# A THEORY OF TURBULENCE
## Part I: Towards Solutions of the Navier-Stokes Equations

**Trinh, Khanh Tuoc**

K.T.Trinh@massey.ac.nz

### Abstract

In this visualisation of turbulence the instantaneous local velocity is expressed in terms of four components to capture the development of and interactions between coherent structures in turbulent flows. It is then possible to isolate the terms linked with each major type of structure and identify corresponding subsets of the Navier-Stokes equations that are easier to solve. Each of these subsets applies to a domain in the flow field not a flow regime. The traditional statistics of turbulence are shown to be not specific to turbulent flow. In particular the Reynolds stresses, the probability density function, the dissipation, production and the energy spectrum are obtained from one subset associated to unsteady state "laminar" flow. The evidence also indicates that the Kolmogorov scale does not represent an eddy size but a conceptual point in the flow field and more importantly that it does not require the assumption of local isotropy.

Applications of this visualisation include a proposal for a theoretical closure of the Reynolds equations, the collapse of all velocity profiles for Newtonian and non-Newtonian flows of all geometries into a unique master curve and the collapse of non-Newtonian pipe flow data onto the Newtonian master curve.

A new partial derivative is proposed that decouples the effect of diffusion and convection thus greatly simplifying the mathematical solution of subsets of the NS equations with implications for a more efficient generation of computer fluid dynamics packages. Finally the implications to turbulent heat and mass transfer are discussed.

## 1    Introduction

Turbulence is a complex time dependent three-dimensional motion widely believed to be governed by equations[1] established independently by Navier and Stokes more than 150 years ago

$$\frac{\partial}{\partial t}(\rho u_i) = -\frac{\partial}{\partial x_i}(\rho u_i u_j) - \frac{\partial}{\partial x_i}p - \frac{\partial}{\partial x_i}\tau_{ij} + \rho g_i \qquad (1)$$

and the equation of continuity

$$\frac{\partial}{\partial x_i}(\rho u_i) = 0 \qquad (2)$$

This fascinating problem has occupied some of the best scientific minds of the last century and a half but a formal solution is yet to be published.

The omnipresence of turbulence in many areas of interest such as aerodynamics, meteorology and process engineering, to name only a few, has nonetheless led to a voluminous literature based on semi-theoretical and empirical solutions and investigations of selected aspects of turbulence structure and mechanisms. According to the Web of Science electronic database, over 3500 papers were published last year alone. It is a challenge to simply keep abreast of the information!

Herbert (1988), a well-known author in the field of transition flow, noted that "different reviews on shear-flow instability (and turbulence)[2] may have little in common and a zero-overlap of cited literature. This curious fact illustrates the many facets of the overall problem, the multitude of views, concepts, and methods, and the need to remain open minded. It also grants me the right to present my own view supported by a selection of references that I know is far from complete".

My interest in this subject started when I was as a postgraduate student 40 years ago and remained active, even during 15 years under the then Stalinist regime of communist Viet Nam that could boast of the most isolated scientific community in the world. This theory was first presented to colleagues in Australasia when I emigrated to New Zealand and wrote up my twenty years of thinking down into a thesis (Trinh, 1992) but the need to start life anew at fifty with five children of school age meant that the theory has been largely unpublished for the last 15 years. There was the occasional presentation in international conferences and congresses but no comprehensive overview. Yet I remain convinced that my work can contribute to the understanding of turbulence, not least because my enforced isolation has allowed me to develop concepts quite independent of the mainstream literature.

---

[1] The suffices i and j in this paper refer to standard vector notation.

[2] Author's addition

I have chosen this forum because it allows me to present my views informally with a hope to engage a dialogue with colleagues who have similarly devoted their life to the understanding of this so common and yet elusive phenomenon.

The aim of this first paper is simply to analyse the likely form of a solution of the Navier-Stokes that would capture the essence of turbulence, in particular to identify the terms in the equations that relate to all the main structures and their interactions. A second goal is to illustrate the application of these concepts to practical problems.

## 2 Velocity fluctuations and Reynolds stresses

Most of the interest in turbulence modelling from a practical engineering view point was originally based on the time averaged parameters of the steady state flow field. Reynolds (1895) has proposed that the instantaneous velocity $u_i$ at any point may be decomposed into a long-time average value $U_i$ and a fluctuating term $U'_i$.

$$u_i = U_i + U'_i \tag{3}$$

with

$$U_i = \lim_{t \to \infty} \int_0^t u_i \, dt \tag{3.1}$$

$$\int_0^\infty U'_i \, dt = 0 \tag{3.2}$$

For simplicity, we will consider the case when
    1. The pressure gradient and the body forces can be neglected
    2. The fluid is incompressible ($\rho$ is constant).

Substituting equation (3) into (1) and taking account of the continuity equation (2) gives:

$$U_i \frac{\partial U_j}{\partial x_j} = \nu \frac{\partial^2 U_i}{\partial x_i} - \frac{\partial \overline{U'_i U'_j}}{\partial x_i} \qquad (4)$$

These are the famous Reynolds equations (Schlichting, 1960, p. 529) also called Reynolds-Averaged-Navier-Stokes equations RANS (Gatski & Rumsey, 2002; Hanjalić & Jakirlić, 2002). The long-time-averaged products $\overline{U'_i U'_j}$ arise from the non-linearity of the Navier-Stokes equations. They have the dimensions of stress and are known as the Reynolds stresses. They are absent in steady laminar flow and form the distinguishing features of turbulence.

The writer proposes that the traditional picture implied by the RANS is an oversimplification and that more information about the Reynolds stresses can be obtained by a more detailed analysis.

## 2.1  Decomposition of the Reynolds Stresses

The derivation of equation (4) implies a velocity trace with a stationary long-time average as shown in Figure 1. Reynolds further imagined the fluctuating components $U'_i$ to be random.

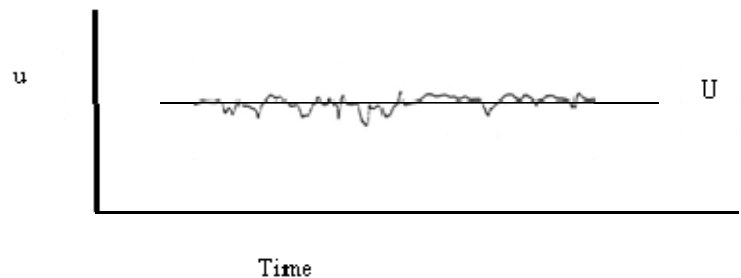

Figure 1 Decomposition of the streamwise component of the instantaneous velocity according to Reynolds (1895), Data of Antonia et al. (1990).

The advance in measuring techniques of the last fifty years have shown conclusively that the instantaneous velocity traces of flow close to a wall show two types of fluctuations: fast and slow. Figure 2 shows a typical trace of streamwise velocity near the wall, redrawn after the measurements of (Antonia, Bisset, & Browne, 1990)

If we draw a smooth line through this velocity trace so that there are no secondary peaks within the typical timescale of the flow $t_v$, we define a locus of smoothed velocity $\tilde{u}_i$ and fast fluctuations $u'_i$ of period $t_f$ relative this base line.

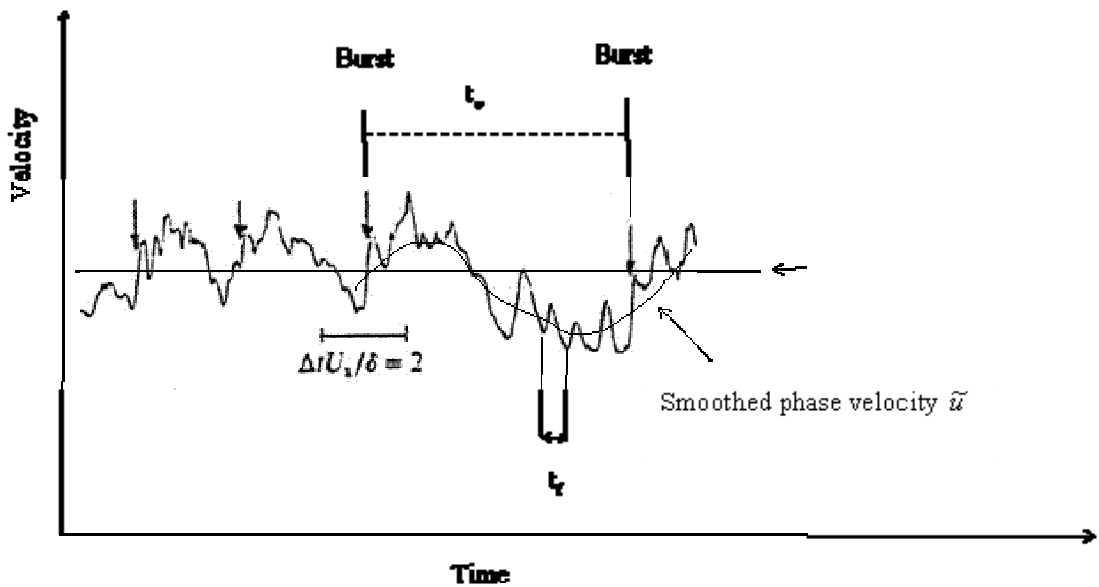

Figure 2 Trace of instantaneous streamwise velocity after measurements by Antonia et al. (1990).

Then we may write

$$\int_0^\infty u'_i dt = 0 \qquad (5)$$

H. T. Kim, Kline, & Reynolds (1971), for example, have obtained the distribution of the smoothed instantaneous streamwise velocities near the wall by conditional sampling at various phases of the bursting cycle (Figure 3).

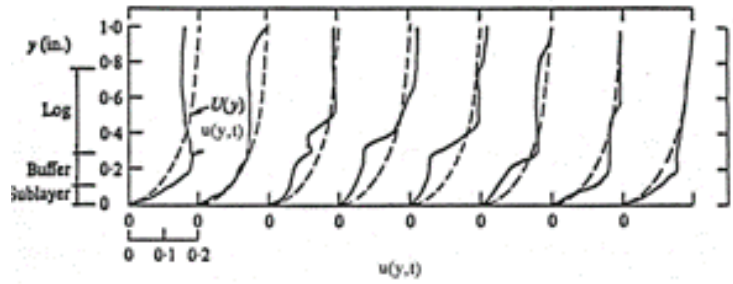

Figure 3 Smoothed phase velocity in a bursting cycle according to Kim et al. (1971).

Mankbadi (1992) also defines the conditional average in the same way as the phase average:

$$\tilde{u}(y,t) = \lim_{N \to \infty} N^{-1} \sum_{n=0}^{N} u(y, t + nt_v) \qquad (6)$$

It is of course necessary to detect first the beginning of an event and determine its characteristic time scale $t_v$. Thus Antonia (1980) defines the conditionally averaged velocity as:

$$\tilde{u}_i = \lim_{t_v \to \infty} \left[ \frac{\int_0^t u_i\, c(t)\, dt}{\int_0^t c(t)\, dt} \right] \qquad (7)$$

It is a measure of the smoothed instantaneous velocity at a particular phase of an event. To avoid confusion in the nomenclature, we will call this "the smoothed phase velocity". Antonia (1980) discusses various detection schemes used to define the function $c(t)$ that presumably locks the sampling onto a special feature associated with the coherent structure.

We draw two conclusions from the work of Kim et al:
    1. The fast fluctuations are eliminated by the conditional sampling process.
    2. The long-time-averaged velocity profile monitored by the Reynolds equations does not correspond to the smoothed phase velocity at any instant in time.

Thus some information is lost in the method of velocity decomposition proposed by Reynolds.

The decomposition of the velocity into fast and slow fluctuations brings out more readily the transient structures of the flow and is crucial to the success of the large eddy simulations LES, direct numerical simulations DNS and the variable-interval time-averaging technique, VITA, of Blackwelder & Kaplan (1976)

The instantaneous velocity may be decomposed in an alternate manner as:

$$u_i = \tilde{u}_i + u'_i \tag{8}$$

Comparing equations (2) and (8) shows that

$$U_i = \frac{1}{t_v} \int_0^{t_v} \tilde{u}_i \, dt \tag{9}$$

and

$$U'_i = \tilde{U}'_i + u'_i \tag{10}$$

where

$$\tilde{U}'_i = \tilde{u}_i - U_i \tag{11}$$

then

$$u_i = U_i + \tilde{U}'_i + u'_i \tag{12}$$

We may average the Navier-Stokes equations over the period $t_f$ of the fast fluctuations. Bird, Stewart, & Lightfoot (1960, p.158) give the results as

$$\frac{\partial(\rho \tilde{u}_i)}{\partial t} = -\frac{\partial p}{\partial x_i} + \mu \frac{\partial^2 \tilde{u}_i}{\partial x_j^2} - \frac{\partial \tilde{u}_i \tilde{u}_j}{\partial x_j} - \frac{\partial \overline{u'_i u'_j}}{\partial x_j} \tag{13}$$

Equation (13) defines a second set of Reynolds stresses $\overline{u'_i u'_j}$ which we will call "fast" Reynolds stresses to differentiate them from the standard Reynolds stresses $\overline{U'_i U'_j}$. In general $u'_i < U'_i$ and the fast Reynolds stresses are smaller in magnitude than the standard Reynolds stresses.

To the writer's knowledge experimental investigations of turbulence, up to the time he first presented this theory to colleagues in Australasia (Trinh, 1992), all targeted the standard Reynolds stresses and no separate measurements existed for the fast Reynolds stresses. Considerations of this second set of Reynolds stresses gives a much better overall picture of the problem, in particular of the causal relationships in the study of the flow structure.

## 2.2 Properties of the Reynolds Stresses

Within a period $t_\nu$, the smoothed velocity $\tilde{u}_i$ varies slowly with time but the fluctuations $u'_i$ may be assumed to be periodic with a timescale $t_f$. In the particular case of steady laminar flow, $\tilde{u}_i = U_i$ and $\tilde{U}'_i = 0$: only the fast fluctuations remain. These are typically remnants of disturbances introduced at the pipe entrance or leading edge of a flat plate by conditions upstream.

We may write the fast fluctuations in the form

$$u'_i = u_{0,i}\left(e^{i\omega t} + e^{-i\omega t}\right) \tag{14}$$

The fast Reynolds stresses $u'_i u'_j$ become

$$u'_i u'_j = u_{0,i} u_{0,j} \left(e^{2i\omega t} + e^{-2i\omega t}\right) + 2 u_{0,i} u_{0,j} \tag{15}$$

Equation (15) shows that the fluctuating periodic motion $u'_i$ generates two components of the "fast" Reynolds stresses: one is oscillating and cancels out upon long-time-averaging, the other, $u_{0,i} u_{0,j}$ is persistent in the sense that it does not depend on the period $t_f$. The term $u_{0,i} u_{0,j}$ indicates the startling possibility that a purely oscillating motion can generate a steady motion which is not aligned in the direction of the oscillations. The qualification steady must be understood as independent of the frequency ω of the fast fluctuations. If the flow is averaged over a longer time than the period $t_\nu$ of the bursting process, the term $u_{0,i} u_{0,j}$ must be understood as transient but non-oscillating. This term indicates the presence of transient shear layers embedded in turbulent flow fields and not aligned in the stream wise direction similar to those associated with the streaming flow in oscillating laminar boundary layers (Schneck & Walburn, 1976; Tetlionis, 1981).

Oblique shear layers have been observed near the wall and upstream of large scale structures by (Blackwelder & Kovasznay, 1972; Hedley & Keffer, 1974; Nychas, Hershey, & Brodkey, 1973; G. L. Brown & Thomas, 1977; Chen & Blackwelder, 1978; Falco, 1977; Spina & Smits, 1987 and Antonia, Browne, & Bisset, 1989). These structures are characteristic of patches of fluid that move within turbulent flow fields. An extraordinary number of these structures have been identified in the past five decades prompting one researcher to say (Fiedler, 1988) "When studying the literature on boundary layers, one is soon lost in a zoo of structures, e.g. horseshoe- and hairpin-eddies, pancake- and surfboard-eddies, typical eddies, vortex rings, mushroom-eddies, arrowhead-eddies, etc…" It is not clear from literature reports whether different observations refer exactly to the same phenomenon and what effects the different methods of event detection have on the results.

In fact the first observations of coherent structures date from (Reynolds, 1883). Interest in these structures was reignited by the classic work of Kline et al. (1967). Using hydrogen bubbles as tracers, they observed inrushes of high-speed fluid from the outer region towards the wall, followed by longitudinal sweeps along the wall. During the sweep phase, the structure of the wall layer shows alternate streaks of high and low-speed fluid. The low-speed streaks become unstable, lift and oscillate until they are eventually ejected into the outer region in a violent burst. Kline et al. (op. cit.) observed that the hydrogen bubble lines in their experiments became contorted during the ejection phase indicating a break-up of the flow into small scales. They refer to the wall-layer process at this point as bursting. Most of turbulent stresses in the wall layer are produced during this short bursting phase compared with the much longer sweep phase. The work of Kline et al. work highlighted the transient nature of the wall layer process and the existence of a secondary stream when most of the turbulent stresses were produced.

Because of the importance of the wall region as highlighted by the work of Kline et al., a large amount of effort has been devoted to its study focussing mainly on the hairpin vortex, the most identifiable coherent structure in that region. Work before 1990 were well reviewed, for example by Cantwell (1981) and Robinson (1991). There have been both physical experiments e.g. (Corino & Brodkey, 1969; Willmarth & Lu, 1972; Blackwelder & Kaplan, 1976; A. A. Townsend, 1979; Head &

Bandhyopadhyay, 1981; Bogard & Tiederman, 1986; Luchak & Tiederman, 1987; Tardu, 1995; (Meinhart & Adrian, 1995; Carlier & Stanislas, 2005), including efforts to induce artificially the creation of a hairpin vortex by injecting a jet of low momentum fluid into a laminar flow field (Arcalar & Smith, 1987b, 1987a; Haidari & Smith, 1994; Gad-el-Hak & Hussain, 1986). With the advent of better computing facilities, direct numerical simulations DNS have been used increasingly to conduct 'numerical experiments" e.g. (Spalart, 1988; Kim, Moin, & Moser, 1987; Jimenez & Pinelli, 1999).

Much more temporal detail can be deduced from numerical experiments. For example, Johansson, Alfresson, & Kim (1991) analysed the data base provided by the DNS of Kim, Moin and Moser (1987) to obtain the conditionally averaged production of turbulent kinetic energy $\widetilde{P}$ which they write as

$$\widetilde{P} = \overline{U'V'}\frac{dU}{dy} - \overline{U'V'}\left(\frac{\partial \widetilde{U}'}{\partial x} + \frac{\partial \widetilde{V}'}{\partial y}\right) - \overline{V'^2}\frac{\partial \widetilde{V}'}{\partial y} - \overline{W'^2}\frac{\partial \widetilde{W}'}{\partial z} \qquad (16)$$

The first term on the right-hand side of equation (16) is the only one that remains in the long-time averaged sense. It is shown in Figure 4b. The total conditionally averaged production $\widetilde{P}$ is substantially higher as seen in Figure 4a. The difference between these two terms is shown in Figure 4c. It points to the existence of an important transient contribution weakly slanted with respect to the wall and which can be attributed to strong gradients in the x- and y- directions of the conditionally averaged streamwise velocity.

Johansson et al. confirm that the Reynolds stresses contribution from the downstream side of the shear layers is spatially spotty but they could follow the associated <U'V'> peaks for distances up to 1000 wall units. Furthermore, they found no signs of oscillatory motions or violent break-up in conjunction with these shear layers which, they believe, indicate a persistent motion of low-speed fluid away from the wall. The writer observes that this motion is very similar to the streaming process observed in laminar oscillating boundary layers.

There has been a slow build up of view that the destabilisation of a laminar flow field cannot be simply explain in terms of growth of periodic disturbances alone as originally investigated by many authors e.g. Tollmien (1929), Schubauer & Skramstad (1943), Schlichting (1960) and must involve a second mechanism e.g. Trefethen, Trefethen, Reddy, & Driscoll (1993); Schoppa & Hussain (2002). Schoppa and Hussain (2002) have analysed their DNS data base to argue that sinusoidal velocity fluctuations led to the production of intense shear layers associated with the streaming flow, that they call transient stress growth TSG. They attribute the lifting of the longitudinal wall vortex into the head of a hairpin vortex directly to the action of the TSG.

Rather than rely on very detailed and complex arguments based on the analysis of vorticity patterns obtained from DNS, PIV (particle imaging velocimetry) or velocity probe measurements with different detection schemes, the writer prefers to use a technique borrowed from the study of laminar oscillating boundary layers (Trinh, 1992) to identify the different terms in the Navier-Stokes equations related to different structures and their interaction.

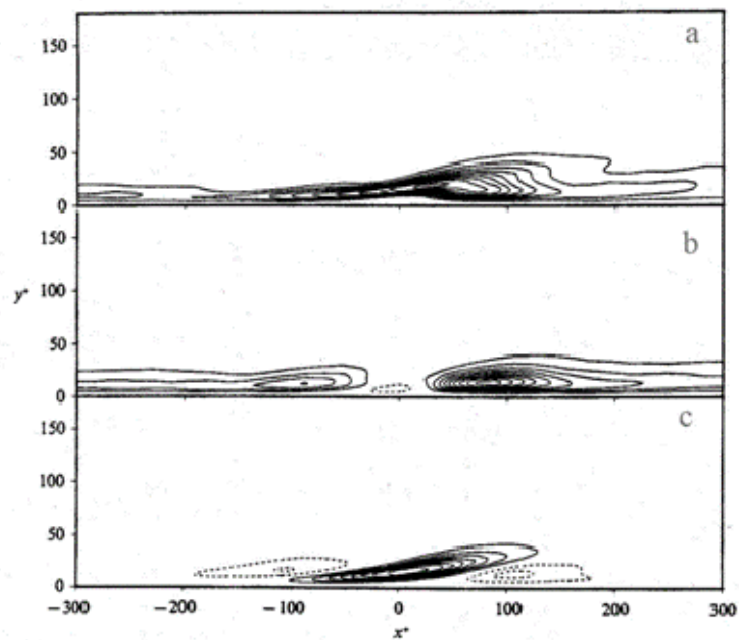

Figure 4 Production of turbulence near the wall. (a) $\widetilde{P}$, (b) $\overline{U'V'}(dU/dy)$, (c) $\widetilde{P} - \overline{U'V'}(dU/dy)$. After Johansson, Alfresson, & Kim (1991)

## 3  Oscillating Laminar Boundary Layers

The analysis of oscillating laminar boundary layers also begins with equation (1). The velocity is decomposed into steady and periodic components. These conditions are exactly the same as those adopted in the DNS (J. Kim, et al., 1987; Spalart, 1988; Laurien & Kleiser, 1989) and the writer believes that techniques developed in the former field of research may be transposed to the study of turbulence. The case of oscillating flow with a zero-mean velocity is particularly interesting since the basic velocity fluctuations imposed by external means do not grow with time because there is no mean motion along the wall. One may thus investigate the effect of the amplitude and frequency of the fluctuations separately. The following treatment of the problem is taken from the excellent book of (Tetlionis, 1981).

We define a stream function $\psi$ such that

$$u = \frac{\partial \psi}{\partial y} \qquad v = \frac{\partial \psi}{\partial x} \tag{17}$$

The basic variables are made non-dimensional

$$x^* = \frac{x}{L} \qquad y^* = \frac{y}{\sqrt{2\nu/\omega}} \qquad t^* = t\omega \tag{18}$$

$$U_e^*(x,t) = \frac{U_e}{U_\infty}(x,t) \qquad \psi^* = \psi \left( U_\infty \sqrt{\frac{2\nu}{\omega}} \right)^{-1} \tag{19}$$

where $U_\infty$ is the approach velocity for $x \to \infty$, $U_e$ is the local mainstream velocity and L is a characteristic dimension of the body. The system of coordinates x, y is attached to the body. The Navier-Stokes equation (1) may be transformed as:

$$\frac{\partial^2 \psi^*}{\partial y^* \partial t^*} - \frac{1}{2}\frac{\partial^3 \psi^*}{\partial y^{*3}} - \frac{\partial U_e^*}{\partial t^*} = \frac{U_e}{L\omega}\left( -\frac{\partial \psi^*}{\partial y^*}\frac{\partial^2 \psi^*}{\partial y^* \partial x^*} + \frac{\partial \psi^*}{\partial x^*}\frac{\partial^2 \psi^*}{\partial y^{*2}} + U_e^* \frac{\partial U_e^*}{\partial x^*} \right) \tag{20}$$

with boundary conditions

$$\psi^{*'} = \frac{\partial \psi^*}{\partial y^*} = 0 \qquad y^* = 0 \tag{21}$$

For large frequencies, the RHS of equation (20) can be neglected since

$$\varepsilon = \frac{U_e}{L\omega} \ll 1 \qquad (22)$$

In this case, Tetlionis reports the solution of equation (20) as:

$$\psi^* = \left[\frac{U_0^*(x^*)}{2}(1-i)[1-e^{(1+i)y^*}] + \frac{U_0^* y^*}{2}\right]e^{it^*} + C \qquad (23)$$

Tetlionis (op. cit. p. 157) points out that equation (23) may be regarded as a generalisation of Stokes' solution (1851) for an oscillating flat plate. This latter solution describes an oscillating flow called the Stokes layer which is often found embedded in other flow fields and has properties almost independent of the host field. Since Stokes also produced a solution for a flat plate started impulsively, often referred to as Stokes' first problem, the oscillating plate will be referred to as the Stokes solution2 for clarity. Van Driest (1956) has used the Stokes solution2 to model the damping function in Prandtl' mixing-length theory (1935) near the wall.

Equation (23) is accurate only to an error of order $\varepsilon$. Tetlionis reports a more accurate solution for the case when $\varepsilon$ cannot be neglected (i.e. for lower frequencies):

$$\psi^* = \frac{U_0^*(x^*)}{2}[\psi_0^*(y^*)e^{it^*} \overline{\psi_0^*(y^*)}e^{-it^*}] + \varepsilon[\psi_1^*(x^*,y^*)e^{2it^*} + \overline{\psi_0^*}e^{-2it^*}] + O(\varepsilon^2) \qquad (24)$$

where $\psi_0$ and $\psi_1$ are the components of the stream function of order $\varepsilon^0$ and $\varepsilon$. Substituting this more accurate solution into equation (20), we find that the multiplication of coefficients of $e^{it^*}$ and $e^{-it^*}$ forms terms that are independent of the oscillating frequency, ω, imposed on the flow field and were not anticipated in equation (24). Thus the full solution of equation (20) is normally written (Stuart, 1966; Tetlionis, 1981) as

$$\psi^* = \frac{U_0^*(x^*)}{2}[\psi_0^*(y^*)e^{it^*} + \overline{\psi_0^*(y^*)}e^{-it^*}]$$
$$+ \varepsilon[\psi_{st}^* + [\psi_1^*(x^*,y^*)e^{2it^*} + \overline{\psi_1^*(x^*,y^*)}e^{-2it^*}] + O(\varepsilon^2) \qquad (25)$$

where the overbar denotes the complex conjugate and $\psi_{st}^*$ results from cancelling of

$e^{it^*}$ and $e^{-it^*}$ terms.

The quantity $\psi_{st}^*$ shows that the interaction of convected inertial effects of forced oscillations with viscous effects near a wall results in a non-oscillating motion that is referred to in the literature as "Streaming". The problem has been known for over a century (Faraday, 1831; Dvorak, 1874; Rayleigh, 1880, 1884; Carriere, 1929; Andrade, 1931; Schlichting, 1932) and studied theoretically (Riley, 1967; Schlichting, 1960; Stuart, 1966; Tetlionis, 1981). The existence of this streaming flow, even in this absence of any mainstream flow, is clearly demonstrated in Figure 5.

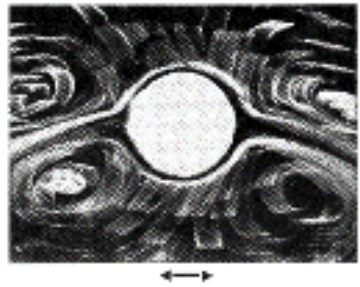

Figure 5 Streaming flow near a vibrating cylinder. After Schlichting (1960).

The governing equation for the streaming function may be extracted from the original Navier-Stokes equations and analysed separately. This is achieved by substituting equation (25) into (20) and collecting the steady terms of order $\varepsilon$. Tetlionis gives the result as

$$-\frac{\partial^3 \psi_{st}^*}{\partial y^{*3}} = -U_0^* \frac{dU_0^*}{dx^*} \overline{\frac{\partial \psi_0^*}{\partial y^*} \frac{\partial \psi_0^*}{\partial y^*}} + \frac{1}{2} U_0^* \frac{dU_0^*}{dx^*} \left( \overline{\psi_0^* \frac{\partial^2 \psi_0^*}{\partial y^{*2}}} + \overline{\psi_0^* \frac{\partial^2 \psi_0^*}{\partial y^{*2}}} \right) + U_0^* \frac{dU_0^*}{dx^*} \quad (26)$$

The boundary conditions imposed in early analyses were:

$$\psi_1^* = \psi_{st}^* = 0 \quad at \quad y^* = 0 \tag{27}$$

$$\frac{\partial \psi_1^*}{\partial y^*} = \frac{\partial \psi_{st}^*}{\partial y^*} = 0 \quad at \quad y^* = 0 \tag{28}$$

$$\frac{\partial \psi_1^*}{\partial y^*} = \frac{\partial \psi_{st}^*}{\partial y^*} = 0 \quad \text{at} \quad y^* \to \infty \tag{29}$$

Similarly the governing equation for $\psi_1$ is obtained by collecting the oscillating terms of order $\varepsilon$. The terms of order $\varepsilon^0$ give

$$\frac{\partial^3 \psi_0^*}{\partial y^{*3}} - 2i \frac{\partial \psi_0^*}{\partial y^*} = -i \tag{30}$$

The solution for the main oscillating component $\psi_0$ is the same as equation (23) and may be arranged as

$$\psi_0^* = -\frac{1}{2}(1-i)\left[1 - e^{-(1+i)y^*}\right] + y^* \tag{31}$$

Stuart (1966) has noted that the complementary function of equation (31) is $\left(A + By^* + Cy^{*2}\right)$ where A, B and C are functions of $y^*$. In order to satisfy the boundary condition in equation (29), it is necessary to put both B and C equal to zero. But then the boundary conditions at the wall cannot be satisfied. Stuart proposes that this anomaly can be remedied by assuming that the derivative $\partial \psi_{st}^* / \partial y^*$ does not reach zero at the outer edge of the Stokes layer but remains finite. Then, assuming C = 0, we obtain

$$\psi_{st}^* = U_0^* \frac{dU_0^*}{dx^*}\left(\frac{13}{8} - \frac{3}{4}y^* - \frac{1}{8}e^{-2y^*} - \frac{3}{2}e^{-y^*}\cos y^* - e^{-y^*}\sin y^* - \frac{1}{2}y^* e^{-y^*}\sin y^*\right) \tag{32}$$

This means that there exist two boundary layers: an oscillating Stokes layer $\delta_s$ and a second layer $\delta_{st}$ created by the intrusion of the streaming flow into the outer inviscid region. Tetlionis estimates the order of magnitude of these two layers as

$$\delta_{st} \approx \frac{L\sqrt{\omega \nu}}{U_\infty} \tag{33}$$

and

$$\delta_s \approx \sqrt{\frac{\nu}{\omega}} \tag{34}$$

$$\frac{\delta_{st}}{\delta_s} \approx \frac{L\omega}{U_\infty} \approx \frac{1}{\varepsilon} \qquad (35)$$

Since ε is small, the streaming layer $\delta_{st}$ is much thicker than the Stokes layer $\delta_s$ as shown in Figure 6.

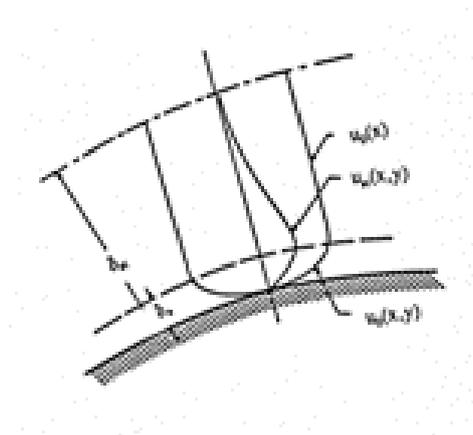

Figure 6 Stokes and Streaming layers. After Tetlionis (1981).

The most important observation is that the streaming flow reaches well beyond the Stokes layer i.e. into the inviscid outer region. This eruption of an unsteady laminar boundary layer is called by various names in kernel studies that attempt to model the wall process of turbulent flow. Peridier, Smith, & Walker (1991) call it viscous-inviscid interaction. These kernel studies arise from the observation that a vortex moving above a wall will induce a laminar sub-boundary layer underneath its path by viscous diffusion of momentum, even if the vortex is introduced into a fluid which was originally at rest (Smith, Walker, Haidari, & Sobrun, 1991). The vortex impresses a periodic disturbance onto the laminar sub-boundary layer underneath. The problem is thus very similar to that discussed by Tetlionis. In these kernel studies the configuration of the vortex must be specified a priori. In the work of Walker (1978) it is a rectilinear vortex, in Chu and Falco (1988) ring vortices, in Liu et al. (1991) hairpin vortices, in Swearingen and Blackwelder (1987), streamwise Goertler vortices. But recently in their numerical simulation Suponitsky, Cohen, & Bar-Yoseph (2005) have shown that vortical disturbances evolve into a hairpin vortex independently of their original geometry over a wide range of orientations.

The investigation of the flow field outside the Stokes layer has been performed by Stuart (1966) and Riley (1967) using asymptotic expansions. The analyses of Stuart and Riley have the advantage that no assumption need be made about the source of the velocity fluctuations. To order $\varepsilon^0$, the flow in this outer layer is inviscid but the interactions of higher orders are not. The problem is very complex and both workers have introduced an essential simplification: they assume that the streaming flow and the potential flow do not interact. In order to express this simplification mathematically, Stuart has rewritten the stream function in the form

$$\psi = \psi_0(x,t) + y U_e(x,t) + \psi_a(x,y,t) \tag{36}$$

where $(\psi_0 + y U_e)$ is the periodic potential flow (including a displacement effect) and $\psi_a$ is an additional flow of which we are especially interested in the steady part. The boundary-layer theory is assumed to be valid and the potential flow balances the given pressure gradient. Then equation (20) becomes

$$U_e \frac{\partial^2 \psi_a}{\partial x \partial y} + \frac{\partial U_e}{\partial x} \frac{\partial \psi_a}{\partial y} - \left( \frac{\partial \psi_0}{\partial x} + \frac{y \partial U_e}{\partial x} \right) + \frac{\partial^2 \psi_a}{\partial t \partial y} + \frac{\partial \psi_a}{\partial y} \frac{\partial^2 \psi_a}{\partial x \partial y} - \frac{\partial \psi_a}{\partial x} \frac{\partial^2 \psi_a}{\partial y^2} = \nu \frac{\partial^3 \psi_a}{\partial y^3} \tag{37}$$

Equation (37) is then averaged with respect to time. The average of $\psi_a$ is denoted by $\psi_{st}^*$

$$\psi_a = \psi_{st} + \psi_t \tag{38}$$

where $\psi_t$ is the time-dependent part of $\psi_a$. Then we have

$$J + \frac{\partial \psi_{st}}{\partial y} \frac{\partial^2 \psi_{st}}{\partial x \partial y} - \frac{\partial \psi_{st}}{\partial x} \frac{\partial^2 \psi_{st}}{\partial y^2} = \nu \frac{\partial^3 \psi_{st}}{\partial y^3} \tag{39}$$

where

$$J = \overline{U_e \frac{\partial^2 \psi_t}{\partial x \partial y}} + \overline{\frac{\partial U_e}{\partial x} \frac{\partial \psi_t}{\partial y}} - \overline{\left( \frac{\partial \psi_0}{\partial x} + y \frac{\partial U_e}{\partial x} \right) \frac{\partial^2 \psi_t}{\partial y^2}} + \overline{\frac{\partial \psi_t}{\partial y} \frac{\partial^2 \psi_t}{\partial x \partial y}} - \overline{\frac{\partial \psi_t}{\partial x} \frac{\partial^2 \psi_t}{\partial y^2}} \tag{40}$$

where the overbar denotes an average with respect to time.

Stuart (op. cit.) has assumed that the function J may be neglected giving

$$\frac{\partial \psi_{st}}{\partial y}\frac{\partial^2 \psi_{st}}{\partial x \partial y} - \frac{\partial \psi_{st}}{\partial x}\frac{\partial^2 \psi_{st}}{\partial y^2} = \nu \frac{\partial^3 \psi_{st}}{\partial y^3} \qquad (41)$$

This linearisation has allowed him to obtain a solution for the streaming layer. This solution is in qualitative agreement with the experiments of Schlichting (op. cit.) for a vibrating cylinder.

### 3.1 The streaming flow

In my view, the relationships in equations (15) and (25) indicate how some of the energy in the main flow is extracted and stored as wave energy, evidenced in the fluctuations of $u_i$, then transformed into kinetic energy $u'_{0,i} u'_{0,j}$ which is released through the creation of a streaming flow, a key element in the mechanism of turbulence production (Trinh, 1992). In the remainder of this paper, it will be called the streaming process. However these observations are not sufficient to answer the other question most often asked: how does turbulence sustain itself against the constant pressure of viscous dissipation? To answer this question, we need to examine the characteristics of the streaming flow and its interactions with the main flow.

The ejections associated with bursting have been compared to jets of fluid essentially in cross flow to the main stream (Grass (1971); Townsend, 1970; Trinh, 1992). The first difference to note is that unlike smoke plumes often studied as steady jets in cross flow, ejections from the wall layer are transient. The reason here is simple: the jets take fluid from the Stokes layer into the outer stream and therefore interrupt the source of the velocity fluctuations that feed the streaming flow. Therefore the cause of the periodic inrush of fast fluid from the outer stream towards the wall is a consequence of the term $u_{i,0} u_{j,0}$ (Trinh, 1992) and not directly dependent of the term $\omega$. This non-oscillatory nature of the streaming flow is supported by the results of Schlichting (1960). As noted before, Johansson et al. (1991) also found the ejections from the wall in turbulent flow are persistent and non-oscillatory.

The main crossflow deflects the wall ejection in a streamwise direction. This is the first interaction. The jet in crossflow has been divided into three zones as shown in Figure 7. The near-field region is jet dominated in the sense that the effects of the crossflow on the jet are not yet significant. In the curvilinear region, the initial jet momentum and the momentum extracted from the crossflow have comparable effects on the jet characteristics. In the far-field region, the effects of the crossflow predominate and the jet is aligned in the direction of the crossflow.

The zones associated with a path of an intermittent streaming jet in a turbulent flow are slightly different. Firstly there is a weak but important interaction with the laminar flow in the Stokes layer. Immediately outside the wall layer, the jet follows a quasi linear path before it becomes deflected into a curvilinear pattern. Since the fluid in the jet moves as coherent structure as evidenced in the analysis of both the DNS (Johansson et al 1991) and probe measurements (G. L. Brown & Thomas, 1977) the fluid velocities in this jet are correlated.

One may therefore express the change in velocity over an axial distance $\lambda$ as a Taylor series

$$\Delta u = \lambda \frac{\partial u}{\partial y} + \frac{\lambda^2}{2!} \frac{\partial^2 u}{\partial y^2} + \frac{\lambda^3}{3!} \frac{\partial^3 u}{\partial y^3} \ldots \tag{42}$$

In the linear region of the path, equation (42) reduces to the expression derived by Prandtl (1935) for the mixing length. However, I have two major differences with the arguments of Prandtl

1. $\lambda$ cannot be compared with the mean free path of molecules, it is a scale of a coherent patch of fluid
2. The scale $\lambda$ is not measured in the normal direction from the wall but along the jet axis.

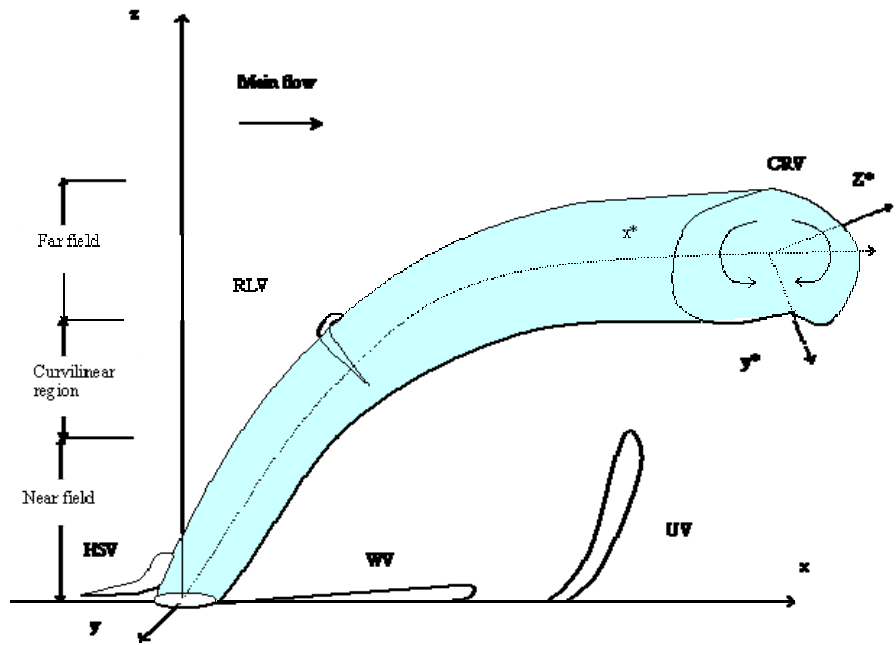

Figure 7 Geometry of a jet in cross flow. CRV Counter-rotating vortex, RLV Ring-like vortex, HSV Horseshoe vortex, WV Wall vortex, UV Upright vortex,, xyz Cartesian coordinates, z*y*z* natural jet coordinates.

Many authors e.g. (Chan, Lin, & Kennedy, 1976; Keffer & Baines, 1963) have found that most parameters of a jet in cross flow like the distribution of lateral velocity correlate better with the natural coordinates of the jet than with fixed Cartesian coordinates. Thus the expression that Prandtl postulated for the mixing-length l

$$l = \kappa y \tag{43}$$

results from a projection of the scale $\lambda$ onto the normal distance and the coefficient $\kappa$, called Karman's universal constant is actually a structural parameter related to the angle of inclination $\alpha$ of the streaming jet

$$\frac{1}{\kappa} = \sin \alpha = \frac{l}{\lambda} \tag{44}$$

Following Prandtl (Prandtl, 1935), we can express the velocity gradient as

$$\frac{dU^+}{dy^+} = \frac{1}{\kappa y^+} \tag{45}$$

and

$$U^+ = \frac{1}{\kappa} \ln y^+ + B \tag{46}$$

which is the famous log-law. This hypothesis was tested against the data of Brown and Thomas (1977) by calculating the value of the scale $l$ at various $y^+$ from well-known data on velocity distributions e.g. (Lawn, 1971; Nikuradse, 1932; Reichardt, 1943), the value of Karman's constant from the log-law and its modified value near the wall from van Driest's damping function (Trinh, 1992; Trinh, 1996; K.T. Trinh, 2005b) and the value of $\alpha$ from equation (44). The scale vectors $\vec{\lambda}$ obtained at different positions $y^+$ were then dovetailed to give a trace of the jet path that fit the data of Brown and Thomas (1977) and Kreplin and Eckelmann (Kreplin & Eckelmann, 1979) almost perfectly as shown in Figure 8 and Figure 9.

The numerical value of $\kappa$ is actually not a universal constant but a function of the relative strengths of the cross flow and the streaming jet as shown in slight differences reported in different geometries : pipe flow (Nikuradse, 1932), boundary layer flow past a flat plate (Clauser, 1954), rotating disks (Dorfman, 1963) but the issue of why it remains reasonably within a very small range of variation requires a much more detailed analysis of the circumstances under which the streaming jet is ejected from

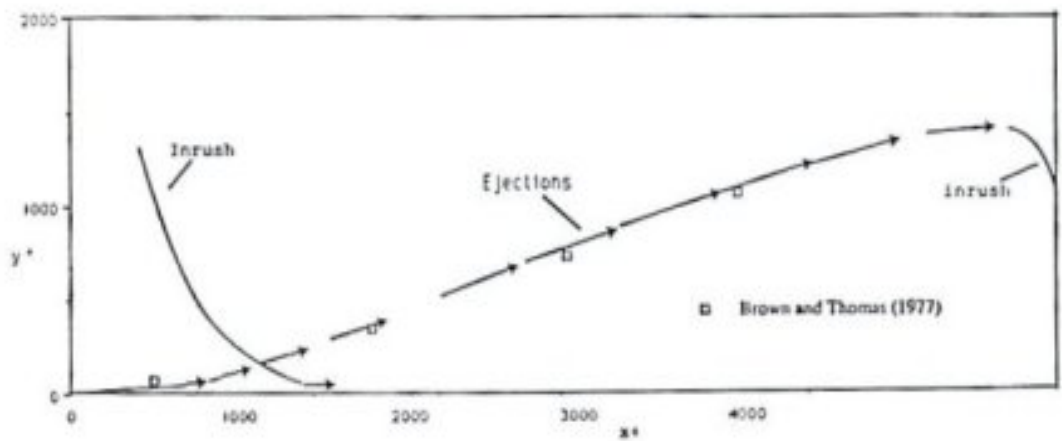

Figure 8 Educed path of streaming jet in the outer region. Data of Brown and Thomas (1977)

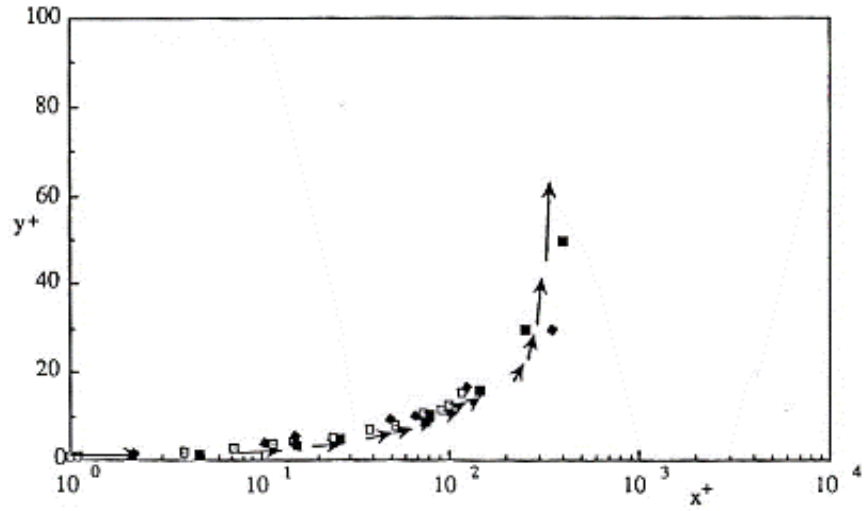

Figure 9 Educed path of streaming jet in the wall layer. Data of Kreplin and Eckelmann (1979)

We should note here that many authors e.g.(Camussi, 2002; Chassaing, George, Claria, & Sananes, 1974) have found that the path of the jet in the far field region follows a power function of the form

$$\frac{y}{D} = A\left(\frac{x}{D}\right)^n \tag{47}$$

The parameter A in equation (47) was found to vary with the $R_j$ the ratio of cross-stream to jet velocity but the exponent n was found by both these workers to be $n \approx 0.385 - 0.39$. Taking the log of equation (47) gives

$$\frac{x}{D} = \frac{1}{0.39}\ln\frac{y}{D} + B \tag{48}$$

Is the correspondence between n and $\kappa$ mere coincidence or does it point to a more universal rule for shear layers induced by jets in cross flow? Some authors e.g. (Pratte & Baines, 1967) quote a different value of n when the distance y is normalised with $R_j D$. More understanding of these jets is required to make a firm call.

When the derivatives of second and higher orders in the Taylor series equation (42) can no longer be neglected, the log-law ceases to apply. A crude estimate of the outer limit of applicability of the log-law is obtained by setting

$$l^+ \frac{dU^+}{dy^+} = \frac{l^{+2}}{2} \frac{d^2U^+}{dy^{+2}} \qquad (49)$$

giving (Doshi & Gill, 1970; Trinh, 1992)

$$l^+ = 2 \frac{dU^+/dy^+}{d^2U^+/dy^{+2}} \qquad (50)$$

which can be recognised as Karman's similarity law (1934). The value of this similarity law is of course that it does not involve the wall parameters, $u_*$ and $\nu$ and equation (50) gives a useful criterion for matching the iterations performed in CFD (computer fluid dynamics) packages with the wall function when this is expressed in terms of a log-law. At the moment, this matching is somewhat arbitrary. This issue is discussed further in section 9.

Ejections of wall fluid imply boundary layer separation and this is only possible with the appearance of a zone of negative pressure behind the streaming jet. Strong shear layers are similarly produced on the upstream side of the jet where the pressure is high as in the forward stagnation region of a cylinder in a flow stream, (Chan, et al., 1976). Johansson et al. (op. cit.) have found that the pressure patterns associated with shear layers near the wall undergo a development where an intense localised high-pressure region around and beneath the centre of the shear layer is found around the stage of maximum strength. At this stage, the maximum amplitude is about $2p_{rms}$ above the mean pressure. Johansson et al. suggest that these strong localised high-pressure regions could be of importance for boundary-layer noise generation.

Johansson et al have also observed that the contribution of the Reynolds stresses to turbulence production in the downstream side of the shear layers is spatially spotty. This is compatible with the existence of a wake behind the ejections. In kernel studies mentioned previously, e.g. Peridier et al (op. cit.), the eruption of the laminar sub-boundary layer underneath the travelling vortex resembles the ejections and represents the intrusion of a stream of low-speed fluid into the outer inviscid region. Peridier et al have shown that a recirculation region exists behind the eruption. Liu, et al. (1991) have shown that the mainstream interacts with hairpin vortices near the wall and produce recirculation regions behind these hairpin vortices. There are other similarities between jets in cross flow and wall ejections. Falco (1977, 1991) has studied coherent structures in a boundary layer with smoke traces with the patterns shown in Figure 10. Falco observed two "typical eddy" forms: a mushroom shape on the back of the large coherent structures and a kidney shape near the edge of the boundary layer, which show striking resemblance to shapes observed with jets in cross-flow (Figure 10). The typical mushroom eddy is evident in the flow visualisation of plumes by Andreopoulos (1989) also reproduced in (Figure 10). The kidney shape represents a cross-section of the jet in the far field region where it is aligned in the direction of main flow. Townsend (1970) postulated that the ejections create roller like structures in the outer region. These have been deduced from probe measurements by Wark & Nagib (1991) who mapped out a recirculation zone associated with the roller-like structure behind the moving ejections (Figure 11) that is strikingly similar to the pattern obtained by Savory, Toy, McGuirk, & Sakellariou (1990) behind jets in crossflow.

The coherent structures created by a jet and the cross flow are in fact more complex and have received a large amount of attention in the last 30 years. Camussi, Guj, & Stella (2002) and Cortelezzi & Karagozian (2001) have summarised these structures, shown in Figure 7 as

1. CRVP (counter-rotating vortex pair) which is evident in the far field region
2. Ring-like vortices which are formed from the upwind shear layer of the jet flow
3. Horseshoe vortices formed upstream of the jet and close to the wall (very similar to the horseshoe vortices in the wall layer before the ejections)
4. WV, wall vortices which develop downstream of the jet orifice and close to the wall identified by McMahon, Hester, & Palfery, (1971) and Fric & Roshko (1994)
5. UV, upright vortices that Fric and Roshko describe as "burst" of the boundary layer fluid.

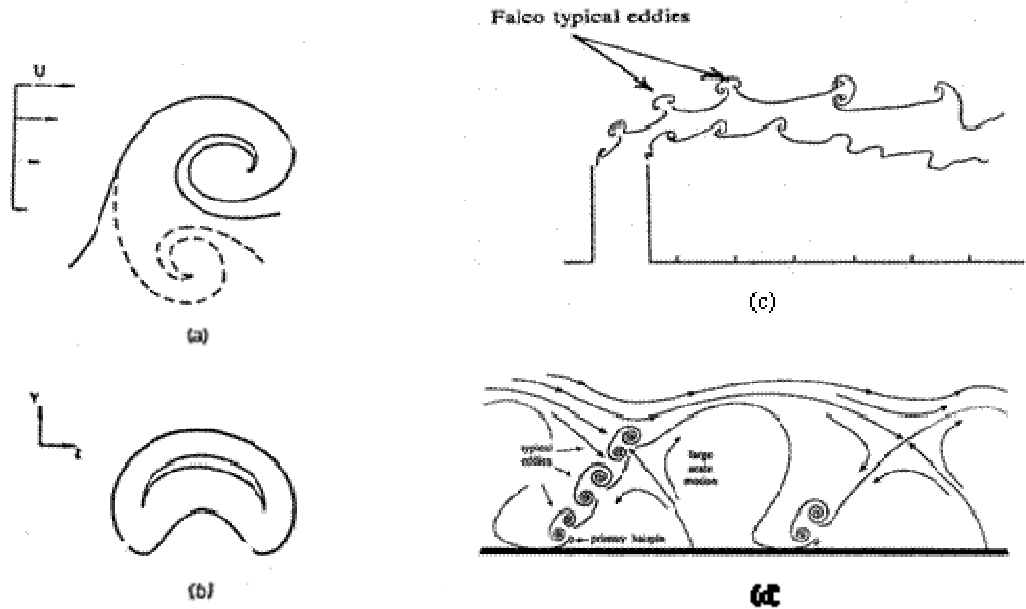

Figure 10 (a),(b) Typical eddies in turbulent boundary layers observed by Falco (1979), (c) Structure of plumes sketch of plumes after Andreopoulos (1989), (d) Sketch of ejection and typical eddies after Falco (1991)

## 4   Influence of the interaction terms

It is clear by now that in the study of turbulent flows we cannot ignore the effect of the interaction function J (equation (39) as Stuart (op. cit.) did in the study of oscillating laminar flow. If anything it should be further detailed. The writer submits that any study of turbulence that neglects this interaction effect will fail to reproduce the fine scale turbulence and require some sort of empirical closure model. So far only the DNS, which do not attempt to linearise the NS equations in that sense, have reproduced this interaction.

Nonetheless we can infer the character of these interactions without a formal analytical solution of the interaction function J. The effect of the interaction terms depends both on the inclination of the streaming jet that changes continuously with distance y from the wall, and the streamwise velocity that increases with distance y.

In the far field of the outer region when the jet path is aligned in the streamwise direction, the CRVP originates as an effect of the bending of the jet itself (Camussi et

25/218

al., Cortelezzi et al. op.cit.) Because the jets in turbulent flow are intermittent, they represent unattached patches of fluids that would be more easily deflected in the streamwise direction. Presumably the CRVP of the ejections would have less interaction with the wake region that derives its vorticity mainly from the cross flow boundary layer, not the jet (Fric & Roshko, 1994). Smith (1998) observed that jet fluid does not flow into the wake before a ratio of jet to cross flow velocities of 10.

The behaviour of the near wall region of jets in cross flow has been documented by many authors (Fric & Roshko, 1994; Kelso & Smits, 1995; Krothapalli, Lourenco, & Buchlin, 1990; McMahon, et al., 1971; Moussa, Trischka, & Eskinazi, 1977). Fric and Roshko (1994) argued that the near-wall flow around a transverse jet does not separate from the jet and shed vortices in the wake like the vortex shedding phenomenon from solid bluff bodies. They observed horseshoe vortices on the upstream side of the jet and argued that the vorticity in the wake region originates from the wall boundary layer flow which wraps around the jet and separates on its lee side creating wall vortices leading eventually to upright vortices that they describe as bursts. The horseshoe vortices are coupled with the periods of vortices that form in the jet wake (Krothapalli et al 1990, Fric and Roshko, 1994, Kelso and Smits 1995).

The horseshoe vortices upstream of the jet are strongly reminiscent of hairpin vortices widely observed in turbulent boundary layers and the wall vortices are similar to the longitudinal vortices in the sweep phase. I believe that the phenomena described here explain the key mechanism of self-sustenance of fully turbulent flows and can be captured with an adequate analysis of the interaction terms. I also agree with Schoppa and Hussain (2002) that the transition from laminar to turbulent flow does not necessarily require the original presence of a parent vortex and may be induced by travelling periodic waves. Nonetheless, once a streamwise horseshoe vortex has been formed, it will generate streaming flow and infant vortices downstream.

Of course there is a limit to the analogy between the jets in cross flow studied extensively in the literature and the ejections because the latter are intermittent. Thus the fluid at the wall in turbulent flows is not continuously supplied from upstream but must rush in from the log-law region to satisfy the equation of continuity as the ejections take the fluid in the low speed streaks away. Nonetheless, the streaming

26/218

flow bursts are not instantaneous and there appears to be an overlap period when the adverse pressure created by the departing ejections can roll up the inrush fluid into streamwise vortices as observed by Kline et al. (1967).
.

The wall layer must be really divided into two separate sub-zones. Very close to the wall, the cross flow velocity is very low and the Reynolds number based on the approach velocity and the jet diameter is small. Under these circumstances there is no wake behind the jet. The Strouhal number for a cylinder increases very rapidly with the Reynolds number and regular Karman vortex streets are only observed above a value of 60 (Schlichting, 1979, p31-32). Thus there is a thin region near the wall where the shear stress is completely defined by viscous diffusion; form drag and convection effects are negligible. Unfortunately since we do not have measurements of the jet diameter, which can only be deduced from detailed analysis of DNS databases that I had no access to, an approximate analysis must be made with data from spherical obstacles.

Nikuradse (1933) created roughness on pipe walls by gluing sand grains of different sizes. He observed that when the equivalent roughness non-dimensional diameter was smaller than 5, the thickness of a laminar sub-layer postulated by Prandtl (1935), the flow behaves as if the wall were smooth. When the roughness diameter d increased further, the wall shear stress increased dramatically because of the contribution of form drag. Gardner and Keey (1958) reported that a wake formed behind a sphere at a particle Reynolds number of

$$\frac{u_d d}{v} = 20 \qquad (51)$$

Taking the velocity $u_r$ at the normal distance d as the approach velocity we get

$$\frac{d u_d}{v} = d^+ u_d^+ = 20 \qquad (52)$$

As shown in sections 7 and 8, in this region

$$u^+ = y^+ \qquad (53)$$

Hence

$$d^+ = \sqrt{20} = 4.5 \qquad (54)$$

Thus when $y^+ < 4.5$ the shear stress is based only on skin friction but above this threshold value, which is very close to Nikuradse's admissible roughness value of 5, this "viscous sub-layer" would behave slightly differently because some form drag is imposed by the presence of the streaming flow.

The wall layer flow before the advent of bursting obeys essentially the solution of order $\varepsilon^0$ and because it persists much longer than the bursting phase, in general the wall layer can be well described by a solution for an unsteady viscous sub-boundary layer as shown in section 7.5.

The region between the wall layer and the far field is dominated, as argued before, by the log-law. The velocity patterns of Wark and Nagib (op.cit.), shown in Figure 11, indicate that there is a wake on the lee side that moves with the ejection as it moves away from the wall. There seems little indication of vortex shedding into this wake region. Many authors (e.g. Keffer and Baines, op.cit.) describe the flow in the wake of jets in cross flow as similar to small scale turbulence. On the upwind side of the jet there is a region of high pressure that creates a strong shear layer with eddies very similar to the "typical eddies" observed by Falco (Figure 10).

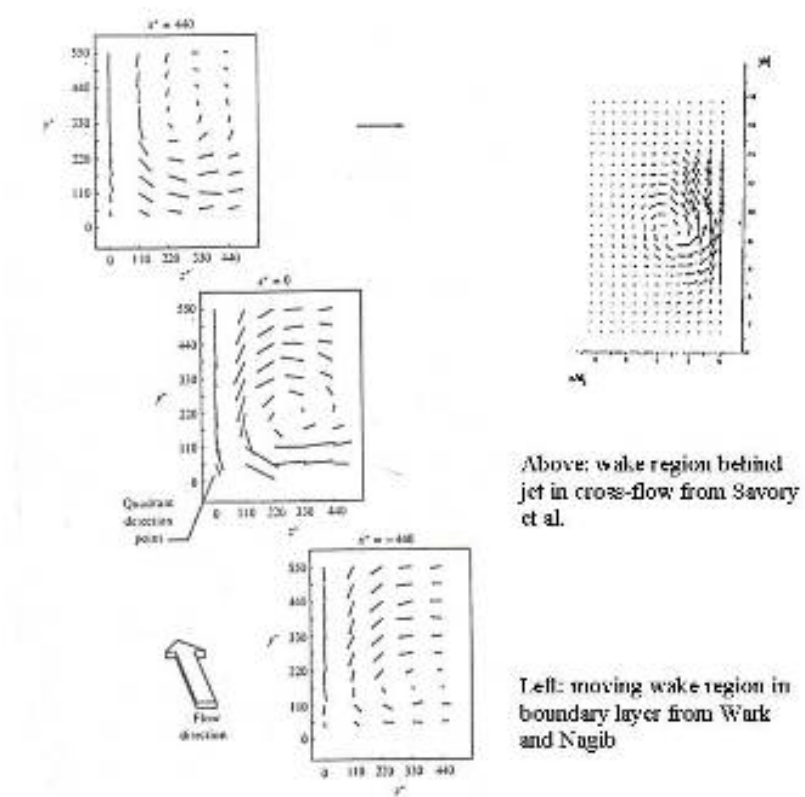

Figure 11 Roller-like structure in wake region behind a fixed jet in cross flow (Savory, et al., 1990) and moving with ejection (Wark and Nagib, 1991)

The transient ejections should be visualised as relatively short cylinders of fluid in cross-flow and I am not sure that we can rule out the possible creation of vortical structures at the ends of these cylinders. Adamorola, Sumner, & Bergstrom (2007) have shown that they can be created at high ratios of cross wind to jet velocity.

In my view, even if vortices are shed from the ejections outside the wall layer, in the linear (log-law) and curvilinear zones of the stream jet path they will still not contribute to turbulence production. This is because the vortices are too far removed from the wall to induce a transient laminar sub-boundary layer. Without the effect of viscosity, there is no way for the flow stream to store the wave energy produced by the velocity fluctuations generated by the rotating vortex and to convert it into the kinetic energy required for streaming. Schlichting (1960, p 431) states more clearly that "in the second approximation (solution of order $\varepsilon$) there appears a term which is not periodic and which represents steady streaming superimposed on the oscillatory

motion…It can also be stated that secondary flow has its origin in the convective terms (The terms on the RHS of equation (20)[3] and is due to the interaction between inertia and viscosity".

The first indication that I saw to support that hypothesis came from estimates of changes in the thickness of the log-law region and wall layer, with Reynolds number and flow configuration. As shown in Figure 13, the dimensionless thickness of the wall layer is independent of the Reynolds number whereas the thickness of the log-law region increases steadily. This would indicate that the behaviour of the log-law region does not impact on the extent of the wall layer. Moreover, the velocity profile of the wall layer is the same for flows at all Reynolds numbers and for all flow configurations (K.T. Trinh, 2005a) as shown in Figure 36.

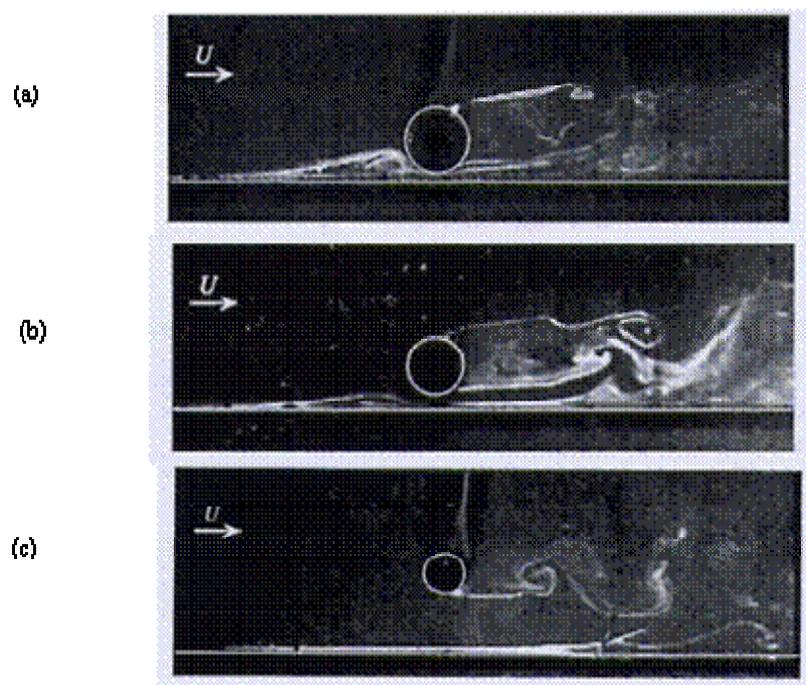

Figure 12 Interaction between a cylinder in cross flow and the wall after Price et al. (2002). (a) y/D=0.125, (b) y/D =0.25, (c) y/D = 1.5

More compelling evidence emerges from the study of works on cylinders in cross flow near a wall (Cigada, Malavasi, & Vanali, 2006; Price, Sumner, Smith, Leong, &

---

[3] Author's addition

Paidoussis, 2002). It is well known that a laminar boundary layer can be induced into turbulent flow by the attachment of a trip wire on the wall (Schlichting, 1979 p 41, 539). This technique has been used by many researchers. Interestingly there is no evidence of vortex shedding on the lee side of the trip wire as shown in Figure 12 (a) reproduced from Price et al (op.cit.). There is only vortex shedding on the far side of the cylinder but this vortex does not appear to interact strongly with the near wall fluid. Instead the pattern next to the wall looks very similar to the wall (longitudinal) vortices observed behind jets in cross-flow, As the cylinder is moved away from the wall its interaction with the wall fluid diminishes until we see clearly vortex shedding on both sides of the cylinder (Figure 12 b and c) but these have no effect on the wall fluid. Jiménez and Pinelli (1999) have recently conducted "numerical experiments" and indicate that perturbations fed from the outer flow have little effect on events in the wall layer.

We note that there is no evidence of vortex shedding behind the jets in the data of Savory et al. and Wark and Nagib in Figure 11, which tends to support the observation of Chan et al. (1976) that the cross flow breaks down into small scale turbulence in the wake.

The reader may by now recognise a strong parallel between the division of the flow field in this discussion with the four zones of constant momentum identified in the PIV experiments of Adrian, Meinhart, & Tomkins (2000). Fife, Wei, Klewicki, & McMurtry (2005) have recently proposed a new scaling scheme based on a hierarchy of scales lined with these four regions.

## 5      Subsets of the Navier-Stokes Equations

### 5.1     Speculations on the form of a solution

It is evident by now that a full analytical solution of the Navier-Stokes equations is extremely difficult because of their non-linearity. Consider the requirements that such a solution must comply with. At any moment, the flow field is populated by a multitude of eddies of different ages, sizes and shapes. This is evident in published pictures educed from both DNS and PIV databases. One would expect an adequate solution to describe each of these individual eddies and its motion. In addition the solution would also need to identify the events and interactions that arise from characteristic features in these eddies. A process that is now regarded as critically important by most serious workers in the field of turbulence, and which has received considerable attention in the last fifty years, occurs in the wall layer. Even the analysis of this very thin layer presents major difficulties. As shown in sections 2, 3 and 4, even a basic solution would require us to express the instantaneous velocity in terms of at least 4 components:

$$u_i = U_i + \tilde{u}_i + u'_i(\omega t) + u_{i,st} \qquad (55)$$

not just 3 as argued by Mankbadi (1992) and Schoppa and Hussain (2002). In fact there may be more components to add in the general case. For example, the fine scale turbulence observed in studies of jet in cross flow (Chan et al, op.cit.) may be captured by another velocity term that Schoppa and Hussain (op.cit.) call "incoherent" velocity in the sense that it does not lead to the formation of a new coherent structure. Now the simple analysis made in sections 2 and 3 allowed us to clearly identify the key features in the wall process and link them to particular terms in the NS. The simplicity with which we can extract a picture of what happens is its main virtue but the solution itself is only accurate to order $\varepsilon$. If we decide to include the terms of order $\varepsilon^2$ in equation (24), the mathematical analysis becomes much more complex. Schlichting (1960) has discussed streaming flow in terms of the method of successive approximations in the study of non-steady boundary layers but does not even bother discussing third–and-higher approximations except to mention that the mathematical difficulties increase exponentially. Nevertheless one would suspect that this neglect of higher order terms immediately come at the cost of missing out on higher interactions between secondary and may be even tertiary and higher level structures. A true solution of the NS equations can surely not do that. Thus we should expect that a solution would take the form of a single, or more likely a sum of, infinite series.

Because of the regeneration properties of self sustaining turbulence, as opposed to decaying turbulence as happens behind a grid for example, we might expect a solution of the form

$$u(t) = u(t + nT) \qquad (56)$$

While one may determine statistically the average scale of the wall layer where most of turbulence is produced by a method of correlation (Meek & Baer, 1970), other methods of burst detection e.g. (Antonia, et al., 1990) show that the time scale of individual streaks in the wall layer is much less well behaved and can vary significantly. This is probably due to the fact that the periodic fluctuations are not based on a single frequency (and its harmonics) but may reflect a non-harmonic spectrum of frequencies. This would be reasonable if one accepts that the velocity fluctuations in the wall layer result from the influence of many vortical structures passing above it giving a spectrum of fluctuations of different frequencies and strengths. Thus the fluctuating velocity term in equation (55) may be more realistically expressed as $\sum_m u'_i(\omega_m t)$ and as a consequence there may be more than one streaming flow component which should also be expressed as $\sum_m u_{ist,m}$. There is evidence that a hairpin vortex in the wall layer may generate "infant vortices" ( e.g. (Arcalar & Smith, 1987 a &b). These would then impress further induced fluctuations on the velocity of the low speed streaks. Thus a burst normally comes from the detachment of many streaming jets, not just one.  Since any one of these jets may trigger the onset of a bursting period, the wall layer cannot be considered as periodic but only intermittent with a complex variation in its period and equation (56) would have to be modified accordingly. Thus any attempt to make the model solution more physically realistic greatly increases the mathematical complexity of the problem.

After this arduous exercise one would then make an ensemble average to give time-averaged velocity distributions and drag coefficients that have been so well documented and used by engineers. The exercise described above is daunting. It is comparable to describing the behaviour of China by studying the life of each of its billion or so individuals and then averaging the characteristics observed. Is it even possible, can one actually obtain a general analytical solution of the NS equations? Without delving into the complex arguments that mathematicians make to prove that a solution does exist, we may take the success of the DNS as an indication that a reliable and detailed picture can be obtained albeit by numerical iterations. There is a parallel between the successive iterations of the DNS and the method of successive approximations discussed earlier. The major difference is that the method of analysis by successive approximations is based on a global scale while the method of numerical iterations is based on quite simple discretisations at the local level. The DNS have only been possible because of the large improvement in the speed of computers that need to carry out the billions of calculation necessary. In a sense, each of the elements in a DNS grid is analogous to an individual in the Chinese example raised previously. No such advance in the tools required is as yet available for an analytical approach. Nonetheless it is hard to see how the iterations could converge if a solution of the DNS equations did not exist. Of course, the fact that the iterations converge does not guarantee that the solution is accurate. But the picture given by the DNS has been found quite compatible with the detailed experimental results made possible by advances in visualisation techniques such as PIV. The DNS have the added advantage over the PIV in that it not only give a picture of the flow field but allows access to local and transient stresses and pressures.

The picture provided by the data bases generated by computer and visual experiments does not necessarily allow us to understand the mechanisms of turbulence: one must extract information about events and causal relationships, which is not a trivial task considering the huge size of these databases, For this one must inevitably have recourse to some statistical tools. I was struck with admiration for the work of Taylor, who practically invented statistical turbulence and was present at each of its major developments before such luminaries like Karman and Batchelor joined in. However, like many other scientists, I have always had some reservation about the blind use of statistics. I will paraphrase here a well known Indian colleague and extend his warnings. Let us suppose that in order to summarise the behaviour of Chinese individuals we assign a value +1 to males and -1 to females, a crude mathematical expression of the yin and yang. Then averaging over the population of a billion plus gives us a result of practically zero, say $+\varepsilon$! Does that mean that the average Chinese is asexist? The slightly positive score of the Chinese probably reflects the one child policy of its government and the fact that the Chinese have a tendency to keep only male children. But then the score of India is more like $-\varepsilon$ if one is to judge from the news reels that are available. An imprudent researcher might actually conclude that the Chinese are more manly that the Indians, and thus make sure that he never travel safely to India! The point of the story is that statistical, indeed any type of, averaging has the troublesome tendency of loosing critical information, even distort superficial impressions. Because of the averaging exercise that resulted in the Reynolds equations, crucial distinction between fast and slow Reynolds stresses that I pointed out in equation (13) has not figured in discussions of turbulence for the best part of the century that followed Reynolds.

Gallup was the first to demonstrate that good assessment of voting trends could be achieved not by polling large random samples of a population but representative samples. The representative sample contains in the right proportion all the major factions of population that matter for the questions asked. In a sense this is also what is needed to obtain a picture of turbulence: an understanding of what structures, events and interactions matter and a suitable weighing of their contribution, including a time scale and location since many of them are transient. This is the approach I started to take since the 1980's first by identifying and analysing the key elements of the puzzle and then assembling them into models that are useful from a practical point of view.

## 5.2    Subsets of the NS equations

The solution of the NS equations is facilitated when various terms are omitted in a process called linearisation. For example, when the Reynolds stresses are neglected from the time-averaged equations, the solution describes conventional laminar flow.

However, the success of the DNS, which do not require any simplification to the original Navier-Stokes equations, suggests that all the terms in the Navier-Stokes equations must be considered in turbulence research. The neglect of any term results in some loss of information. The impact of this loss depends on the particular term being neglected. The following is a summary of my assessment of the impact caused by various simplifications to the Navier-Stokes equations.

### 5.2.1        Fast Fluctuating Components of the Velocity

According to the present analysis, the fast fluctuations are the trigger for the streaming flow (or ejections) from the wall. Without them the secondary streams would not exist and the flow would not be turbulent.

It is interesting to note that all the successful DNS (e.g. Laurien and Kleiser 1989,

Spalart 1988, Kim, Moin and Moser 1987) have included fluctuating components of the velocity. Spalart has noted that the use of an imposed fluctuating velocity in conjunction with spectral methods helps with faster convergence in the iteration process.

### 5.2.2  The Interaction Terms

When the Navier-Stokes equations are solved in their original unsteady form for boundary and initial conditions with fluctuating velocity components, the solution may potentially include all the features of turbulence. However, if the interaction terms J in equation (39) are neglected, the solution will still fail to reproduce the small scale features. The velocity field shows only a laminar oscillating boundary layer with a secondary streaming flow but no wake formation on its downstream side with consequent breakup of the streamwise flow into small scale.

### 5.2.3  The Streaming Function

A more drastic simplification consists in assuming that the streaming function is negligible (the factor $\varepsilon$ in equation (21) is small). The solution then reproduces a fluctuating laminar boundary layer without secondary streaming flow. This kind of solution is typified by a laminar boundary layer with small velocity disturbances as encountered in the study of boundary layer stability or in the low-speed streak phase of the wall layer.

### 5.2.4  The Reynolds Stresses

The neglect of the Reynolds stresses in the time-averaged boundary-layer equations gives the "laminar" boundary-layer equations solved for example by Blasius for flow past a flat plate (Blasius, 1908).

We must differentiate here between two types of subsets of the NS equations. The first is traditionally linked with an explanation in terms of flow regimes and is typified by letting some terms tend to zero and leads to laminar flow solutions if the Reynolds stresses are ignored or potential flow if the viscosity decreases to zero. We may call these asymptotic solutions of the NS equations. The second is based on extracting particular terms from the NS to obtain equations that help analyse particular features of the flow. Orr (1907) and Sommerfield (1908) obtained in that way the governing equations for the development of small perturbations (the fast fluctuations) in the theory of stability. The attraction of these methods is that the subsets are often much easier to solve than the original Navier-Stokes equations.

## 6      Additive Layers in Turbulent Flow near a Wall

The writer proposes a different interpretation for the asymptotic solutions of the NS equations. The analysis described in sections 2 and 3 indicate that the subset of the Navier-Stokes equations where the terms of order $\varepsilon$ are neglected apply to a region near the wall where the effect of the ejections on the long-time average velocity field has not yet become important. It does not actually require that there are no velocity fluctuations, only that they are small enough for their effect on the smoothed phase velocity $\tilde{u}_i$ to be negligible. Thus the asymptotic subsets of the NS equations may be viewed as applicable to particular domains rather than regimes of flow. We will come back to discuss how that observation allows us to model the wall layer in section 7.

As the Reynolds number increases, the magnitude of the velocity fluctuations grows ($\varepsilon$ increases) according to well-known analyses of stability of laminar flows e.g. (Dryden, 1934, 1936; Schiller, 1922; Schlichting, 1932, 1933, 1935; Schubauer & Skramstad, 1943; Tollmien, 1929) and eventually a streaming flow appears. The structure of the flow field begins to change when the ejections start to disturb the outer quasi-inviscid region beyond the wall layer. At Reynolds numbers just above the critical value, e.g. Re =2100 for pipe flow, only the far field section of the intermittent jets penetrates the outer region. The disturbance to the previously "quasi-potential" flow may be compared with that of a wall-parallel jet since the ejections are here aligned in the direction of main flow. This region has been described by Cole's law of the wake (Coles, 1956). In the present visualisation, upon transition, the first layer to be added to the wall layer is the law-of-the-wake region as shown in Figure 13.

As the Reynolds number increases, so does the intensity of the ejections which reach further into the outer region. Eventually, the linear region of the jet path begins to intrude into the outer region. At that Reynolds number the main stream begin to interact with the jet to create a wake; the logarithmic law-of-the-wall makes its appearance. I believe that the dramatic increase in the boundary layer thickness may be explained in terms of this penetration of the ejections into the outer hitherto "inviscid" region. Figure 13 has been compiled from velocity measurements of boundary layer flow by (Klebanoff, 1954) and pipe flow by various authors (Bogue, 1961; Eckelmann, 1974; Laufer, 1954; Lawn, 1971; Nikuradse, 1932; Senecal & Rothfus, 1953). It shows how the law-of-the-wall, (hereafter called the log-law) and law-of-the-wake regions are added to the wall layer.

The advance of the ejections beyond the wall layer may be seen by monitoring the growth of the log-law region. This is best observed when the experimental velocity profiles are reduced to a common basis. I have not found any experimental measurements that could illustrate the penetration of the ejections into the outer region within a bursting period of turbulent flow. Some information can be obtained from studies of oscillating pipe flow (Akhavan, Kamm, & Saphiro, 1991; Hino, Kawashiwayanagi, Nakayama, & Hara, 1983). Schneck and Walburn (1976) have argued in their study of pulsatile blood flow that the secondary streaming flow results from a tendency of viscous forces to resist the reversal of flow imposed by the oscillating motion of the main stream. This is demonstrated more clearly in the experiments of Gad-el-Hak, Davis, McMurray, & Orszag (1983) who generated an artificial bursting process in a laminar boundary layer on a flat plate by decelerating it. The magnitude of the deceleration and the corresponding adverse pressure gradient must be sufficient to induce separation and ejection of low-speed fluid from the wall. This is evident in the oscillatory experiments of Akhavan et al (op.cit.) who found that turbulence begins at $Re_\omega$ = 500-550. The velocity profiles at different phases of a cycle are shown in Figure 14 for $Re_\omega$ = 1080. Except for the regions of oscillating potential flow, the similarity transformation adopted in section 8 collapses them onto the velocity profiles for "steady" turbulent pipe flow. The acceleration phase, where the pressure gradient is favourable, is laminar. The velocity profile here exhibits only two regions: (a) a wall layer which coincides very well with the profiles for laminar boundary layer flow and those for the wall layer of steady turbulent pipe flow, and (b) a fluctuating potential flow in the outer region. The data illustrates clearly the growth of the log-law region in the deceleration phase of the cycle.

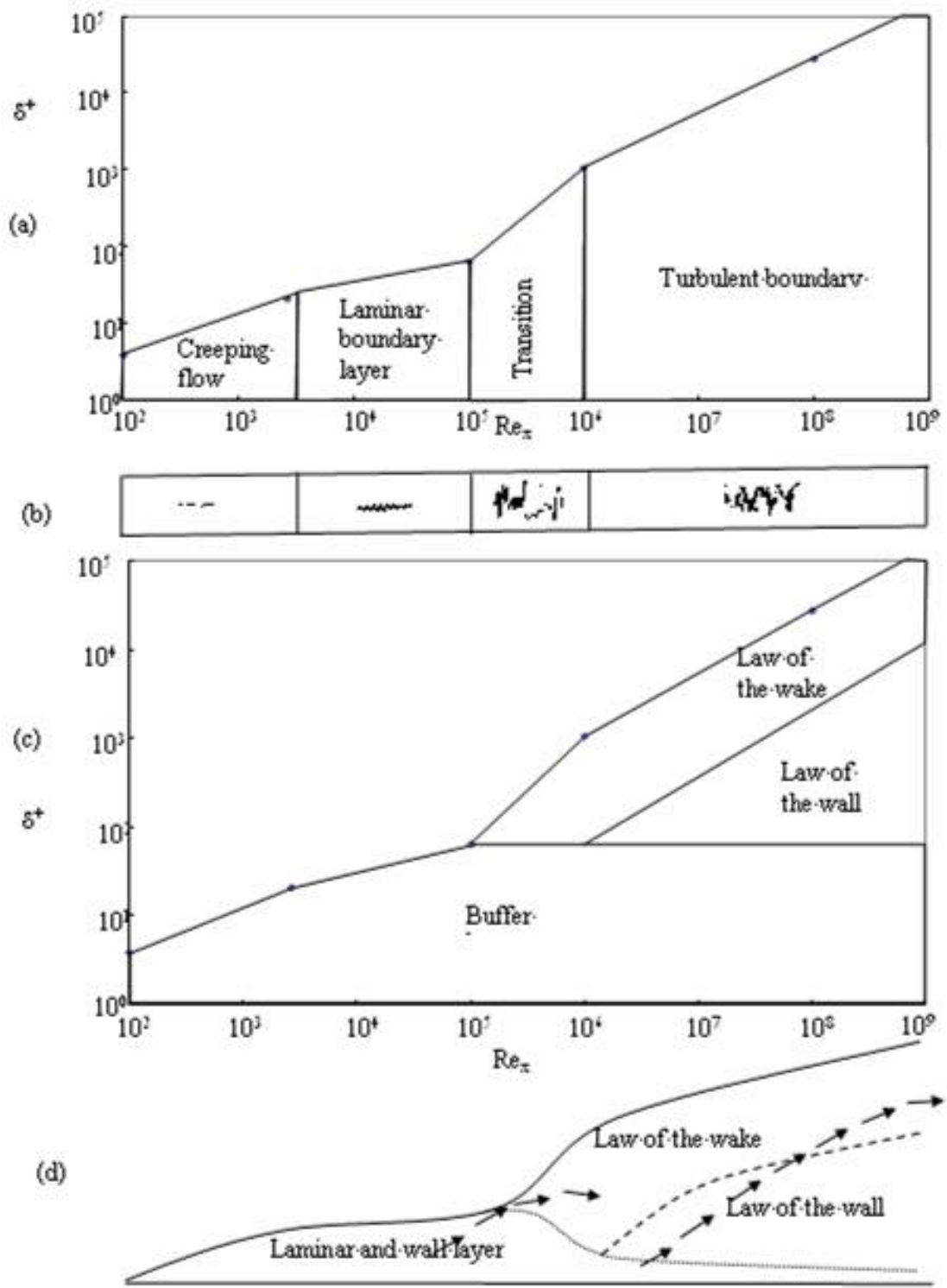

Figure 13 Representations of flow regimes (a) Reynolds flow regimes, (b) Representative velocity traces (after Schubauer and Skramstadt (1947), (c) Additive layers in normalised dimensions, (d) Physical thickness of layers (not to scale) and path of ejections in transition and fully turbulent flow.

The division of a turbulent boundary layer into three regions is well known in the literature (Bradshaw, 1971; Cecebi & Smith, 1974; Hinze, 1959). The physical interpretation presented here is new. Figure 13 suggests that it is useful to turn the picture by 90° and investigate the evolution of turbulence with scaling parameters based on the friction velocity and the normal distance rather than concentrate on the streamwise velocity as suggested by Reynolds (1883).

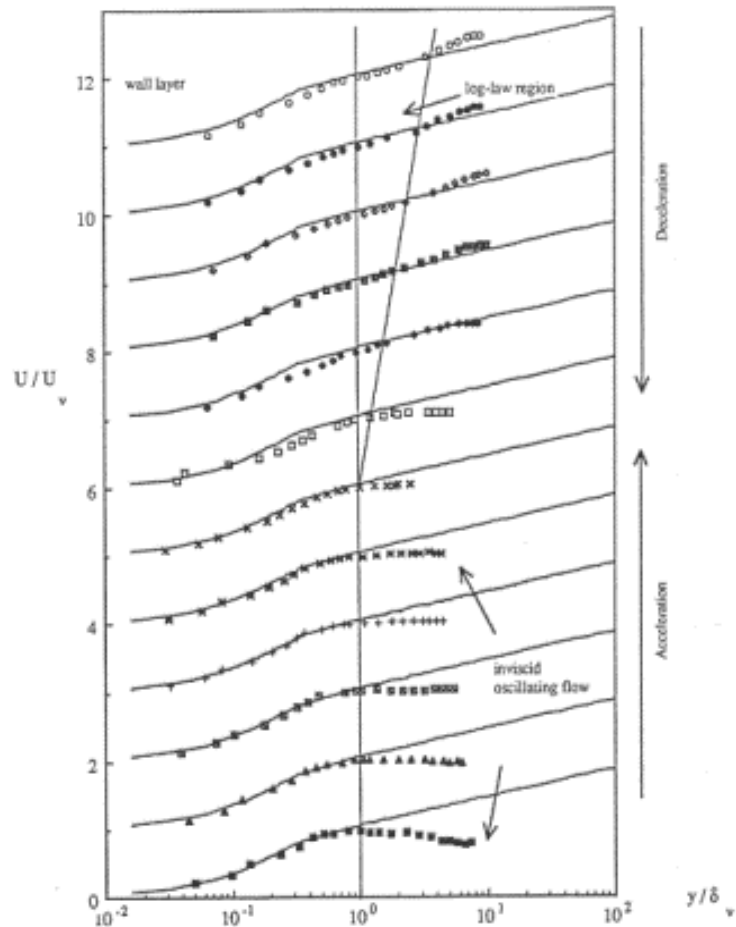

Figure 14 Penetration of the log-law into the outer region during a cycle oscillating pipe flow. From Trinh (1992). Data of Akhavan et al. (1991)

## 7    Further deductions from this visualisation

Sometime in the middle to late 1980s, I decided to give up on a formal general solution of the NS equations based on a multi-component decomposition of the instantaneous velocity and leave that to better mathematicians. I concentrated much more on a conditionally sampled picture of the flow field to obtain results that are of more immediate use to practical applications

This is achieved by two further steps:
> 1. A description of each region of the flow field, preferably through simple mathematical models obtained from the analysis of the subsets of the NS equations,
>
> 2. Matching criteria to link these models.

These two problems may be analysed at various levels of complexity. I have aimed only at the simplest.

The following is a brief outline of the major results I obtained over the years. The details will be published separately.

## 7.1　Energy flow in turbulence

In this visualisation, the flow of energy in a turbulent field may be described as follows. Energy is extracted from the main flow through periodic fluctuations, however they may be induced. The writer agrees with the arguments of Schoppa and Hussain (op.cit.) that the existence of a streamwise horseshoe vortex is not necessary to induce transition to turbulence. Even acoustic vibrations have been shown to induce streaming (e.g. Schlichting 1960, p 431; Frater, 1967). However, once self-sustenance of turbulent flow has been achieved, the main source of periodic fluctuations will be vortices travelling above the wall. The energy is stored in the sweep phase of the wall layer process in the form of growing wave fluctuations. This wave energy is then transformed into kinetic energy through the streaming process expressed in mathematical terms by the fast Reynolds stresses. When the magnitude of the fast Reynolds stresses reaches a critical threshold, fluid is ejected from the wall layer into the outer flow bringing with it the energy contained in the streaming flow. Fluid rushes into the wall layer after a burst to satisfy the law of conservation of mass. The energy contained in the streaming jets is dissipated eventually in the far field region through viscous interactions. Energy is also extracted from the cross flow to break it up when it impinges on the streaming jets. Some of that energy is dissipated as small scale turbulence but some is returned through the formation of "infant" vortices. In the wall layer, the infant vortices start a new sweep phase and perpetuate the process of turbulence generation.

## 7.2   A Definition of Turbulence

To a beginner, the study of turbulence is immediately hampered by the surprising lack of a clear and concise definition of the physical process. Tsinober (2001) has published a long list of attempts at a definition by some of the most noted researchers in turbulence. The most common descriptions are vague: "a motion in which an irregular fluctuation (mixing, or eddying motion) is superimposed on the main stream" (Schlichting 1960), "a fluid motion of complex and irregular character" (Bayly, Orszag, & Herbert, 1988) or negative as in the breakdown of laminar flow (Reynolds' experiment 1883). Some of the definitions are quite controversial like Saffman's (1981) "One of the best definition of turbulence is that it is a field of random chaotic vorticity" because the words random and chaotic would imply that a formal mathematical solution, which is necessarily deterministic, does not exist. Perhaps the most accurate definition can be attributed to Bradshaw (1971) "The only short but satisfactory answer to the question "what is turbulence" is that it is the general-solution of the Navier-Stokes equation". This definition cannot be argued with but it is singularly unhelpful since no general solution of the NS yet exists 160 years after they were formulated.

The writer proposes that turbulence should be defined as "a system with a main cross flow containing secondary intermittent streaming, at some angle to the direction of the main flow and with which it interacts".

## 7.3   Basic Elements of a Turbulent Flow Field

In the present work, the dynamics of turbulence near a wall are viewed in terms of three flow components:

   1. A main wall stream which is seen as a non-steady laminar boundary layer flow. Small perturbations in the layer grow into waves. In the stability theory, these are called primary disturbances. In the present study the vortical structures created within this layer are still attached to the wall and their Reynolds stresses are not considered characteristic of turbulence but of an unsteady Blasius layer (1908) with a fluctuating component in its approach velocity.

2. A first secondary stream which is modelled after an intermittent jet in crossflow. The writer identifies this stream with the ejections observed by Kline et al (1967). In the statistical theory of turbulence, they are called the primary eddies but they correspond to the secondary instabilities in the theory of stability. Kline et al (op. cit.) and Grass (1971) also refer to them as eddies and have mapped their path close to the wall. In this study, the primary eddies are seen as the main source of turbulence production and their shedding frequency determines the time scale of the bursting process.

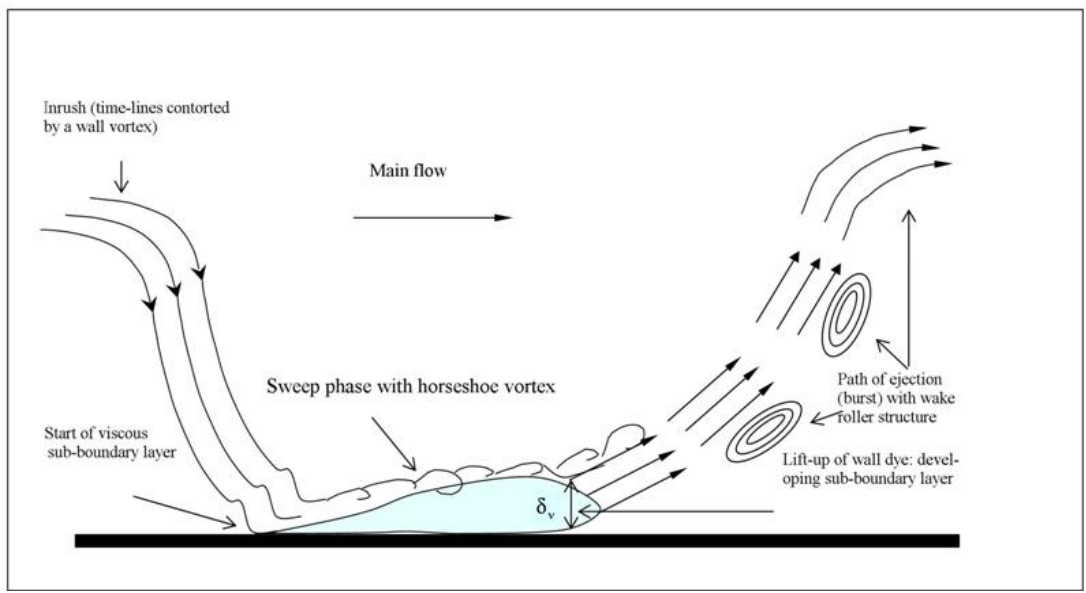

Figure 15 Basic visualisation of wall turbulence.

3. A second secondary stream which results from the interaction between the first two flow components. It is viewed in terms of a wake on the lee side of the intermittent jet. The structures behind this jet are called secondary eddies or tertiary instabilities. The theory of stability differentiates only between the primary and secondary instabilities. In this study, the secondary eddies (tertiary instabilities) are seen as responsible for the fine scale turbulence in the log-law region and for turbulence regeneration in the wall layer.

Turbulence thus results from the very complex interactions between the three flows as shown in Figure 15.

## 7.4 A criterion for transition

The first major concept presented to students of fluid mechanics is Reynolds' use of a dimensionless number to identify the transition from laminar to turbulent flow (Reynolds, 1883). It is widely accepted that laminar flow in a pipe ends when the dimensionless group

$$\mathrm{Re} = \frac{DV}{\nu} \tag{57}$$

reaches a value in the vicinity of 2100. The visualization illustrated in Figure 13 suggests that the flow field ceases to be completely laminar when the solution of order $\varepsilon$ protrudes from the layer described by the solution of order $\varepsilon^0$ in equation (23). This event depends on the value of the factor

$$\varepsilon = \frac{U_e}{L\omega} = \left(\frac{U_e L}{\nu}\right)\left(\frac{\nu}{L^2 \omega}\right) \tag{58}$$

which is, in my view, a more appropriate criterion. It still contains the Reynolds number when one equates the typical length L with the diameter D but states that transition is also dependent on the frequency of the original periodic disturbances.

In fact, it has been known for a long time that one can preserve laminar flow for much longer than suggested by Reynolds if the pipe entrance is carefully designed to reduce disturbances. For example, Ekman (1910) has shown that laminar pipe flow can be sustained up to a Reynolds number of 40,000. On the other hand Prandtl and his colleagues often hasten the development of turbulent boundary layers in wind tunnels by introducing disturbances through a trip wire (Schlichting 1960, p. 39).

While the parameter $\varepsilon$ is a theoretically correct parameter for determining transition, the problem is that the characteristics of small disturbances are mostly unknown in situations of practical interest. When calculating the size of the different domains where solutions of different orders in $\varepsilon$ apply, as discussed in section 8, I noted that the non-dimensional thickness $\delta_\nu^+ = \delta u_*/\nu$, where $u_* = \sqrt{\tau_w/\rho}$ is the shear velocity, $\tau_w$ the wall shear stress, $\rho$ and $\nu$ the fluid density and kinematic viscosity, of the layer obeying the solution of order $\varepsilon^0$ reached a maximum value approximately 64-65 at the end of the laminar flow regime and stayed subsequently constant. This value is the same for many flows for which I could obtain data from the literature: pipe flow, boundary layer on a flat plate, and flow between parallel plates. The corresponding thickness of the Karman buffer layer shown in Figure 13 follows the same trend. The relation between these two thicknesses is explained in section 8. Thus the value of $\delta_v^+$ is a convenient parameter to delimit flow regimes.

## 7.5 The wall layer

It will be noted from the analysis made in sections 2 and 2.1 that the solution of order $\varepsilon^0$ is independent of the solution of order $\varepsilon$. The question is where does it apply? Figure 13 suggests that it applies to all laminar flows and to the sweep phase of turbulent wall layers. Because the sweep phase lasts so much longer than the bursting phase e.g. (Walker, Abbott, Scharnhorst, & Weigand, 1989) it dominates the average velocity distribution in the wall layer.

This is the wall layer which can be modelled by any equation describing an unsteady state (developing) laminar boundary layer. Einstein and Li (1956) were the first to attempt modelling the development of the local velocity with Stokes' solution1. The idea was quickly picked up by others (Hanratty, 1956; Meek & Baer, 1970) but any other model for an unsteady state laminar boundary layer can be used.

The governing equation for the Stokes solution1 (1851) applied to turbulent flow is

$$\frac{\partial \tilde{u}}{\partial t} = \nu \frac{\partial^2 \tilde{u}}{\partial y^2} \qquad (59)$$

Stokes has solved this equation for the conditions:

| IC  | t = 0 | all y | $\tilde{u} = U_v$ |
| BC1 | t > 0 | y = 0 | u = 0 |
| BC2 | t > 0 | y = ∞ | $\tilde{u} = U_v$ |

where $U_v$ is the approach velocity for this sub-boundary layer. The velocity at any time t after the start of a period is given by:

$$\frac{\tilde{u}}{U_v} = \text{erf}(\eta_s) \qquad (60)$$

where $\eta_s = \dfrac{y}{\sqrt{4\nu t}}$

The average wall-shear stress is

$$\tau_w = \frac{\mu U_\nu}{t_\nu} \int_0^{t_\nu} \left(\frac{\partial \tilde{u}}{\partial y}\right)_{y=0} dt = \frac{\mu U_\nu}{t_\nu \sqrt{\pi}} \int_0^{t_\nu} \frac{1}{\sqrt{\nu t}} dt \tag{61}$$

Equation (61) may be rearranged as

$$t_\nu^+ = \frac{2}{\sqrt{\pi}} U_\nu^+ \tag{62}$$

The time-averaged velocity profile near the wall may be obtained by rearranging equation (62) as

$$\frac{U^+}{U_\nu^+} = \int \mathrm{erf}\left(\frac{y^+}{4 U_\nu^+ \sqrt{t/t_\nu}}\right) d\left(\frac{t}{t_\nu}\right) \tag{63}$$

### 7.5.1 The time-averaged velocity profile of the wall layer

Einstein and Li (op.cit.) and Hanratty (op.cit.) have applied equation (63) between the wall and the edge of the buffer layer (Karman, 1934), where $U_b^+ = 13.5$ and $\delta_b^+ = 30$, and obtained good agreement with measured velocity profiles but the relationship of this model to various streamwise positions x along the wall is ill-defined.

Meek and Baer (op.cit.) endeavoured to define more formally the domain of application of equation (63) by taking the edge of the wall layer as the position where $u/U_\nu = 0.99$, which corresponds to $y = \delta_\nu$ and $\eta_s = 1.87$. Substituting these values into equation (60) gives

$$\delta_\nu^+ = 4.16 U_\nu^+ \tag{64}$$

Back-substitution of equation (64) into (62) gives

$$\delta_\nu^+ = 3.78 t_\nu^+ \tag{65}$$

Meek and Baer matched equation (64) with Prandtl's law of the wall, equation (46) and obtained $U_\nu^+ = 14.9$, $\delta_\nu^+ = 64$. Substituting this new criterion into equation (63) gave good predictions for the time-averaged velocity over the whole wall layer, $0 < y^+ < 64$. Meek and Baer also showed that equation (62) gave a very good prediction of the time scale of the wall layer process as shown in Figure 16.

Many authors have subsequently used the Einstein-Li approach with further refinements to model the wall layer e.g. (Hanratty, 1956, 1989; Black, 1969). Reichardt (1971) has included the effect of the pressure gradient into equation (56). After this spade of work in the sixties and early seventies, the use of Stokes solution1 lapsed in the literature, except for attempts to use the concept in penetration theories of heat and mass transfer discussed in section 13. I realised after a while that many researchers could not reconcile the concept of a laminar sub-layer, even intermittent, with the intense activity that Kline et al. (1967) first identified in the wall layer and confirmed by many others (Corino & Brodkey, 1969; H. T. Kim, et al., 1971; Offen & Kline, 1974). In particular the Stokes solution1 looked incompatible with the coherent structures that dominated the studies of turbulence in the last fifty years e.g.(Cantwell, 1981; Robinson, 1991; Adrian, et al., 2000; Carlier & Stanislas, 2005; Jeong & Hussain, 1995; Jeong, Hussain, Schoppa, & Kim, 1997; Smith & Walker, 1995; Swearingen & Blackwelder, 1987). Indeed even a cursory search in The Web of Knowledge database returned thousands of papers devoted to coherent structures and entire books have been devoted to their understanding e.g. (Holmes, Lumley, & Berkooz, 1998) and one researcher (McNaugton, 2008) stated the common "hope that understanding these 'coherent structures' will give insight into the mechanism of turbulence, and so useful information for explaining phenomena and formulating models." In the same breadth he acknowledged that "Unfortunately little of practical value has been achieved in the 50 years of research into turbulence structure because of the very complexity of turbulence, so that there is still no accepted explanation of what the observed structures are and how they are formed, evolve and interact."

Particular attention has been paid to the horseshoe or hairpin vortices that have been seen by many as crucial to an understanding of wall turbulence e.g. (Arcalar & Smith, 1987b; Gad-el-Hak & Hussain, 1986; Schoppa & Hussain, 2000; Suponitsky, et al., 2005). Clearly much more convincing evidence must be presented to make the Einstein-Li approach acceptable.

### 7.5.2 "Turbulence" statistics of the wall layer

My first argument is based on a fundamentally different view for the physical

48/218

interpretation of the exercise. Einstein and Li viewed the wall layer as a laminar flow which is disrupted because it is periodically bombarded by eddies coming from the outer region. During the quiescent intervals when the faster fluid is in contact with the wall, momentum is exchanged with the outer flow by a diffusion process.

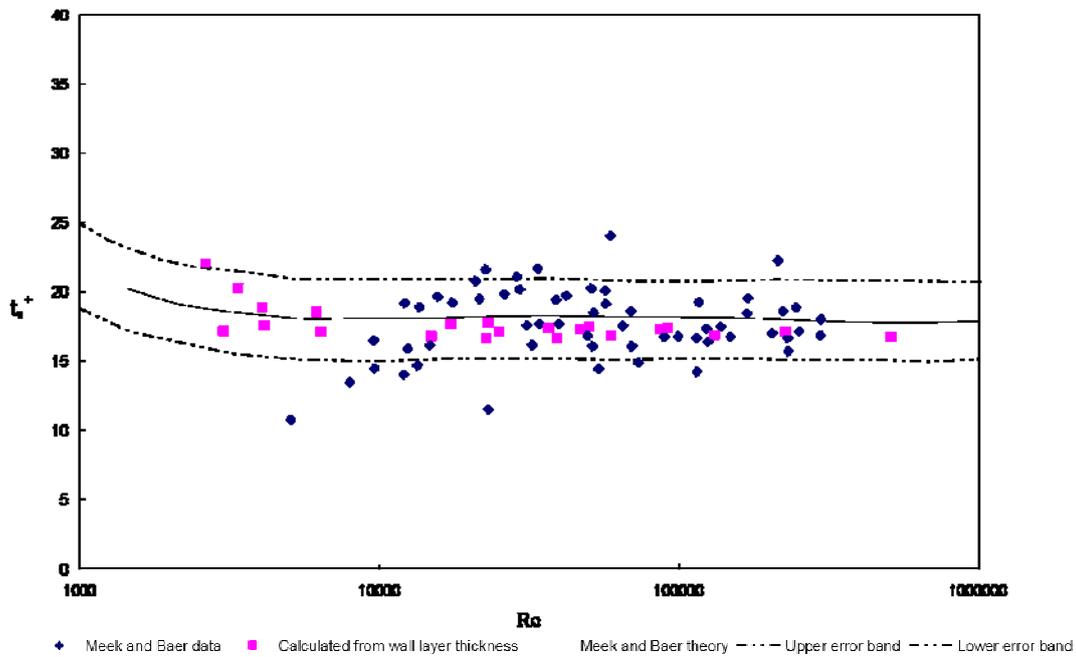

Figure 16 The time scale of the wall layer process according to Meek and Baer (1970)

The derivation in sections 2 and 3 is based on the contrary view that turbulence is mainly produced in the wall layer, not the outer region. It is clear from this derivation that the Stokes solution1, a form of the solution of order $\varepsilon^0$, does not imply that fast velocity fluctuations are not present in the wall layer. This solution only gives a distribution of the smoothed phase velocity $\tilde{u}_i$ in the sweep phase of the wall layer and gives no information about the fast fluctuations $u'_i$ that exist, as is well shown by many experimental measurements in the last 5 decades. The form of the solution depends on the geometry used. For pipe flow Szymanski's derivation (Szymanski, 1932) is more appropriate because it takes account of curvature, for flow between parallel plates the solution takes the form of a Fourier series (Bird, Stewart, & Lightfoot, 1979; Theodore, 1971), for spherical coordinates we obtain a Legendre series. I have applied these derivations, with suitable modifications, of initial conditions and obtained quite satisfactory correlations with experimentally available profiles of time-averaged properties (section 12). But I still prefer to use the Stokes solution1 because of its mathematical simplicity noting that it applies well whenever the surface curvature can be neglected, i.e. when the wall layer thickness is very small compared with the thickness of the turbulent boundary layer.

The solution of order $\varepsilon^0$ describes essentially how the wall retards the velocity of the adjacent fluid through diffusion of viscous momentum and its thickness represents the maximum penetration of viscous momentum into the outer flow; to paraphrase Prantl's brilliant concept for a laminar boundary layer (Prandtl, 1904). Many more deductions can be made about the properties and usefulness of the Stokes solution1 as a model for the sweep phase than were explored by its early proponents.

**7.5.2.1 Probability density distribution of the streamwise velocity**

For example, I calculated the probability density distribution (pdf) in the wall layer by generating two thousand data points from equation (60) and sampling them for frequency of occurrence and compared the results with the experimental pdf of Kreplin and Eckelmann (1979) in Figure 17.

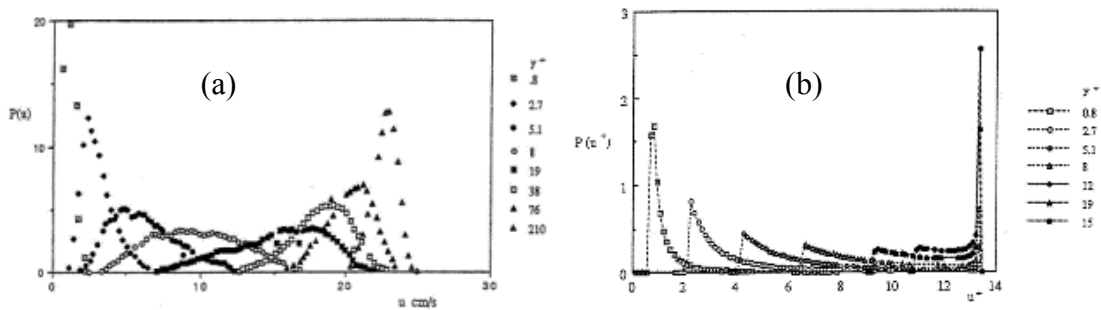

Figure 17 (a) Probability density function educed from the Stokes solution1, (b) Data of Kreplin and Eckelman (1974). Reproduced from Trinh (1992, 2005)

There is a remarkable resemblance between Figures 17 (a) and (b). In particular for each $y^+$, the peak on the two Figure 17 (a) and (b) occurs at the same values of $U^+$. This indicates that the Stokes solution1 correctly estimates the value of the dominant wavelengths. As the distance from the wall $y^+$ increases from 0 to 13, the relative size of the peaks diminish in the proportions measured. The pdf becomes more widely spread.

Two basic differences between the simulation and the experiment of Eckelmann must be stressed:

1. Eckelmann's velocity population spanned the whole range of possible values across the channel, *i.e.* from 0 to $U_m$ at the axis. In this simulation, velocities were limited to $U_v^+$ ($\approx 2/3\ U_m^+$), that at the edge of the buffer layer. Thus the generated population is an underestimate of the real population over which the instantaneous velocity was sampled. This condition makes the estimated probability density distribution peak higher than the measured value.

2. The Stokes solution assumes a uniform bonding approach velocity U. The simulation therefore forces the velocity at the edge of the wall layer to a probability of 1. In reality the velocity fluctuations persist right to the channel axis.

For values of $y^+$ below 12, the maximum in the profile of fluctuating velocity, the curve is skewed to the left. For the remainder it is skewed to the right. The pdf for $y^+ = 13$ has not been measured by Eckelmann but can be shown to be a minimum in this simulation. This trend can be observed on the predicted plot but a definite bias to the left is strongly apparent even though some skewing to the right occurs near the edge of the buffer layer. This is probably because the model assumes that at the end of the period the initial uniform flow profile is restored abruptly. This is not borne out in reality.

The agreement between Figure 17 (a) and (b) is better for values of $y^+ < 12$ than in the remainder of the buffer layer. This suggests that the Stokes solution models best the front portion of the sweeps where the Reynolds stresses are minimal and the disturbing influence of the ejections is least.

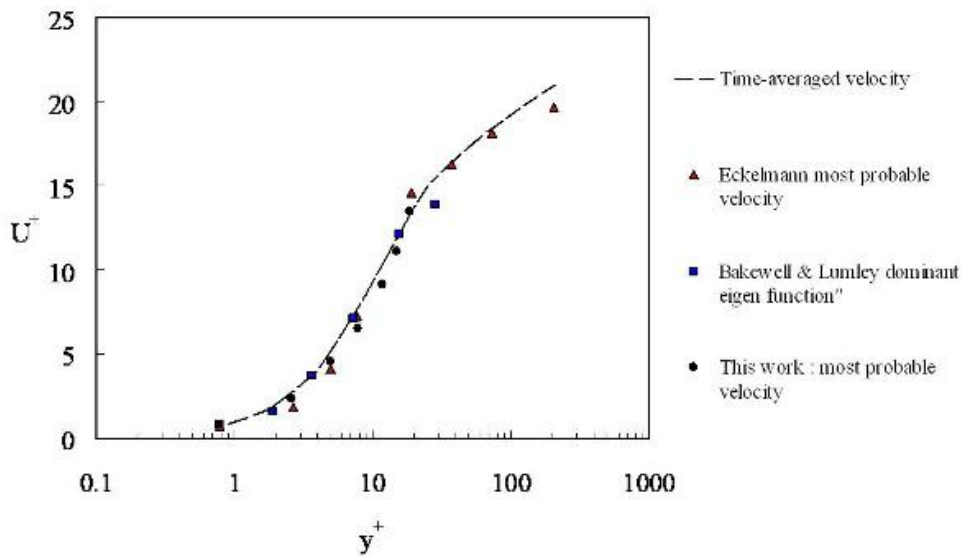

Figure 18 Dominant velocities compared to time averaged velocity profile. Data from (Bakewell & Lumley, 1967; Eckelmann, 1974; Trinh, 1992)

When one overlays the values of the most probable velocities at various position $y^+$ obtained by Eckelmann with the values predicted in Figure 17 (a), the correspondence is remarkable (Figure 18). The data from Bakewell & Lumley (1967) also plotted was obtained by an orthogonal decomposition of the instantaneous velocity traces, a tool that has also been used successfully in other flow geometries, e.g. by Takeda (1999) for the study of turbulent Taylor vortex flow. Such decompositions were supposed to identify the most important velocity fluctuations in turbulent flow fields.

Considering the crudity of the model, the agreement is already enlightening. A better model is obtained by noting that the ejections result in a non-uniform velocity outside the wall layer and the approach velocity in the Stokes solution1 needs to be modified accordingly for a more rigorous analysis.

### 7.5.2.2 The moving front of turbulence

The Stokes solution1 cannot show how the unsteady viscous state sub-boundary layer behaves in the x direction. A picture can be obtained by using a time-space transformation (Trinh & Keey, 1992a; 1992b). Essentially the Stokes solution1 is transformed into the Blasius solution for a laminar boundary layer on a flat plate (Blasius, 1908) with an extended form of Taylor's hypothesis which yields

$$\delta_v = 4.96\sqrt{\frac{x_v \nu}{U_v}} \quad (66)$$

We will come back to discuss the implications of this transformation in section 9.2. For the moment, we note that if we assume that slow speed streaks of all ages t (and therefore lengths) have an equal probability of passing a fixed probe in the flow field, then we need to average over all x to obtain a statistical average for the sub-boundary layer thickness giving

$$\delta_b = 3.31\sqrt{\frac{x_v \nu}{U_v}} \quad (67)$$

which may be rearranged as (Trinh 1992)

$$\delta_b^+ = 1.28\sqrt{x_v^+} \quad (68)$$

The trace of this average sub-boundary layer thickness given by equation (68) also fits the 'moving front of turbulence' of Kreplin and Eckelmann (1979) perfectly as shown in Figure 19.

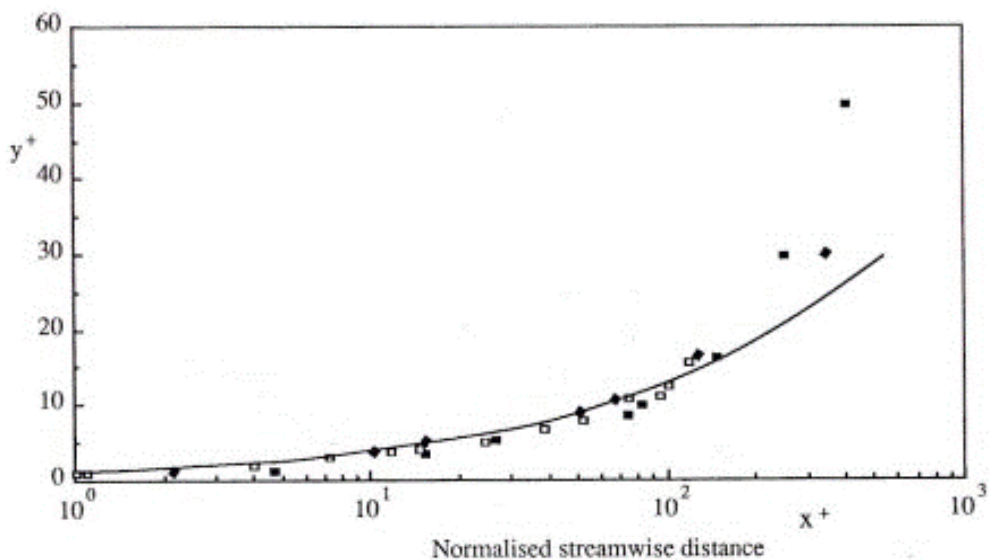

Figure 19 "Moving front of turbulence" and statistically averaged edge of the solution of order $\varepsilon^0$ using equation (65). Data of Kreplin and Eckelmann (1979).

Thus the shape of the hairpin vortex can be explained by two completely separate events. The legs of the vortex are shaped in the sweep phase because the viscous sub-boundary layer that is induced under the travelling vortex growth in thickness as the diffusion of viscous momentum penetrates into the main flow pushing the vortex further away. Then in the following bursting phase, the streaming ejection, whose path is well captured by the log-law as shown in Figure 8 and Figure 9, lifts the head to a much steeper angle.

### 7.5.2.3 The fluctuating velocity

The r.m.s. fluctuating velocity at any point $y^+$ is given by sampling over the distribution of low-speed streaks, using:

$$\frac{\sqrt{\overline{u'^2}}}{u_*} = \frac{\sqrt{(\tilde{u}^+ - U^+)^2 dt^+}}{t_v^+} \qquad (69)$$

where the streamwise velocity is calculated from equation (59) . The trends compare well with the measurements of Eckelmann (1974) and Laufer (Laufer, 1954), in Figure 20 but the predicted peak occurs sooner than in the experimental data and the predicted fluctuating velocity in falls off more rapidly. This is because the simple model proposed has assumed a constant approach velocity $U_v^+$ at the edge of the wall layer whereas it is in reality fluctuating. A better model would use an approach velocity consisting of a time averaged value $U_v^+$ and a fluctuating component $u_v'^+$.

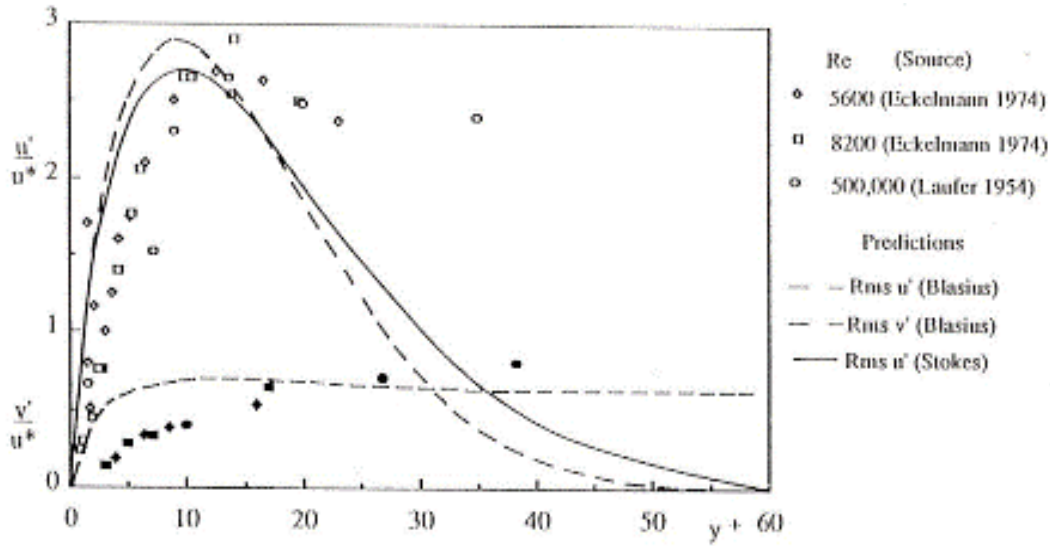

Figure 20. Longitudinal and normal fluctuating velocities near the wall at different Reynolds numbers and predictions from the Stokes solution1 and Blasius solution.

A similar prediction can be made by averaging the velocity fluctuations from the Blasius solution over all possible lengths $x$ as shown in Figure 20.

**7.5.2.4 Correlation function of the wall shear stress**

The wall shear stress can be calculated according to the Blasius equation and the correlation coefficient obtained according to the definition

$$f(x) = \frac{\overline{(du/dy)_0 (du/dy)_x}}{\overline{(du/dy)_0^2}} \qquad (70)$$

It is compared with the measurements of Kreplin and Eckelmann (Figure 21), Again agreement is good for the first part of the curve but the correlation function (crf) is overestimated at the tail end of the curve, presumably for the same reason as the fluctuating velocity was underestimated.

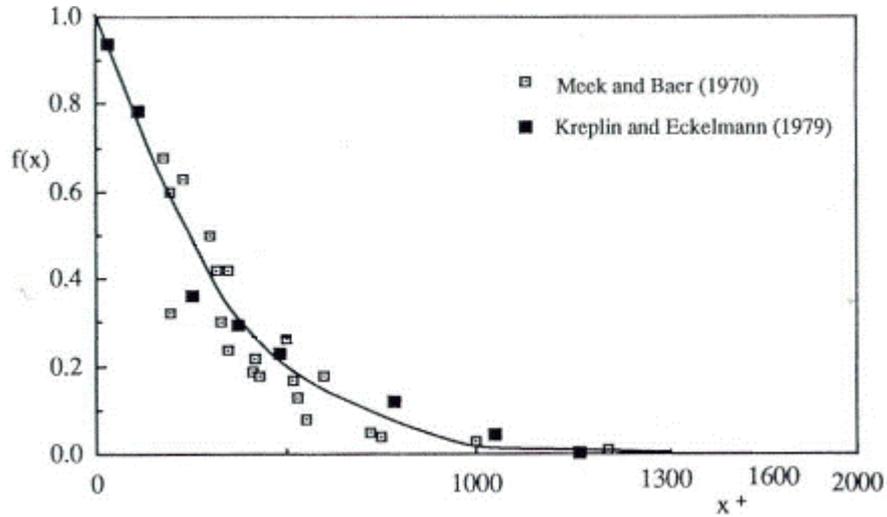

Figure 21 Correlation function for the wall shear stress. Line represents equation (67). Data of Meek and Baer (1970) and Kreplin and Eckelmann (1979).

### 7.5.2.5 The production of turbulence

The production of turbulent energy is defined as

$$P = \overline{u'v'}\frac{dU}{dy} \tag{71}$$

In pipe flow, the local shear stress is (Bird, Stewart and Lightfoot 1960, p. 162)

$$\tau = \tau_v + \overline{u'v'} = \tau_w\left(1-\frac{y}{R}\right) \tag{72}$$

The viscous shear stress is defined as

$$\tau_v = \mu\frac{dU}{dy} \tag{73}$$

Therefore we may write

$$\frac{\overline{u'v'}}{\tau} = \frac{\tau_t}{\tau} = 1 - \frac{dU^+/dy^+}{1 - y^+/R^+} \tag{74}$$

where $U^+ = U/u_*$. Combining equations (74) and (71) and rearranging gives

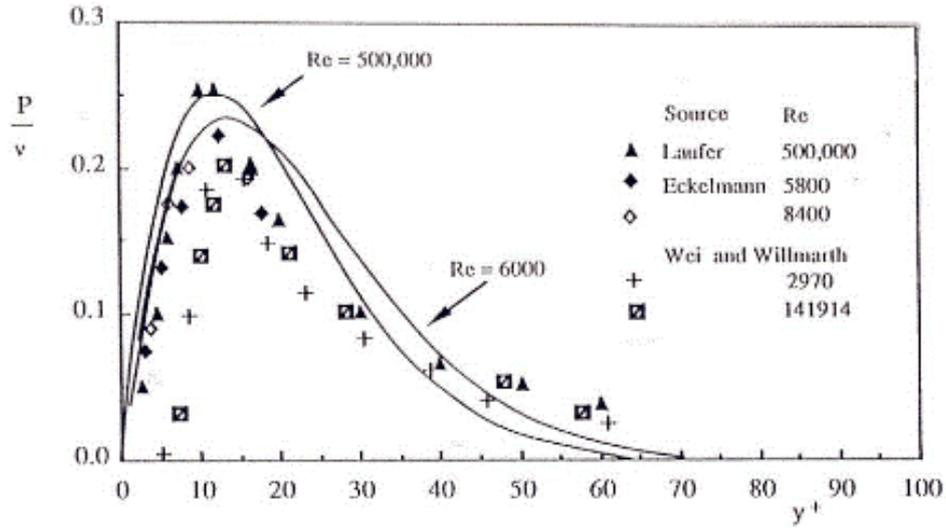

Figure 22 Production of turbulence predicted from the Stokes solution1.

$$\frac{P}{\nu} = \left(1 - \frac{y^+}{R^+}\right)\frac{dU^+}{dy^+} - \left(\frac{dU^+}{dy^+}\right)^2 \tag{75}$$

by using values of the time averaged velocity profile obtained from equation (63), we can estimate the profile of turbulence production as shown in Figure 22 against the data of Laufer (op. cit.) and Eckelmann (op. cit.). Note that normalising the production term P with $U_\nu^+$ and $\delta_\nu^+$ does not result in a single curve because of the presence of the factor $(1-y^+/R^+)$.

Since time and normal distance are related in the Stokes solution1, we can plot the production of turbulence predicted against normalized time with a crude transformation based on

$$\overline{t^+} \cong \overline{y^+} \tag{76}$$

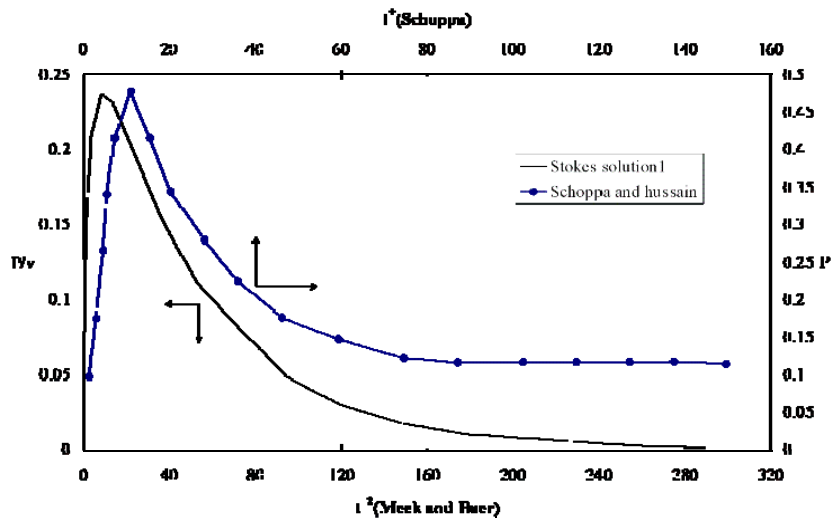

Figure 23 Production of turbulence against time. Data of Schoppa and Hussain (2002)

Note that the dimensionless time used by Schoppa and Hussain and Kreplin and Eckelmann is the square of the dimensionless time used by Meek and Baer. The evolution of the production with time is plotted against the data of Schoppa and Hussain
Figure 23. Schoppa and Hussain normalized their production data "by the kinetic energy E" not the kinematic viscosity $\nu$ thus a direct comparison of numerical values is not possible. In addition, the transformation between time and space in equation (76) is at best crude. Nonetheless, the trend in their data is well captured qualitatively by the prediction of turbulence production from the Stokes solution1. I am not quite sure that I understand fully how Schoppa and Hussain extracted P from their database but it appears to me that they did not sample conditionally in the way that Johansson et al. (1987) did (Figure 4), which would be the only way to capture the contribution of the streaming jet (or STG) and that their P data simply reflects the contribution of the solution of order $\varepsilon^0$.

I wish to record here my respect for the contributions of those pillars of turbulence research, Reynolds, Prandtl, Taylor, Kolmogorov, to name only a few, whose ideas have given us a foundation to work on. We stand on the shoulders of these giants to obtain a glimpse of the true physical nature of the phenomenon. As we follow the pathway they have cleared in the jungle that is turbulence, we have the further advantage of a huge data base of experimental evidence gathered painstakingly with often great ingenuity by researchers no less worthy of our praise. It is however incumbent upon us, the latter generation, to modify existing concepts, even those put forward by geniuses of their times, in the light of new experimental information. I will thus be devoting much of this paper to introduce a number of areas in turbulence research where I respectfully diverge from the views of my illustrious predecessors or where I believe that their approach can be expanded fruitfully.

By now the reader may have come to the same startling conclusion I made in the 1980's that the traditional statistics of turbulence can in fact be derived completely from the solution of order $\varepsilon^0$ that <u>looks like</u> the solution for an unsteady laminar viscous boundary layer! The only difference is that the solution of order $\varepsilon^0$ here deals with the smoothed phase velocity $\tilde{u}$ not the instantaneous velocity $u$. This observation has major implications and my first reaction was to double check the calculations. My second reaction was amazement at my blindness for having stared at this data for almost 14 years without seeing these simple patterns. This affliction is nicely described by Dan Brown in The Da Vinci Code (2003) as "Scotoma[4], the mind sees what it wishes to see". We are so convinced that these measurements, derived directly from complex turbulent velocity fluctuations, must be characteristic of turbulence that we refuse to see any alternative interpretation. In hindsight, the explanation is of course not difficult. All of the results presented above are based on long time statistical averages that not only do not capture fast quasi-periodic fluctuations but also smudge out the contributions of the transient fast streaming flow. As shown in Figure 18, the time averaged velocity profile can be very well approximated by the dominant value of the instantaneous velocity be it derived from direct experimental sampling (Eckelmann, 1974) or orthogonal decomposition of velocity traces (Bakewell and Lumley, 1967), neither of which assume viscous flow a priori. But as shown by the contribution calculated from the Stokes solution1, this dominant contribution cannot be interpreted as the imprint of a dominant *eddy* or disturbance.

---

[4] Scotoma is a medical condition defined as "a blind spot in an otherwise normal field of vision". Dan Brown used the concept to describe a comparable mental condition.

The most serious consequence of this observation was that none of the classic statistics of turbulence, that I, like many others, had been exposed to in my early training and readings, could be used to differentiate between laminar and turbulent flow regimes so neatly made by Reynolds (1883). I realized then why turbulence had been so difficult to understand: all the theories that had been developed had been inspired by and validated with the wrong data! This became my primary incentive in the next few years to look for another phenomenon that could distinguish these two flow regimes and, as shown in sections 3, I believe that I found it in the streaming flow and its interactions with the main boundary layer flow.

Schoppa and Hussain (2002) are the only authors who also identified that the streaming flow driven by the fast Reynolds stresses, that they call transient stress gradient TSG, is an important component of turbulence production. In particular, their detailed analysis of vorticity patterns from their DNS highlighted the important role of the velocity fluctuations in the Z direction in the formation of the TSG. However I disagree with their argument that the peak in
Figure 23 indicates the time when the streaming jet is produced. The analysis in this work indicates that this peak is found in the sweep phase and the streaming jet belongs in the next burst phase which comes later.

In addition, the emphasis of their work was to explain how the horseshoe vortices came to be so shaped and in fact many of the analyses in the last half centuries have focused on the (vortical) coherent structures in turbulent flow. I believe on the other hand that the primary action in the wall layer is found in the low speed streaks underneath the hairpin vortices because this is were the streaming flow has its roots.

The only way to illustrate the contribution of the fast Reynolds stresses is by conditional sampling as shown in Figure 4. Unfortunately experimental data describing the details of this phenomenon is still scarce.

## 7.6 The energy spectrum and the Kolmogorov scale

The most powerful data used in turbulence studies is probably the energy spectrum derived from the correlation function, crf. The correlation for the velocity can be obtained from measurements of the instantaneous velocity by, for example, hot anemometers. The auto correlation is given by:

$$f(\tau) = \frac{\overline{u(t)u(t+\tau)}}{\overline{u(t)^2}} \quad (77)$$

where $\tau$ is here a time delay. The two point spatial correlation is

$$f(r) = \frac{\overline{u(x,t)(u(x+r,t)}}{\overline{u(x,t)^2}} \quad (78)$$

It is used to estimate a macroscale $L_x$ (Bradshaw, 1971) which is traditionally interpreted as the size of the eddy passing the probe.

$$L_x = \int f(r)dr \quad (79)$$

and a (Taylor) microscale (Bradshaw, 1971; Lesieur, 2008), an estimate of the size of an eddy where viscous dissipation occurs

$$l_t = \frac{15\overline{u'^2}\nu}{\zeta} \quad (80)$$

where $\zeta$ is the energy per unit volume[5]. This scale can be determined by fitting a vertex parabola to the correlation function. It is traditionally argued that large eddies contain mainly kinetic energy, are unstable and breakdown to smaller and smaller eddies. If the difference between the large and small eddies is large (e.g. at high Reynolds numbers) a wide spectrum of intermediate eddies exist which contain kinetic energy and dissipate little but during the process of degeneration the anisotropic characteristic of the large eddies is lost. Kolmogorov (1941b) concluded that the properties of the smallest eddies are statistically independent of the primary eddies and are determined only by the rate of dissipation per unit mass. Thus in a small volume the components of the fluctuating velocity are equal. This situation is called local isotropy and does not require that the bulk stream itself be isotropic.

---

[5] I have used the symbol $\zeta$ for the energy per unit volume rather than $\varepsilon$ used in conventional texts on turbulence between the symbol $\varepsilon$ has already been used in sections 2 and 3 with respect to oscillating flow.

Bradshaw (1971) illustrated how vortex stretching at different scales leads to local isotropy with a "family tree" shown in Figure 24. Frisch (U. Frisch, 1995; U Frisch, Sulem, & Nelkin, 1978) has illustrated the concept of an energy cascade first postulated by Richardson (1922) by the breakdown of large unstable eddies with no loss of energy until the smallest eddies are reached (Figure 24).

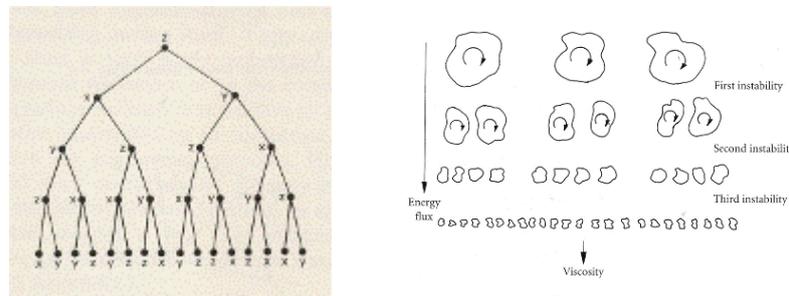

Figure 24 Left: Bradshaw (1971) representation of "family tree for local isotropy. Right: Frisch (1978, 1995) representation of Richardson's (1922) energy cascade.

The energy spectrum describes the flow of energy between different scales and can be obtained from a Fourier transform of the crf

$$\frac{E(k)}{\overline{u'^2}} = \frac{2}{\pi} \int_{-\infty}^{\infty} f(r) \exp(-ikr) dr \qquad (81)$$

Kolmogorov (1941a) argued that the smallest eddies where all the remaining turbulent energy is dissipated must scale with the (kinetic) energy per unit volume $\zeta$ and the viscosity and obtained by dimensional analysis

$$l_k = \frac{\mu^{3/4}}{\rho^{1/2}} \varsigma^{-1/4} \qquad (82)$$

Kolmogorov also argued that in the range of unstable eddies, called the inertial sub-range, the only relevant variable for the spectrum is the energy per unit volume and showed by dimensional analysis that

$$E(k) = C(K) k^{-5/3} \varsigma^{2/3} \qquad (83)$$

The slope of -5/3 has been widely observed e.g. (Lesieur, 2008).

We can obtain the energy spectrum by taking the Fourier transform of the crf shown in Figure 21 but because the curve predicted by the solution of order $\varepsilon^0$ overestimates the crf at large values of $x^+$ the spectrum obtained will not be accurate for a critical range of wavenumbers of interest. There is another intriguing way of obtaining it from time averaged parameters that allow us to discuss the physical implication of this spectrum.

The pdf predicted in section 7.5.2.1 shows that the velocity field passing a probe situated at a distance $y^+$ in composed of a distribution of velocities, each of which must contribute to the total fluid energy at that position. By energy, we mean mainly kinetic energy, which has the dimension of stress. Figure 18 also shows that the contribution at each position $y^+$ reflects the contribution of a dominant eigenmode. We suspect the reverse to be true: The time-averaged turbulent shear stress at each position $y^+$ may be used as a good estimate of the contribution of the dominant wave length. It is thus edifying to plot out the variation of turbulent stress with distance. In this exercise, we start from the equation

$$\tau = \tau_v + \tau_t \tag{84}$$

The shear stress in a pipe at y is given by

$$\tau = \tau_w (1 - \frac{y}{R}) = \tau_w (1 - \frac{y^+}{R^+}) \tag{85}$$

The laminar contribution can be calculated from Newton's law of viscosity

$$\tau_v = \mu \frac{dU}{dy} \tag{86}$$

Combining equations (84), (85) and (86) gives

$$\frac{\tau_t}{\tau} = 1 - \frac{dU^+/dy^+}{1 - y^+/R^+} \tag{87}$$

It is useful for comparison purposes to define here a dimensionless wavenumber based on the pipe diameter D and the distance $x^+$ travelled by the dominant Eigen mode and calculated from Taylor's hypothesis and the time scale of the wall layer measured by Meek and Baer (1970)

$$x^+ = U^+ t_v^{+^2} \qquad (88)$$

$$k_x D = \frac{2\pi D^+}{x^+} \qquad (89)$$

$U^+$ is the time averaged velocity at position $y^+$, which of course is also equal to the velocity of the dominant Eigen mode. Figure 25 shows a plot of the turbulent shear stress spectrum in a pipe measured by Lawn (1971) at $Re = 9.10^4$. Included are points of $(\tau_t/\tau)/(2\pi)$ against $kR$ calculated with equations (87) and (89) respectively using velocity data of Laufer (op.cit.) and Eckelmann (op.cit.) near the wall. Estimates from the Stokes solution1 are also included.

The changes in $E(kR)$ and $(\tau_t/\tau)/(2\pi)$ with $kR$ show very similar trends. In particular for wavelengths smaller than $kR \approx 3.5$ both variables tend to level out to a constant value. The drop in of $(\tau_t/\tau)/(2\pi)$ for wavelengths in the range $10 \leq kR \leq 3.5$ is strikingly similar to the drop in $E(kR)$ despite much scatter in the data for $(\tau_t/\tau)/(2\pi)$. This scatter arises from the difficulty of making measurements of velocity very close to the wall. In principle, we can by-pass that difficulty by taking direct measurements of $\overline{u'v'}$ such as those of Eckelmann (1979). These data confirm that equation (87) gives good predictions of $\rho\overline{u'v'}/\tau$ for the range $5 < y^+ < 100$ but since the size of the probe itself is equal to $d_p^+ = 2$, measurements below $y^+ = 5$ show a bias. Calculations from the velocity profile involve numerical differentiation, an inaccurate exercise in itself, but further involve very small differences between two very similar numbers. For example, in order to obtain a value of $E(kR) \approx 10^{-3}$ we must get a slope $dU^+/dy^+$ accurate to four decimal points. Given these difficulties, it is all the more remarkable that the two sets of variables agree so well and that the plot of $(\tau_t/\tau)/(2\pi)$ is able to reproduce so clearly the famous $-5/3$ slope predicted by Kolmogorov.

Yet more physical insight can be obtained by looking at the dissipation scales implied.

If we neglect $\tau_t$ in equation (84) we assume that all the energy is dissipative and a rearrangement of equation (85) after integration gives

$$U^+ = y^+ \qquad (90)$$

If on the other hand we neglect the term $\tau_v$ in equation (81) and follow Prandtl's derivation we get the log law in equation (46). Thus one might expect that the viscous and kinetic contributions to the local shear stress will be equal at the intersection of equations (46) and (90) which occurs for pipe Reynolds numbers between 6,000 and $10^6$ at an average value of $y_k^+ = 11.8$ according to experimental data (e.g. Nikuradse 1931).

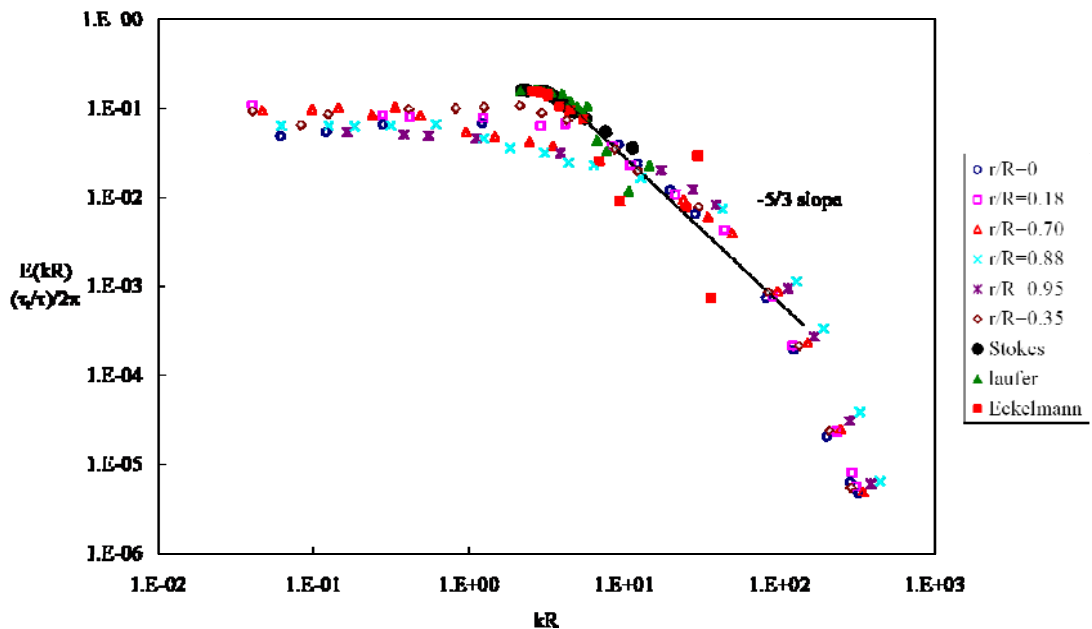

Figure 25 Shear stress and turbulent stress spectra. Spectral data of Lawn (1971), Re=921000, calculated shear stress distributions based on velocity data of Laufer (1954), Eckelmann (1974) and the Stokes solution1.

The reader may verify that indeed at that position $y_k^+ = 11.8$ we obtain the equality $\tau_v = \tau_t$, as shown for example in Figure 34. That equilibrium between kinetic and viscous energies is characteristic of the equilibrium range of wavenumbers and also represents the defining condition of the Kolmogorov scale (equation (82).

It is very interesting to note that the gradient of turbulent stress spectrum in Figure 25 is -5/3 for wavenumbers in the range $4 < kR < 50$ including the point $kR \approx 5$ equivalent to $y_k^+ = 11.8$. We note here that equation (82) implies that the Kolmogorov scale is found at the scale where the turbulent and viscous energies are equal and it is defined by the point where E (or) has dropped to half of its value as low kR which is equal to $kR \approx 5$. The slope of -5/3 applies, in Kolmogorov's argument to the inertial sub-range where the energy spectrum is only dependent on $\zeta$ whereas it is found in the turbulent stress spectrum to straddle the interface of this inertial subrange and the beginning of the dissipation sub range.

Kolmogorov was impressive for the elegance of his reasoning and the simplicity of his dissipative eddy scale that has been perhaps the most used concept in the analysis of transport phenomena as stated in the Royal Society (London) 50 years memorial issue (1991). Yet the evidence presented here led me to two blasphemic disagreements with Kolmogorov's views:

1. The Kolmogorov scale is not necessarily an eddy scale; it just captures well the position where the kinetic and viscous energies in the flow field are in balance.
2. The Kolmogorov scale applies in regions that are strongly anisotropic and does not require the assumption of local isotropy.

Permit me to detail further these two arguments. The advantage of the concept of local isotropy is to allow Kolmogorov to argue that there exists a universal form of the spectrum that is independent of the strongly anisotropic characteristics of the main flow from which is it generated. But Kolmogorov stated (1941b) that this hypothesis of local isotropy would apply to "sufficiently small domains G of the four dimensional space ($x_1, x_2, x_3, t$) *not lying near the boundary of the flow or its other singularities*". This statement of Kolmogorov is at odds with the spectral representations of the wall shear stress that show a wave number space where there does exist a range of dissipative scales e.g. Lawn (1991) shown in Figure 25. It implies that viscous dissipation at the wall cannot be explained in terms of a scale represented by equation (82) and is better represented by equation (90) which led Prandtl (1935) to postulate the existence of (steady state) "laminar sub-layer". However, we know from the classic work of Kline et al. (1967), which has been validated by many others e.g. (Corino & Brodkey, 1969; Offen & Kline, 1974) that the wall layer is transient and highly active and cannot described by the concept of a *steady state* laminar sub-layer where turbulent shear stresses are absent. This paradox is resolved when we note that the error function (erf) obtained in the Stokes solution1 degenerates to a linear relationship between the stress and shear rate for very small values of $y^+$ at all values of *t* and that the time averaged relationship does coincide with equation (90) . Popovich and Hummel (1967) showed that this linear relationship exists and Figure 36 shows that the erf does reduce to equation (90) very near the wall.

The mathematical formulation of the Kolmogorov scale equation (82) is based on his first similarity hypothesis that "the distributions $F_n$ are uniquely determined by the quantities $\nu$ and $\zeta$ " and obtained by dimensional analysis. Kolmogorov himself never argued that the scale bearing his name represented an eddy, even when he used equation (82) in his analysis of energy dissipation in locally isotropic turbulence (Kolmogorov, 1941a) although he described it as "the scale of the finest pulsations". The interpretation that the Kolmogorov scales represent the "smallest, dissipative, turbulent eddies" appears in subsequent papers that seek to apply Kolmogorov' ideas to predictive theories of turbulent flow e.g. Wilson & Thomas (1985). I have never found published experimental data able to identify Kolmogorov *eddies*, even from reported DNS analyses, in the wall layer of turbulent flow. Yet most statistical measurements point to this region as having the highest level of energy dissipation e.g. (Lawn, 1971).

The requirement for a universal equilibrium range is based on the fact that only the local viscosity and energy content are relevant. This hypothesis allowed Kolmogorov to ignore two cumbersome variables L and U that are characteristic of the geometry and Reynolds number of the main flow. If they are included, all solutions become specific to the geometry and flow conditions of the problem under consideration and loose any claim to general applicability. This was the great quest in Kolmogorov's time. It started with Taylor's invention of statistical turbulence (G.I. Taylor, 1921, 1935). The first study of turbulence mechanism were restricted to homogeneous and isotropic flows (Batchelor, 1960) and first experimental works focused on turbulence behind grids e.g. Comte-Bellot & Corrsin (1966) which unfortunately were only dissipative and could not reveal how turbulence preserved itself. The argument I have set forth in this paper is that the solutions the NS equations of different orders of approximation, really solutions of subsets of the NS equations based on omission of different terms, are only constrained by their domain of application, and the solution of order $\varepsilon^0$ can be found embedded in fully turbulent flows as shown in Figure 36, without the requirement of isotropy. In so far as it is only dependent on the magnitude of $\varepsilon$, this form of solution is independent of the larger scale of the flow field or its geometry.

The other point that I am making here in comparing the stress distribution obtained from the Stokes solution1 and the measured stress spectrum is that the wavenumber does not automatically translate to an eddy scale but can be interpreted as the distance travelled by a velocity front. The Stokes solution1 shows how a uniform velocity front degenerates through the penetration of viscous retardation from the wall and this can be well described by a correlation function since we are analyzing velocity behaviour within the same coherent body of fluid. Another point is worth making here.

The experimental characterisation of steady turbulent flow at a macroscopic level, say in a pipe, is relatively easily done using the similarity principles of Reynolds and others but the detailed statistical characterisation has to be based on local instantaneous velocity fields. In the classic statistical approach one follows the cascade of energy through a hierarchy of scales and the local picture is a solution of order n where n is the number of scales in the hierarchy. Thus the solution of order n is the most complex and has only been possible with many assumptions, the most potent being local isotropy. I started to realize at some point in the development of my views that I was adopting strategically and philosophically an approach in the reverse direction. The solution of order $\varepsilon^0$ is the simplest and most elementary and applies at the local level. As we include higher order terms, the solutions of higher order become more complex and include more structures and interactions until at some level n one captures most of the elements that populate the overall flow field. This approach delivers us from the tyranny of isotropy whether global or local, because the solution of order $\varepsilon^0$, which has universal application as shown in the zonal similarity analysis (Figure 36, section 8), applies in shear layers that exist at solid and fluid interfaces. By their nature, these shear layers cannot be isotropic.

One disturbing question remained in my mind for a long time about this interpretation of the energy spectrum: it has been measured from experimental correlations of instantaneous velocity outside the wall layer where the turbulent stresses predominate completely. Lawn (1971) for example, has shown the spectra measured at many Reynolds number and many positions in the radial direction, some very close to the wall, some very close to the pipe axis, pretty much collapse into a single 'universal" curve. The spectrum of turbulent stress calculated in Figure 25 is fully based on the variations in the wall layer only. Why should we believe that it can match the spectra near the pipe axis?

To answer that question, we observe that the velocity decomposition in equation (12) is not confined to the wall layer alone. Consider a velocity probe in the outer region, it will react to the passage of a number of coherent structures such as, for example a rotating vortex. The component $U_i$ is closest to the velocity with which the structure moves with the main stream. The component $\widetilde{u}_i$ is the difference between the time averaged velocity and the smoothed phase velocity of the circulation flow inside the vortex. The component $u'_i$ arises from periodic fluctuations impressed on the circulation flow by vortical structures passing nearby. Equation (13) still applies but equation (20) now needs to be solved with different boundary conditions: the velocity at the boundary of the vortex $(y = 0)$ is not zero. Instead the boundary condition stipulates that the velocity gradients in the two streams on either side of the fluid interface between the main flow and the vortex must be equal. While the form of the solutions of order $\varepsilon^0$ and $\varepsilon$ will change with boundary conditions, the existence of a streaming flow is not negated. Thus the probe can pick up velocity fluctuations due to an originally transient laminar circulating flow. The breakdown of the circulation flow and therefore of the large eddy follows essentially the same interaction between the cross and streaming flows observed in the wall layer. The breakdown stops when the eddy size is so small that the circulating flow is unable to produce a streaming flow. The value of this scale may be defined by the path length of the circulating flow (and therefore its local Reynolds number) and related to the curve of neutral stability for growth of disturbances e.g. {Schubauer, 1943). In addition, it will be increasingly more difficult for the main stream to induce circulation as the vortex size decreases.

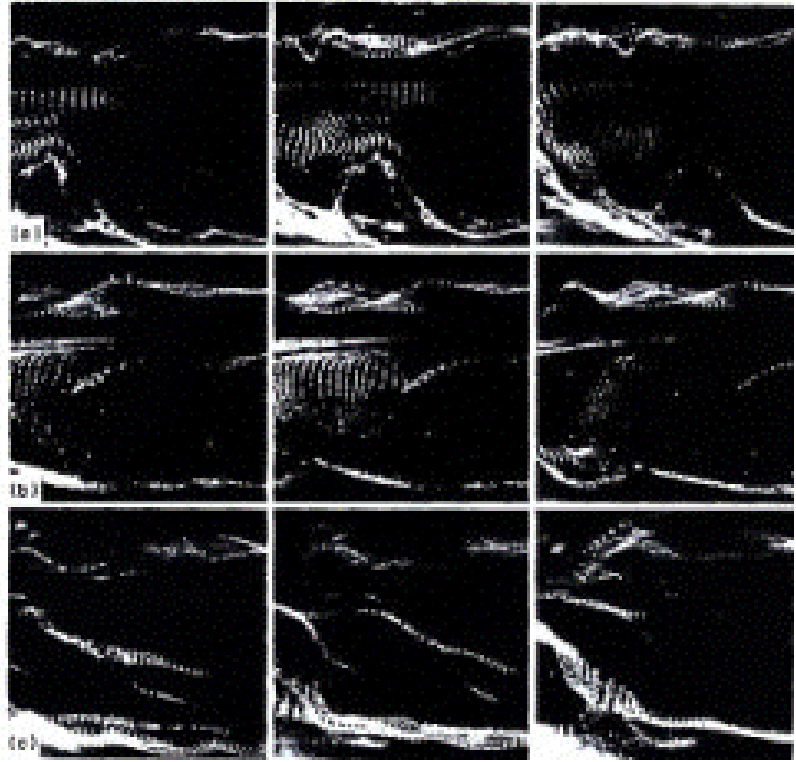

Figure 26 Sweep-burst cycle at fluid interfaces from (Turney & Banerjee, 2008) Sequential pictures of bursts near wall and near sheared gas-liquid interfaces. (a) near wall: shear free interface; (b) near interface: countercurrently sheared interface; (c) near interface: cocurrently sheared interface. The lower part of each picture is a side view and the upper part is a plan view of the same structure,

A similar situation is found with surface circulation of gas bubbles dispersed in a fluid. At small enough sizes, circulation in the bubble interface stops and the bubble can be treated as a solid sphere (Calderbank, 1967). In that sense, I agree that there *may* exist eddies that are purely dissipative and not productive of turbulence if this is what Kolmogorov implied when he said "$\lambda$ (or $l_k$) is the scale of the finest pulsations". But the scale $\lambda$ cannot be regarded as the wavelength of the fast velocity fluctuations $u'_i$; it should be regarded as the scale of the circulating flow described by

$\tilde{u}_i$.

So where are we most likely to observe these energy dissipating vortices? In my view in regions where there is no regeneration of turbulence particularly behind the streaming jets especially in the log law.

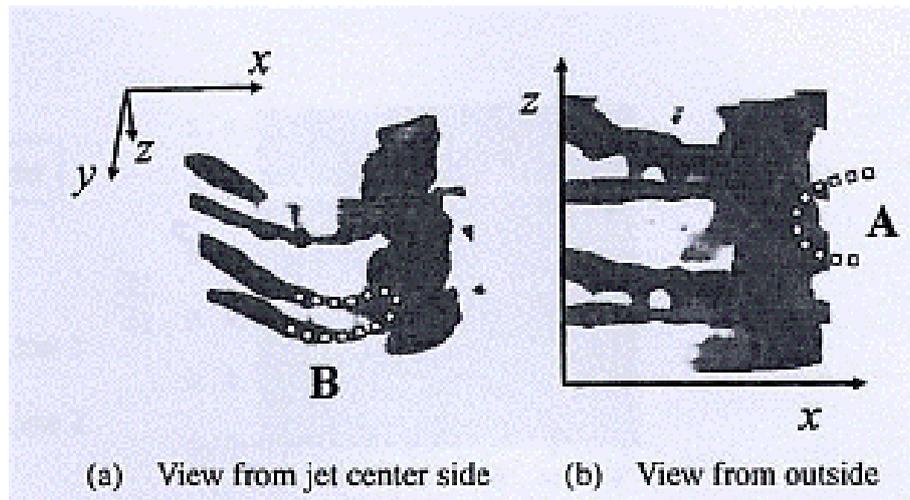

Figure 27 Hairpin vortex embedded in an axisymetry jet from Toyoda & Hiramoto, (2006)

There are tantalising descriptions such as Keffer and Baines (1963) of "small scale turbulence" behind jets in cross flow or Johansson et al (1991) "spatially spotty turbulence". But I have not had the opportunity to find numerical data to analyse properly. The two components decomposition of the instantaneous velocity is most likely two apply in these regions of small scale turbulence.

Is there evidence that a streaming process can occur without the presence of a solid wall? For many years I searched unsuccessfully for that evidence but the availability of web data bases has dramatically brought much work to my attention recently. Banerjee and his colleagues (Rashidi M, Hetsroni G, & S., 1991; Turney & Banerjee, 2008) have shown conclusively that sweep and burst sequences do occur at fluid interfaces away from the wall as shown in Figure 26. In fact we do not even need an interface between two different fluids: Toyoda and Hiramoto (2006) have created hairpin vortices within axisymetric jets (Figure 27). Clearly streaming jets exist away from solid walls as long as shear layers are produced, even at liquid-liquid interfaces.

Since it is extremely difficult to document the breakdown a large eddy because of its transient nature, there is no experimental documentation that is available to my knowledge. The most potent "proof" presented in the classical literature has been the energy cascade deduced by an interpretation of the energy spectrum as a distribution between eddies of different scales. One situation has been studied extensively: Taylor vortices because they are relatively steady in space and therefore easy to follow. Following the investigation of Coles (1965) the flow between rotating cylinders has been used to investigate the transition from laminar to turbulent fluid motion and over 2000 experimental, numerical and theoretical studies have been performed for this geometry (Caton, Janiaud, & Hopfinger, 2000)

Taylor (1923) was the first to show that small disturbances could destabilise the Couette flow of a viscous fluid between two rotating cylinders and lead to the appearance secondary flows called Taylor vortices that are originally laminar. In this regime called laminar Taylor Vortex flow (R4 illustration in Figure 28), which starts with a critical Taylor number of

$$Ta = \frac{U_i d}{\nu}\sqrt{\frac{d}{R_i}} \approx 41.3 \qquad (91)$$

where $d$ is the gap between the cylinders, $R_i$ the inner cylinder radius and $U_i$ the peripheral velocity of the inner cylinder; the inflow and outflow boundaries between laminar vortices are flat and perpendicular to the cylinder axis. As the Reynolds number increases a second critical Reynolds number is reached, $R = 5.5$ in the work of Wang, Olsen, & Vigil (2005), where TVF gives way to wavy vortex flow (WVF) which is characterized by large amplitude travelling azimuthal waves superimposed on the boundaries of the Taylor vortices (illustration at R=16 in Figure 28). Caton et al. (1999) showed that the primary instability that leads to the formation of the wavy pattern is a direct bifurcation. The second bifurcation leads to a pattern of drifting non-axisymmetric vortices is a subcritical Hopf bifurcation.

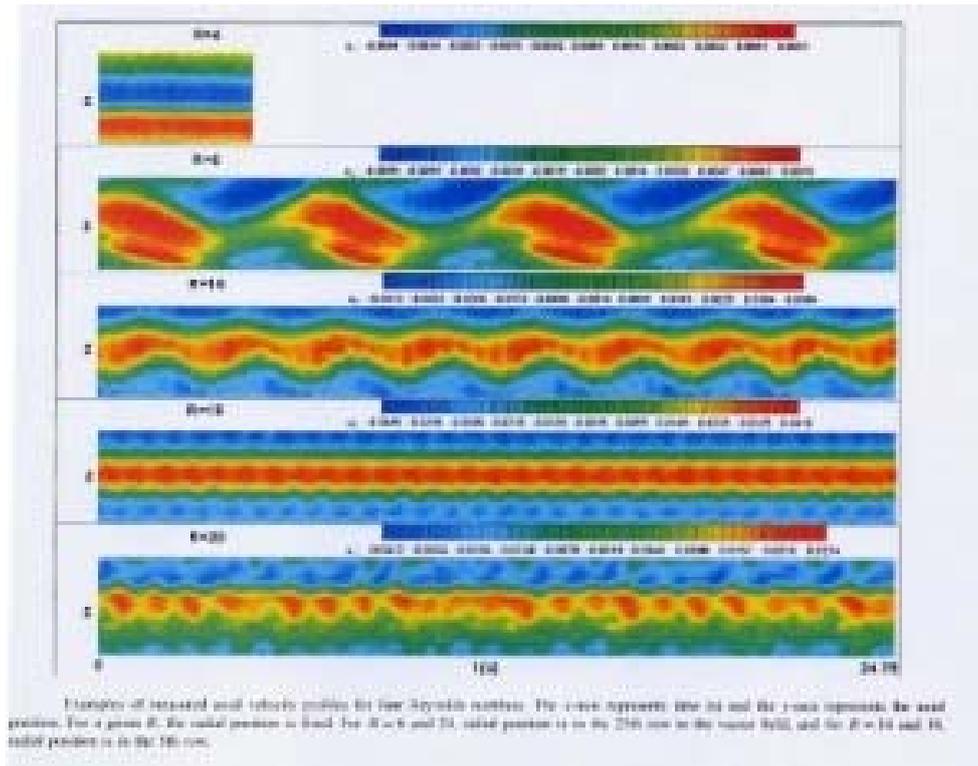

Figure 28 Evolution of regimes in Taylor vortex flow from Wang et al. (2005) R is ratio of current Reynolds number to critical Reynolds number where Taylor vortices start to appear.

As Re increases further, the amplitude of the azimuthal waves vary with time giving rise to-quasi periodic regimes known as modulated wavy vortex flow (MWVF). Some authors (Gorman & Swinney, 1979; Zhang & Swinney, 1985) distinguish two modes of MWV flow, GS and ZS. A subtle transition further occurs around $R \approx 12$ where a broad peak appears in the velocity power spectra with a rise in background noise level. Brandstater and Sweeney (1987) argued that the flow dynamics seem to fit the description of deterministic chaos, even though the transition itself seems not understood mathematically and this regime is called chaotic vortex flow (CVF).
Not all authors report detection of this regime, e.g. (Wang, et al., 2005; Wereley & Lueptow, 1998). A further increase in Reynolds number leads to a disappearance of the azimuthal waves and fully turbulent flow TTVF), even though the Taylor vortex structure remains. Wang et al. (op.cit.) put the critical Reynolds number for transition to turbulence at $R = 18$ then report a reappearance of periodic velocity fluctuations at $R = 20$ (Figure 28).

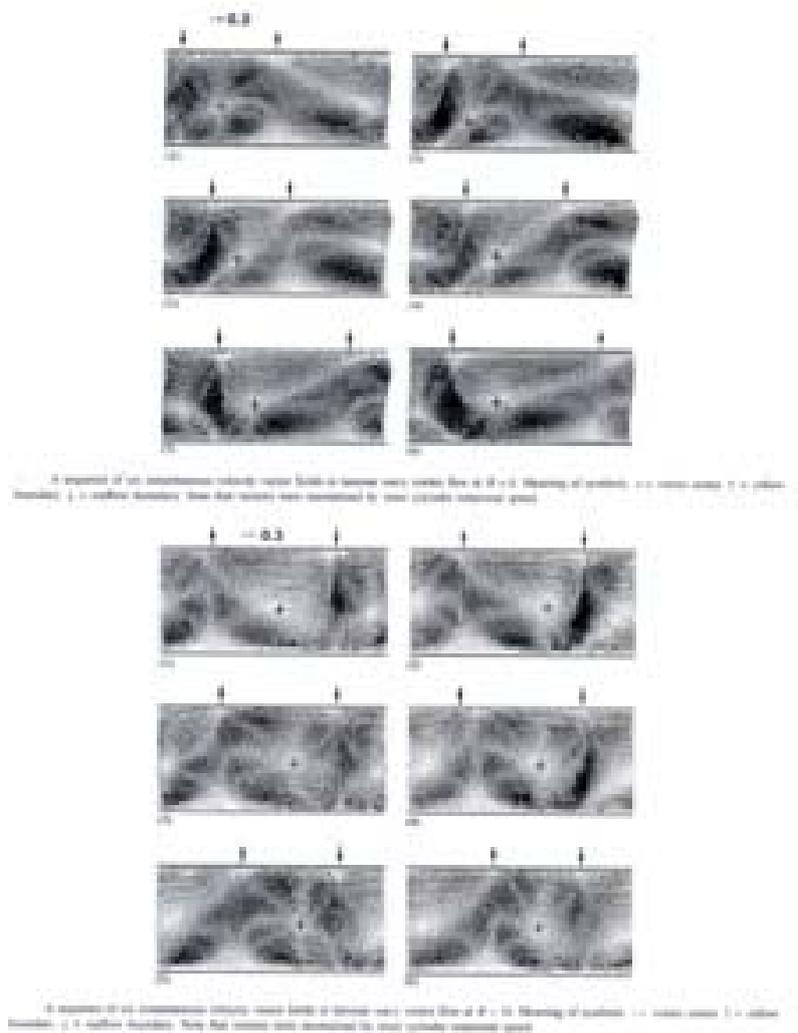

Figure 29 Oscillation and non-oscillation of Taylor vortices. From Wang et al. (2005)

In wavy flow the boundaries of the Taylor vortex oscillate whereas in stationary vortex flow they remain constant as shown by the velocity patterns presented by Wang et al. for $R = 6$ and $R = 18$ in Figure 29.

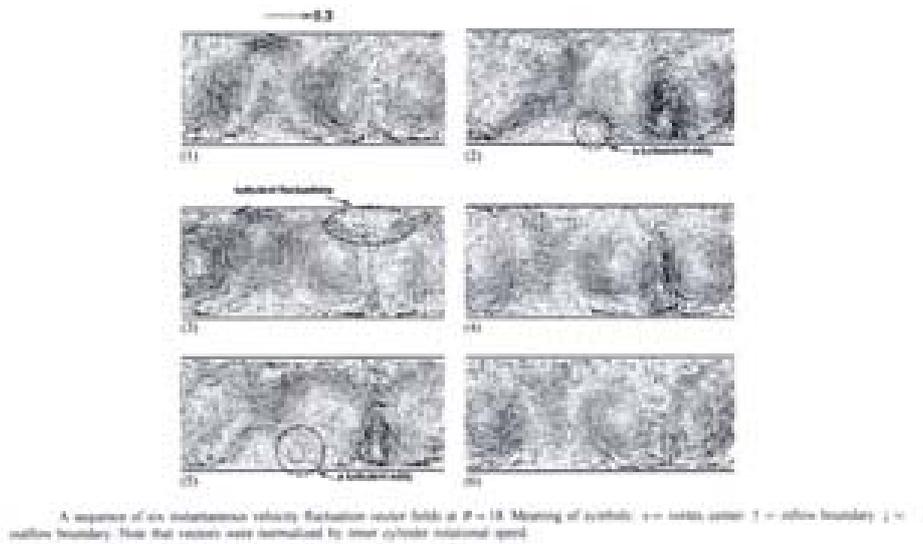

Figure 30 Turbulent eddy in turbulent Taylor vortex flow. From Wang et al. (2005)

Wang et al. reported that the correlation length of the azimuthal velocity decrease sharply at $R = 18$ and simultaneously no contribution to the instantaneous velocity can be attributed to the travelling azimuthal waves that have disappeared and all velocity fluctuations are dominated by turbulent "eddies". They also report at $R = 18$ the presence of turbulent vortices (Figure 30) but interestingly these are not found within the Taylor vortices, which do not break down, but in the spaces between adjacent Taylor vortices and between those Taylor vortices and the wall. At $R = 20$ the azimuthal waves reappear and the correlation length increases again until it reaches a small peak at $R = 30$ after which the waves once again decay. Wang et al. correctly attribute, in my view, the variations in the correlation length to the scale of the azimuthal waves and do not interpret this correlation as an eddy scale, even in the TTVF regime.

Wereley and Lueptow (Wereley & Lueptow, 1998) observed that as the Reynolds number of the inner cylinder increases, the vortices become stronger and the outflow between pairs of vortices becomes increasingly jet-like. The patterns of fluctuations of the radial velocity

$$\frac{\langle u' \rangle}{U} = \frac{\sqrt{\langle u'^2 \rangle}}{R_i \omega} \tag{92}$$

at the boundaries of a Taylor vortex where the fluid flows from the inner to the outer cylinder (outflow) and from the outer to the inner cylinder (inflow) show some interesting patterns (Figure 31).

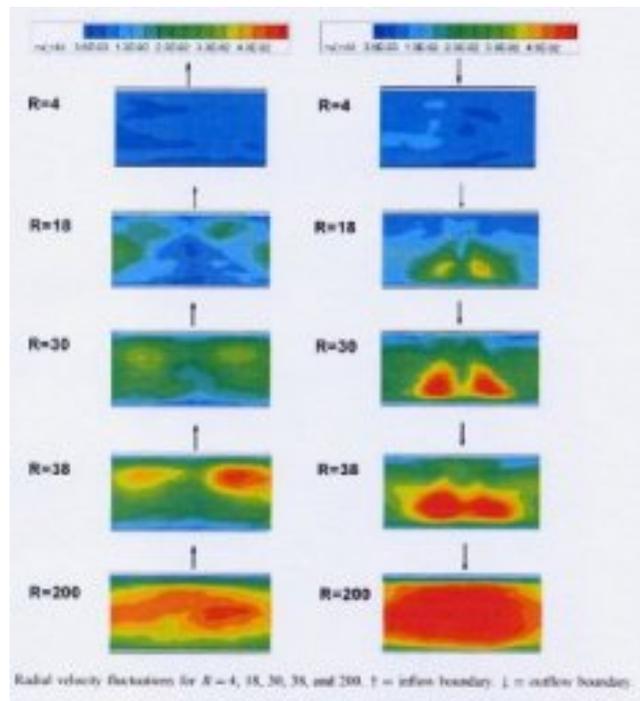

Figure 31 Development of zones of radial velocity fluctuations with Reynolds number in TVF. From Wang et al. (2005). The top boundary represents the inner cylinder wall, the bottom represents the outer cylinder wall.

For $R = 4$ (LTVF) they are close to zero in the entire field. At $R = 18$ when transition to TTVF begins non-zero values are observed but the fluctuations show a strong spatial dependence being strongest near the outer cylinder for the outflow and the inner cylinder for the inflow boundary. Only when R is large (i.e. $R = 200$) do the fluctuations become homogeneously distributed. Thus the fluctuations develop in the boundary flow around the Taylor vortex and eventually propagate into the TV but there is no evidence that the vortex itself breaks down into an increasing number of smaller and smaller vortices with increasing Reynolds numbers as one might expect from the schematic of Frisch in Figure 24. Wang et al. identify two different contributions to the velocity fluctuations: those due to the large azimuthal wave motion and those due to a shorter correlation length that they attribute to turbulence.

By combining information about correlation length and velocity fluctuations, they identify that zones at the start of radial flow, for example the inner cylinder wall in outflow, are dominated by the contribution of the azimuthal waves, and the reverse is true at the end of the radial flow. This sort of analysis further highlights the argument made in sections 7.5 that long time velocity fluctuations may be created by unsteady laminar flow completed unrelated to turbulence. In fact even the long term averages of velocity fluctuations based on the short correlation lengths that Wang et al. attribute to turbulence do not give any indication about the streaming flow or bursting process well accepted as characteristic of turbulence. Thus one may not quite agree with the statement of Atkhen, Fontaine, & Wesfreid (2000) that "Couette-Taylor instabilities are prototypes for general studies in hydrodynamic stability and transition to turbulence".

Takeda (1999) was able to extract more information in his study of transition between the MWV-CWV and TTV flow regimes. One dimensional Fourier analysis of a data set acquired with an ultrasonic velocity profile monitor yielded space-dependent power spectra and time-dependent energy spectra. The spectra at $R = 13.5$ showed the normal general decrease in value with increasing frequency or wavenumber but there are multiple peaks superimposed on the base line. He identified each peak in the power spectrum either as a fundamental, a harmonic or a linear combination of two or more fundamental frequencies and found three intrinsic wave modes for the MWV regime: the WVF, GS and ZS modes. The WVF mode disappears at $R = 21$ in his work. Takeda used a two-dimensional Fourier transform to study the nature of the quasi-periodic state, which follows the MWV regime, quantitatively since various spatial modes contribute to the same peak in the power spectrum. Beyond the flow regime of the quasi-periodic state, the flow structure becomes less periodic and Takeda switches to orthogonal decomposition of a time series velocity profile stored in a $M \times N = 1024(time) \times 128(space)$ matrix since Fourier analysis is no longer as effective. The eigenvectors obtained are elemental spatial patterns of the fluctuation velocity field and the eigenvalue yields information about the contribution of the corresponding elemental patterns to the velocity field. The eigenmode of index 0 is the time averaged velocity profile (corresponding to the basic TVF mode). The eigenvalue spectrum shows a very sharp decrease for the initial modes, implying that only a few modes are necessary to describe the velocity flow field.

In the two dimensional Fourier analysis Takeda distinguished three basic coherent modes. The first mode prevails at the lowest values $R < 10$ studied and has fairly constant intensity up to $R = 21$. It is identified as the WVF mode since it is the only mode appearing at $R < 10$. The second mode increases logarithmically between $10 < R < 17$ after which it remains constant and corresponds to one of two kinds of MWV modes, GS or ZS. Both disappear suddenly at $R \approx 21-22$ corresponding to the disappearance of the azimuthal wave motion. But a third mode, apparently first identified by Takeda, appears at $R = 23$ and disappears at $R = 36$. The frequency of this mode is much higher than those of the WVF and MWV modes and Takeda identified this as a fast azimuthal wave. The amplitude of that mode is localized near the inflow and outflow boundaries. Is this mode, which has not been studied extensively, be related to a streaming flow? Much more work is needed.

Thus the transition to turbulence in Taylor vortex flow is much more complex than in shear flow past a plane surface because of the superimposition of different wave modes. The evidence shows however that even when the TTVF regime can be clearly identified, e.g. by measurements showing a torque coefficient-Taylor number relationship (Taylor 1923) of

$$C_M \sim T_a^{-0.2} \tag{93}$$

the Taylor vortex remains intact. This indicates to me that we need to revise the argument that the energy spectrum shows that the large "eddies" or coherent structures are intrinsically unstable and break down to smaller and smaller eddies. I have strong misgivings about the interpretation of the energy spectrum as an energy cascade which implies that large eddies break up to many smaller eddies because of some yet undefined internal forces. Figure 25 suggests that the energy spectrum can be interpreted as a distribution of the dominant contribution to shear stress in real space, not necessarily between eddies, although I am quite open to new experimental evidence that can show that this is actually happening. It seems to me that large eddies can only be broken by external forces such as collision with another transient coherent structure. In this case the coherent structure called the original Taylor vortex does not break down to several smaller vortices but the internal flow within its boundaries shows an internal breakdown to small scale turbulence as can be seen by even old photographs e.g. by Schultz-Grunow and Hein (Schlichting p.527).

In a large number of practical applications, the most relevant variables sought are the steady state time averaged velocity distributions and friction losses. In these situations the loss of detail incurred during the averaging process is not necessarily a disadvantage because small approximations or errors do not have a major impact on the results. For example, both Prandtl (1935) and Karman (1934) have been able to produce good correlations between the turbulent pipe friction factor and the Reynolds from a logarithmic representation of the velocity profile thus neglecting completely that in the outer region (near the pipe axis) the actual velocity distribution obeys Coles' law of the wake.

Operations and processes involving turbulence can be treated in four ways:
- A black box or empirical approach that does not require any understanding of turbulence mechanisms and is not based on solutions of the NS equations even in linearised form. This empirical approach typically only requires dimensional analysis to identify the major dimensionless the dimensionless groups of variables involved and relies on experimental data to quantify the parameters involved in the correlations for example (Blasius, 1913)

$$f = \frac{0.079}{\mathrm{Re}^{1/4}} \tag{94}$$

- At the other extreme, we find a transparent box or fully mathematical approach where every detail of the turbulent flow field is known through a formal solution of the NS equation. This would be like knowing where each of the billion Chinese is and what he or she is doing at any time. At the moment this is not within reach.

- A white box or theoretical approach based on formal solutions of subsets of the NS equations, giving a theoretical description (with suitable approximations) of the key parameters relevant to the particular problem considered without attempting a complete description of all turbulent parameters. We are hopefully progressing towards the white box.

- Most present work take a gray box approach where a general form of correlation is derived, usually from the Reynolds equations RANS, complemented by assumptions about missing information such as the form of the Reynolds stresses. Here we have different shades of gray: light when less experimental data is used to supplement the theoretical understanding (semi-theoretical approach) and darker when the theoretical solution is in very broad general form and much more experimental data is used to fill in the gaps (semi-empirical approach). An example is Prandtl's logarithmic correlation

$$\frac{1}{\sqrt{f}} = 4.0 \log \operatorname{Re} \sqrt{f} - 0.4 \qquad (95)$$

  where the coefficients 4.0 and 0.4 were obtained experimentally.

In the following we take advantage of the fact that the solution of order $\varepsilon_0$, an analytical solution of a clearly defined subset of the NS equations, is able to capture the turbulence characteristics of the wall layer, because the contribution of the streaming flow in the wall layer can be neglected in this domain of application and also make use of the well known log-law to analyse common problems of interest. In this sense our approach is between a white box and a gray box – a lighter shade of gray but hopefully the significant reduction in experimental parameters will inspire greater confidence in the solution obtained.

## 8  A master curve for the time averaged velocity profile

Since the early days of turbulence research, many have been looking for a master curve for the velocity profile that would apply to all situations. Prandtl and his student Nikuradse thought that such a curve could be obtained by normalising the velocity and normal distance with the wall parameters $u_*$ and $v$ and called it the universal velocity profile (Figure 32).

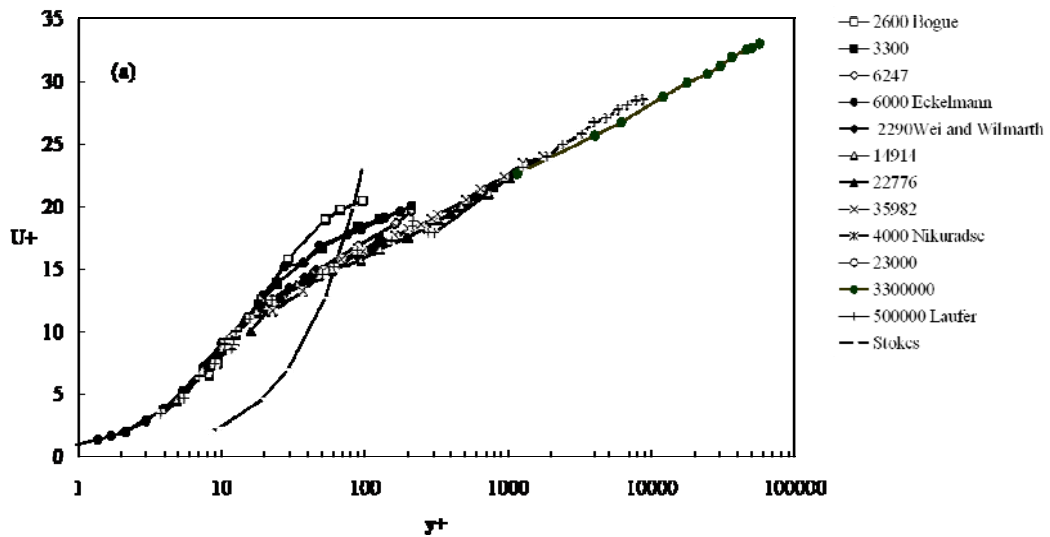

Figure 32 Prandtl-Nikuradse universal velocity profile for turbulent pipe flow. Data from Bogue (D.C. Bogue, 1961; Eckelmann, 1974; Laufer, 1954; Nikuradse, 1932; Wei & Willmarth, 1989),

Their (and others') experimental data show however that the profile normalised with the wall parameters still had a (small) Reynolds number dependence.

The analysis in sections 2, 3 and 4 indicate that the wall layer, the log-law region and the far field region are based on completely different subsets of the NS equations reflecting different flow mechanism. It appears unreasonable then to expect that all three regions would scale with the wall parameters, especially since the log law and the far field regions have no direct contact with the wall. Better parameters can be found *at the interfaces between these regions since they would be common to both adjacent regions*. Thus for the inner region, the thickness of the wall region $\delta_v$ and the velocity at its boundary $U_v$ looked much more attractive candidates. These parameters can be determined from measured velocity profile in several ways.

1. The first method is to fit a straight line to log-normal plot of the normalised velocity profile and take the lower limit of applicability of the log-law (Figure 33) as the defining point for $U_v$ and $\delta_v$. This method is not highly accurate.

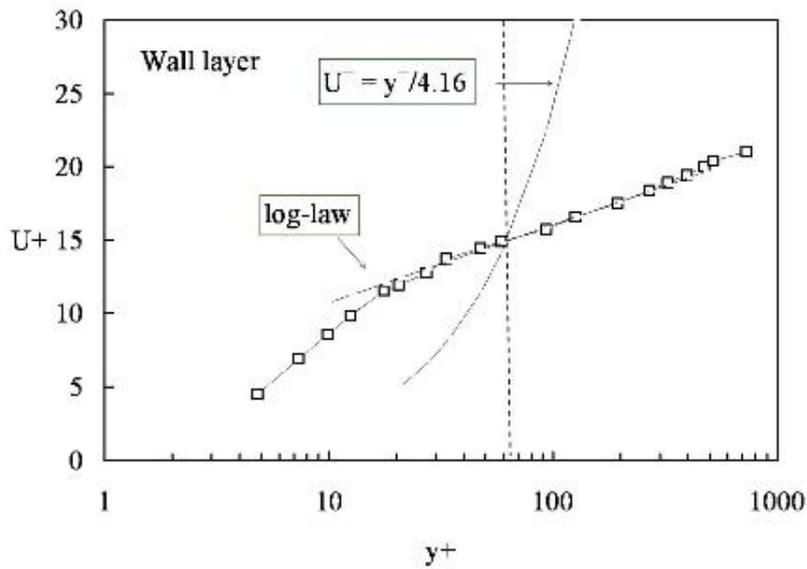

Figure 33 Edge of wall layer determined from slope of log-law and by intersection with the Stokes solution1. Data of Wei and Wilmarth (1989), Re = 14914.

2. The second method is to plot the profile of the viscous stress contribution $\tau_v/\tau$ (Figure 34).

   Because the viscous contribution falls asymptotically a cut-off point must be taken arbitrarily. In this work a value of 4% is chosen because it coincides with the point where the power spectrum in Figure 25 begins to flatten out.

3. The third method relies on the fact that the solution of order $\varepsilon^0$ is a boundary layer solution and the edge of this layer represents the maximum distance of penetration of diffusion of viscous momentum from the wall. The thickness of layer can therefore be defined by a solution such as equation (64) . Its intersection with measured velocity profiles (Figure 33) gives values of $U_v$ and $\delta_v$ in full agreement with the two first methods. But the intersection is easier to see and the determination of this edge is sharper and easier to determine.

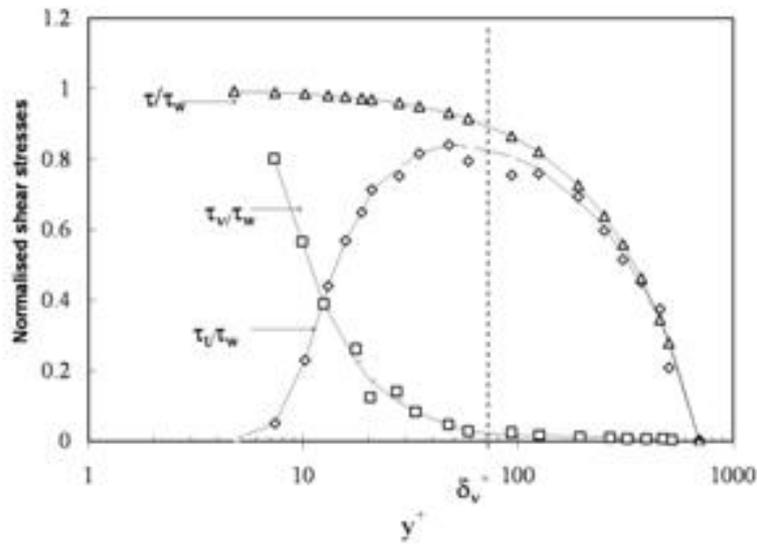

Figure 34 Edge of wall layer determined from normalised viscous stress distribution. Data of Wei and Wilmarth (1989), Re = 14914.

Each of these three techniques (which essentially return the same results within experimental uncertainty) allows us to obtain the wall layer thickness, the maximum distance for penetration of wall viscous momentum into the main flow. As shown in Figure 35, this wall layer thickness decreases rapidly from a value of $R^+$, the normalised pipe radius for laminar flow, to a stable value of approximately 64 to 67 depending on different authors as the Reynolds number increases.

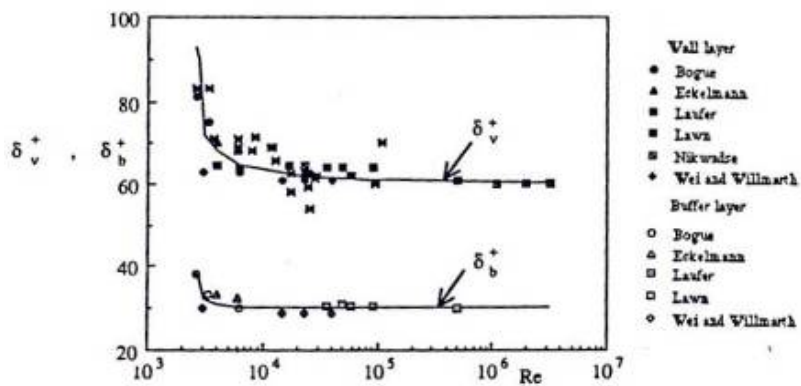

Figure 35 Wall and buffer layer thicknesses in pipe flow.

Also shown is the average distance $\delta_b$ of penetration of wall viscous momentum which coincides with Karman's definition of the edge of the buffer layer. The advantage of the first two methods is that they do not involve any mathematical model of the velocity profile.

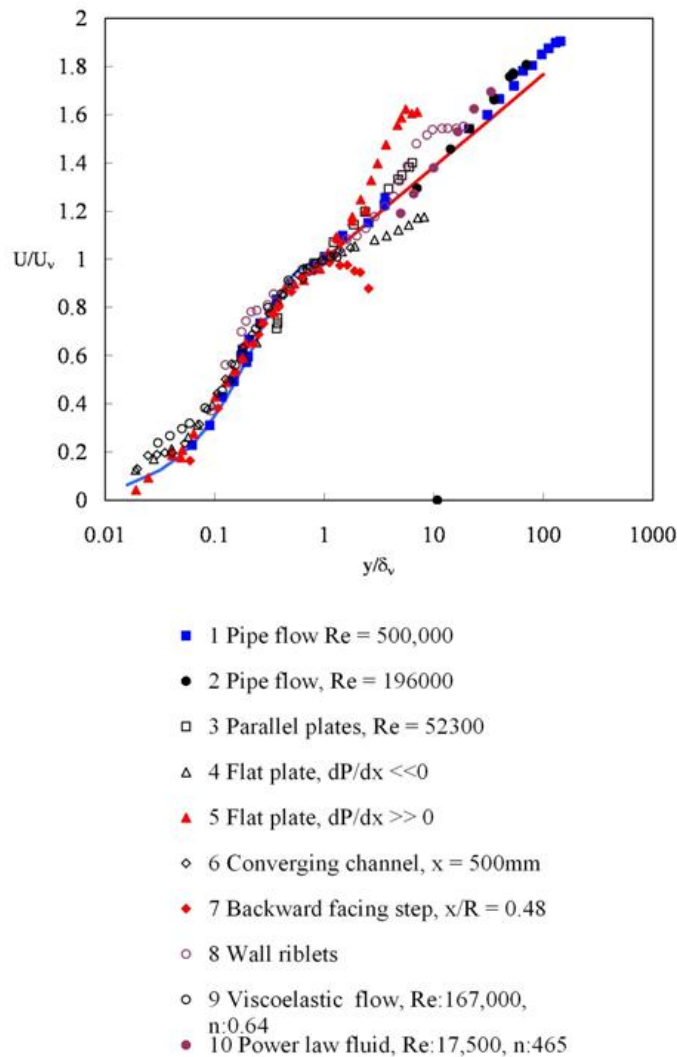

- ■ 1 Pipe flow Re = 500,000
- ● 2 Pipe flow, Re = 196000
- □ 3 Parallel plates, Re = 52300
- △ 4 Flat plate, dP/dx <<0
- ▲ 5 Flat plate, dP/dx >> 0
- ◇ 6 Converging channel, x = 500mm
- ◆ 7 Backward facing step, x/R = 0.48
- ○ 8 Wall riblets
- ○ 9 Viscoelastic flow, Re:167,000, n:0.64
- ● 10 Power law fluid, Re:17,500, n:465

Figure 36 Zonal similar velocity profile for different types of fluids and flow configurations from Trinh (2005). Data from [1] Laufer (1954), [2, 10] Bogue (1962), [3] Schlinder & Sage (1953)), [4,5] Kline, et al. (1967),[6] Tanaka & Yabuki (1986), [7] Devenport & Sutton (1991), [8] Bandyopadhyay (1986), [9] Pinho & Whitelaw (1990)

The velocity profile normalised with the parameters $U_v$ and $\delta_v$ for a large range of Reynolds numbers, different flow configurations and fluid rheological behaviours is shown in Figure 36. To differentiate it from the traditional "universal velocity profile" representation I have called it the zonal similarity analysis.

Several points can made from the zonal similarity analysis shown in Figure 36.
1. There does exist a unique master curve for the inner region of the velocity field past smooth surfaces but it must be normalised with parameters shared by the two adjacent zones; the wall layer and the log law region. This curve applies to all Reynolds numbers, all types of fluids and all flow geometries, including recirculation regions.

    All the data in the wall layer convincing collapse into a single curve. I have a shown a few examples from the hundreds of experimental curves I used to test the methodology and I stopped testing for more when I became satisfied that there was likely to be no exception to this pattern (except for profiles on rough walls). This deals with the question of universality which Biferale, Calzavarini, Lanotte, Toschi, & Tripiccione (2004) pointed out has been a "central matter of investigation in the scientific community". However, Biferale et al. follow the basic arguments of Kolmogorov that "a strong requirement for universality to hold is that large scale anisotropic fluctuations become more and more negligible going to smaller and smaller scales". In this analysis I consider that universality is satisfied when the same normalised inner region profile applies to all flows. The decoupling of events in the main stream (identified by many with large anisotropic fluctuations) and event in the dissipation zone (identified by many with small scale fluctuations) is only a necessary condition for the well documented universality of the high wave number portion of the power strum but does not in itself guarantee that universality.

    The reason for my belief in this truly universal master curve in the wall layer is simpler: The solution of order $\varepsilon^0$ that describes the sweep phase (the dominant contribution to the time-averaged velocity profile in the wall layer) is independent of the terms of order $\varepsilon$ and higher as seen in section 3. Therefore the solution of order $\varepsilon^0$ has the same form irrespective of the

Reynolds number and the geometry of the system but the numerical values of its parameters are not. The effects of geometry and Reynolds number are captured in the values of $U_v$ and $\delta_v$ that are determined for each particular system. Hence, using them as normalising parameters in effect has accounted for the influence of Reynolds number and flow geometry.

Thus any formulation for the solution of order $\varepsilon^0$ can be averaged to fit the zonal velocity profile in the wall layer but I have preferred to use the Stokes solution1 as shown in Figure 34 because it is the simplest and easiest to manipulate. The similar profile predicted by the time-averaged Stokes solution1 agrees very well with experimental data but a small discrepancy exists very near the edge of the wall layer because the Stokes solution1 assumes a zero velocity gradient at the wall layer edge, which is clearly not true in turbulence flows.

This method does not require any empirical assumption about the form of the Reynolds stresses or about the characteristics of the coherent structures that populate the flow field

2. Data in the log-law region also collapse for all geometries and Reynolds number but the extent of application of the log-law varies greatly between the systems considered.

3. The profiles in the outer (or far field) region do not collapse into this similarity representation for two reasons
    - The normalising parameters used are not shared by the outer region
    - The pressure term in the NS equations dominate in the outer region which requires a different method of scaling. The matching point between the log-law and outer regions is also highly dependent on the Reynolds number and the flow geometry.
4. The ability of the zonal similarity analysis to collapse data from non-Newtonian fluids both purely viscous and viscoelastic onto the Newtonian master curve suggests that the mechanism of turbulence production in non-Newtonian fluids is not different than the mechanism in Newtonian fluids, as suggested by a number of authors e.g. (McComb, 1991; Tennekes, 1966) This issue is discussed in more details in section 10.

5. The ability of the master curve to represent velocity profiles at all Reynolds numbers indicate that there is no need for a separate wall function for low Reynolds numbers e.g. (Patel, Rodi, & Scheuerer, 1985), for recirculation flow e.g. (Cruz, Batista, & Bortolus, 2000) or for heat transfer where details very close to the wall are required e.g. (Herrero, Grau, Grifoll, & Giralt, 1991). I believe that instead of focusing our efforts into the development of different formulations for different Reynolds numbers and geometries, we should direct our efforts into the development of better matching criteria for different asymptotic solutions of the NS equations, which are relatively easy to obtain. This led me to the study of a number of closure methods.

Table 1 shows how the thickness $\delta_v^+$ of the wall layer under different situations. The variation of this layer can be quite complex, for example in the recirculation region behind a backward facing step (Figure 37). But its value in fully developed turbulent pipe flow and boundary layer flow past flat plate is quite constant.

| Type of fluid | n | Re | $\delta_v^+$ | Source | Remarks |
|---|---|---|---|---|---|
| Newtonian | 1 | 2600 | 83.2 | Bogue (1962) | |
| | | 3300 | 75.0 | | |
| Viscous, non-Newtonian | 0.745 | 3660 | 80.1 | | Metzner-Reed (1955) generalised Reynolds number |
| | 0.70 | 11700 | 80.05 | | |
| | 0.59 | 6100 | 85.0 | | |
| | 0.53 | 17400 | 79.85 | | |
| | 0.465 | 7880 | 80.0 | | |
| Viscoelastic | 1 | 16700 | 60.0 | Pinho & Whitelaw | Reynolds number based on |

| | 0.90 | 16700 | 105.0 | (1990) | the non-Newtonian viscosity at the average wall shear stress |
| | 0.75 | 16700 | 155.0 | | |
| | 0.64 | 16700 | 180.0 | | |
| | | 459000 | 60 | Wells (1968) | Reynolds number based on the solvent viscosity |
| | | 98000 | 74 | | |
| | | 38700 | 82 | | |
| | | 211000 | 60 | | |
| | | 69900 | 75.6 | | |
| | | 13300 | 88 | | |

Table 1 Dimensionless wall layer thickness in pipe flow for different Reynolds numbers and fluid properties

The analysis of flow in recirculation regions is among the most challenging in flow dynamics. Figure 36 and Figure 37 indicate that the zonal similarity analysis performs well and gives useful thickness parameters that can be used for matching with other estimates of flow outside the circulation zone by ,say, CFD packages.

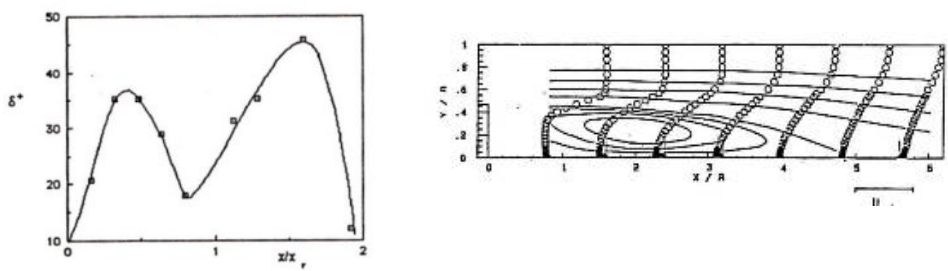

Figure 37 Streamlines and velocity profiles behind a backward facing step (Devenport & Sutton, 1991) and estimated thickness of wall layer (Trinh 1992, 2005)

# 9 Closure theorems

## 9.1 A theoretical closure for the Reynolds equations

The NS equations and the equation of continuity form a closed set that can be solved in principle, even though no general solution has been obtained in the last 160 years because of the great difficulties arising from the non linear terms. When Reynolds averaged the NS equations to give the famous Reynolds equations RANS, a degree of freedom is lost and there is no longer sufficient information to solve this new set of equations. This is the famous closure problem in turbulence. There are many ways to address this closure problem (Hinze, 1959; Lesieur, 2008; McComb, 1991; Schlichting, 1960; Tsinober, 2001).

One the first attempts was made by Prandtl (Prandtl, 1925, 1935) which resulted in his formulation of the logarithmic law of the wall. The mathematical success of his approach, verified with the data of Nikuradse (1932) was marred however by his unfortunate analogy of the scale $l$ with the free path of molecules and called 'mixing-length'. While the physical interpretation of this scale is now widely discredited, the mathematical estimate from a truncated Taylor series does not depend on it and the law of the wall remains one of the most useful tools in turbulence predictions. In Prandtl's approach the Reynolds stresses are not modelled directly but through a turbulent eddy viscosity, a popular concept introduced by Boussinesq (1877). The critical step in Prandtl's analysis is the derivation of a turbulent length scale that allowed him then to estimate the eddy viscosity. The postulation of an algebraic relationship for the length scale ready for use with the RANS is referred to in the literature as a zero-order closure model (Gatski & Rumsey, 2002). Prandtl's mixing length was postulated for the plane shear flow with unidirectional mean flow. In addition, Prandtl focused only on the log-law region and made no attempt to model the outer region or the wall layer. Prandtl did assume that there was a thin region near the wall $0 < y^+ < 5$, called the laminar sublayer where viscous forces would completely dominate resulting, in his view, to complete damping of turbulent fluctuations.

Many other workers have extended the zero order analysis to two-layer mixing-length models. Cecebi & Smith (1974) essentially expressed Prandtl's and Cole's laws in multidimensional terms and Baldwin and Lomax (1978) expressed the mixing length in terms of vorticity.

The considerable difficulties linked with analytical solutions for flow in complex geometries were by-passed with the introduction of computational fluid dynamics CFD in the early seventies (B.E. Launder & Spalding, 1974).The problem with zero order models is that the parameters of the model e.g. the boundary layer thickness in the Cecebi-Smith model must be evaluated by searching along grid lines in the normal direction. One equation models such as that of Spalart & Allmaras (1994) are local and can be used with any type of grid. The approach here is to calculate the eddy viscosity through the formulation of a transport equation. The next development was to calculate the eddy viscosity from two local quantities both estimated from transport equations.

$$E_v = C_\mu \frac{k^2}{\varepsilon} \tag{96}$$

Where the symbols here are

$E_v$    Turbulent kinematic viscosity

$k$    Turbulent kinetic energy

$\varepsilon$    Turbulent energy dissipation rate

$C_\mu$    Model coefficient

This is the famous and popular $k - \varepsilon$ model. An alternative is the $k - \omega$ model where $\omega$ is the dissipation rate of kinetic energy per unit $k$. The coefficients of the terms in the transport equations for these models are determined empirically from experimental observations. One well-known challenge of the $k - \varepsilon$ models is the treatment of the inner region. Bradshaw, Launder, & Lumley (1991) tested CFD packages developed by authors around the world and found that any model which invokes the logarithmic law-of-the-wall gave reasonable predictions of the velocity field irrespective of the model for the outer flow. But the so-called standard $k - \varepsilon$ models, those derived for high Reynolds numbers, do not give an accurate representation of recirculation regions and of the near wall at transition and low Reynolds numbers. Thus there is an extensive list of modifications of Prandtl's law of the wall to deal with these situations e.g. (Gavrilakis, 1992) which has also been shown to apply to two dimensional flow (Zanoun & Durst, 2003). A further weakness of the $k-\varepsilon$ models is the unrealistic isotropic nature of the eddy viscosity. This has led to the development of so-called non-linear eddy viscosity models such as the algebraic stress models.

The latest development is the introduction of differential second-moment turbulence closure models DSM (B. E. Launder & Sandham, 2002) that are based on transport equations for the turbulent stresses and turbulent fluxes. The advantage of the DSM is in the exact treatment of the turbulence production term and of anisotropy of the turbulent stress field. Hanjalíc and Jakirlić (2002) believe that the DSM will eventually replace the present popular $k-\varepsilon$ model but admit that "despite more than three decades of development and significant progress, these models are still viewed… as a …target than as a proven and mature technique". The modelling of the $\overline{u_i u_j}$ and $\varepsilon$ equations is still based on the a characteristic turbulence time scale $\tau = k/\varepsilon$ and a length scale $L = k^{3/2}/\varepsilon$ but one now has the opportunity to model two new important terms: the pressure-strain term $\varphi_{ij}$ and the stress dissipation rate $\varepsilon_{ij}$.

I have included this very succinct overview of closure techniques only to put my own views, described below, in context. It have not attempted to capture adequately, even in the most general manner, the huge diversity of approaches resulting from the avalanche of computer modelling work of the last forty years. For a more detailed introduction, the reader may consult excellent books and reviews such as that of Launder and Sandham (op. cit.) or (Patel, et al., 1985; Rodi, 1980) for earlier models.

The striking feature of all existing closure models is the empirical nature of coefficients used, which are more and more numerous as the models increase in complexity to give adequate descriptions of complex industrial applications. There were of course considerations of basic theoretical understanding, particularly of the energy cascade introduced by Kolmogorov, but the fundamental empirical nature simply reflects the state of poor understanding of turbulence mechanisms and, in my view, the constraints of the RANS used as a starting point.

In the 1980's I began to work on a method of effecting closure theoretically by matching two separate descriptions of the velocity field: one based on a time averaged solution using subsets of the RANS equations, the other based on an unsteady state solution of subsets of the NS equations. We begin with the all important inner region where I saw the opportunity to by-pass altogether the modelling of the Reynolds stresses.

Consider the log-law region. We start our modelling by referencing Millikan's similarity argument (Millikan, 1939). The outer (far field) region must scale with the outer variables and the wall layer must scale with the wall parameters as shown by the solution of order $\varepsilon^0$. Therefore dimensional considerations tell us that there exists an intermediary region these two different scaling laws overlap. Millikan showed that this overlap requires the velocity profile in this intermediary region to take a semi-logarithmic form. The attraction of Millikan's analysis is that it does not require any assumption about the physical structures that populate the log-law region, unlike Prandtl's analysis (op. cit.) that involves assumptions about the link between the Reynolds stresses and the mixing-length, even though both arrive to the same log-law. Of course the Millikan analysis, which is purely mathematical, only gives the form of the solution, not the values of the coefficients involved.

We now attempt to derive a formulation compatible with Millikan's analysis, using a zero equation model to highlight the basic considerations without dealing with the added complexities of higher models.

The local time-averaged shear stress at a distance y is given by

$$\tau = (\mu + \rho E_v)\frac{dU}{dy} \qquad (97)$$

Rearranging in dimensionless form gives

$$U^+ = \int_0^{y^+} \frac{\tau/\tau_w}{1 + E_v/\nu} dy^+ \qquad (98)$$

To estimate the eddy diffusivity, we expand the turbulent shear stress into a Taylor series of the velocity gradient equation (42) but only keep the first order derivative.

Following Prandtl, we use a normal projection of this correlation length

$$\tau_t = \left(\rho l \frac{dU}{dy}\right)^2 = \rho E_v \frac{dU}{dy} \tag{99}$$

Hence

$$E_v = l^2 \frac{dU}{dy} = l\sqrt{\frac{\tau_t}{\rho}} \tag{100}$$

Equation (98) may be rearranged as

$$\frac{dU^+}{dy^+} = \frac{\tau/\tau_w}{1 + E_v/\nu} \tag{101}$$

In the log law region, the viscous contribution is small (Figure 34) and $1 << E_v/\nu$ and we may write

$$\frac{dU^+}{dy^+} \cong \frac{\tau/\tau_w}{E_v/\nu} = \frac{\tau/\tau_w}{(l/\nu)\sqrt{\tau_t/\rho}} = \frac{\tau/\tau_w}{(lu_*/\nu)\sqrt{\tau_t/\tau}\sqrt{\tau/\tau_w}} \tag{102}$$

Substituting for equation (43)

$$\frac{dU^+}{dy^+} = \frac{\sqrt{\tau/(\tau_w \tau_t)}}{\kappa y^+} \tag{103}$$

which does not lead to a log-law. To obtain the log law from equation (103) Prandtl had to assume further that

$$\tau = \tau_w = \tau_t \tag{104}$$

This assumption is clearly at odds with reality particularly at low Reynolds number and we therefore will not adopt it. In addition, as pointed out in section 3.1 we do not interpret $l$ as a mixing-length, a distance that eddies at y travel before they loose their identity in analogy to the mean free path of gas molecules but as a typical scale of the streaming jet. The Karman constant $\kappa$ is interpreted as the angle of inclination between the jet path and the normal direction (section 4). Since the path of a jet in cross flow can be approximated by an arc, I thought that Karman's constant could be regarded as a mutated value of $\pi$

$$\frac{1}{\kappa} = \pi f(\frac{\tau_t}{\tau}) \tag{105}$$

and named this derivation the eddy path theory (Trinh, 1969) which went through many modifications. The function f in equation (105) arises from the expectation that the path of the eddy will be dependent on the relative strengths of the streaming flow and the cross flow. For the moment we simply accept the widely quoted value $\kappa = 0.4$ to avoid complicating the formulation any further.

In order to extend the application of equation (103) right to the wall, we introduce a new factor F such that

$$l = \kappa y \sqrt{\frac{\tau}{\tau_w}} F \tag{106}$$

And

$$\frac{E_v}{v} = \kappa y^+ \left(\frac{\tau}{\tau_w}\right) \sqrt{\frac{\tau}{\tau}} F \tag{107}$$

Van driest (1956) was the first author I met who proposed a similar factor that he called a damping factor. As introduced in section 3.1, he argued that the Stokes solution2 can be used to describe how eddies from the outer region would be damped by viscous resistance near the wall. As discussed in section 4, I believe that the log-law is related to the streaming jet outside the wall layer. The inclusion of a damping factor F is only to make equation (107) compatible with the solution of order $\varepsilon^0$ (the Stokes solution1 not solution2) and give a single formulation for the inner region. It has no physical significance, and certainly I do not believe that the function F represents a damping of turbulent eddies from the outer flow that bombard the wall layer, a view that was made obsolete by the ground breaking work of (Kline, et al., 1967).

The form of this factor may be inferred from the velocity gradient in the Stokes solution1

$$\frac{\partial u}{\partial y} = \frac{2}{\sqrt{\pi}} e^{-\eta_s^2} \tag{108}$$

A form of the damping function compatible with the error function is

$$F = 1 - e^{-b\left[\frac{y^+(\tau/\tau_w)}{\delta_v^+}\right]^2} \qquad (109)$$

The damping function must satisfy two limits

$F = 0$      at      $y^+ = 0$

$F \to 1$      as      $y^+ \to \delta_v^+$

As with all boundary layer solutions, an arbitrary cut-off value must be taken. In the present formulation a value F=0.99999 at $y^+ = \delta_v^+$ was found most suitable. At high Reynolds numbers, the wall layer is thin and $\delta_v^+ \ll R^+$ and $\tau/\tau_w \cong 1$, equation (109) gives $b = 11.2$. Then for pipe flow

$$U^+ = \int_0^{y^+} \frac{(1 - y^+/R^+)dy^+}{1 + 0.4y^+\left(1 - \frac{y^+}{R^+}\right)\sqrt{\frac{E_v}{E_v + v}}\left(1 - e^{-11.2\left[\frac{y^+(1-y^+/R^+)}{\delta_v^+}\right]^2}\right)} \qquad (110)$$

Equation (110) must pass by the point $(\delta_v^+, U_v^+)$.

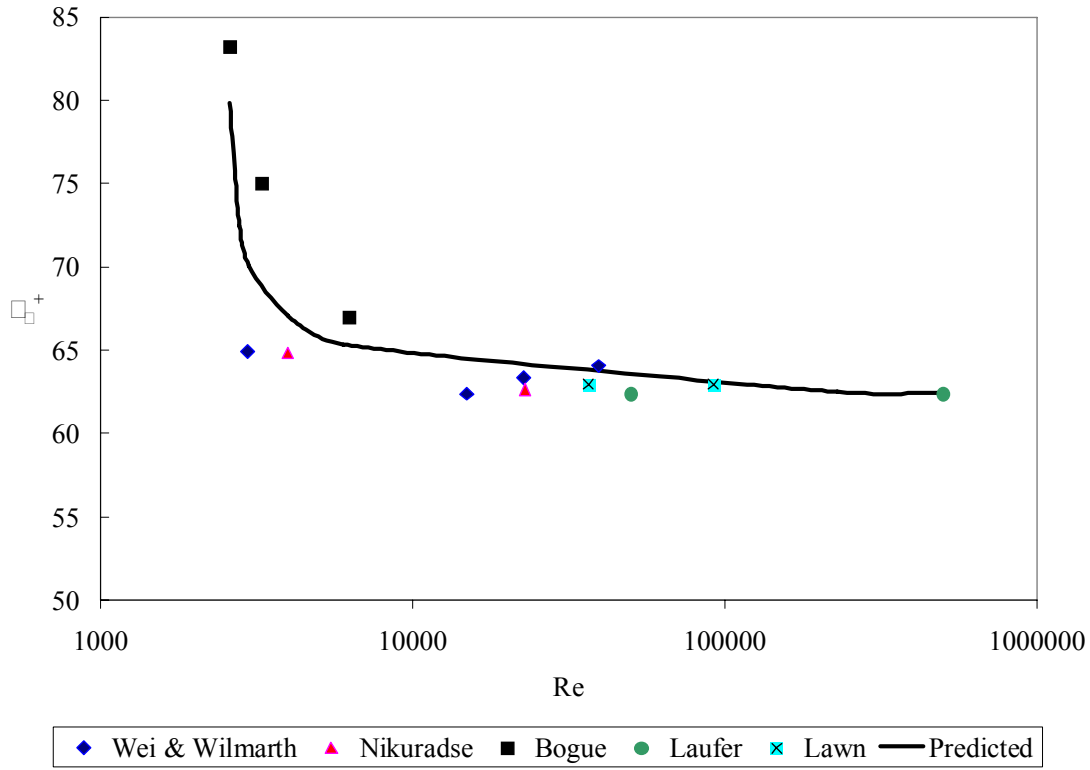

Figure 38 Wall layer thickness predicted by equation (111).

Here we note that we have derived a solution of order $\varepsilon^0$ from the original NS equations for the wall layer which must also pass through the point $(\delta_v^+, U_v^+)$

$$\delta_v^+ = 4.16 U_v^+ \tag{61}$$

Matching equations (110) and (61) gives

$$\delta_v^+ = 4.16 \int_0^{\delta_v^+} \frac{(1 - y^+/R^+)\,dy^+}{1 + 0.4 y^+ \left(1 - \dfrac{y^+}{R^+}\right)\sqrt{\dfrac{E_v}{1+E_v}}\left(1 - e^{-11.2\left[\frac{y^+(1-y^+/R^+)}{\delta_v^+}\right]^2}\right)} \tag{111}$$

This equation can be solved iteratively for $\delta_{v+}$ for each value of $R^+$. The effect of the Reynolds number is introduced through the dimensionless radius $R^+ = \dfrac{\mathrm{Re}\sqrt{f}}{2\sqrt{2}}$. The wall layer thicknesses predicted by equation (111) are shown against experimental data in Figure 38. These predictions are fully compatible with the determination of the wall layer thickness by the use of the Stokes1 solution1 shown in Figure 35.

The velocity profiles predicted from equation (110) are plotted against literature data in Figure 39. The predictions fit the data in the inner region (wall layer + log law) but not in the pipe core which requires a different correlation than the log-law as discussed in section 4. This derivation also gives quite good predictions of the turbulent shear stress distribution (Figure 40), the dissipation (Figure 41) and the production of turbulence (Figure 42).

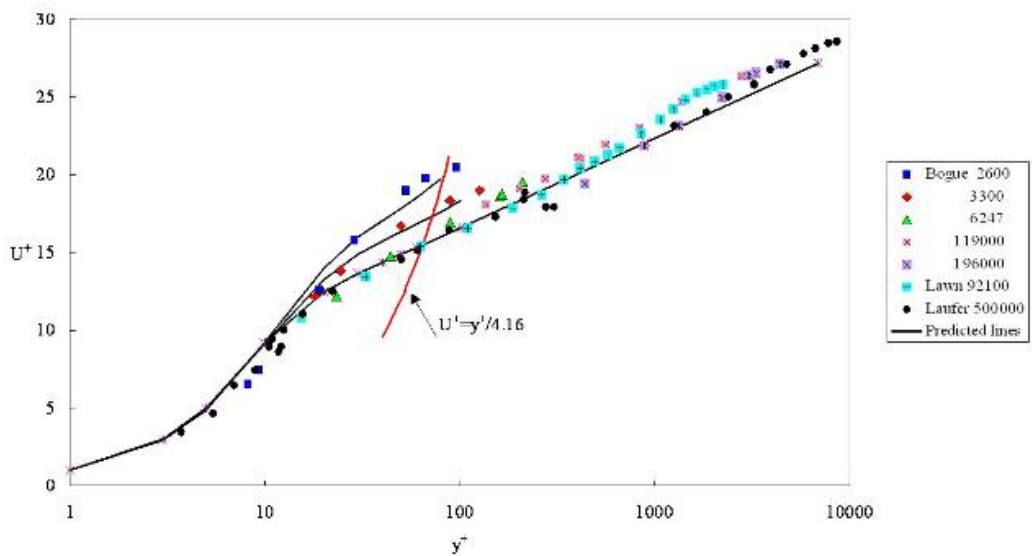

Figure 39 Velocity profile predicted by equation (110). Data of Bogue (1961), Lawn (1971) and Laufer (1954)

We could have used other formulations for the damping function such as those of van Driest (1956), Khishinevski and Korninenko (1967), or Deissler (1955) see (Trinh, 1992). But these would only have given an adequate prediction for higher Reynolds number because they do not include the parameter $R^+$ which allows the present formulation to deal more adequately with the effect of low Reynolds numbers in the transition region.

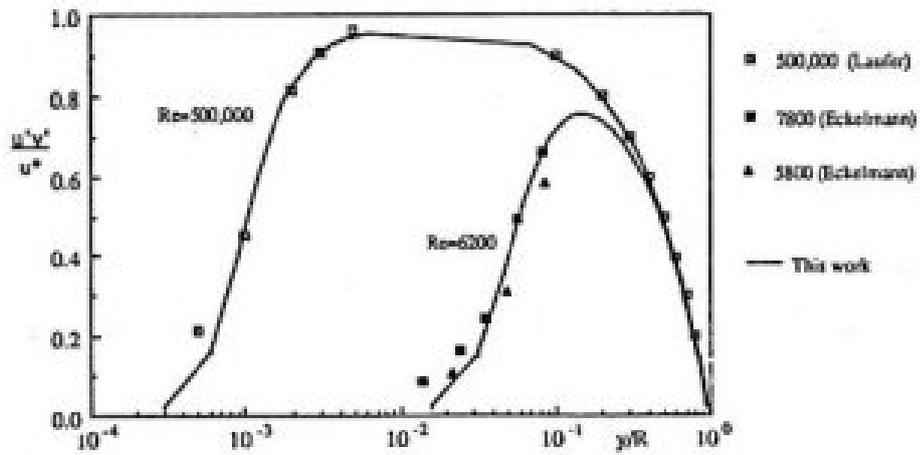

Figure 40 Reynolds stress distribution according to theoretical closure of eddy path theory. Data of Laufer (op.cit.) and Eckelmann (op.cit.)

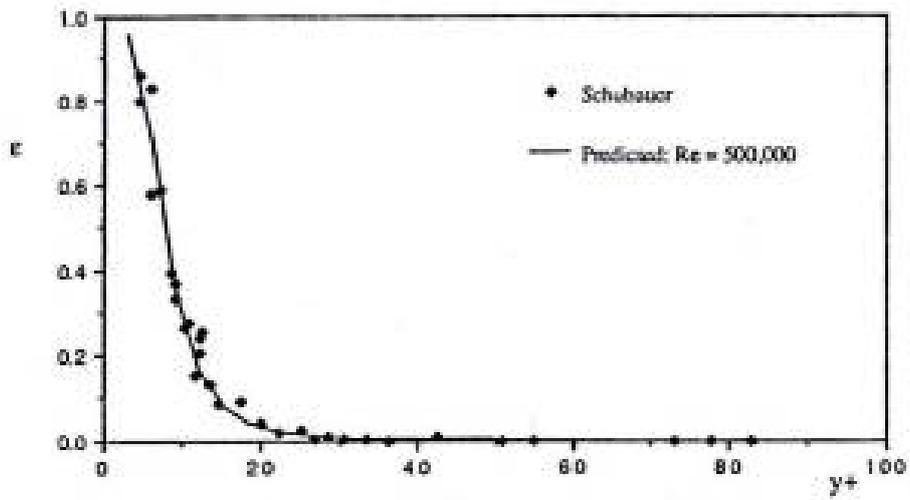

Figure 41 Dissipation predicted by eddy-path theory. Data of (Schubauer, 1954)

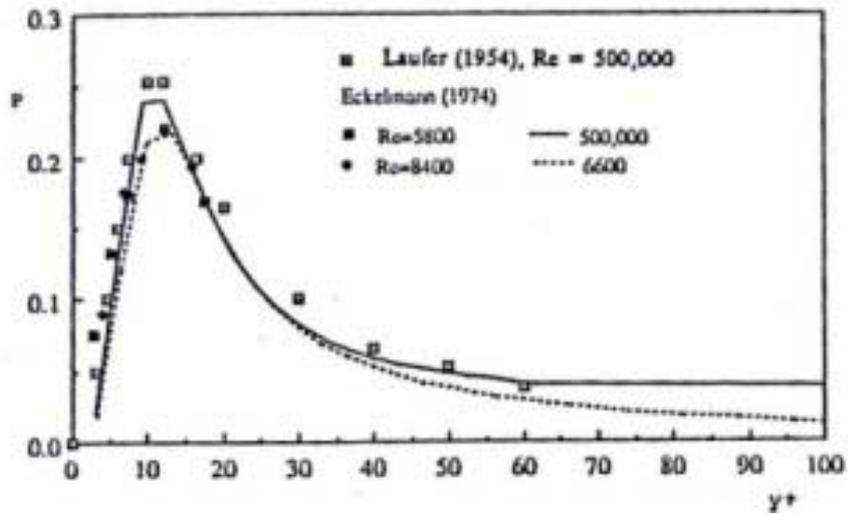

Figure 42 Production of turbulence predicted by eddy path theory. Data of Eckelmann (op.cit.) and Laufer (op.cit.)

We could have of course started equation (45), which results from Millikan's analysis and modify with a damping function without reference to the mixing length and obtained similar results. It is more interesting to look at possible forms of equation (106) where the original formulation of the mixing length equation (43) was modified with a damping function to give a formulation that is compatible with the solution of order $\varepsilon_0$ near the wall. The ratio $\sqrt{\tau/\tau_w}$ was introduced simply for convenience of mathematical computation. Dimensionally speaking, any other ratio of stresses could have been chosen. For example we may have considered

$$l = \kappa y F \sqrt{\frac{\tau}{\tau_t}} \qquad (112)$$

which leads to

$$\frac{E_v}{\nu} = \kappa y^+ F \sqrt{\frac{\tau}{\tau_w}} \qquad (113)$$

Or

$$l = \kappa y F \sqrt{\frac{\tau_w}{\tau_t}} \qquad (114)$$

$$\frac{E_\nu}{\nu} = \kappa y^+ F \tag{115}$$

Or

$$l = \kappa y F \tag{116}$$

$$\frac{E_\nu}{\nu} = \kappa y^+ F \sqrt{\frac{\tau_t}{\tau_w}} \tag{117}$$

With each of these formulations, the constant $b$ in equation (110) needs to be modified for best fit against the experimental data then all formulations give the same type of prediction shown in Figure 39. Real differences in the predictions of the eddy viscosity only become apparent when $y^+ \leq 0.1$ (Trinh, unpublished work, 1986; 1991) but these have negligible impact on the time averaged velocity profile or the friction factor predicted. A plot of the turbulent stress distribution overlapped on the energy spectrum such as shown in Figure 25 shows that all formulation give good agreement with energy spectrum up to $kR \approx 20$ but equation (106) for example gives too high values at higher wave numbers whereas equation (116) gives too low values (Trinh, 1992). Since it is impossible to make experimental measurements of turbulent stresses or even velocity as such small distances from the wall, the energy spectrum gives us an alternative for studying the behaviour of the Reynolds stresses at very small values of $y^+$.

### 9.1.1 A closed solution for external boundary layers

The damping function used in equation (110) requires the stipulation of the dimensionless pipe radius $R^+$, or in the case of boundary layer flow, the boundary layer thickness $\delta^+$. The diameter of a pipe limits the size of both the length and time scales. The thickness of an external boundary layer can, on the other hand, grow indefinitely. Application of the log-law to flow past flat plates e.g. (Schultz-Grunow, 1941) correlates the velocity profile near the wall but does not give the friction factor because the thickness of the turbulent boundary layer in this approach is undefined and the problem is not closed. The traditional method of dealing with this difficulty has been to solve the integral momentum equation numerically, using Prandtl's velocity distribution, to give the boundary layer thickness and the friction factor (Schlichting 1960, p 601). A second way of describing the velocity profile in turbulent boundary layer flow originated from the empirical power law relationship introduced by Blasius (1913). Nikuradse (1932) used both the log-law and power law correlations for the velocity profiles in pipe flow to obtain alternate derivations for the friction factor (Schlichting, 1960). Recently, Zagarola, Perry and Smits (1997) have argued from new, more careful measurements of pipe flow data that both the log-law and the power law in the region apply in the inner region but give much more restrictive limits to these so-called overlap regions, particularly for the log-law

Neglecting, the small but real inconsistencies that these laws have with real velocity profiles, I applied the principle of matching these two separate formulations for the velocity profile in external boundary layers in 1982 to develop a second closure criterion.

The friction factor on a flat plate may be written in the general power law form

$$f = \frac{\alpha}{\text{Re}_\delta^\beta} \qquad (118)$$

The velocity profile may be also written as

$$\frac{U}{U_\infty} = \left(\frac{y}{\delta}\right)^p \qquad (119)$$

The indices $p$ and $\beta$ are related (Schlichting 1960)

$$p = \frac{\beta}{2-\beta} \qquad (120)$$

A derivation by standard methods (Schlichting p.598) gives the boundary layer thickness as

$$\frac{\delta}{x} = \left[\frac{2\alpha(\beta+1)(\beta+1)}{2-\beta}\right]^{1/(\beta+1)} \left(\frac{\nu}{xU_\infty}\right)^{\beta/(\beta+1)} \qquad 121$$

The power index p is obtained here by forcing equation (119) through the edge of the wall layer $(\delta_v^+, U_v^+)$ If we take these values as 16 and 64 respectively, the coefficients α and β may be obtained (Trinh 1992) as

$$\beta = \frac{5\ln U_\infty^+ - 13.86}{U_\infty^+ + 2.5\ln U_\infty^+ - 25.00} \tag{122}$$

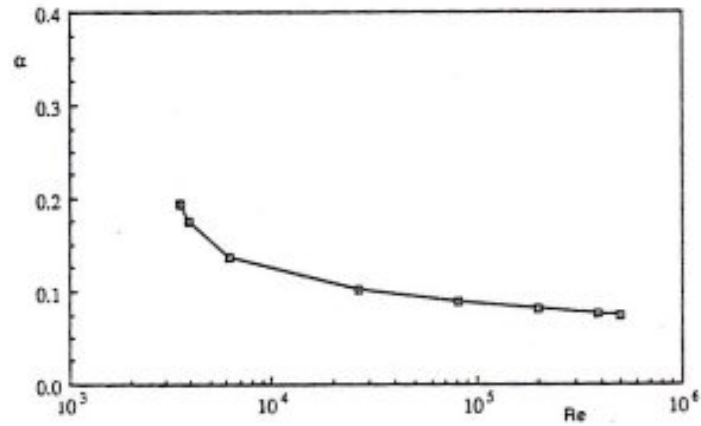

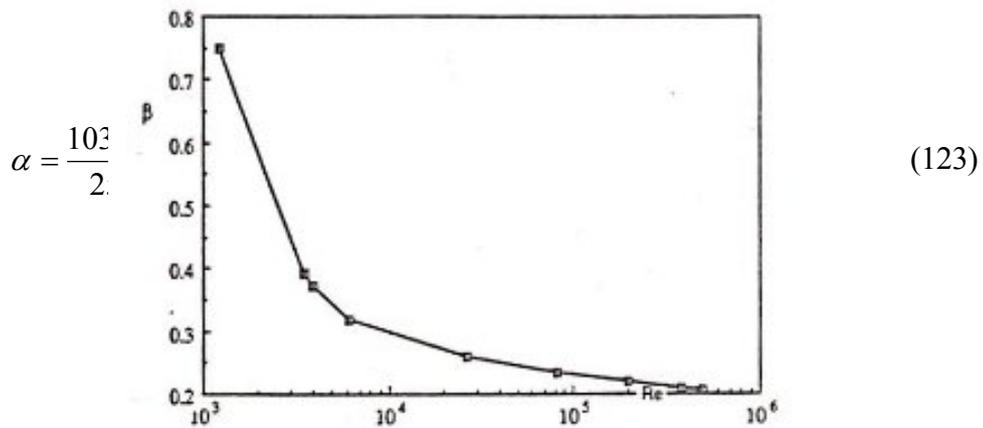

$$\alpha = \frac{10?}{2.} \tag{123}$$

Figure 43 Variations of factors $\alpha$ with Reynolds number

And

Figure 44 Variations of factors $\beta$ with Reynolds number

Figure 43 and Figure 44 show that the coefficients $\alpha$ and $\beta$ are functions of the Reynolds number, rather than true constants as one might conclude from the experimental formulation of Blasius. Nikuradse (in Schlichting, 1960) showed that $p$, which is related to $\alpha$ and $\beta$, does change with Reynolds number.

The estimate of the boundary layer thickness in equation (121) is now fed back into the log-law to give the local friction factor. After some lengthy but straightforward mathematical manipulations, the average friction factor over a length L is given by

$$\frac{1}{\sqrt{f}} = A\ln\left(L^+ \operatorname{Re}_L^{-\beta/\beta+1}\right) + \frac{A}{\beta+1}\ln\left[\frac{2\alpha(\beta+1)(\beta+2)}{\beta(2-\beta)}\right] + 3.05 \qquad (124)$$

The results are shown in Figure 45 against the experimental data from several authors reported by Schlichting (1960).

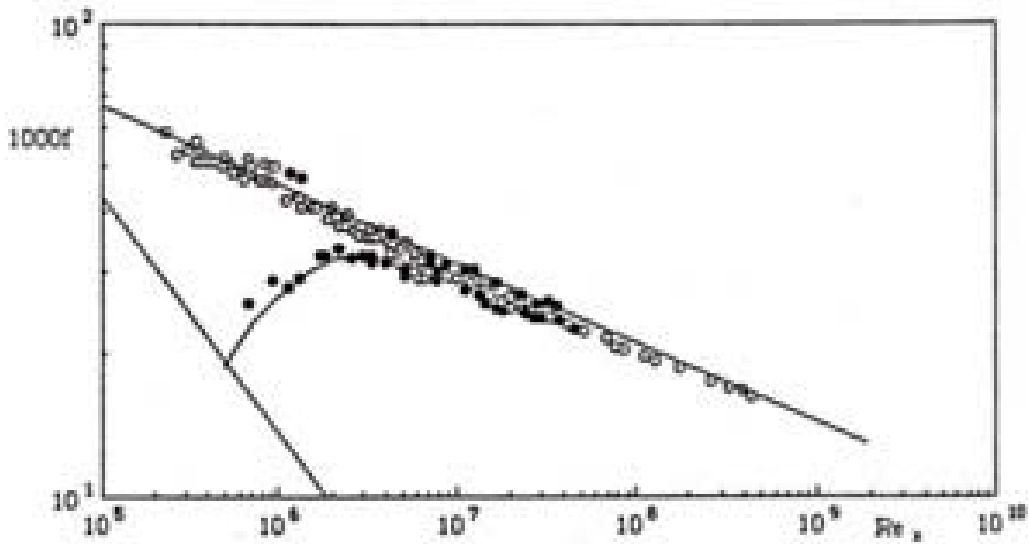

Figure 45 Friction factor for turbulent boundary layer flow past a flat plate predicted by equation (124).

The advantage of the present technique is that it gives a closed solution for quick estimates of the friction factor in external boundary layers. It also allows us to implement the zero-order closure model in CFD packages more efficiently because the thickness of the boundary layer can be coupled with the parameters of the log-law. Others (Schlichting, 1960; L. C. Thomas & Al-Sharif, 1981; F. M. White, Christoph, & Lessmann, 1973) have used the integral momentum and energy theorems in conjunction with either the logarithmic or power law distributions of velocity and temperature separately. The simultaneous use of two types of distributions can be made only after a relationship has been established between the coefficients in the power law and those in the log-law.

**9.1.2    Some further thoughts about higher order closure approaches**

The closure technique and the visualisation of turbulence proposed here have significant implications to the modelling of turbulent flows. Unfortunately, I have not really had the opportunity to develop computer modelling of practical systems[6]. Nonetheless, I wish to summarise here the basic consequences as I see them.

1. The TANS equations and their implications

The first issue lays with RANS equations themselves, the governing equations universally used in commercial CFDs. They are based on a two component decomposition of the instantaneous velocity. When we use a four-component decomposition equation (55) we obtain a new set of time-averaged NS equations that I will call temporarily TANS (Tuoc Averaged Navier Stokes equations) for lack of a better nomenclature. The procedure for averaging is the same as Reynolds' (1895). It is not difficult, just lengthy but clearly show extra terms that did not figure in the RANS.

Critically, the traditional Reynolds stresses RS are now separated into slow Reynolds stresses SRS that arise from a difference between the smoothed phase velocity of the coherent structures and the long time averaged velocity and fast Reynolds stresses FRS due to the streaming flow. These behave differently and modelling them separately will capture more adequately the physics of turbulent flows.

---

[6] I did try to write a CFD program for several months in the early 1990's. Unfortunately, after twenty years from 1970 to 1990 without access to computers in the isolation of communist Viet Nam, my computer skills acquired mainly in the 1960's had become quite rusty. Despite upgrading them with an excellent book on FORTRAN programming and Patankar's excellent introduction to CFD (Patankar, 1980), I did not have the skills to write proper graphical presentations of the results. The discretisation and solver codes were not an issue. In fact before 1982, all my calculations were made with pen, paper and a slide rule and in 1983-84 I had the extraordinary luck- and money- to pick up a TI programmable calculator, my most sophisticated tool that facilitated the iterations required for my closure technique.

There will also be new interaction terms involving the smoothed phase and streaming flow velocities, see equation (40). The all important pressure strain terms are also significantly modified. In second moment closure models such as, for example, the popular Gibson-Launder model, the kinematic Reynolds stresses are obtained from (B.E. Launder & Shima, 1989)[7]

$$\frac{D\overline{u_i u_j}}{Dt} = P_{ij} - \varepsilon_{ij} + \varphi_{ij} + d_{ij} + \partial\left(\frac{v\partial\overline{u_i u_j}}{\partial x_k}\right)\bigg/\partial x_k \qquad (125)$$

Where the vector terms have the usual significance in CFD modelling

$P_{ij}$     Turbulence production

$\varepsilon_{ij}$     Dissipation

$\varphi$     Pressure-strain

$d_{ij}$     Turbulence diffusion

The pressure strain term denotes the averaged product of the fluctuating pressure and strain fields

$$\varphi_{ij} = \overline{\frac{P}{\rho}\left(\frac{\partial u_i}{\partial x_j} + \frac{\partial u_j}{\partial x_i}\right)} \qquad (126)$$

Four distinct contributions are traditionally associated with this term

---

[7] The symbols in this section are those of Launder and Shima. In the presentation of a body of work spanning over 25 years, with another 15 pondering the issues sporadically, one quickly runs out of symbols for individual variable parameters. Instead of introducing a complex and messy system of subscripts and superscripts I have decided to assign temporary significance to symbols that match different quoted authors in selected sections of this script. Thus the symbol $\varepsilon$ in general describes the Strouhal-Reynolds number product (equation 58) after the nomenclature of Tetlionis for oscillating boundary layer flow but refers to the energy dissipation rate in the nomenclature of Launder and Sharma.

$$\varphi_{ij} = \varphi_{ij1} + \varphi_{ij2} + \varphi_{ij1}^{w} + \varphi_{ij2}^{w} \qquad (127)$$

The first term $\varphi_{ij1}$ relates to the anisotropic part of the Reynolds stress, the second $\varphi_{ij2}$ to the production of turbulence and the last two $\varphi_{ij1}^{w}, \varphi_{ij2}^{w}$ to the redistribution of velocity fluctuations in the direction normal to the wall to those parallel to it by the first two[8].

The TANS equations allow us to model the pressure strain term in a different way. One would expect that the pressure will be greatly affected by the streaming flow in the log-law region because it is a cross flow with respect to the main flow direction. On the other hand in the outer region, it is essentially aligned with the main flow and should have only minor effects on the local pressure, as one would expect from a coherent body of fluid convected essentially with a velocity equal to 80% of the maximum boundary layer velocity (Adrian, et al., 2000). In the wall layer the term again diminishes in importance, mainly because of the relatively low magnitude of the streamwise velocity component.

In many CFDs, $k$ and $\varepsilon$ are related through a typical time scale of turbulence $\tau$

$$\tau = \frac{k}{\varepsilon} \qquad (128)$$

While this equation is dimensionally correct, it does not provide a full picture of the critical events.

Meek and Baer (1973) successfully normalised the time scale in the wall layer of turbulent pipe flow with the friction velocity and the kinematic viscosity, the so called wall parameters (Figure 16). However, Rao, Narasimha, & Narayanan (1971) showed the average period of velocity fluctuations in the outer region could not be correlated in the same way and found that there data must be normalised with the outer flow parameters, the thickness of the boundary layer and the approach velocity.

---

[8] Note that the terms $\varphi_{ij1}$ is sometimes called the slow term and $\varphi_{ij2}$ the fast term in the pressure strain terminology (Hanjalic & Jakirlic, 2002). The terms slow and fast here are used in a different context than my slow and fast Reynolds stresses

In my view, there is no controversy here, because these authors were modelling two different events with different time scales. The Meek and Baer time scale is related to the duration of the sweep phase of the wall cycle. But we know that the bursting cycle is of much shorter duration and follows necessarily a different time scale. In addition, once the ejections or bursts are detached from the wall, it makes little sense to scale the velocity fluctuations that they induce with the wall parameters. Since the average thickness of the wall layer is roughly equal to Karman's buffer layer thickness Figure 35 andFigure 38, the ejections are often observed to emanate from the distance $y^+ = 30$ e.g. (Fife, et al., 2005). The normalised time scale linked with the fluid contained in the low-speed streaks which move as a coherent structure once it detaches from the wall, is similar to the Strouhal number of jets and cylinders in cross flow ad scales with the local (outer) stream parameters as shown by Rao et.al.

Thus modelling these two separate and distinct phenomena with a unique time scale must surely present major mathematical difficulties.

The zero order models attempt to propose algebraic relationships that can be directly substituted into the RANS. As such as they are "global" in nature i.e. the formulation applies to the entire profile as a whole. The modern use of transport equations allows us to model each parameter locally and can be used flexibly for example with unstructured grids. The problem with local models is that the solution must be matched with the non-slip condition at the wall. Near the wall it requires a modelling of the small scales. The traditional literature is dominated there by the concept of energy cascade and locally isotropic dissipation scale of Kolmogorov which unfortunately do not shed any light on the sweep burst cycle observed by (Kline, et al., 1967) and others.

Many authors decided to match the iterated solution with the law of the wall thus providing it a simpler set of effective "boundary conditions", as Launder and Shima nicely put it. Unfortunately this match point, the first grid point for the iterated solution is "largely empirical, even for attached flows…and this method is problematical for separated flows" (Schwab, 1998), p.88

2. An alternate closure strategy

The solution of order $\varepsilon^0$ is an exact solution of the subset of the NS equations that applies to the region where the diffusion of negative momentum from the wall affects the flow field. The subset itself is based on the neglect of the fast velocity fluctuations impressed by the vortex travelling above the wall and the streaming flow. It turned out that the dominant mode in the orthogonal decomposition of the instantaneous velocity coincides very well with to the most probable velocity and predictions of the phase velocity of the solution of order $\varepsilon^0$ (Figure 18). This indicates that in the wall layer, the contribution of the streaming flow to the profile of the streamwise velocity profile can be neglected, essentially because the bursting phase is much shorter than the sweep phase. Of course the same may not be true of the normal and azimuthal profiles because the velocity v of the streaming flow is quite substantial. That point should be checked and kept in mind in three dimensionally modeling.

Nonetheless the solution of order $\varepsilon^0$ reproduced very well all the "turbulence" parameters of usual concern to CFD modellists: Reynolds stresses, production and dissipation of turbulence etc… section 7.5). It is immediately apparent that in the wall the traditional Reynolds stresses are identified with the slow Reynolds stresses in my decomposition and that the fast Reynolds stresses do not make a large contribution. This allows to nicely side step the thorny issue of modeling the Reynolds stresses in the wall layer that has bothered many authors.

Since the solution of order $\varepsilon^0$ is unsteady state, it cannot be incorporated directly into CFDs based on time averaged NS equations but is can be used to interface with solutions in the outer flow where the strength of the CFD lies, unless details of the wall layer itself are sought. Therefore the strategy of CFD modeling I propose shifts from modeling the terms in the RANS or TANS in the inner region to one of determining a proper position for this interface. This is the exercise presented in section 9.1. It defines the inner limit of applicability of the log-law. Applying the Prandtl-Nikuradse version of the law of the wall (e.g. in standard $k-\varepsilon$ model (B.E. Launder & Spalding, 1974) is tantamount to assuming that the wall layer has a fixed thickness of $\delta^+ = 64$. Clearly this will not capture the effect of transition Reynolds numbers (Figure 38) or even less the significant variations of wall layer thickness in the recirculation regions e.g. behind a backward facing step (table 1, Figure 37) This is what pushed many workers to develop special models to deal with low Re or special circumstances. The present technique can be applied equally to higher order closure models.

The solution of order $\varepsilon^0$ is relatively easy because it is based essentially on viscous diffusion of momentum. The use of the solution of order $\varepsilon^0$ provides a particularly useful tool for special conditions that have encouraged the development of the more complex closure models, e.g. variable density fluids, rotating systems because it is relatively easily derived. An example for non-Newtonian fluids is presented in section 10.

3. Outer limit of the log-law

Actually, the profile in the outer region is relatively insensitive to the model used (e.g. Bradshaw et. al. 1991) as long as the effect of the pressure term is reasonably well accounted for.

It therefore makes sense to start the iteration process in the bulk flow and then move towards the wall by linking with the log-law region. In older standard $k-\varepsilon$ models the switch from the iteration model to the law of the wall is performed quite arbitrarily. Figure 36 shows that the outer limit of the log-law varies considerably with the geometry and conditions of the outer flow. In some transitional flows, it may not even exist and the outer flow model should be coupled directly to the wall layer (e.g. Figure 36). Therefore a fixed arbitrary interface will add significant uncertainties to the solution obtained.

I argued that the log-law ceases to apply when the path of the ejected ceases to be linear and therefore equation (50) defines the outer limit of the log-law. It should be used as the defining condition for interfacing the outer flow iterations with the log-law.

**9.2    Matching the steady and unsteady state solutions**

The usefulness of juxtaposition of two separate formulations of the same problem led me naturally to matching Lagrangian and Eulerian descriptions of transport processes in the mid 1980s. My first attempt was made with heat transfer in a laminar boundary layer.

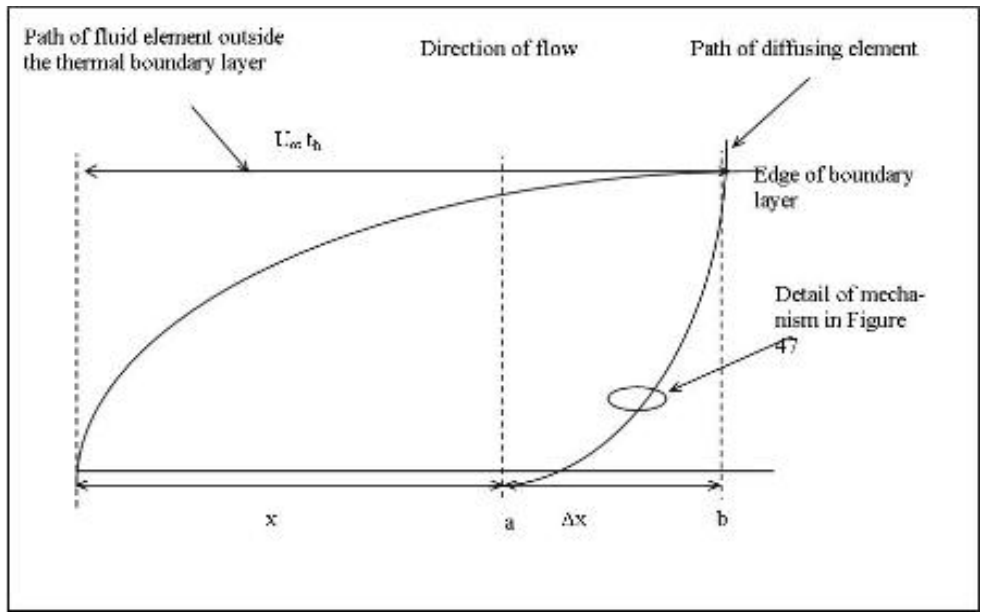

Figure 46 Diffusion path and convection velocities of an entity of heat.

When we think of it, the flow of heat can be described physically as an unsteady state phenomenon, even when the momentum and thermal boundary layers profiles for laminar flow are steady. For a flat plate, each element of fluid moves from the leading edge only once and will experience a change in momentum along a streamline due to the effect of diffusion of viscous momentum from the wall. Similarly each thermal front generated from the wall also moves across the boundary layer only once. For convenience we will refer to a small portion of this front as an "entity" of heat. Thus the motions of these elements of fluid and entities of heat can be described with unsteady state equations. The appearance of steady state profile profiles is perceived in an Eulerian framework because of an endless repetition of unsteady state movements of elements that must be described in a Lagrangian framework. The movement of the elements of fluid and entities of heat are in different directions

(Figure 46).

Consider next the nature of forces acting on an entity of heat. At time t, the entity of heat enters an element of fluid at position (x, y), drawn in full line and coloured red in Figure 47, which has velocities, u and v. At time (t + δt), the element of fluid has moved by convection to a new position, not shown in Figure 47, and the thermal entity has diffused to an adjacent element at (x + δx, y + δy), drawn in dotted lines and coloured orange with a brick pattern. *Since the thermal entity is a scalar property, it has no mass*, feels the effect of diffusion forces only and convects with the velocity of the fluid element where it temporarily resides. *The convective forces act on the host element of fluid, not on the entity of heat.*

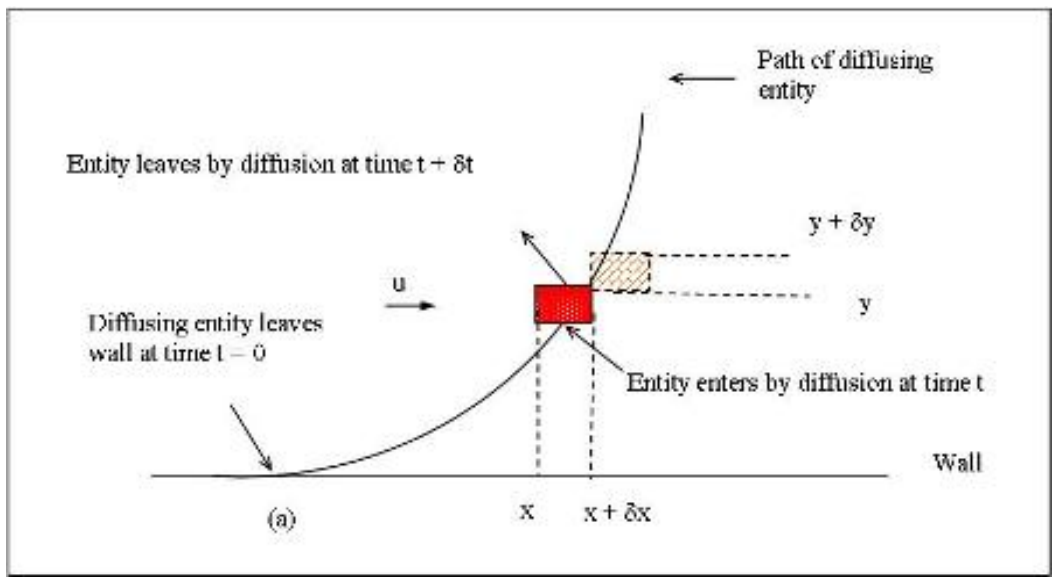

Figure 47 Convection and diffusion of an entity of heat

This physical analysis shows that the elements of fluid and heat move in different directions owing to different driving forces and it should be possible to separate the effects of diffusion and convection in the mathematical analysis. Consider now the mathematical description of this physical visualisation.

We begin with the unsteady state equation for heat diffusion

$$\frac{\partial \theta}{\partial t} = \alpha \frac{\partial^2 \theta}{\partial y^2} \tag{129}$$

Where $\theta$ is the temperature. Then assuming a third order polynomial temperature profile

$$\frac{\theta - \theta_w}{\theta_m - \theta_w} = 1.5\eta_h - \frac{1}{2}\eta_h^3 \tag{130}$$

where

$$\eta_h = \frac{y}{\delta_h(t)} \tag{131}$$

The thermal boundary layer thickness at time t is given by

$$\delta_h(t) = 2(2\alpha t)^{1/2} \tag{132}$$

The rate of heat transfer is

$$q_w = 0.53 k \Delta\theta_\infty (\alpha t)^{1/2} \tag{133}$$

Trinh and Keey (1992 a) proposed that the time scale $t_d$ of the diffusion process must be equal to the time that an entity of heat starting at x on the wall takes to move *across* the thermal boundary layer whereas previous proponents of the penetration theories (Higbie, 1935) based their time scale on the (convective) motion *along* the boundary layer

$$t = \frac{x}{U_\infty} \tag{134}$$

Trinh and Keey defined

$$t_d = \frac{x}{U_\infty f(u)} \tag{135}$$

where the variables x and $U_\infty$ have been retained only for convenience since they are easily measured. The reader is referred to the Trinh and Keey paper for the detailed calculation of the function f(u) and examples of its applications in several well known problems. Only the main steps are shown here. The function f(u) can be estimated from the integral energy equation as

$$f(u) = \frac{8}{3} M \frac{U_\infty}{U_d} \tag{136}$$

Where

$$M = \frac{\delta_d^*}{\delta_d} \tag{137}$$

is called usually a shape factor (Schlichting, 1979), 208), $\delta_d^*$ is the integral energy thickness, $\delta_d$ the boundary layer thickness and $U_\infty$ and $U_d$ are the velocities at the edge of the viscous and thermal boundary layers respectively. The ratio $U_\infty/U_d$ is a function of the ratio of the viscous and thermal boundary layer thicknesses $\sigma = \delta_d/\delta_v$.

For $Pr \gg 1$

$$f(u) = \frac{2}{5}\sigma - \frac{1}{35}\sigma^3 \tag{138}$$

and

$$Nu_x = 0.53\, Re_x^{1/2}\, Pr^{1/2} \sqrt{\frac{2}{5}Pr^{-b} - \frac{1}{35}Pr^{-3b}} \tag{139}$$

where the index b can be calculated from the implicit relation

$$Pr^{b-0.5} = 1.636\sqrt{\frac{2}{5}Pr^{-b} - \frac{1}{35}Pr^{-3b}} \tag{140}$$

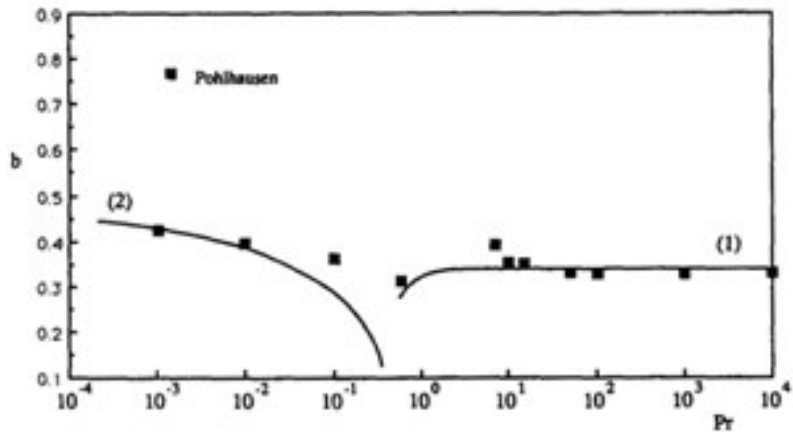

Figure 48 Variation of index b (From Trinh and Keey 1992a)

For Pr<1 a separate correlation must be derived for b because the thermal layer is now thicker than the momentum layer. (Trinh and Keey 1992a) as shown in line (2) in Figure 48. The derivation is very inaccurate for $0.1 < Pr < 1$ because it involves taking logarithms near the value zero.

$$Pr^{b-1/2} = 1.636\sqrt{1 + \frac{2}{3}Pr^b - \frac{8}{5}Pr^{2b} + \frac{32}{105}Pr^{4b}} \tag{141}$$

Figure 48 shows the predictions of *b* against the experimental data of Polhausen (Polhausen, 1921; H. B. Squire, 1942). For $Pr \gg 10$, $b \approx 1/3$ and we obtain the Polhausen equation

$$Nu_x = 0.324 \, Re_x^{1/2} \, Pr^{1/3} \qquad \qquad 142$$

Figure 49 shows a comparison between Nusselt numbers predicted by Trinh and Keey for a flat plate and the results from the well-known Polhausen analysis.

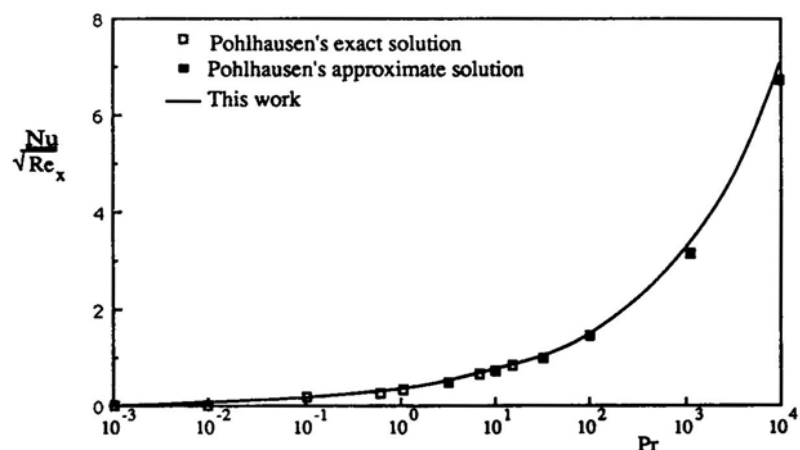

Figure 49  Comparison between the Nusselt numbers in the solutions of Polhausen (1921) and this work.

### 9.3   A new partial derivative for transport dynamics

My incursion into the matching of unsteady state and steady state solutions of the NS and RANS equations began with reading the work of Einstein-Li which I found truly inspiring, and then Meek and Baer. However, I could not agree with their physical visualisation of eddies from the outer flow bombarding a transient viscous sub-boundary layer given the work of Kline et al. Though I quickly found that the traditional parameters measured directly from velocity fluctuations in turbulent flow fields could be coordinated by the Stokes solution1 (section 7.5) there remained many uneasy and nagging questions.

My first concern lay with the Eulerian framework of the Stokes solution. Since the ground breaking work of Kline et al., we have known that the low speed streaks occur at seemingly random points in time and space. Yet the Stokes solution1 used by Einstein and Li and others after them is based on an Eulerian governing equation (59) . Is that reasonable? Mathematically speaking, a Lagrangian solution of the NS equation, even a subset, seemed to me to offer attractive advantages. In order to understand the behaviour of even a single coherent structure we need to follow it in time and space. Presumably, if the field of vision is large enough one can also extract the Lagrangian history of an unattached coherent structure from say a PIV experiment but there is scarcely any publication of that type as far as I know.

My second concern lay with the way the Stokes solution1 itself was derived. The governing equation for unsteady flow pat a flat plate with a zero pressure gradient is

$$\frac{\partial u}{\partial t} + u\frac{\partial u}{\partial x} + v\frac{\partial u}{\partial y} = \nu\frac{\partial^2 u}{\partial y^2} \tag{143}$$

Stokes neglected the convection terms to obtain equation (59) which was to apply to a flat plate suddenly set in motion. Applying this solution to the sweep phase is not without controversy since the time scale is about 700 microseconds according to the data of Meek and Baer and the convection terms not quite negligible. They certainly are not when the Stokes solution1 is transformed into the Blasius steady state solution for laminar boundary layer flow past flat plates (Trinh & Keey, 1992b).

The underlying principle there is that the decoupling of the diffusion and convection forces in convective transport as discussed in section 9.2. There was a need to formalise the mathematical expression of this decoupling. We start with the governing equation for heat transfer

$$\rho C_p \left( \frac{\partial \theta}{\partial t} + v_x\frac{\partial \theta}{\partial x} + v_y\frac{\partial \theta}{\partial y} + v_z\frac{\partial \theta}{\partial z} \right) = k\nabla^2\theta \tag{144}$$

where $\rho$ is the density, k the thermal conductivity and $C_p$ the thermal capacity of the fluid. For a laminar boundary layer past a flat plate, equation (144) can be simplified to

$$\frac{\partial \theta}{\partial t} + u\frac{\partial \theta}{\partial x} + v\frac{\partial \theta}{\partial y} = \alpha\frac{\partial^2 \theta}{\partial y^2} \tag{145}$$

Equation (145) may be rewritten as

$$\frac{D\theta}{Dt} = \alpha \frac{D^2\theta}{Dy^2} \tag{146}$$

where

$$\frac{D\theta}{Dt} = \frac{\partial \theta}{\partial t} + u\frac{\partial \theta}{\partial x} + v\frac{\partial \theta}{\partial y} \tag{147}$$

is called the substantial derivative.

Bird *et al.* (1960, p.73) illustrate the difference between the Eulerian partial derivative ∂θ/∂t and the Lagrangian substantial derivative Dθ/Dt with the following example. Suppose you want to count the fish population in a river. The Eulerian partial derivative gives the rate of change in fish concentration at a fixed point (x,y) in the river as seen by an observer standing on the shore. An observer in a boat drifting with the current will see the change in fish concentration on the side of the boat as given by the substantial derivative.

However, a fish swimming in the river will have a different perception of the fish population, which not described by either of these derivatives. Clearly a Lagrangian derivative is required but the convection velocities u and v are no longer relevant to this case; the velocity and path of the fish are. The introduction of this new kind of derivative, which is clearly needed, lies at the centre of this discussion (Trinh, 2002).

The fish in the previous example is represented here by the entity of heat illustrated in Figure 47. An observer attached to the *entity of heat* moving across the boundary layer will perceive the changes in temperature according to an equation similar to (Higbie, 1935)

$$\frac{\partial \theta}{\partial t} = \alpha \frac{\partial^2 \theta}{\partial y^2} \tag{129}$$

but the frame of reference in the penetration theory is not attached to the wall as implied by Highbie. A new symbol $\mathcal{D}/\mathcal{D}t$ must be introduced for this new Lagrangian derivative *along the path of diffusion*. For a laminar boundary layer on a flat plate we obtain:

$$\frac{D\theta}{Dt} = \alpha \frac{D^2\theta}{Dy^2} \tag{148}$$

This equation has the same form as the Fourier equation for heat transfer e.g. (Incropera, Dewitt, Bergman, & Lavine, 2007) that applies strictly only to stationary media, but equation (148) can monitor the diffusion of heat into a convection stream along the path of thermal diffusion and give a more formal basis to penetration theories.

The decoupling of the flow paths of heat and fluid elements in unit operations is evident in the analysis of some well established unit operations. Consider for example the working of a multiple effect evaporator (K. T. Trinh, 2005, Figure 50). Saturated fresh steam enters the steam chest of first effect outside the tube bundle and condensed. It condenses and releases its latent heat which passes through the tube wall to the milk stream inside the tubes. The water in the milk, called cow water, evaporates. This cow water vapour is separated from the concentrated milk at the bottom of the first calandria through a separator and flows into the vapour chest of the next effect. There it condenses and releases its latent heat to further evaporate cow water from the concentrated milk line. Very clearly the system re-uses the latent heat fed in with the fresh steam not the fluid in the fresh steam. In fact the condensate in the second effect (dotted green line) is collected in a line physically separate from the fresh steam condensate (continuous green line) which is cleaner and can be returned to the boiler.

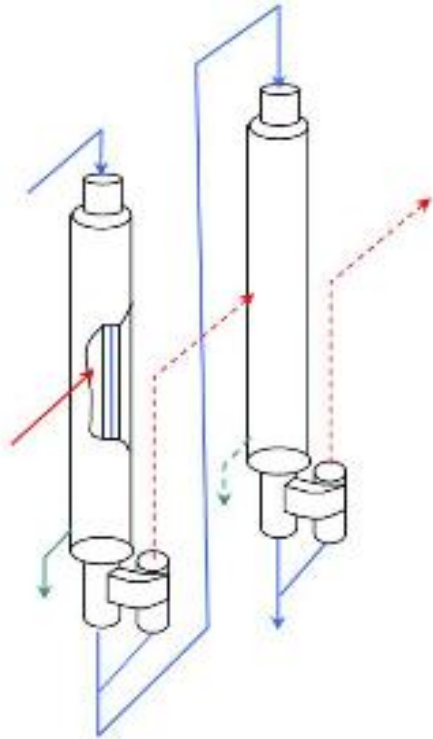

Figure 50 Heat and fluid flow in a two effect evaporator for milk. Red continuous line: fresh steam, red dotted line: vapour from milk cow water, blue line: milk line, green line: condensate from steam and cow water vapour

In my introductory lecture on turbulent transport for newcomers to the field, I illustrate the mechanism with a video showing the transport of potassium permanganate from a small cell into a beaker of water.

In Figure 51(a) the colour of the permanganate spreads slowly by diffusion over the time scale of six consecutive frames. In Figure 51(b) the colour spreads over the entire beaker in the same time scale because the source cell is moved (convected) but if we look carefully, at each new position the diffusion process has not changed.

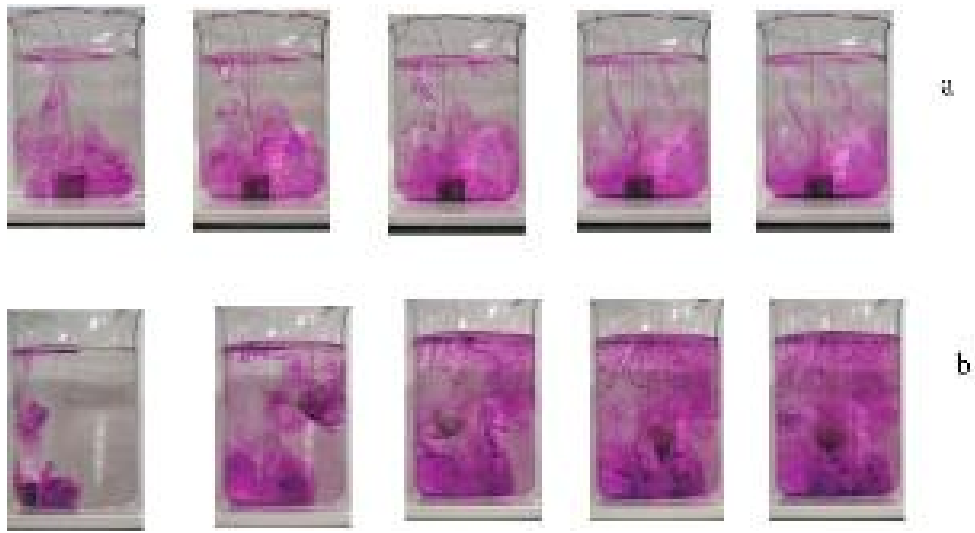

Figure 51 Diffusion of potassium permanganate from a fixed source (a) and a moving source (b)

We may thus explain convective transport with two distinct mechanisms Figure 52. In laminar flow, heat is transferred through the wall to the convective stream travels by diffusion (equation (129) across the boundary layer. The convection forces do not act directly on the heat elements themselves but remove the heat together with the fluid elements convected in the streamwise direction to maintain a high temperature gradient. Of course, in a stagnant fluid without a heat sink, the temperature eventually becomes uniform and heat transfer ceases.

In turbulent transport, heat is first transferred through the wall into the low-speed streaks by diffusion. These are then ejected as coherent structures and because they have higher temperature than the outer quasi-inviscid flow, they represent moving heat sources from which heat is again diffused into the new surroundings. Because the same ejected eddies contain simultaneously heat, mass and momentum, the laws that govern heat, mass and momentum in the log-law and outer region must naturally have analogous forms.

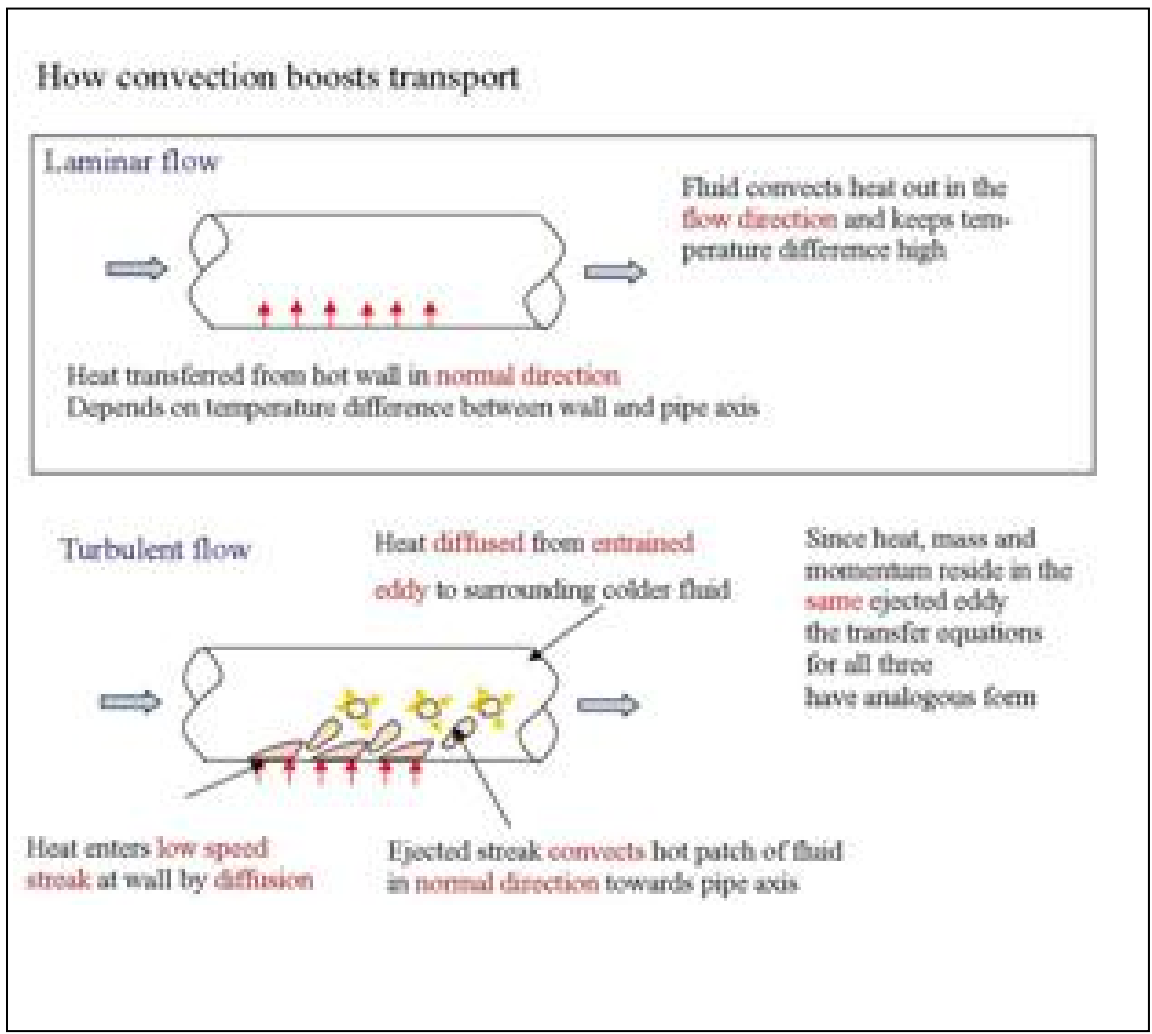

Figure 52 Effect of convection on transport dynamics

Of course the introduction of a new partial derivative to the literature of mathematics is not a trivial exercise. I was hoping that my mathematician colleagues would not view this as an impudent encroachment into their field. In fact I am very poor at mathematics and simply could not deal with the legendary non-linearity of the NS equations so took the opportunity to remove the convection terms from the solution. I was behaving like any half decent engineer faced with an impossibly complex task, be it building the longest suspension bridge across a highly turbulent sea span, the fastest undetectable deep sea submarine or a theoretical solution to a confusing and little understood phenomenon. I simplified the problem by inventing a new tool and taking a different approach.

Thus in 1994 I floated my concept in a discussion with our professor of applied mathematics Graeme Wake. His muted response was "It rings true but what are the rules of this new derivative?" I floated the concept to a larger audience in 2002 (Trinh, 2002) and again the response was mixed. Though there was no disagreement with the physical visualisation behind this new derivative and enthusiastic support from some, notably the session chairman, the average response was somewhat muted bewilderment as the audience tried to digest the implications to transport analysis. Clearly more details were needed.

It must be remembered that the temperature at any point $(x_L, y_L, t)$ in the solution to equation (148) does not correspond to the value of the steady state solution obtained, for example by Polhausen (op.cit.) because the set $(x_E, y_E)$ in Polhausen's solution refers to a fixed point in space whereas for the set $(x_L, y_L)$ that can be derived in this Lagrangian solution refers to an element of heat at time t. Lagrangian solutions are useful in providing the history of targeted diffusing elements and therefore insight into transport mechanisms, but they cannot be used directly in practical engineering applications that must necessarily be developed in an Eulerian framework.

It must be clearly understood that this new partial derivative along the diffusion path of transport quantities is not the same as the traditional Lagrangian derivative (Bennett, 2006) that follow the path of convection. Such Lagrangian derivatives capture the changes to the fluid elements along their convection path.

One way of validating the existence of this new partial derivative is of course to show that it can predict accurately well known transport problems. I chose to compare the predictions of the new Lagrangian solutions with well known Eulerian solutions that are widely accepted, rather than compare directly with experimental measurements. We illustrate this exercise with two simple examples.

We first try to estimate the thickness of the laminar boundary layer on a flat plate. It represents maximum penetration of viscous momentum from the wall, a physical phenomenon which is independent from the frame of reference of the mathematical analysis. In the classic solution of Blasius (1908) it is given by

$$\delta_v = 5\sqrt{x\nu} \tag{149}$$

Equation (149) is plotted against the Reynolds number

$$\mathrm{Re}_x = \frac{xU_\infty}{\nu} \tag{150}$$

in Figure 53 for a specific example detailed in Table 2.

Table 2. Parameters used in example of laminar flow and heat transfer past a flat plate

| Test fluid used | Water | unit |
|---|---|---|
| Temperature | 20 | °C |
| Viscosity | 0.001002 | kg/ms |
| Density | 998 | kg/m$^3$ |
| Heat capacity | 4.182 | kJ/kgK |
| Thermal conductivity | 0.603 | W/mK |
| Prandtl number | 6.935296 | |
| Approach velocity | 0.2 | m/s |

The physical properties were taken from (Bayley, Owen, & Turner 1972)

The thickness of the boundary layer in the Lagrangian solution is given by (64) and can be rearranged as.

$$\delta_s = 3.97\sqrt{4\nu t} \tag{151}$$

To transform time into distance, we call on Taylor's hypothesis

$$x = \int u\,dt = \overline{U} t_v \tag{152}$$

The average velocity $\overline{U}$ is simply calculated from the Blasius solution according to the method of Trinh and Keey (op.cit.) and sketched in section 9.2. The time scale $t_v$ is then substituted into equation (152). The boundary layer thickness calculated from the Stokes solution1 is 6.5% smaller than the value obtained by Blasius. However, when we use the approximate solution to equation (59) (Bird Stewart Lightfoot, 1960) and the velocity profile of the Polhausen (1921), the results coincide almost exactly with those of Polhausen as shown in Figure 53.

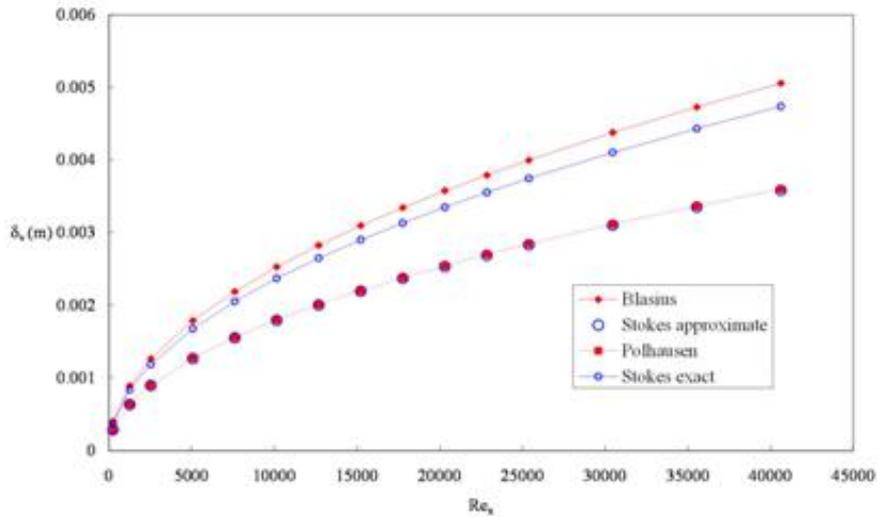

Figure 53 Edge of a laminar boundary layer on a flat plate from Eulerian and Lagrangian solutions

In the second example, we follow an element of heat through a thermal boundary layer. We keep the flow conditions in Table 2 and raise the plate temperature from a position $x_0 = 0.05m$ to 25°C for a 5°C driving force. We monitor a heat element starting at the plate at the position $x = 0.10m$.

We divide the flow field into a rectangular grid. The velocity at each of the node points of that grid is calculated from the Polhausen approximate solution for laminar flow past a flat plate. The velocity associated with this rectangular element of fluid is simply taken in this example as the arithmetic average of the velocities at the four corners of the grid point

$$\bar{u} = \left( \frac{u(x,y) + u(x, y+\delta y) + u(x+\delta x, y+\delta y) + u(x+\delta x, y)}{4} \right) \qquad (153)$$

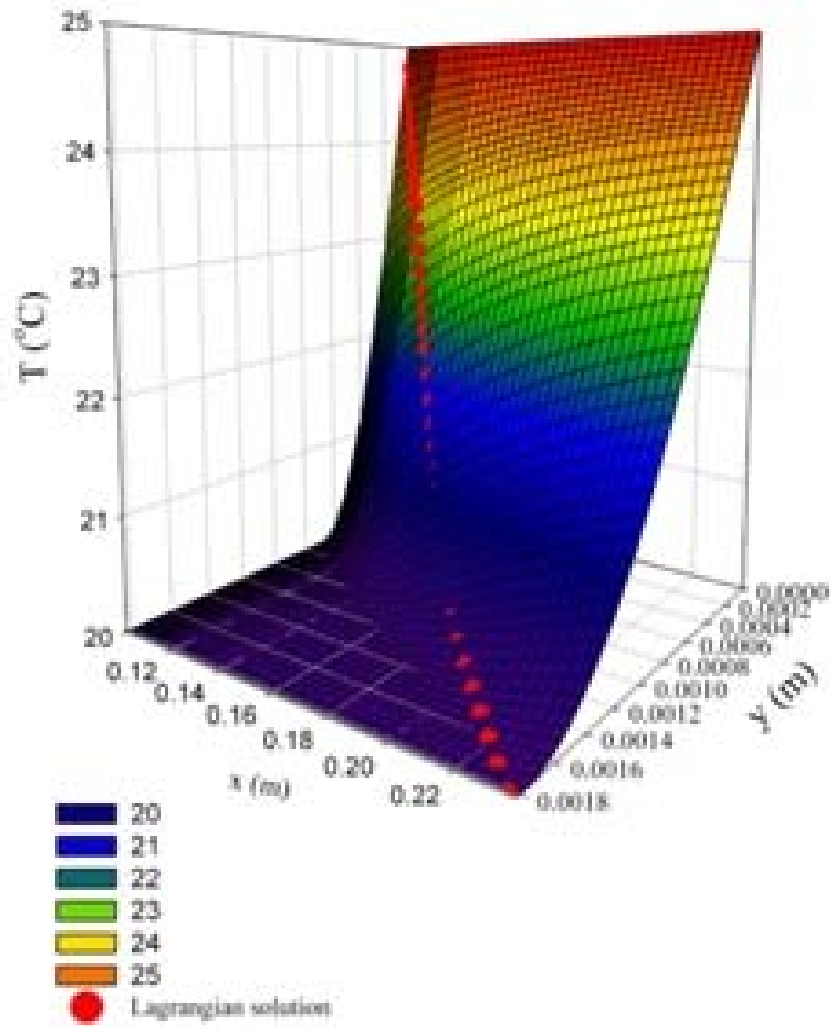

Figure 54 Diffusion path for a heat element starting at the wall at x=0.1 m and comparison with the Polhausen approximate solution for heat transfer in a laminar boundary layer

We then calculate the time elapsed

$$\delta t = \frac{\delta x}{u} \qquad (154)$$

Thus for the first cell with $\delta x = 0.00125m$, $\delta y = 1.64337E-05$ the convection velocity of the fluid element is 0.003738359m/s, the time increment associated with diffusion of heat between points (0.10000,0) and (0.10125, 1.64377E-05) is 0.334371309s. The temperature at position (0.10125, 1.64377E-05) is then obtained from the solution of equation (148)

$$\frac{\theta_w - \theta}{\theta_w - \theta_\infty} = erf\left(\frac{y}{\sqrt{4\alpha t}}\right) \qquad (155)$$

Giving θ =24.78933904°C, The next element with corners (0.10125, 1.64377E-05), (0.1025, 1.64377E-05), (0.10125, 3.28754E-05) and (0.1025, 3.28754E-05) has velocity 0.011145875m/s with a time increment 0.112149s. Thus point (0.1025, 3.28754E-05) is associated with a time elapsed of (0.334371309 + 0.112149 =0.446520419s) giving a temperature of 24.63574535°C. This procedure allows us to follow the diffusion of heat between adjacent fluid elements.

The results of the Polhausen approximate solution for steady state (Eulerian) heat transfer in a laminar flat plate boundary layer (Polhausen, 1921; H. B. Squire, 1942) are presented in a 3D mesh plot in Figure 54. The results from equation (155) are overlaid as a 3D scatter plot. They fall almost exactly on the response surface of the mesh plot. In fact the standard deviation between the 38 points generated in the Lagrangian solution (scatter plot) and the corresponding sets of (x,y) in the Polhausen solution is 0.01%. The edge of the boundary layer in the mesh plot is clearly seen where the response surface flattens to a horizontal plane ($\theta = 20^oC$ coloured dark purple). The red scatter plot shows that the Lagrangian solution does indeed predict the boundary layer thickness accurately but supports and illustrate the argument of Trinh and Keey (op.cit.) that the thickness of thermal boundary which began at $x_o = 0.05m$ in this example, is not situated vertically above the position where the element of heat left the plate ($x_1 = 0.10m$) in this example but at a distance much further downstream ($x = 0.235m$) in this case. Simultaneously, we obtain an estimate of the period of for the diffusion process ($t_d = 1.207464\,s$) and can calculate the rate of penetration of the thermal front, which is not available in the Eulerian solution. This extra information can be useful in many applications, for example in mixing processes.

### 9.3.1 Further observations on this new derivative and its potential applications

1. While equations (146), (129) and (148) have similar forms, only $\partial\theta/\partial t$ and $D\theta/Dt$ are truly partial derivatives of the variable $\theta$ with respect to time. The substantial derivative $D\theta/Dt$ is not. Thus similar techniques of analysis may be applied to equations (129) and (148) provided one remembers that the frames of reference for the solutions are different. The same is true for the traditional Lagrangian analysis.
2. This derivative allows us to extract from the NS equations subsets that are much more solvable because of the reduced problems of non linearity as the convection terms are omitted. As I tried to formalise the description and proof of this derivative I could envisage a new approach to the solution of the NS equations: breaking them into manageable subsets and reconnecting the individual solutions to obtain the global one. Unfortunately I have never could carve out the time from my bread-and-butter duties to think seriously about this possibility.
3. Computer aided solutions are much easier to envisage. A similar analysis for momentum diffusion that turns the Stokes solution1 into the Blasius laminar boundary layer solution for flat plates will be presented separately. Trinh and Keey (1992b) have already introduced that procedure in general terms and it was used to describe the "moving front of turbulence" in the wall layer in section 7.5.2.2
4. The evidence in Figure 36 shows that equation (60) adequately correlates the velocity profile in the wall layer of flows in all geometries and types of fluid. We may thus use the present transformation to effect closure in CFD's in an approach similar to that presented in section 9.1: we match the edge of the unsteady and steady state wall layers to define the inner limit of the log-law.
5. This transformation exercise is not a traditional CFD process based on discretisation of the original partial differential equations and iterations. It starts with the matching of the integrated solutions and therefore is much less demanding of computer time.
6. It is useful to use a fine grid close to the wall to minimise the effect of the

singularity introduced by the no-slip condition at the wall but we can switch to much coarser grid elements afterwards without great loss of accuracy. This has been applied in Figure 54.

7. The decoupling of the diffusion and convection processes suggests another method of CFD modelling similar to the large eddy simulation LES technique but instead of modelling the large scales only and using the experimental statistical data for the fine scales, we model the large convection patterns with the traditional CFD and overlay it with the diffusion analysis. It is useful here to point out that there is a fundamental difference between the diffusion transport from the wall to the fluid in the low speed streak and the transport from the ejected streak to its new surroundings. In the former situation, the source of the transport quantity is constant whether it is heat, mass or momentum; in the latter situation, there is no longer any renewal of the heat, mass or viscous momentum in the ejected eddy which now becomes a decaying source.

8. In fact the same decoupling and recoupling approach may allow is to significantly reduce the resource requirements of DNS making them eventually applicable to more complex industrial problems than the present DNS that are constrained because of computer time to relatively low Reynolds numbers and simple geometries.

## 10   Non-Newtonian flow

We continue to explore the implications of the present approach to turbulence studies. The next point is best made by studying non-Newtonian fluid flow.

### 10.1   Purely viscous power law fluids

Consider now the solution of order $\varepsilon^0$ for a power law fluid

$$\tau = K\,\dot{\gamma}^n \tag{156}$$

as a case study for non-Newtonian turbulent flow (Trinh, 1994; Trinh, 1999).

    K is called the consistency coefficient,

    n        the flow behaviour index, and

$\dot{\gamma}$      the shear rate

The governing equation for Stokes' law is

$$\frac{\partial \tilde{u}}{\partial t} = -\frac{1}{\rho}\frac{\partial \tau}{\partial y} \tag{157}$$

Substituting equation (156) into (157) gives

$$\rho\frac{\partial \tilde{u}}{\partial t} = n\,K\left(\frac{\partial \tilde{u}}{\partial y}\right)^{n-1}\frac{\partial^2 \tilde{u}}{\partial y^2} \tag{158}$$

The relative distance $\eta(t,y)$ into the viscous sub-boundary layer near the wall $\delta_i(t)$ is defined as

$$\eta(t,y) = \frac{y}{\delta_i(t)} \tag{159}$$

and the velocity $\phi$ relative to the velocity $U_v$ at the edge of the wall layer is

$$\phi = \frac{\tilde{u}}{\tilde{U}_v} \tag{160}$$

Substituting these new variables into equation (158) and integrating with respect to $\eta$ gives

$$\delta_i^n \frac{d\delta_i}{dt} = \frac{K}{\rho}\tilde{U}_v^{n-1}\frac{N}{M} \tag{161}$$

in which

$$N = \int_0^1 n(\phi')^{n-1}(\phi'')d\eta(t,y) = [(\phi')^n]_0^1 \tag{162a}$$

and

$$M = \int_0^1 (\phi')\eta(t,y)d\eta(t,y) = 1 - \int_0^1 \phi\,d\eta(t,y) \tag{162b}$$

where the primes denote derivatives with respect to $\eta$.

Equation (161) can now be integrated separately with respect to the variables $\delta_i$ and t over a characteristic time T:

$$\int_0^{\delta_e} \delta_i^n\,d\delta_i = \int_0^T \frac{K}{\rho}\tilde{U}_v^{n-1}\frac{N}{M}dt \tag{163}$$

The instantaneous wall-layer thickness at time $t_v$, which coincides with the onset of ejection, is

$$\delta_v = \delta_i(t_v) = \left[\frac{K}{\rho}(n+1)\tilde{U}_v^{n-1}\frac{N}{M}t_v\right]^{1/(n+1)} \tag{164}$$

The instantaneous wall shear stress is:

$$\tau_{w,i} = K\left[-\left(\frac{\partial \tilde{u}}{\partial y}\right)_{w,i}\right]^n \tag{165}$$

$$\tau_{w,i} = K\left[\tilde{U}_v(\phi')_w \frac{\partial \eta}{\partial y}\right]^n \tag{166}$$

$$\tau_{w,i} = KN\frac{U_v^n}{\left[\delta_i(t)\right]^n} \tag{167}$$

The time-averaged wall shear stress is given by

$$\tau_w = \frac{1}{t_v}\int_0^{t_v} \tau_{w,i}\,dt \tag{168}$$

$$\tau_w = K\,N\,(n+1)\frac{\tilde{U}_v^n}{\tilde{\delta}_v^n} \tag{169}$$

$$\tau_w = (n+1)\tau_e \tag{170}$$

where $\tau_e = \tau_{w,t_v}$ is the instantaneous wall-shear stress at time $t_v$ which coincides with the end of the low-speed-streak phase and the onset of ejection. Henceforth, this wall shear stress will be called the *critical local instantaneous wall-shear stress at the point of ejection* or simply the *critical shear stress*. The suffix e has been introduced as an alternative to $w, t_v$ for convenience and to highlight the fact that the variables associated with that suffix are estimated at the time just before ejection. Similarly $\tilde{U}_e = \tilde{U}_v$ and $\delta_e = \delta_v$.

The thickness of the Stokes layer $\delta_v$ at the end of the period $t_v$ is related to the critical approach velocity $U_v$ by putting $t = t_v$ in equation (167) and rearranging:

$$\delta_v = \left[\frac{KN}{\tau_e}\right]^{1/n} \widetilde{U}_v \tag{171}$$

The time-averaged shear velocity is usually given the symbol $u_*$ and defined as:

$$u_* = \sqrt{\frac{\tau_w}{\rho}} \tag{172}$$

We define a new normalising parameter, the *critical* shear velocity $u_{e*}$

$$u_{e*} = \sqrt{\tau_e/\rho} = \sqrt{\frac{\tau_w}{\rho(n+1)}} = \frac{u_*}{\sqrt{n+1}} \tag{173}$$

The thickness of the Stokes layer may be normalised with the critical wall shear stress as:

$$\delta_e^+ = \frac{\delta_v u_{e*}}{\nu_e} = \frac{\delta_e u_{e*}^{2/n-1} \rho^{1/n}}{K^{1/n}} \tag{174}$$

The normalised critical approach velocity at the point of bursting is

$$\widetilde{U}_e^+ = \frac{\widetilde{U}_e}{u_{e*}} \tag{175}$$

Combining equations (171), (173), (174) and (175) gives

$$\delta_e^+ = N^{1/n} \widetilde{U}_e^+ \tag{176}$$

The coefficients M and N can be determined once the relative velocity $\phi$ as a function of $\eta$ is known. Following (Polhausen, 1921) and (Bird, Stewart, & Lightfoot, 1960), we assume that the velocity profile can be described approximately by a third-order polynomial:

$$\phi = 1.5\eta - 0.5\eta^3 \tag{177}$$

Back-substitution into equations (162a) and (162b) respectively yields the unknown coefficients:

$$N = (3/2)^n \tag{178}$$

and

$$M = 3/8 \tag{179}$$

Substitution of equation (178) into (176) gives

$$\delta_e^+ = 1.5 \widetilde{U}_e^+ \tag{180}$$

Equation (180) shows clearly that the relation between the wall layer thickness and the critical approach velocity of the Stokes layer at the point of bursting is independent of the fluid rheology, specifically the flow behaviour index, when normalised with the critical instantaneous shear velocity.

We should note that for Newtonian fluids, n=1, equation (178) gives

$$N_{n=1} = 3/2 = (N)^{1/n} \tag{181}$$

Therefore equation (160) may be written as:

$$\delta_e^+ = N_{n=1} \widetilde{U}_e^+ \tag{182}$$

Thus the previous conclusion is not dependent on the form of the velocity profile assumed. For example, the same derivation leading to equation (182 may be performed with a fourth-order polynomial also proposed by (Polhausen, 1921). Of course the numerical value of the coefficient N changes with the velocity profile assumed but equation (180) does not. In the exact Stokes solution (Bird, 1959)

$$N_{n=1} = 2.08 \tag{183}$$

We recall that the zonal similarity analysis collapses both Newtonian and power law fluids onto the same velocity profile (Figure 36). Data in the wall layer correlated well with the time averaged Stokes solution1 for Newtonian fluids. In fact if we take the trouble of integrating numerically the velocity profile in the exact solution for power law fluids (Bird, 1959), we find that this profile does not differ much from the integrated erf in equation (60) . In addition the time average of the solution of equation (177) also fits the data.

Because instantaneous shear and velocity profiles are very difficult to measure, it is convenient to re-express these instability criteria in terms of time-averaged shear stresses through the use of equation (170) . Since the edge of the wall layer is defined by the maximum penetration of viscous momentum from the wall then the time averaged wall layer thickness is equal to the thickness of the transient viscous sub-boundary layer at the point of ejection $\delta_v = \delta_e$ and $\widetilde{U}_v = \widetilde{U}_e$. Then the velocity at the edge of the wall layer, normalised with the time-averaged shear velocity, becomes

$$\widetilde{U}_v^+ = \frac{\widetilde{U}_e}{\sqrt{\tau_w/\rho}} = \frac{\widetilde{U}_e^+}{\sqrt{n+1}} \tag{184}$$

Similarly

$$\delta_v^+ = \frac{\delta_e u_*^{2/n-1} \rho^{1/n}}{K^{1/n}} = \delta_e^+ (n+1)^{2-n/2n} \quad (185)$$

Combining equations (182), (184) and (185) gives

$$\delta_v^+ = 2.08(n+1)^{1/n} \widetilde{U}_v^+ \quad (186)$$

Equations (182) and (186) indicate that the apparent thickening of the wall layer, seen in traditional velocity plots normalised with the time-averaged shear velocity $u_*$ such as those of (Bogue, 1961), are not real. This apparent thickening is the consequence of an integration process, which relates the critical instantaneous wall shear stress, the normalising parameter with physical significance, to the time-averaged wall shear stress, which is traditionally more easily measured.

The critical apparent (non-Newtonian) kinematic viscosity at the wall $\nu_e$ is defined as

$$\nu_e = \frac{\tau_{w,T}}{\rho(-\partial u/\partial y)_e} = K^{1/n} \tau_e^{(n-1)/n} \quad (187)$$

The critical instantaneous Reynolds number becomes

$$\text{Re}_e = \frac{DV}{\nu_e} = \frac{DV}{K^{1/n} \tau_e^{(n-1)/n}} \quad (188)$$

and the critical instantaneous friction factor is

$$f_e = \frac{2\tau_e}{\rho V^2} \quad (189)$$

These definitions can be compared with the more conventional definitions of the time-averaged friction factor

$$f = \frac{2\tau_w}{\rho V^2} \quad (190)$$

and the generalised Metzner-Reed Reynolds number (Metzner & Reed, 1955)

$$\text{Re}_g = \frac{D^n V^{2-n} \rho}{K 8^{n-1} \left(\frac{3n+1}{4n}\right)^n} \quad (191)$$

The relation between the critical instantaneous Reynolds number and the generalised Metzner-Reed Reynolds number can be derived from equations (170), (187), (188), (190) and (191):

$$\text{Re}_e = \left(\text{Re}_g\, f^{1-n}\right)^{\frac{1}{n}} 2^{\frac{5(n-1)}{n}} \left(\frac{3n+1}{4n}\right)\left(\frac{n+1}{2}\right)^{\frac{n-1}{n}} \qquad (192)$$

The friction factor in viscous non-Newtonian pipe flow has been measured by (Dodge, 1959), Bogue (1961) and (Yoo, 1974). Figure 55 shows a plot of time-averaged friction factor against the Metzner-Dodge Reynolds number.

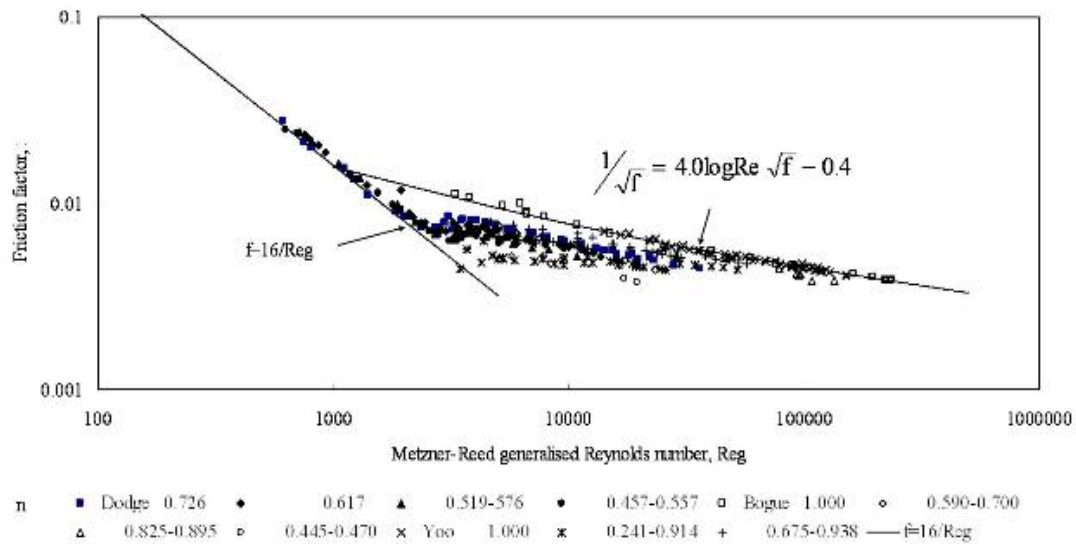

Figure 55 Plot of time-averaged friction factor against generalized Metzner-Reed Reynolds number.

The data for different values of the flow behaviour index, n, fall on different lines. Figure 56 shows a plot of the critical instantaneous friction factor against the critical instantaneous Reynolds number, defined in terms of the critical wall shear stress. All the data falls on a unique plot.

For Newtonian fluids, the critical friction velocity is related to the time averaged friction velocity by putting n = 1 in equation (173) and:

$$u_{e*} = u_* \sqrt{2} \qquad (193)$$

Similarly equation (170) gives

$$\tau_w = 2\tau_e \qquad (194)$$

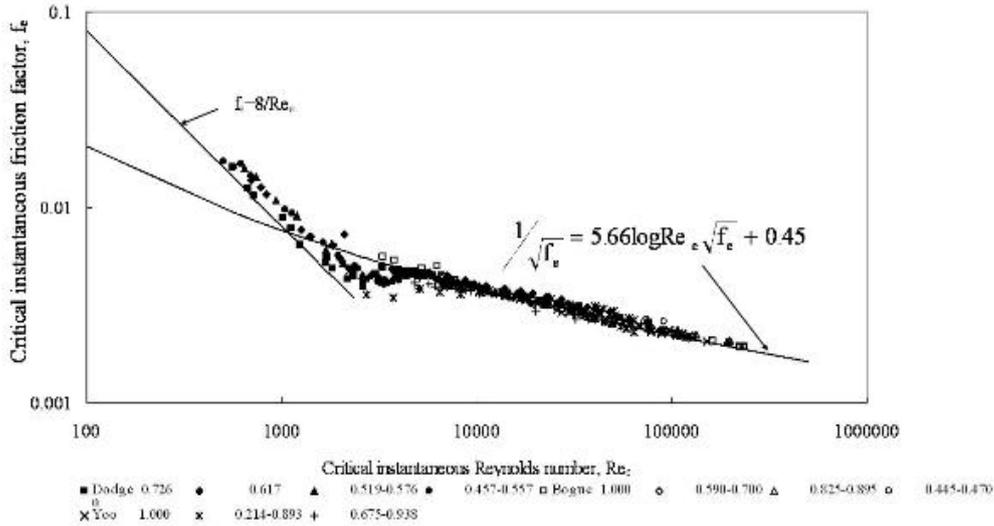

Figure 56. Similarity plot of instantaneous critical friction factor against Reynolds number.

$$f = 2f_e \qquad (195)$$

Prandtl's logarithmic law (Prandtl, 1935); (Nikuradse, 1932)

$$\frac{1}{\sqrt{f}} = 4.0 \log(Re \sqrt{f}) - 0.4 \qquad (196)$$

may be rewritten in terms of the critical wall shear stress

$$\frac{1}{\sqrt{f_e}} = 5.66 \log(Re_e \sqrt{f_e}) + 0.45 \qquad (197)$$

Equation (197) fits the data of Dodge, Bogue and Yoo quite closely, as shown in Figure 56. The agreement is not perfect because the wall layer analysis has been made for flow past a flat surface. The application of this analysis to circular pipe flow implies that the curvature effects can be neglected, which is true only at high Reynolds numbers when the wall layer is very thin compared to the pipe radius.

Similarly, the Blasius power law (Blasius, 1913)

$$f = \frac{0.079}{Re^{1/4}} \qquad (94)$$

may be rewritten as

$$f_e = \frac{0.04}{Re_e^{1/4}} \qquad (198)$$

Equation (198) also fits the data presented in Figure 56.

The apparent viscosity of non-Newtonian fluids changes with the applied shear stress or shear rate. Therefore, it is important for engineering correlations to identify exactly the shear stress (or shear rate) at which the viscosity used to normalise the velocity $U$, the distance y and at which the Reynolds number is calculated. Experimental data shows that both the velocity profile e.g. (Bogue & Metzner, 1963) and the friction factor-Reynolds number curves (Figure 55) of power law fluids are shifted when the velocity and distance are normalised with the viscosity calculated at the time averaged value of the wall shear stress. The data of Bogue (1961) and others (K.T. Trinh, 2005a) show that the wall thickness $\delta_v^+$, normalised with the time averaged wall shear stress, becomes thicker as the behaviour index n decreases. In an effort to collapse the non-Newtonian data onto their Newtonian counterparts, many authors have resorted to different definitions of effective viscosity (e.g. Alves et al., Metzner and Reed, Edwards and Smith op. cit.). The present paper argues that since turbulence is a time-dependent phenomenon, the viscosity should be estimated at the value of the local instantaneous wall shear stress, not the time averaged value. When this is done, there is no need to define an effective viscosity.

The more important point to stress again here is that there is a danger in starting our analyses from the RANS equations because the implied constant of integration has misled a number of authors to argue that the mechanism of turbulence is different in Newtonian and non-Newtonian fluids e.g.(Tennekes, 1966) I argue that they are not.

**10.2   Other fluid models**

The derivation in section 10.1 can easily be applied to fluids with other rheological equations of state such as Bingham Plastic and Herschel-Bulkey fluids (Trinh, 1994 unpublished). But a more useful approach can be taken for the general case.

The collapse of literature data into a single curve when the friction factor and Reynolds numbers are expressed in terms of the instantaneous wall shear stress at the point of ejection is not perfect because the Stokes solution applies only to a flat surface and curvature cannot be neglected. We can remedy that situation by using the Szymanski analysis (Szymanski, 1932) for unsteady state pipe flow discussed in section 12.3.2. For the moment let us accept that this solution yields the wall layer thickness in power fluids as

$$\delta_v^+ = \left(\delta_v^+\right)_{n=1} \left(\frac{3n+1}{4n}\right) \tag{199}$$

The reader may recognise the product $\left([3n+1]/4n\right)$ as the difference between the shear rate at the pipe wall for power law fluids and the shear rate for Newtonian fluids at the same flowrate (Skelland, 1967)

$$\frac{3n+1}{4n} = \frac{\dot{\gamma}_w}{\left(\dot{\gamma}_w\right)_{n=1}} = \frac{\dot{\gamma}_w}{8V/D} \tag{200}$$

where V is the average velocity and D the pipe diameter. In fact we can use a more general form of equation (200) that is applicable to all non-Newtonian fluid rheological model by using the behaviour index $n'$ which is the slope of the log-log plot of $\tau_w$ and $8V/D$ as shown by the derivation of Mooney and Rabinowitsch (Skelland op.cit)

$$\frac{3n'+1}{4n'} = \frac{\dot{\gamma}_w}{\left(\dot{\gamma}_w\right)_{n=1}} = \frac{\dot{\gamma}_w}{8V/D} \tag{201}$$

As discussed in section 9.1, the log law must pass through the point $U_v^+, \delta_v^+$ which allows obtain a velocity profile and an estimate of the friction factor using the well known methods of Prandtl (1935) and Karman(1934). Instead of using the point $U_v^+, \delta_v^+$ as a forcing constant for the log law, we can use the point $U_k^+, y_k^+$ (Trinh, 1969) discussed in section 7.6 because equation (90) conveniently gives

$$U_k^+ = y_k^+ \tag{202}$$

To a good approximation the Kolmogorov scale in non-Newtonian fluids is also related to its value in Newtonian fluids in the same proportion as $\delta_v^+$ as can be seen by plotting equation (202) against the data of Bogue (1961) used in Table 1. We should note that because of experimental errors not all the data of Bogue show an apparent thickening of the wall layer.

$$y_k^+ = (y_k^+)_{n=1} \left( \frac{3n'+1}{4n'} \right) \qquad (203)$$

Then at high Reynolds numbers, when $(y_k^+)_{n=1} = 11.8$ the log-law becomes conveniently

$$U^+ = 2.5 \ln y^+ + 9 - 2.5 \ln \left( \frac{3n'+1}{4n'} \right) + \frac{3}{n'} \qquad (204)$$

and the friction factor is:

$$\frac{1}{\sqrt{f}} = \frac{4.06}{n'} \log \left( \text{Re}_g f^{1-n'/2} \right) + 2.16 - \frac{2.78}{n'} \qquad (205)$$

Equation (205) is shown against the data of Dodge and Bogue in Figure 57.

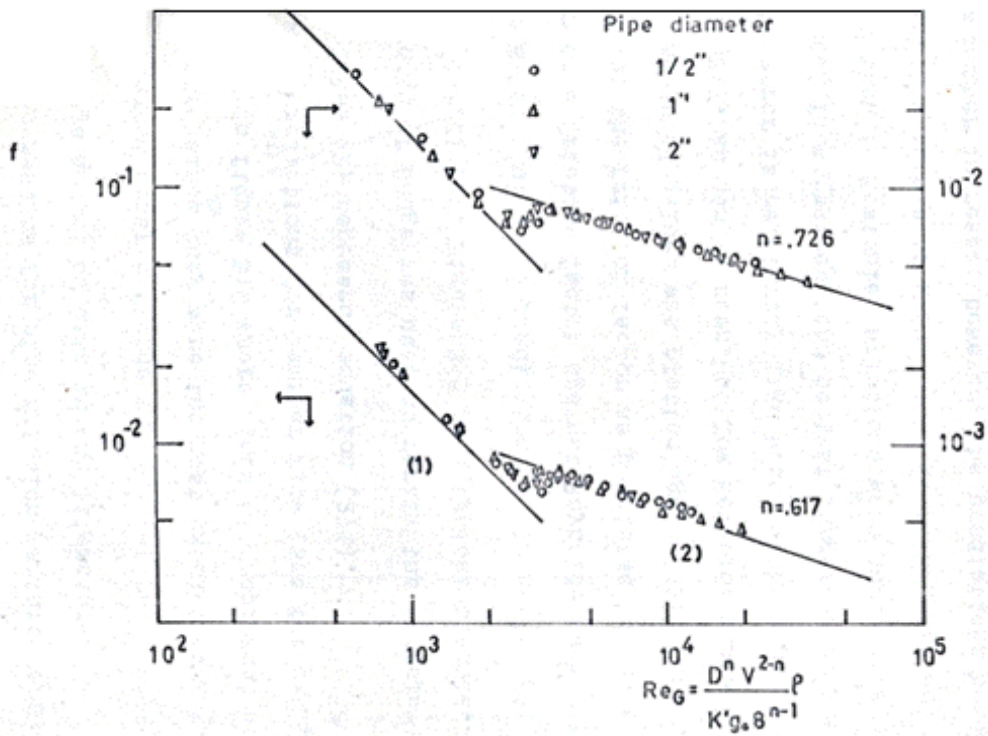

Figure 57 Friction factor for power law fluids in pipe flow predicted by equation (205). Data of Dodge (1959) an Bogue (1961)

We may also express the function $([3n'+1]/4n')$ in terms of parameters in different rheological models. For Bingham plastics, the friction factor becomes

$$\frac{1}{\sqrt{f}} = 4.07 \log\left(\frac{DV\rho}{\mu_B}\right)\sqrt{f}\left(1 - \frac{4}{3}\alpha_\tau + \frac{1}{3}\alpha_\tau^4\right) + \frac{8.35(1-\alpha_\tau)}{1 - \frac{4}{3}\alpha_\tau + \frac{1}{3}\alpha_\tau^4} - 9.06 \qquad (206)$$

where $\mu_B$ is the plastic viscosity, $\alpha_\tau = \tau/\tau_y$ and $\tau_y$ is the yield stress. Equation (206) is plotted against the data of Thomas (1960) in Figure 58.

We can also develop power law relationships similar to Blasius (1913)

$$f = \frac{\alpha}{\text{Re}_g^\beta} \qquad (207)$$

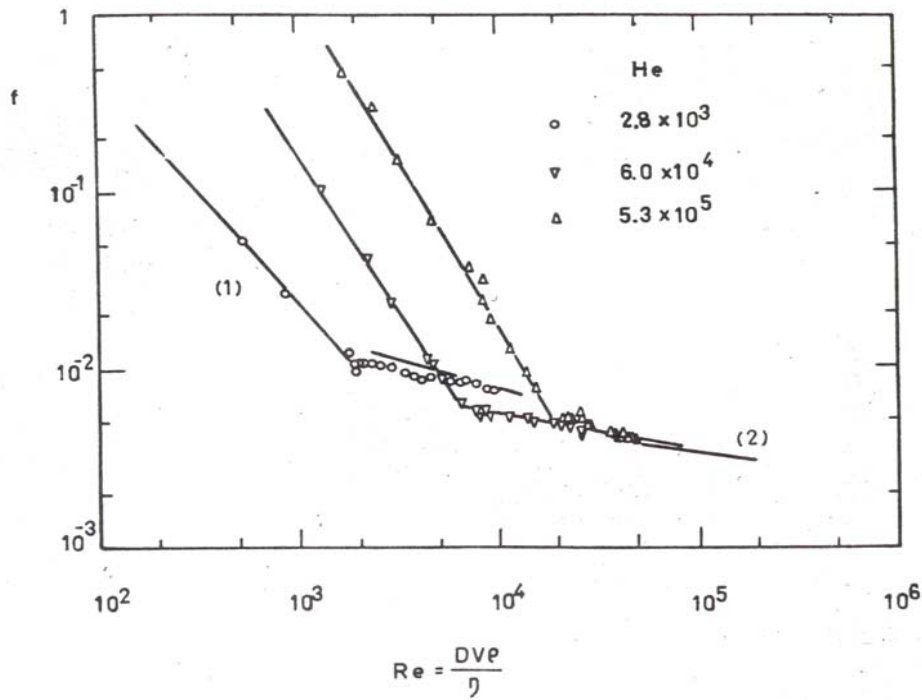

Figure 58 Friction factor for Bingham plastic fluids in pipe flow. Data of Thomas (1960). Line 1: Laminar solution (Skelland (1967), line 2: Equation (206)

This corresponds to a power law velocity profile of the form (Skelland, 1967)

$$\frac{U}{U_\infty} = \left(\frac{y}{R}\right)^{\frac{\beta n}{2-\beta(2-n)}} \qquad (208)$$

Equating the power index in equation (208) with the one seventh power law velocity profile (Schlichting 1960) gives

$$\beta = \frac{1}{3n+1} \quad (209)$$

Forcing equation (208) through the point $\left(U_k^+ = y_k^+ = 11.8((3n+1)/4n)\right)$ gives

$$\alpha = \frac{11.8^{-\frac{6n}{3n+1}} \left(\frac{3n+1}{4n}\right)^{-\frac{7n}{3n+1}} 2^{\frac{4+n}{3n+1}}}{\phi^{\frac{7n}{3n+1}}} \quad (210)$$

where $\phi = V/U_\infty$ is the ratio of the average velocity to the maximum velocity at the pipe axis. Equation (207) also fits the data of Dodge quite well (Trinh, 1993). The same derivation can also be applied easily to other fluid models.

Equation (207) allows us to estimate shear rate at the wall in turbulent flow (Trinh, 1969), which I have found surprisingly enough is available nowhere else

$$\dot{\gamma}_w = \left[ \frac{\alpha V^{2-2\beta+n\beta} \rho^{1-\beta} K^{\beta-1} 8^{\beta(1-n)}}{2D^{\beta n} \left(\frac{3n+1}{4n}\right)^{\beta n}} \right]^{1/n} \quad (211)$$

Figure 59 shows that the log-log plots of $\tau_w$ vs. $8V/D$ results in different lines for laminar and turbulent flows as first pointed out by Bowen (1961), but a curve of $\tau_w$ vs. $\dot{\gamma}_w$ results in a single line. The ability to estimate the shear rate in turbulent flow is important because rheological measurements of non-Newtonian fluids cannot safely be extrapolated outside their range of measurement as argued by Metzner in his discussion of the results of Clapp (Clapp, 1961) and therefore it is important to know the range at which the measurement should be taken.

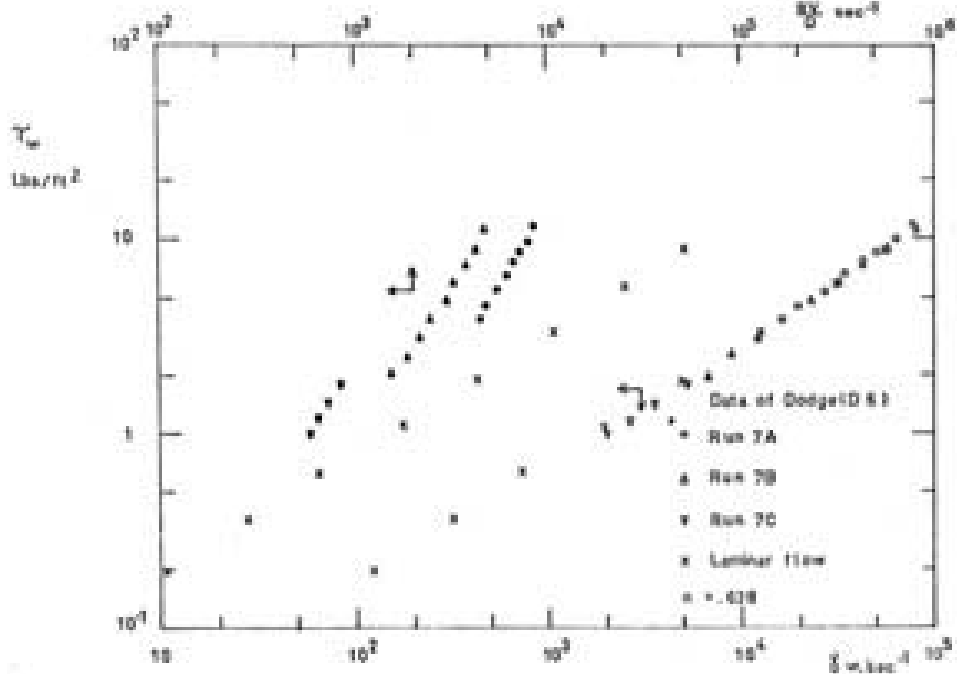

Figure 59  Wall shear rate in power law turbulent flow. Data of Dodge (1959*)*

**10.3    Drag reduction**

**10.3.1.1        Viscoelastic turbulent flows**

The phenomenon of drag reduction DR was discovered accidentally by Toms (1949) who found that addition of a small amount of polymers to water reduced the turbulent friction factor substantially. The most notable application is in the Alaska Pipeline System where it helps decrease power requirements and increase design throughput (Burger, Chorn, & Perkins, 1980).  and friction drag on vessel hulls ((National Research Council, 1997). Kawaguchi, Li, Yu, & Wei (2007) report that 70% of the pumping power used to drive hot water in primary pipelines or district heating systems was saved by adding only a few hundred ppm of surfactant into the circulating water. The significant potential of this technology in the reduction of energy usage in the face of dwindling resources as well as reduction in environmental pollution has encouraged a large number of studies on viscoelastic flows. These studies are also made to add further insight into the mechanics of turbulence and polymer behaviour. Excellent reviews have been published in the last 40 years (Gyr & Bewersdorff, 1995; Hoyt, 1972; Landahl, 1973; Liaw, Zakin, & Patterson, 1971; J. Lumley, 1969; J. L. Lumley, 1973; McComb, 1991; Nieuwstadt & Den Toonder, 2001; Virk, 1975; C. M. White & Mungal, 2008).

A number of physical phenomena have been observed in DR flows. All authors agree that there is a substantial thickening of the wall layer. Thus the mean velocity distribution is modified and the shear in the boundary layer is redistributed (White & Mungal, 2008). At low DR the Prandtl-Nikuradse log law is pushed further away from the wall and therefore the slope of the Karman buffer layer (the region between the Prandtl laminar sub-layer and the log law) is increased. There is a limiting universal velocity profile called Virk's asymptote (Virk, Mickley, & Smith, 1970). The near wall structure is also altered significantly. Sibilla and Beretta (2005) report from their DNS analysis that the mean vortex in the polymer flow is weaker less tilted and inclined, while its length and radius are higher. Cross-flow velocity fluctuations near the wall drop consistently. Similar observations are found in the work of White, Somandepalli, & Mungal (2004) and Dubief, Terrapon, Shaqfeh, Moin, & Lele (2004)

Polymer solutions are known to be viscoelastic and many authors naturally sought to explain DR in terms of the polymer relaxation time. Darby & Chang (1984) provided new scaling criteria by including a relaxation time and collapsed viscoelastic and non-elastic friction factors. Dimensional analysis introduced two parameters, the Deborah number (Seyer & Metzner, 1967) and the Weissenberg number. Both are dimensionless ratio of time scales: a typical polymer response time, usually the relaxation time which for linear polymers can be related to the size and number of monomers (Flory, 1971) and a representative time scale of the turbulent process. The difficulty lies in the proper choice of this time scale. Some authors use $T = v/u_*^2$ (White & Mungal, 2008).

More detailed analyses of the mechanics of turbulent viscoelastic flow involve considerations of the energy required to stretch the polymer. Most authors agree that this process occurs in the buffer layer and alters the flow dynamics. White and Mungal (op.cit.) identify two schools of thoughts. The first focuses on viscous effects and the second on elastic effects. The first school proposes that the effect polymer stretching is to increase the effective viscosity of the solution which in turn suppresses turbulent fluctuations. The second school argues that the onset of DR occurs when the elastic energy stored by the partially stretched polymer becomes comparable to the kinetic energy in the buffer layer at some scale larger than the Kolmogorov scale.

The forces linked with polymer stretching are extracted from laser velocimetry, PIV and DNS experiments by expressing the total local averaged stress as

$$\tau = \mu \frac{\partial U}{dy} - \rho \overline{u'v'} + \tau_{xy}^p \qquad (212)$$

White and Mungal state that "the increasing role of the polymer shear stress is perhaps the most interesting and controversial characteristic of HDR flows".

DR can also be achieved by the addition of cationic surfactants. Myska & Zakin (1997) and Gasljevic, Aguilar, & Matthys (2001) point out that the mechanisms of DR in solutions with polymer and surfactant show clear differences but the later is less intensively researched.

### 10.3.1.2    Wall riblets, compliant surfaces and surface bubbles

Polymer addition is not the only way to achieve DR. It has been known for some time that the drag force of a Newtonian fluid in turbulent flow past a surface can be reduced significantly by the addition of small riblets on the wall in the streamwise direction e.g. (Bandyopadhyay, 1986; Bechert & Hage, 2006). The principle is copied from observations of shark skin and now widely applied, for example in the design of swimsuits to improve swimmers performance.

Another technique which is copied from the swimming techniques of dolphins is to use compliant surfaces (Bandyopadhyay, 1986; Bandyopadhyay, et al., 2005; Choi, et al., 1997). The resultant wall motion is a uniform wave travelling downstream (Fukagata, Kern, Chatelain, Koumoutsakos, & Kasagi, 2008). In wind tunnel measurements Lee, Fisher, & Schwarz (1995) showed that the flow-induced surface displacements resulted in a reduction in the growth rates of unstable Tollmien-Schlichting waves and a delay in the onset of turbulence.

The generation or injection of bubbles near the surface is another exciting technique for drag reduction e.g. (McCormick & Bhattacharyya, 1973; Merkle & Deutsch, 1989; Mohanarangam, Cheung, Tu, & Chen, 2009; Sanders, Winkel, Dowling, Perlin, & Ceccio, 2006). The effect is attributed to the deformable properties of bubble surfaces e.g. (Oishi, Murai, Tasaka, & Yasushi, 2009). Interestingly, Sanders, et al. (2006) found that "skin-friction drag reduction was observed when the bubbly mixture was closer to the plate surface than 300 wall units of the boundary-layer flow without air injection...Skin-friction drag reduction was lost when the near-wall shear induced the bubbles to migrate from the plate surface". This effect appears compatible with the observation presented in section 4 and Figure 12 that moving vortices only affect the wall layer process when they are situated within the zone of the solution of order $\varepsilon^0$. It also raises the intriguing issue whether the hydrogen bubbles generated in wall layer visualisation (e.g. Kline et al., Corino and Brodey op.cit.) would have altered the statistics of the low speed streaks. Recently Fukuda, et al. (2000) showed significant DR by coating the wall surface with a super water repellent paint SWR and introducing air which forms a thin lubricating film.

These technologies are being actively studied applied in both air and water transports e.g. (Bandyopadhyay, 2005; Bushnell, 2003; Choi, et al., 1997; Davies, Carpenter, Ali, & Lockerby, 2006; McCormick & Bhattacharyya, 1973; Reneaux, 2004). Riblet technology has been applied to the design of sports swim suits (Orlando news, 2004) creating controversy at international sports events.

**10.3.2 My views**

The study of viscoelastic behaviour can be confusing and frustrating because it is approached from so many different angles, that often do not overlap well: Physical chemists look for formulations that explain the structure of polymers, often restricted in practical applications because they are confined to linear models, hence low shear rates or because they require a large number of experimentally measured parameters which are a challenge to determine in themselves. Physicists and mathematicians look for challenges in non-linear solutions and engineers are interested in process operations that involve high shear and non-linear models. The famous, and probably most studied, the Oldroy model (Bird, et al., 1979) requires eight parameters! Since drag reduction can be realised by widely different means, different types of additives as well as very different surface modifications, it is reasonable to analyse the fluid dynamics process and the polymer behaviour separately. I believe that we need to identify the fundamental mechanism of turbulence reduction before trying to see how polymer elasticity feeds into that mechanism.

While it is relatively easy to explain the apparent thickening of the wall layer in purely viscous fluids in terms of an integration constant (section 10.1), the phenomenon of drag reduction cannot be explained in that way.

In my view, the experimental evidence supports two firm statements:
1. The time-averaged wall layer in viscoelastic fluids is thicker in real physical terms unlike the apparent thickening in non-elastic fluids which is a consequence of expressing the apparent viscosity in terms of the time averaged wall stress instead of the instantaneous stress which appears in the NS equations.
2. The velocity profile in this wall layer can still be described by a solution of order $\varepsilon^0$. This is clearly evident in Figure 36 for both polymer additives and riblets. Similarly, I dispute the argument set forth by some of my colleagues e.g. Meyer (1966) that a new log-law emerges in viscoelastic flow. Since the solution of order $\varepsilon^0$ is based on the diffusion of viscous momentum, the viscoelasticity of the fluid does not affect the velocity profile. Even the most recent reviews (e.g. White and Mungal, op.cit.) agree that laminar pipe flow of polymer and Newtonian fluids show no significant differences. I have been looking out for any publication that would disagree but am not aware of any. In fact the viscosity of viscoelastic fluids is still measured in current rheometers in the same laminar flow configurations used for Newtonian fluids.
3. Two corollaries are derived from statements 1 and 2.
    a. The greater thickness of the wall layer is coupled with a longer time scale as expressed by equation (65)
    b. The time space transformation discussed in section 9.2 and (K.T. Trinh

& R.B. Keey, 1992) indicates that the low speed streaks will also be longer, which is observed e.g. in DNS studies (Sibilla & Beretta, 2005)

My explanation on the DR process is as follows.

Firstly, the elasticity of the flow medium in viscoelastic fluids dampens the magnitude of the oscillations (fast velocity fluctuations) $\omega$ and slows down their rate of growth. As a consequence the magnitude of the fast Reynolds stresses (strength of the streaming flow) is reduced, the ejection of the streaming flow is delayed and the sweep phase lasts longer. For example CAI, JIN, & Yang (2008) reported experimental observations that "the non-dimensional burst period (for turbulent flows in flexible tubes)[9] increases". The solution of order $\varepsilon^0$ is not affected since it does not involve the fast fluctuations but its time scale is extended. The *instantaneous* edge of this wall layer during a sweep phase is therefore the *same* for Newtonian fluids of equivalent viscosity and DR flows but the **time-averaged** thickness of the wall layer is **greater** in DR flows because of the extended time scale (Table 1). At the same time all the "turbulence" statistics evaluated in section 7.5 are altered because of the period of the wall layer. Indeed comparisons of the standard Reynolds stresses with the polymer stress contribution is in my view unfortunate because the critical variables that control the streaming process are the fast Reynolds stresses, not the standard Reynolds stresses, and these must be obtained by conditional sampling of instantaneous flow patterns. This visualisation is summarised in Figure 60.

---

[9] Author's addition

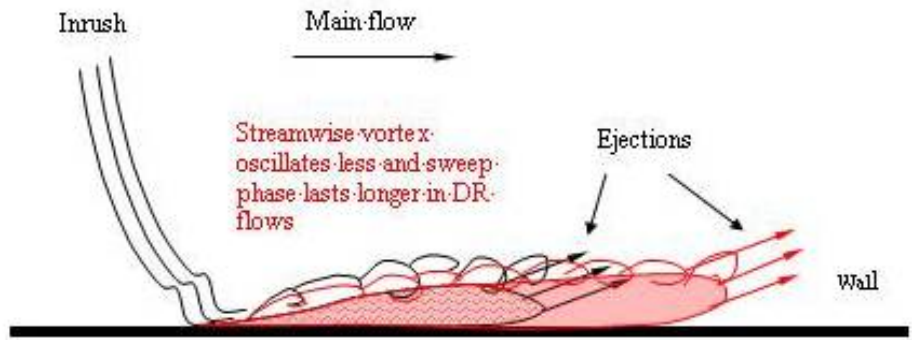

Figure 60 Comparison of Newtonian and drag-reduced turbulent flows

So far the visualisation is based solely on fluid mechanics arguments, experimental data (section 8) and analysis of the basic equations (20), (26) and (37), section 3. The specifics of the polymer attributes and surface modifications as well as other potential manipulations can be worked in later when this general theory is applied to particular situations.

A key determinant of the streaming flow is the magnitude of the fast fluctuations in particular the fluctuations in the z direction that Schoppa and Hussain (op.cit.) have rightly identified, in my view, as a main contributor to the streaming jets. It is evident in the lateral oscillations of the low speed streaks observed by many authors (e.g. Kline et al. and Offen and Kline, op.cit.). Riblets very successfully constrain these oscillations and thus stabilise the horseshoe vortex. Compliant surfaces work on another principle: I believe that they counteract the fast fluctuations by creating complementary waves. Bubbles and air films are another means of providing compliant surfaces. Thus we do not even need physical constraints like riblets. I have long held the belief that we should be able to reduce the fast fluctuations by simply introducing waves, even acoustic waves, of the right frequency. I even planned to test the reverse experiment: enhancing heat transfer without increasing the flow rate by acoustic streaming. This application, if successful, would be useful for thick fluids in the food industry where a turbulent Reynolds number is hard to achieve, particularly in agitated vessels. Unfortunately I never had the time, opportunity or funding to test these ideas. Thus I was very gratified to find that colleagues have acted on similar premises. For example Du & Karniadakis (2000) have shown by DNS that a transverse travelling wave can eliminate the low speed streaks and demonstrated the application with electromagnetic tiles in salt water.

Concentrated viscoelastic solutions have high viscosity which hampers the growth of wave disturbances according to existing stability analyses and to that extent delay the onset of ejection. In the same way of course, an increase in viscosity would reduce the value of the Reynolds number for a given flow rate and thus drive towards more laminar flow. Frater (1967) has studied acoustic streaming in a viscoelastic fluid with two time constants: the first $\lambda_1$ is associated with the stress tensor, the second $\lambda_2$ with the strain tensor. He found that the thickness of the inner layer (solution of order $\varepsilon^0$ increased with increases in the ratio $\lambda_1/\lambda_2$ and the value of the product $\omega\lambda_2$ which the reader may recognise as similar to the Deborah number proposed by Metzner (op.cit.).

We now proceed to apply these concepts to simple DR predictions

### 10.3.2.1 Virk's asymptote

As noted previously, a substantial number of authors have argued that the slope of the log-law close to the wall in viscoelastic fluids is different from that for Newtonian fluids as shown in Figure 61 taken by (McComb, 1991). In fact Virk, et al. (1970) has argued as the concentration of polymers increase this new log law tends towards a limiting profile called Virk's asymptote.

$$U^+ = 11.7 \ln y^+ - 17 \tag{213}$$

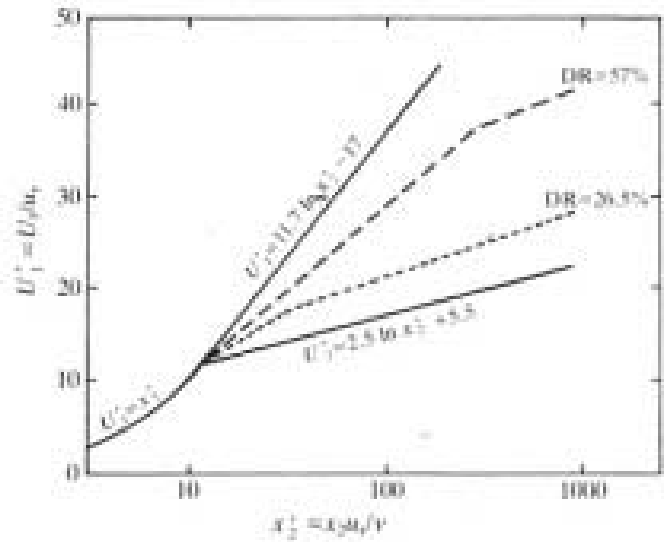

Mean velocity profiles in a pipe during the injection of Polyox WSR301 (R = 3.5 × 10⁴) (data taken from McComb and Rabie 1982).

Figure 61. Log law profiles in viscoelastic fluids (from (McComb & Rabie, 1979))

This visualisation of a different mechanism of turbulence has a flow on effect on how the energy and dissipation terms are modelled and affect the outcome of CFD predictions particularly of heat and mass transfer in viscoelastic media e.g. (Koskinen, et al., 2005)

Figure 36 shows a different picture: the velocity profile of viscoelastic fluids coincide with that of Newtonian fluids when expressed under the zonal similarity analysis. When the velocity profile such as those obtained by Pinho and Whitelaw (1990) is plotted in the Prandtl-Nikuradse presentation, there is indeed a gradual shift of the curve with increased polymer concentration similar to Figure 61. But when it is plotted in zonal similarity format all points collapse onto a unique curve which is well represented by the averaged error function obtained in the Stokes solution1. With increased polymer concentration, the thickness of the wall layer increases and the outer portion of the error function is now situated at distances traditionally attributed to the power law region for Newtonian fluids. In addition, the portion of the error function profile between $0.1 < y^+/\delta_v^+ < 0.8$ can be approximated by a straight line as shown in Figure 36.

$$\frac{U^+}{U_v^+} \approx 0.36\ln\left(\frac{y^+}{\delta_v^+}\right)+1.1 \tag{214}$$

At the end of the laminar region, $\text{Re} = 2100$ $R^+ = 64.7$, $U_v^+ = R^+/2$ and equation (214) becomes

$$U^+ = 11.7\ln y^+ - 18 \tag{215}$$

which almost coincides perfectly with Virk's asymptote.

If we match equation (214) with the edge of the wall layer at high Reynolds numbers $\delta_b^+ = 64.7$, $U_b^+ = 15.6$ (section 7.5.2.2, Figure 38) we obtain

$$U^+ = 5.6\ln y^+ - 6.24 \tag{216}$$

which is very similar to Karman's correlation for the buffer layer.

I believe that some of my colleagues have mistakenly taken this section of the error function profile for a new log-law. In the zonal similarity representation, Virk's asymptote also coincides with the averaged error function which means that when the viscoelasticity is strong enough, the solution of order $\varepsilon$, the streaming function is suppressed and the flow follows the solution of order $\varepsilon^0$: the flow is laminar with small disturbances. There is no argument however that, whereas the apparent thickening of the wall layer and shift of the log-law in purely viscous Newtonian is an artefact created by the process of integration from the NS equations to the Reynolds equation, the thickening created by viscoelastic effects or riblets is real.

### 10.3.2.2 Turbulent viscoelastic pipe flow

The work of Frater (op.cit.) shows that the elastic time scales[10] $\lambda_1$ and $\lambda_2$ modify the behaviour of the fast fluctuations $u_i'$. In fact anything that retards the growth of the fast fluctuations (addition of long chain polymers, cationic surfactants, riblets, compliant surfaces, transverse waves or bubble generation will as argued previously reduce the value of the parameter $\varepsilon$. As a direct consequence, the streaming function obtained in section (3) becomes weaker and it takes longer before ejection of the low speed streak fluid. Assuming a linear relation we write

$$t_{ve} = t_{v,ne} + t_{DR} \tag{217}$$

---

[10] In this section the symbol $\lambda$ refers to relaxation time scales of viscoelastic fluids.

Where $t_{v,ne}$ is the time scale of the wall layer for a non-elastic fluid of the same viscosity

$t_{DR}$ the increase in time scale responsible for drag reduction

$t_{ve}$ the total time scale of the wall layer in the viscoelastic fluid

The wall layer thickness equation (65) becomes
Then

$$\delta_{ve}^+ = 3.78(t_N^+ + t_{DR}^+) \tag{218}$$

The parameter B in the log law, equation (46) is obtained by forcing it through the edge of the wall layer and taking account of equation

$$U^+ = 2.5\ln(\frac{y^+}{\delta_v^+}) - U_v^+ = 2.5\ln(\frac{y^+}{\delta_v^+}) - \frac{\delta_v^+}{4.16} \tag{219}$$

Substituting for $\delta_v^+ = 67$ [11] gives the results of Prandtl-Nikuradse for Newtonian fluids (Nikuradse, op.cit.). The friction factor may be obtained by the classic procedure of Prandtl and Karman (op.cit.).

$$U^+ = 2.5\ln y^+ + 5.5 \tag{220}$$

The difficulty in applying it to drag reduction resides in estimates of the DR time scale $t_{DR}$.

---

[11] There is a slight discrepancy for the value of $\delta_v^+$ required to fit in with logarithmic correlations of velocity profiles among various authors and this work because the former is based on average fits of profiles in a large range of Reynolds numbers, not just asymptotic values of the wall layer thickness.

The situation for polymer solutions can be visualised simply by analysing the response of a Maxwell liquid described by

$$\dot{\gamma} = \frac{1}{G}\frac{d\tau}{dt} + \frac{\tau}{\mu} \qquad (221)$$

to an applied shear stress. G is the modulus of rigidity of the liquid. When the stress is applied slowly, the first term on the RHS is small and the fluid behaves essentially as a viscous liquid. This velocity profile in laminar flow and the solution of order $\varepsilon^0$ is the same for viscoelastic and non-elastic fluids. However when the velocity fluctuations are fast, the first term on the RHS dominate and the fluid behaves like an elastic solid of relaxation time scale

$$\lambda = \frac{\mu_a}{G} \qquad (222)$$

Basic analysis of wave propagation in liquid systems, e.g. (Elmore & Heald, 1969; Squire, 1971), indicates that pressure waves in fluids are related to the volumetric strain induced and the fast fluctuations would necessarily be reduced in viscoelastic fluids because of their solid-like behaviour. Thus clearly $t_{DR}$ must be a function of $\lambda$.

The simplest correlation for the friction factor is based on forcing the log-law through the Kolmogorov point (section 10.2). For a Newtonian fluid, it can be formally obtained by equating (80) and 220)

$$y_K^+ = 2.5\ln y_K^+ + 5.5 = 11.7 \qquad 223$$

When the wall layer in non-Newtonian inelastic fluids is normalised with the apparent viscosity, based on the time-averaged wall shear stress, it is apparently thicker than the wall layer in Newtonian fluids because of the integration process (sections 10.1 and 10.2). For pipe flow the ratio of thicknesses in power law and Newtonian fluids is equal to $(3n+1/4n)$ as argued in section 12. This factor can be introduced into equation 219 which is then equated with (80) to give $y^+_{K,NN,ne}$. The variations of $y^+_{K,NN,ne}$ with $n'$ can be analysed by calculating its value from published friction factor data (D.C. Bogue, 1961; D.W. Dodge, 1959; A. D. Thomas, 1960; Yoo, 1974).

The results confirm the validity of equation (203) that I derived earlier from more heuristic arguments (Trinh, 1969). When we apply the same procedure to viscoelastic polymer solutions, there is a further shift in the Kolmogorov point. From dimensional consideration, the only parameter pertinent to the wall layer that can combined with this the distance shift $\Delta y$ to gives a time scale is the friction velocity $u_*$ then

$$\left[\frac{u_*}{\Delta y}\right] = \left[\frac{\mu_a}{G}\right] \qquad (224)$$

$$\left[\frac{\Delta y u_* \rho}{\mu_a}\right] = \left[\frac{\tau_w}{G}\right] \qquad (225)$$

where the brackets indicate that we are comparing dimensions. Thus the ratio $\tau_w/G$, called the Weissenberg number, represent a shift in dimensionless distance. Because of the success of equation (203) in predicting purely viscous turbulent flows, it was extended to viscoelastic fluids by assuming an additive relationship

$$y_k^+ = 11.8\left(\frac{3n+1}{4n}\right) + a\frac{\tau_w}{G} \qquad (226)$$

To a first approximation the constant $a$ is taken as unity. Then the friction factor is given by

$$\frac{1}{\sqrt{f}} = 4.07\log\left(\mathrm{Re}_g\, f^{1-n/2}\right) + 2.16 + \frac{\tau_w}{G\sqrt{2}} - \frac{2.78}{n} - 4.07\log\left(1 + \frac{\tau_w}{11.8G\left(3n+1/4n\right)}\right) \qquad (227)$$

Considering the crudity of the assumption equation (227) predicted quite well the measurements of Metzner and Park (Figure 62).

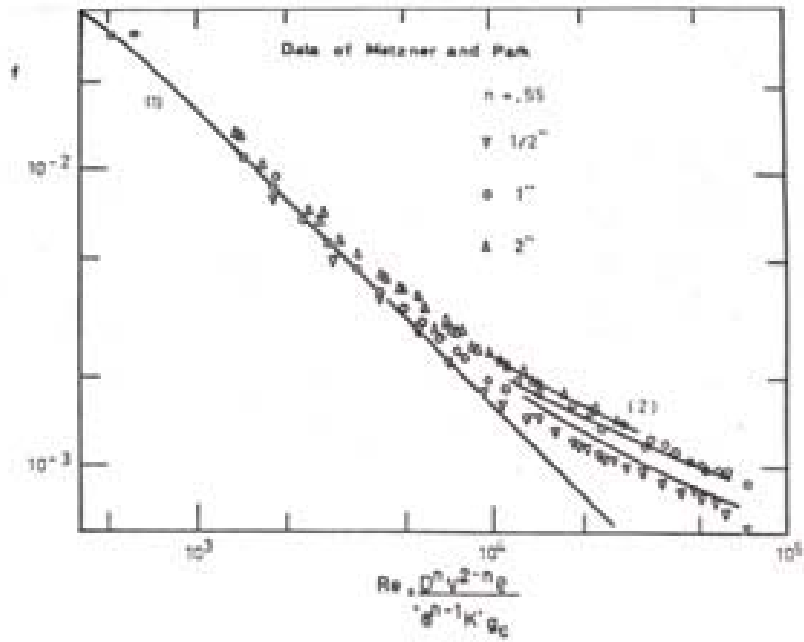

Figure 62 Friction factor in viscoelastic pipe flow (from Trinh 1969.)Lines (2) represent equation (227) with G values calculated from the rheology data of Metzner and Park (1964).

The advantage of this approach is that it relies only on fluid dynamic considerations. The polymer properties are introduced through the experimental measurements of G performed by Metzner and Park at different shear stresses and show that the relaxation time scale is not a single value for each polymer solution but varies with the Reynolds number. The prediction of these time scales from polymer considerations can be treated separately as an exercise in polymer chemistry.

The use of the Kolmogorov point to obtain correlations for the friction factor polymer solutions continues to attract researchers e.g. (Andrade, Petronílio, Edilsonde, Maneschy, & Cruz, 2007). These authors used the Carreau model and the Dodge-Metzner correlation to estimate the Kolmogorov scale in polymer solutions by further assuming that the shear rate at the wall can be obtained by the approximation

$$\dot{\gamma}_w = \frac{u_*}{y_K} \tag{228}$$

## 11     Time scales of turbulence

There are a multitude of time scales that can be identified in turbulent flows and it is important to know which ones should be compared with the viscoelastic time scales of the fluid. This choice affects the significance of the dimensionless number that can be formed. I count at least four major time scales.

The first is the time scale of the all important wall layer which was measured, for example by Meek and Baer. Near the wall it must be really separated into two components: the time scale of the sweep phase and the time scale of the bursting phase. The former is much larger and dominates the wall layer time scale.

I was perplexed when I met the paper of Rao et al. (1971) years after it was published. The time scale they measured in turbulent flow scaled with the outer variables $U_\infty$ and $\delta$, completely contradicting the scaling of Meek and Baer. The matter was resolved in my mind when I realised that the probes of Meek & Baer and Rao et al. captured the signatures of completely different coherent structures the former next to the wall, the other well into the main flow. Strouhal (1878) was the first to scale the periodic shedding of vortices behind a cylinder in cross flow with a dimensionless number

$$\text{Str} = \frac{nd}{U_\infty} \tag{229}$$

now known as the Strouhal number where n is the frequency of vortex shedding. The reader can see the similarity between the original Strouhal number and the normalised Rao time scale, which is measured outside the wall layer; however the latter is more complex. I half agree with the argument of Rao et al. that their time scale and that of Meek and Baer must be connected. While the Rao time scale is based essentially on the signatures of the vortex shed behind the ejections from the wall these are intermittent and the frequency of ejection must impact on the Rao time scale. However the reverse is not true: The Rao events do not affect the wall layer time scale, as we have argued previously in section 7.6.

The least understood time scale is that of the fast fluctuations $u'_i$ because they are lost in the statistical averaging process commonly used to study turbulence. Thus all studies of oscillatory flow must start with some arbitrarily defined initial periodic disturbance. Schneck and Walburn (op. cit.) for example define a wave Strouhal number. In the same way, the initial disturbances introduced in DNS studies of turbulence mechanisms are arbitrary and so are the ones used in the theory of stability. In fact these fast fluctuations cannot just be described in simple sinusoidal form, as shown for example by the decomposition of the velocity trace in Figure 2, even if the original disturbances to laminar flow were sinusoidal. This is presumably because fast Reynolds stresses are continuously produced in oscillatory flow and even though many of them will be weak and eventually dissipated by viscous diffusion they will modify the form of the original disturbances. An understanding of these fast fluctuations is important since their time scale contributes to the parameter $\varepsilon$, equation (58), that can be used to define transition to turbulence.

Of course a number of further time scales are introduced by viscoelastic fluids that intrinsic to their composition and nature.

## 12   Transition between laminar and turbulent flows

### 12.1   An alternate criterion for transition

The difficulties in predicting the various time scales from basic mathematical analysis of the NS equations led me to consider alternative engineering approaches.

The decomposition of a turbulent flow field into additive layers in Figure 13 has stressed that transition from purely laminar flow begins to be noticed when streaming jets emerge from the domain of the solution of order $\varepsilon^0$. This solution is confined by a domain limited in the normal direction by the thickness $\delta_v^+$. Calculations of this parameter $\delta_v^+$ from literature data on velocity profiles as well as from the closure technique in section 9.1 show that this parameter reaches a stable value of about 64 (or 67 depending on the researcher) once the flow field has reached full turbulence, in the conventional sense (Figure 38). This is true for practically all the external boundary layer data that I have been able to analyse but there is a small distortion for flow in closed conduits, such as cylindrical pipes in transitional flow because of the influence of the opposite wall. Therefore the Reynolds number at which the parameter $\delta_v^+$ reaches its asymptotic value defines the end of the laminar region. We note that the domain of application of the solution of order $\varepsilon^0$ is only constant when expressed in terms of normalised units as shown in Figure 13. When one expresses these domains in terms of absolute units of length, then the domain for order $\varepsilon^0$ diminishes in thickness dramatically with Reynolds number as shown for pipe flow in Figure 63. Thus an observer or probe situated at the location Ob in Figure 63 may feel as though the flow at that location has broken down to smaller turbulence scales but in fact the solution of order $\varepsilon^0$ has not been altered; it has simply moved out of the field of vision of the observer.

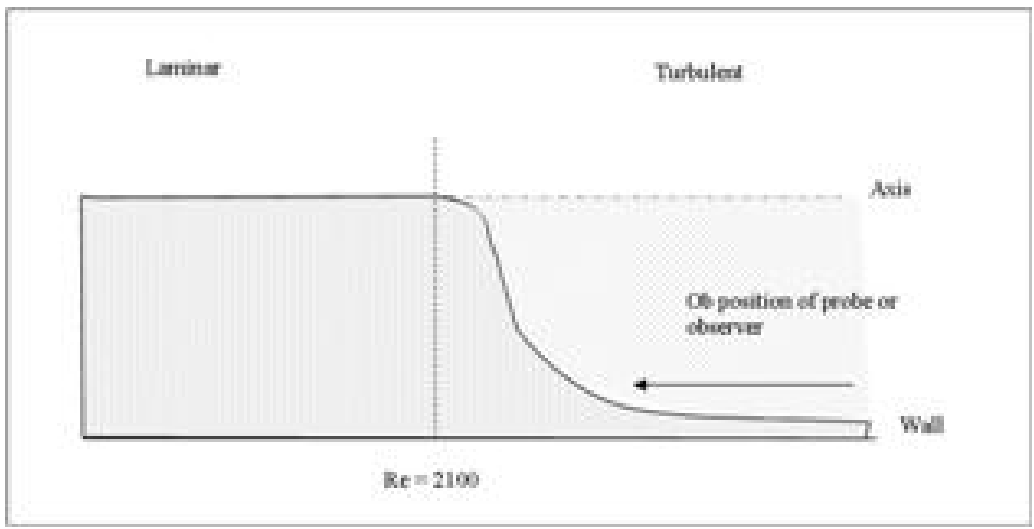

Figure 63  Variation of physical value of wall layer thickness in pipe flow with Reynolds number (not to scale).

Clearly at the end of the laminar flow regime we have

$$\delta_v^+ = R^+ \tag{230}$$

and we wonder whether this equality may be used as an alternate criterion for determining transition to from laminar flow without the need to calculate $\varepsilon$. For pipe flow the relation $R^+ = 64$ simply gives the well-known result $Re = 2100$.

Since we were able to produce a unique master velocity profile for the inner region by using the set $U_v^+, \delta_v^+$ as normalising parameters in the zonal similarity analysis (Figure 36), we wonder whether we can similarly produce a unique for friction factors by using the values of f and Re at the end of the laminar flow regime.

## 12.2 The frictional law of corresponding states

Obot (1993) has independently come to the conclusion that the transition Reynolds number can be used as a scaling factor for flow and heat transfer in channels of different geometries, two phase flow and mixing. He called this similarity plot "the frictional law of corresponding states", Figure 64.

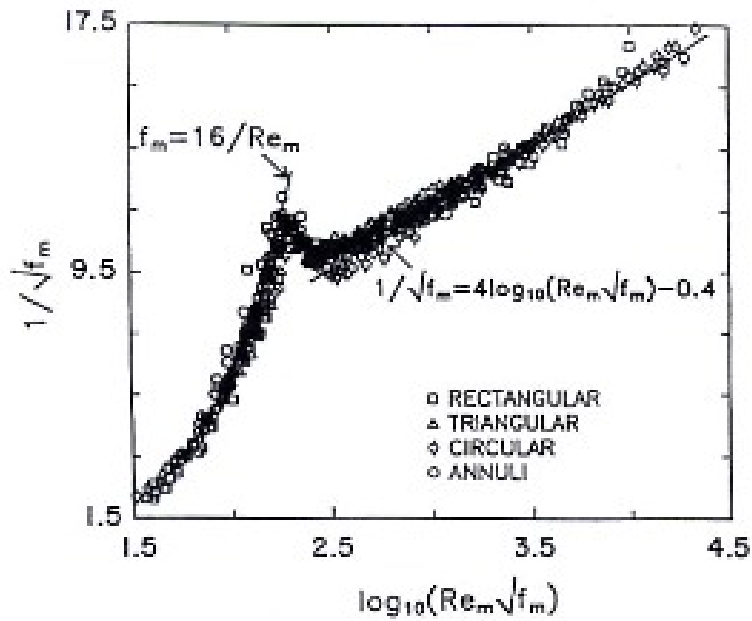

Figure 64 Obot law of corresponding states. $Re_m = Re/Re_c$, $f_m = f/f_c$, $Re_c$ critical Reynolds number at transition.

## 12.3 The retardation of turbulence function

As discussed before in section 10.3.2.1, the wall layer thickness is equal to the pipe radius at the critical Reynolds number at transition and the normalised velocity is equal to the critical friction factor. Thus Figure 36 and Figure 64 are equivalent. However, three differences between the Obot's approach and the present analysis need to be made.

1. Obot did not relate his scaling law to a universal velocity profile as shown in Figure 36. He argued heuristically that the conditions at transition must somehow define the structure of the subsequent turbulent flows.
2. Secondly, the friction factor in conduit flow is often expressed in terms of mass averaged flow velocity V whereas the parameter $U_v^+$ used here represents the velocity at the conduit axis. Since the ratio $\phi = V/U_\infty$ varies between flow geometries an error is introduced when the traditional friction factor is used. Thus the data presented by Obot has a significant scatter about the master curve.
3. Thirdly the similarity breaks down when a new component is introduced in the calculation of the friction factor: form drag which occurs behind bluff bodies, including pipe roughness. In such cases one cannot expect that all portions of the curve would collapse neatly into a single master curve. We encountered a similar problem for the outer region of the velocity profile in section 8.
4. A similar plot can also be constructed for the Nusselt number.

The more interesting issue here is to ask how these concepts affect the prediction of transition in non-Newtonian fluids. In section 10.1 we saw that the wall layer thicknesses are the same for Newtonian and purely viscous power law fluids when normalised with the instantaneous critical wall shear stress but different when normalised with the time-averaged wall shear stress. We immediately suspect that the difference in the transitional Reynolds number will therefore follow the same pattern but were hampered because the Stokes solution1 used as a model for the analysis neglected the curvature of the surface.

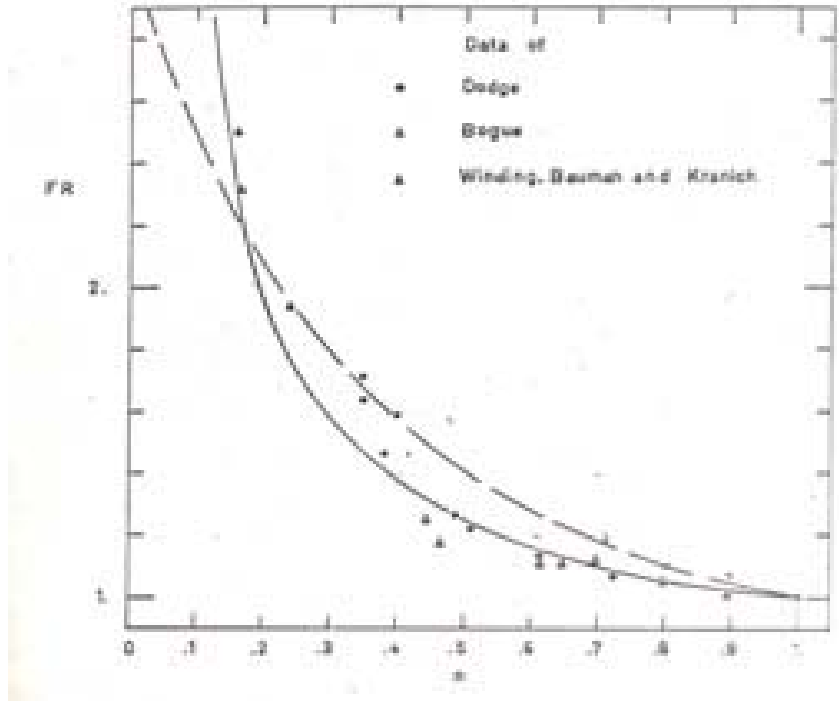

Figure 65 Apparent retardation of turbulence (ratio of critical Reynolds number in power law and Newtonian fluids) after Trinh, 1969. Full line: equation (231), dotted line: prediction by (Hanks & Ricks, 1975)

Equation (199) suggest that the critical apparent Reynolds number in power law fluids can be related to the Newtonian value by

$$\mathrm{Re}_{a,c} = \mathrm{Re}_{c,n=1}\left(\frac{3n+1}{4n}\right) \qquad (231)$$

This equation is shown against data calculated from literature measurements in Figure 65. A similar correlation for Bingham plastic fluids (Trinh 1969) is shown in Figure 66.

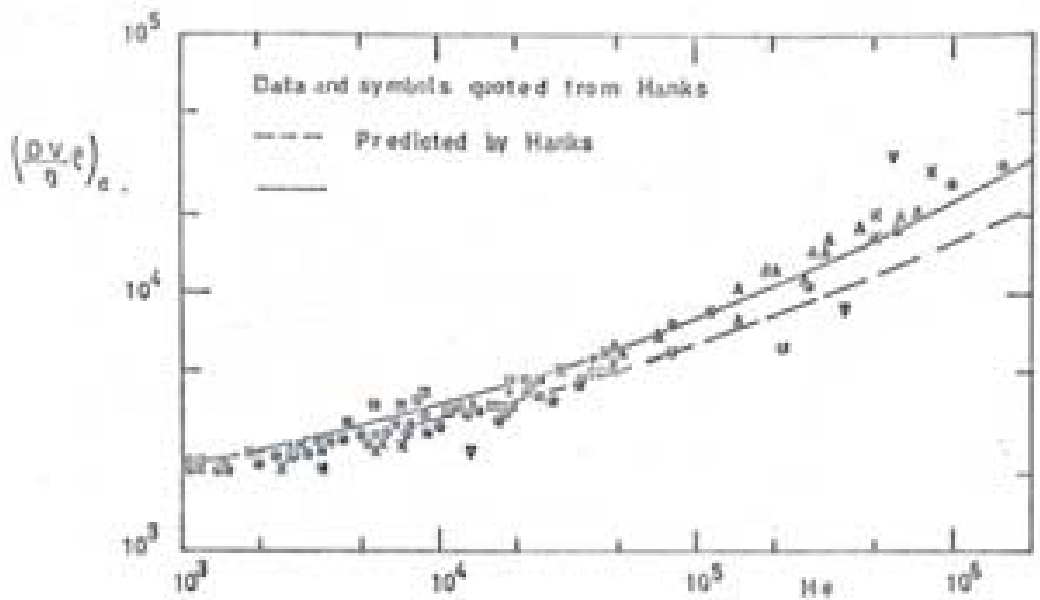

Figure 66 Apparent retardation of turbulence (ratio of critical Reynolds number in Bingham plastic and Newtonian fluids) after Trinh, 1969. Full line: equation (231), dotted line: prediction by (Hanks & Dadia, 1971)

We may now use this theoretical estimate of the critical Reynolds numbers in non-Newtonian fluids to produce a master curve for time averaged friction factors. An example is shown for pipe flow of power law fluids in Figure 68.

Most experimental data in non-Newtonian turbulent flow show significant experimental variations, especially near the transition between laminar and turbulent flow. In fact it is very rare to find measurements that give an exact value for the transition Reynolds number, even in the simplest case of power law pipe flow. It is therefore important to define clearly the way this critical Reynolds number is obtained because the variations of this value, between different schemes of evaluation, are large enough to allow justification of any theoretical argument.

In my work, I draw two straight lines for the laminar and turbulent regimes and take the value of $Re_c$ at the intersection. This is shown with the data of Dodge $n = .726$ in Figure 67

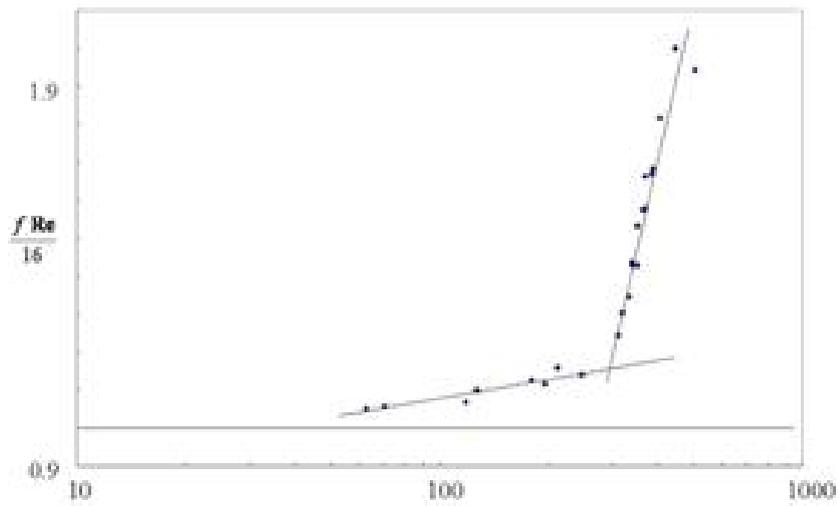

Figure 67 Determination of critical Reynolds number. Data of Dodge (1959)

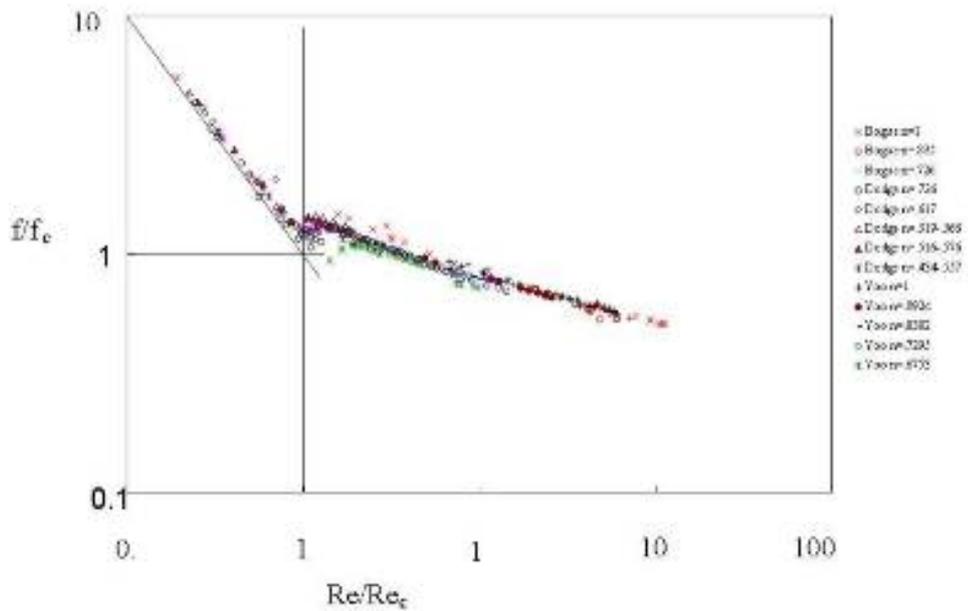

Figure 68 Reduced plot of pipe friction factors for power law fluids

We note here that line (1) in Figure 67 should be horizontal since it describes laminar flow where

$$f = \frac{16}{\text{Re}_g} \tag{232}$$

The fact that this line has a slope illustrates clearly how even widely trusted data in non-Newtonian turbulent can harbour significant measurement uncertainties; The Dodge & Metzner (1959) correlation is the most quoted for purely viscous non-Newtonian pipe flow and figures in all textbooks related to the field.

There is a short third straight line in the transition region between lines (1) and (2) as shown for example in the data of Bowen (1961). A better estimate of $\text{Re}_c$ would be obtained by taking the intersection of that line and line 1 in Figure 67 but most data available in the literature do not have measurements in the transition region.

Finally the friction factor data in the literature is often plotted against the Metzner-Reed generalised Reynolds number $\text{Re}_g$, which is different from the Reynolds number based on the wall apparent shear rate $\text{Re}_a$ by a factor $(3n'+1)/(4n')$ hence

$$\text{Re}_{g,c} = \text{Re}_{c,n=1} \left( \frac{3n+1}{4n} \right)^2 \tag{233}$$

Equation (233) was used to calculate the critical Reynolds number in power law fluids and prepare Figure 68 with very satisfactory results.

## 12.3.1 Turbulent flow between parallel surfaces

This is a classic problem when a fluid is contained between two parallel plates separated by a distance d. At time $t = 0$ the upper plate is moved with velocity $U_d$. In most classic problems of this type e.g. (Bird, et al., 1979; Theodore, 1971), the fluid is originally at rest. To model turbulent flows, we identify time zero with the inrush of fluid from the mainstream at the start of a new low-speed streak. The governing equation is again

$$\frac{\partial \tilde{u}}{\partial t} = \nu \frac{\partial^2 \tilde{u}}{\partial y^2} \tag{59}$$

We again define the dimensionless velocity, distance and time as

$$\phi = \frac{u}{U_d} \tag{234}$$

$$\xi = \frac{y}{d} \tag{235}$$

$$\tau = \frac{vt}{d^2} \tag{236}$$

We seek a solution of the form

$$\phi(\xi,\tau) = \phi_\infty(\xi) - \phi_t(\xi,\tau) \tag{237}$$

Hence equation (59) is solved for the conditions

| | | | |
|---|---|---|---|
| IC  | $\tau = 0$ | $0 \leq \xi \leq 1$ | $\phi = \phi_0$ |
| BC1 | $\tau > 0$ | $\xi = 0$ | $\phi = 0$ |
| BC2 | $\tau > 0$ | $\xi = 1$ | $\phi = 1$ |

At time $t \to \infty$ this becomes laminar Couette flow with a linear velocity profile

$$\phi_\infty = \xi \tag{238}$$

Substituting into equation (59) gives

$$\frac{\partial \phi_t}{\partial \tau} = \frac{\partial^2 \phi_t}{\partial \xi^2} \tag{239}$$

We now use the method of separation of variables, setting

$$\phi_t = \varsigma(\xi) T(\tau) \tag{240}$$

Substituting into (239), rearranging and equating both sides of the equation to a parameter $(-\alpha^2)$

$$\frac{1}{T}\frac{dT}{d\tau} = \frac{1}{\varsigma}\frac{d^2\varsigma}{d\xi^2} = -\alpha^2 \tag{241}$$

Equation (241) then gives

$$T = C_0 e^{-\alpha^2 \tau} \tag{242}$$

$$\varsigma = \left(C_1 \sin \alpha\xi + C_2 \cos \alpha\xi\right) \tag{243}$$

$$\phi = \xi + C_0 e^{-\alpha^2 \tau}\left(C_1 \sin \alpha\xi + C_2 \cos \alpha\xi\right) \tag{244}$$

The constants $C_0, C_1, C_2$ are determined from the IC and BC.

BC1 gives $\quad C_0 C_2 = 0$ Then BC2 gives

$$C_0 C_1 e^{-\alpha^2 \tau} \sin \alpha = 0 \tag{245}$$

Which is satisfied if either $C_0 C_1 = 0$ or $\sin \alpha = 0$. Since $C_0 C_1 \neq 0$ because $\phi$ is a function of $\tau$, we put

$$\alpha = n\pi \quad \text{where} \quad n = 1,2,3,4... \tag{246}$$

and

$$\phi = \xi + \sum_{1}^{\infty} C_n e^{-\alpha^2 \tau} \sin(n\pi\xi) \tag{247}$$

where $C_n = C_0 C_1$.

We now proceed to identify the form of the initial velocity profile $\phi_0$. From equation (247)

$$\phi_0 = \xi + \sum_{1}^{\infty} C_3 \sin(n\pi\xi) \tag{248}$$

The initial velocity profile cannot be assumed to be flat as in traditional start up problems (Bird et al, Theodore op. cit.) because in this case it refers to measurements at the inrush phase where fluid rushes from the log-law region towards the wall. Since the thickness of the wall layer in this phase is zero, we expect a profile made up of a log law that extends straight to the wall and a law of the wake. Following Karman (op.cit.) and Prandtl (op.cit.), I originally ignored the law of the wake to simplify the derivation. I used a simplified form of the law of the wall equation (110)

$$U^+ = \int_0^{y^+} \frac{1}{\frac{1}{1-y^+/R^+} + \kappa y^+} dy^+ \approx \int_0^{y^+} \frac{1}{1+\kappa y^+} dy^+ \qquad (249)$$

Because of the simplifications, a better fit is obtained by putting $1/\kappa = 2.7$ and

$$U^+ = 2.7 \ln\left(1 + \frac{y^+}{R^+}\right) \qquad (250)$$

$$\phi_0 = \frac{U^+}{U_d^+} = \frac{2.7}{U_d^+} \ln\left(1 + \frac{d^+\xi}{2.7}\right) \qquad (251)$$

Equating (248) and (251) gives

$$\xi + \sum_1^\infty C_3 \sin(n\pi\xi) = \frac{2.7}{U_d^+} \ln\left(1 + \frac{d^+\xi}{2.7}\right) \qquad (252)$$

Multiplying both sides by $\sin(m\pi\xi)$ and integrating with respect to $\xi$

$$\int_0^1 \xi \sin(m\pi\xi d\xi) + \sum C_3 \sin(n\pi\xi)\sin(m\pi\xi)d\xi = \int_0^1 \frac{2.7}{U_d^+} \ln\left(1 + \frac{d^+\xi}{2.7}\right)\sin(m\pi\xi)d\xi$$

$$(253)$$

The function

$$C_4 = \int_0^1 \frac{2.7}{U_d^+} \ln\left(1 + \frac{d^+\xi}{2.7}\right) \sin(m\pi\xi) d\xi \qquad (254)$$

is best calculated numerically for each value of $d^+$. Because of its orthogonality, the function $\sin(n\pi\xi)\sin(m\pi\xi)$ only contributes when $m = n$ and we get

$$C_3 = \frac{2}{n\pi}(-1)^n - C_4 \qquad (255)$$

Then

$$\phi = \xi + \sum_1^\infty \left(\frac{2}{n\pi}(-1)^n - C_4\right) e^{-\alpha^2 \tau} \sin(n\pi\xi) \qquad (256)$$

For steady state fully developed turbulent flow, equation (256) is time averaged over the range $0 \leq t^+ \leq t_v^+$ for each value of $d^+$, hence Reynolds number. The period of the streaming flow $t_v^+$ is taken from Figure 16 or as $t_v^+ = 17$ for high Reynolds numbers.

It is not possible to set up an exact experiment where fluid is simply held between two parallel plates but a closely analogous situation is found for fluids between rotating cylinders, especially when the gap between the cylinders is small and curvature can be neglected. In this case, the data of Reichardt (1963) shows that the velocity profile exhibits an inflection point at the gap axis. Equation (256) can be applied to each side of the axis as shown in
Figure 69.

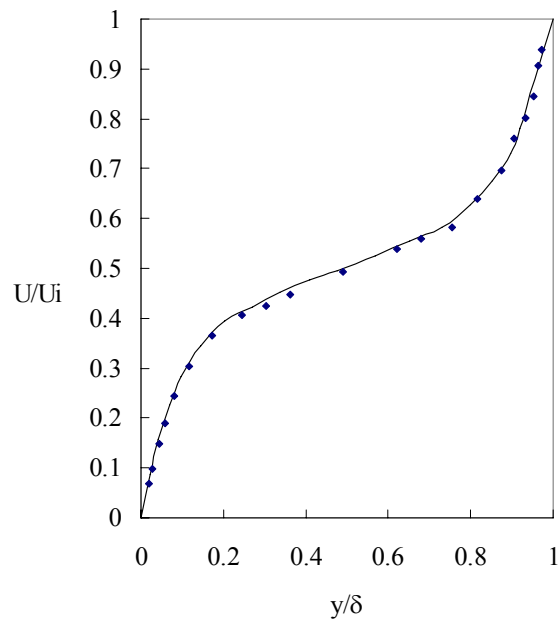

Figure 69 Turbulent flow between rotating cylinders. Data of Reichart (1963). Solid line: time averaged equation (256)

### 12.3.2 Turbulent pipe flow

We start here with the Szymanski solution (1932) which, like the Stokes solution1, is traditionally used to describe start up flow in pipes. The governing equation is

$$\rho \frac{\partial \tilde{u}}{\partial t} = \frac{\Delta P}{L} + \mu \frac{1}{r} \frac{\partial}{\partial r}\left(r \frac{\partial \tilde{u}}{\partial r}\right) \tag{257}$$

Introducing the dimensionless variables

$$\phi = \frac{4\tilde{u}L}{\Delta P R^2} \tag{258}$$

$$\xi = \frac{r}{R} \tag{259}$$

$$\tau = \frac{\nu t}{R^2} \tag{260}$$

Gives

$$\frac{\partial \phi}{\partial \tau} = 4 + \frac{1}{\xi} \frac{\partial}{\partial \xi}\left(\xi \frac{\partial \phi}{\partial \xi}\right) \tag{261}$$

The solution for $t \to \infty$ is obtained by putting $\partial \phi_\infty / \partial \tau = 0$ giving

$$\phi_\infty = 1 - \xi^2 \tag{262}$$

which is simply the Poiseuille laminar parabolic profile. The full solution is thus made up of a steady component and a time-dependent component.

$$\phi = \phi_\infty + \phi_t \tag{263}$$

Substituting into equation (257) gives

$$\frac{\partial \phi_t}{\partial \tau} = \frac{1}{\xi} \frac{\partial}{\partial \xi}\left(\xi \frac{\partial \phi_t}{\partial \xi}\right) \tag{264}$$

which Szymanski solved by the method of separation of variables with initial and boundary conditions

I.C. $\qquad \tau = 0 \qquad \phi_t = 0$

B.C.1 $\qquad \xi = 0 \qquad \phi_t = finite$

B.C.2 $\qquad \xi = 1 \qquad \phi_t = 0$

Putting

$$\phi_t(\xi,\tau) = \varsigma(\xi)T(\tau) \tag{265}$$

Gives

$$\frac{1}{T}\frac{dT}{d\tau} = \frac{1}{\varsigma}\frac{1}{\xi}\frac{d^2\varsigma}{d\xi^2} = -\alpha^2 \tag{266}$$

$$T = C_0 e^{-\alpha^2\tau} \tag{267}$$

$$\varsigma = C_1 J_0(\alpha\xi) + C_2 Y_0(\alpha\xi) \tag{268}$$

Where $C_0, C_1, C_2$ are constants to be determined and $J_0, Y_0$ are Bessel functions of zero order.

From BC1, $\phi_t$ is finite hence $\varsigma$ must also be finite. Because $Y_0(\alpha\xi)$ tends to minus infinity as $\xi$ tends to zero, $C_2$ must be zero.

From BC2 $\phi_t = 0$ at $\xi = 1$. This requires that $\varsigma$ and therefore $J_0(\alpha\xi)$ be zero at $\xi = 1$.

The values of $\alpha$ where the oscillating function $J_0(\alpha) = 0$ is tabulated in most handbooks of mathematics e.g. $\alpha_1 = 2.405, \alpha_2 = 5.520...$ Then

$$\phi = (1-\xi^2) - \sum_{n=1}^{\infty} B_n J_0(\alpha_n \xi) e^{-\alpha_n^2 \tau} \tag{269}$$

where $B_n = C_0 C_{1n}$. Applying the IC, Szymanski obtained

$$B_n = \frac{8}{\alpha_n^3 J_1(\alpha_n)} \tag{270}$$

Other investigations have been made by Balmer and Florina (1980), Gorla and Madden (1984), Letelier and Leutheusser (1967), Otis (1985) and Patience and Methrotra (1989) but most workers have compared their results with developments in the entrance region because measurements of start up flow are not readily available.

Consider the implications of the Szymanski solution. Like the Stokes solution1, it describes how an original flat velocity distribution is altered over time by the effect viscous diffusion of momentum until it degenerates to the Poiseuille solution at $t \to \infty$. However the Stokes solution1 applies, according to the evidence in sections 7.5, to a mathematical domain bounded by the normalised distance $\delta_v^+$ but the Szymanski solution applies to a physical domain in space delimited by the radius $R$. Thus a probe situated at a position $r$ will actually see a change in flow regime with changing characteristic time $\tau$ of the process.

The agreement with experimental measurements of turbulent velocity profiles is improved by noting again that the initial velocity cannot be flat because it refers to measurements at the inrush phase where fluid rushes from the log-law region towards the wall.. I used again the simplified form of the law of the wall equation (250)

$$U^+ = 2.7 \ln\left(1 + \frac{y^+}{R^+}\right) \tag{250}$$

$$\phi_0 = \frac{U^+}{V^+} = \frac{2.7}{V^+}\ln\left(1 + \frac{R^+(1-\xi)}{2.7}\right) \tag{271}$$

The solution is still equation (269) but the factor $B_n$ must be modified

$$B_n = \frac{8}{\alpha_n^3 J_1(\alpha_n)} + \frac{2\int_0^1 \phi_0 J_0(\alpha_n \xi)\xi d\xi}{[J_1(\alpha_n)]^2} \tag{272}$$

Equations (269) and (272) were solved numerically and time averaged for a period $T^+ = 17.6$ given by Meek and Baer (1970). Equation (272) fitted measured velocity profiles in turbulent pipe flow well but equation (269) did not. Unfortunately I lost the graph when I emigrated from Viet Nam and have yet to sit down to replot it again but I did manage to keep the graph for a similar solution: turbulent flow between parallel surfaces discussed in section 12.3.1

While the Szymanki solution is better able to account for curvature than the Stokes solution1, it is less convenient for determining the thickness of the wall layer. It must be estimated by method 2, Figure 34 section 8. But it does highlight more clearly the basic fact that given sufficient time the original log-law profile will be reduced into the Poiseuille parabolic velocity profile. The reason that this does not occur is because the ejections interrupt this process and limit the time scale available.

The more interesting issue here is that for non-Newtonian fluids we have an estimate of the wall layer thickness $\delta_v^+$ in pipe flow which should be equal to $R_c^+$ at the transition point. The relation between this wall layer thickness for power law fluids and for Newtonian fluids in the Szymanski solution clearly has to obey equation (199).

A very similar problem in encountered for flow between two parallel plates.

## 13  Analogy between heat mass and momentum transfer in turbulent flows

The study of heat and mass transfer has been dominated from an early stage by the similar form of the scalar transport equations and the NS equations. Following Boussinesq, the transport flux (e.g. of heat) can be defined in terms of an eddy viscosity

$$q = -(k + \rho E_h)\frac{d\theta}{dy} \tag{273}$$

where  $k$  is the thermal conductivity

$E_h$  the eddy thermal diffusivity

$q$  the rate of heat transfer flux

Equation (273) may be rearranged as

$$\theta^+ = \int_0^{y^+} \frac{q/q_w}{1+\frac{E_h}{k}} dy^+ \tag{274}$$

which is very similar to equation (101)

### 13.1  Reynolds' analogy and its paradoxes

Reynolds (1874) was the first to propose a formal analogy between heat, mass and momentum transfer expressed by

$$\frac{h}{\rho C_p V} = St = \frac{f}{2} \qquad (275)$$

which requires that the velocity and temperature profiles be the same. (Bird, et. A. 1960, p.382)

$$\frac{d\theta^+}{dy^+} = \frac{dU^+}{dy^+} \qquad (276)$$

It is normally assumed that Reynolds analogy implies two conditions

$$q/q_w = \tau/\tau_w \qquad (277)$$

$$\Pr_t = E_h/E_v = 1 \qquad (278)$$

Equations (277) and (278) are at odds with experimental evidence. These are the paradoxes of the Reynolds analogy. The distributions of heat flux and shear stress in turbulent pipe flow are shown in Figure 70 and clearly are not equal. The distribution of shear stress is linear and unique for all Reynolds numbers but the distribution of heat fluxes is dependent on both the Reynolds and Prandtl number. Similarly, many workers have shown that the turbulent Prandtl number $\Pr_t$ is not unity. Blom & deVries (1968) has collected experimental measurements of the turbulent Prandtl number as shown Figure 71. Later data and interpretations of the turbulent Prandtl number can be found in (Churchill, 2002; Kays, 1994; Malhotra & Kang, 1984; McEligot, Pickett, & Taylor, 1976; McEligot & Taylor, 1996).

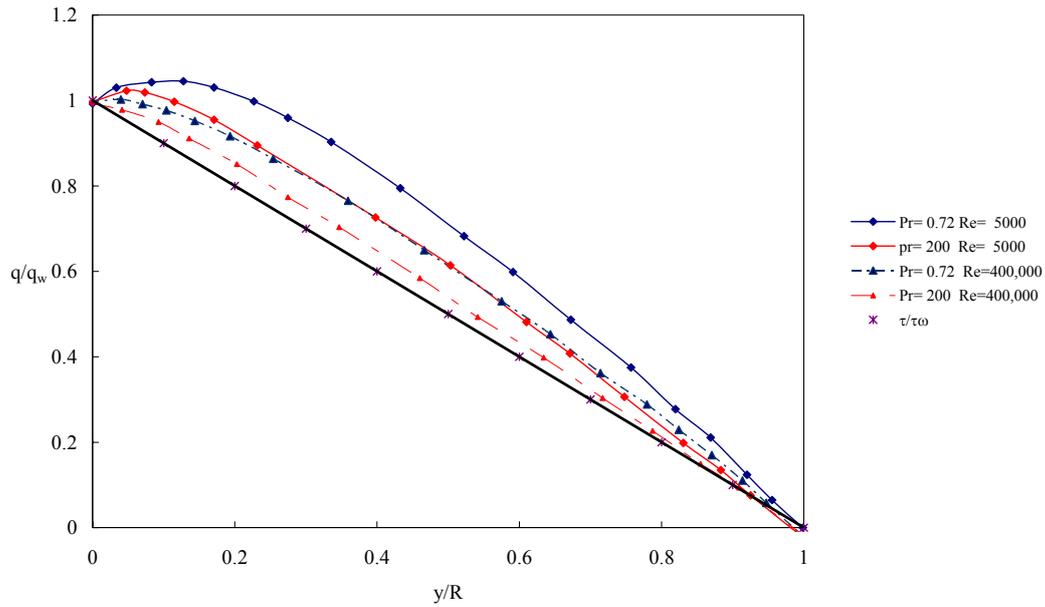

Figure 70. Distribution of shear stress and heat flux in turbulent pipe flow. From Hinze (1959)

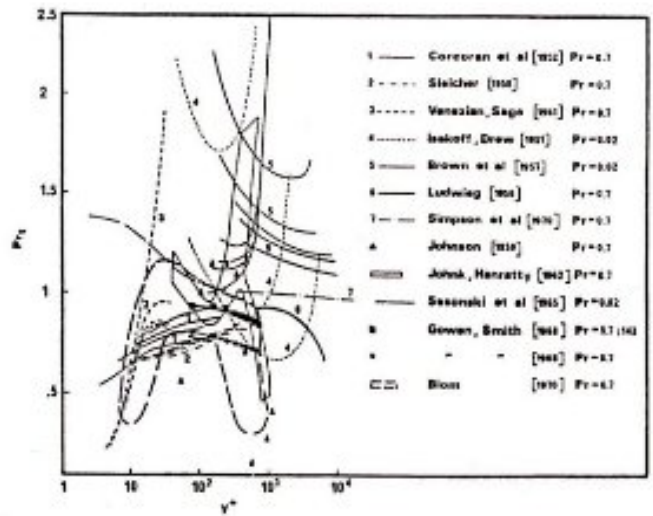

Figure 71, Turbulent Prandtl number $\Pr_t$ from Blom and deVries (1968)

The paradoxes of Reynolds' analogy can easily be explained (Trinh 1969) by comparing equations (98), (274) and (276). Equation (276) can be obtained from equations (98) and (274) without recourse to the assumptions in equations (277) and (278) when we apply a number of simplifications that require

$$\frac{E_h}{k} \gg 1 \quad and \quad \frac{E_v}{v} \gg 1 \qquad \frac{E_h}{E_v} = \frac{q/q_w}{\tau/\tau_w} \qquad (279)$$

As noted in sections 6 and 7 linearisations of the NS equations by approximations or simplifications do not necessarily invalidate their asymptotic solutions but do restrict their domain of applicability. Equation (279) applies only when the diffusive contributions to transport are negligible i.e. when both the velocity and temperature profiles follow a log-law as shown in Figure 72. Thus it was realised very early that Reynolds' analogy only applied to the turbulent core and Prandtl (1910) improved it by assuming that it only applied up to the laminar sub-layer.

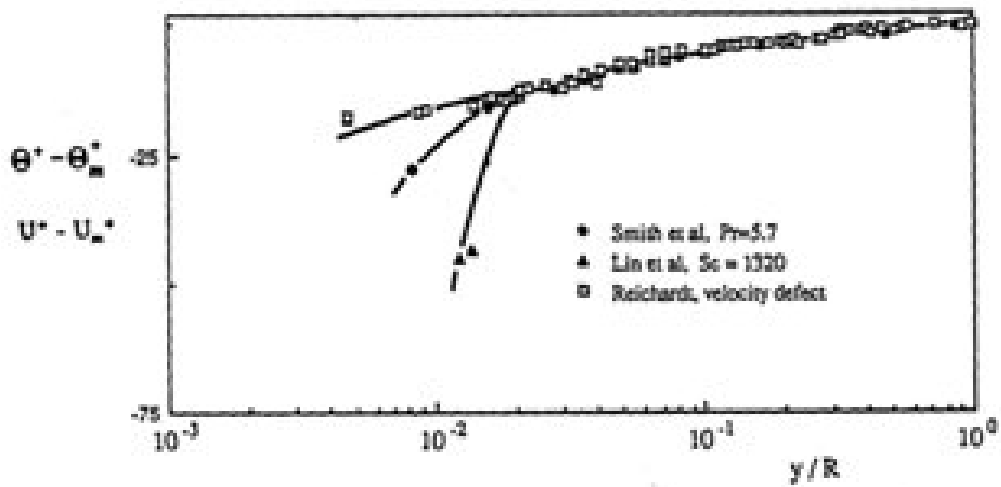

Figure 72 Reynolds' analogy and its range of applicability

### 13.2 Other analogies

Since these early days, many other analogies have been to improve agreement with experimental data. Three main approaches have been adopted:
Empirical correlations the best known being Colburn's analogy (1933)

$$St = \frac{f}{2} \Pr^{-2/3} \qquad (280)$$

Colburn originally formulated it for pipe flow but subsequent experimental verifications show a discrepancy of about 15% as summarised for example in Bird et al. (1960, p. 400). The agreement for external boundary layer flow is better.

Boundary layer theories e.g. (Deissler, 1955; Karman, 1939; Levich, 1962; Martinelli, 1947; Metzner & Friend, 1958; Reichardt, 1961) are based on solutions of equation (274). As in momentum transfer, the closure of the models is based on mathematical or physical postulates about the eddy diffusivity $E_h$. This information is often supplied through some experimental measurement of the turbulent Prandtl number. Because of this, boundary layer theories are still referred to as analogies, even though their form has been obtained in a more theoretical framework than the Colburn analogy. In particular Spalding (1961) successfully expanded the velocity near the wall in a Taylor series and showed that the eddy diffusivity scaled with $y^3$. By analogy the eddy thermal diffusivity is assumed to be $E_h \sim y^3$ in many heat transfer studies (Churchill, 1996, 1997)

Penetration theories originated with Higbie (1935) who used the equation for unsteady conduction to model the transport process in jets and packed columns.

$$\frac{\partial \theta}{\partial t} = \alpha \frac{\partial^2 \theta}{\partial y^2} \qquad (129$$

where $\alpha$ is the thermal diffusivity. The well-known solution is

$$q_w = \frac{1}{\sqrt{\pi}} \frac{\Delta \theta_m}{\sqrt{\alpha t_h}} \qquad (281)$$

Higbie closed the derivation by assuming that the typical time scale over which equation ((281) applies is the contact time

$$t = \frac{x}{U_\infty} \qquad (134$$

where x is the swept length. In an effort to apply Higbie's approach to turbulent transport, Danckwerts (1951) assumed that the surface near the wall is periodically swept clean by eddies penetrating from the bulk stream. The rate of renewal of the surface fluid near the wall is a function of the probability of occurrence of eddies of various frequencies. Danckwerts assumed this probability distribution to be uniform.

Subsequent postulates of the surface renewal distributions have been reviewed by Mathpati & Joshi (2007; Pletcher (1988; Ruckenstein (1987; Sideman & Pinczewski (1975). Many of these postulates do not link the assumed distribution of eddies to the improved understanding of the coherent structures or the wall structure but more recent work does e.g. (Fortuin, Musschenga, & Hamersma, 1992).

Ruckenstein (1968) first attempted to derive a physical model for the distribution function by modelling the eddy as a roll cell which circulates the fluid from the wall to the outer region. The motion close to the wall surface is assumed to obey the laminar transport equation

$$U\frac{\partial \theta}{\partial x}+V\frac{\partial \theta}{\partial y}=\alpha \frac{\partial^2 \theta}{\partial y^2} \qquad (282)$$

Ruckenstein calls this state "pseudo-laminar flow" but does not elaborate about the relation between this state and the bursting phenomenon at the wall. Thomas and Fan (1971) used an eddy cell model proposed by Lamont and Scott (1970) in conjunction with a wall model by Black (1969) and the time scale measured by Meek and Baer (1970) to model the whole process. In both these approaches, the differentiation between the instantaneous fluxes and their time-averaged values is unclear and rough approximations are necessary to effect closure of the solution. Experimental measurements to vindicate these visualisations are difficult to obtain because the wall layer in mass transfer processes is extremely thin. Perhaps the most extensive studies have been attempted by Hanratty and his associates. Their ideas have evolved, along with improved experimental evidence, from a belief that the eddy diffusivity near the wall is proportional to $y^4$ at very high Schmidt numbers (Son & Hanratty, 1967), as predicted by Deissler (1955) to a belief that a more accurate power index is 3.38 (Shaw & Hanratty, 1964, 1977) to an argument that the analogy between heat and mass transfer breaks down completely very close to the wall (Na & Hanratty, 2000). The research of Hanratty showed that the characteristic length scale of mass transfer in the longitudinal direction is equal to that for momentum transfer (Shaw and Hanratty 1964, 1977) but the time scale for mass transfer is much shorter that for momentum transfer.

To explain this perplexing effect, Campbell and Hanratty (1981, 1983) have solved the unsteady mass transfer equations without neglecting the normal component of the convection velocity, which they model as a function of both time and distance. They found that only the low frequency components of the velocity fluctuations affect the mass transfer rates and that the energetic frequencies associated with the bursting process have no effect. In their explanation, the concentration sub-boundary layer acts as a low pass filter for the effect of velocity fluctuations on the mass transport close to the wall. The existence of two time scales in the wall region of heat or mass transfer has been noted by all modern investigators. Their explanation is varied. McLeod and Ponton (1977) differentiate between the renewal period and the transit time which is defined as the average time that an eddy takes to pass over a fixed observer at the wall. Loughlin et al. (1985) and more recently Fortuin et al. (1992) differentiate between the renewal time and the age of an eddy.

### 13.3 This visualisation

The analysis of heat transfer starts with the energy equation for a fluid of constant density

$$\rho C_p \frac{D\theta}{Dt} = -k\nabla^2\theta \tag{283}$$

Where the velocity terms must be taken from the solution of the NS equations. Our analysis in sections 3 and 4 indicates immediately that the solution of equation (283) must be also be linked with velocity fields from two different solutions of order $\varepsilon^0$ and $\varepsilon$.

### 13.3.1 Physical concepts

In the present application to heat and mass transfer (Trinh 1992), the ejected streaming flow brings patches of wall-layer fluid into the outer region by a mechanism of agitation. As Broadwell and Mungal (1991) put it in their review of the large structures in turbulent flow, agitation is the convection and dispersion of bulk fluid from one region to the other. This process is linked with the solution of order $\varepsilon$. The final blending of this convected wall-fluid with the surrounding stream is achieved by molecular diffusion. Since heat, mass and momentum are convected simultaneously by the same ejection the profiles of velocity, temperature and mass change in the same manner. This physical interpretation is fully compatible with the existence of a Reynolds analogy outside the wall layer.

On the other hand the heat transfer linked with the solution of order $\varepsilon^0$ is essentially diffusive. We note that the velocity $\tilde{u}_i$ to be used in equation (283) is the smoothed velocity in the sweep phase of the wall layer, not the instantaneous velocity that will also have a fast fluctuating component $u'_i$.

### 13.3.2 The paradox of time scales

Most modern penetration theories for turbulent transport in boundary layer flow start with some form of a momentum model for the wall layer based on the original proposal by Einstein and Li (1956, op.cit.). This approach is well validated by the observation by the Hanratty group that the characteristic length scale of mass transfer in the longitudinal direction is equal to that for momentum transfer (Shaw and Hanratty 1964, 1977). However when the time of scale the wall layer measured for example by Kline et al (op.cit.) and Meek and Baer (op.cit.) is used in conjunction with equation (129) the solutions are unable to predict accurately the heat or mass transfer rate at the wall. In fact we can define two non-dimensional time scales from equations (59) and (129)

$$t_v^+ = u_* \sqrt{\frac{t_v}{\nu}} \tag{284}$$

$$t_h^+ = u_* \sqrt{\frac{t_h}{\alpha}} \tag{285}$$

The existence of two time scales implies that two different mechanisms exist for momentum and heat (or mass) transfer. But the application of Reynolds' analogy for the turbulent core implies that there is only one common mechanism of agitation. Surprisingly this paradox has not been discussed in the literature.

The only situation when the distinction between these two time scales does not matter is when the process is so short that the diffusions implied in equations (59) and (129) have not reached steady state. In this case the ratio of the thermal and momentum layer thickness scale with $\text{Pr}^{1/2}$ and the only time scale that matters is given by equation (134). This explains the successful prediction of gas absorption in short wetted wall columns by Highbie (op.cit.).

When steady state is reached the heat transfer coefficient scales with $\text{Pr}^{1/3}$ for high Prandtl numbers and Highbies' approach fails. The surface renewal theory of Danckwerts (1955) discussed in section 13.2 dealt with this problem by including a second time scale. As discussed in section 13.2, he postulated that while the fluid resides near the wall eddies from the outer flow regularly come in and renew the surface of this wall layer. Many authors have expanded on the ideas of Danckwerts, mainly by postulating different heuristic functions for the distribution of renewal times. A recent description of these can be found in Mathpati and Joshi (2007) who have also reviewed boundary layer theories of transport.

The differentiation between the residence time of a packet of fluid at the wall and the frequency at which it is renewed by fluid from the bulk flow does not explain, in my view, the difference between the two time scales in equations (284) and (285) because surely the momentum, heat and mass in the wall layer must be renewed simultaneously by the incoming eddies. For example, if one explains the second time scale in the thermal wall layer in terms of the "age" of eddies sweeping the wall (Fortuin et al, 1991) one is faced with the question as to why the age of ejected lumps of fluid from the wall does not affect Reynolds' analogy. Indeed any argument based on convective forces which are specific to heat and mass transfer (i.e. not present in momentum transfer) raises a paradox when one applies Reynolds' analogy.

In the present theory, summarised in Figure 52 there is only one mechanism for agitation in a turbulent boundary layer. This agitation process relies on the intermittent ejection of wall fluid into the outer region and its time scale is $t_\nu$, the time scale of the momentum wall layer. The second time scale $t_h$ reflects a diffusion process within the wall layer and therefore does not relate to the agitation mechanism;

it is related to the diffusion of heat and mass across the wall layer, as argued in section 9.2 and 9.3 and best understood if one analyses equations (59) and (129) in a Lagrangian context. Because of the difference in momentum and thermal diffusivities, the depth of penetration of heat and viscous momentum from the wall differ as shown by the conceptual illustration of a mapping of the velocity and temperature contours in Figure 73.

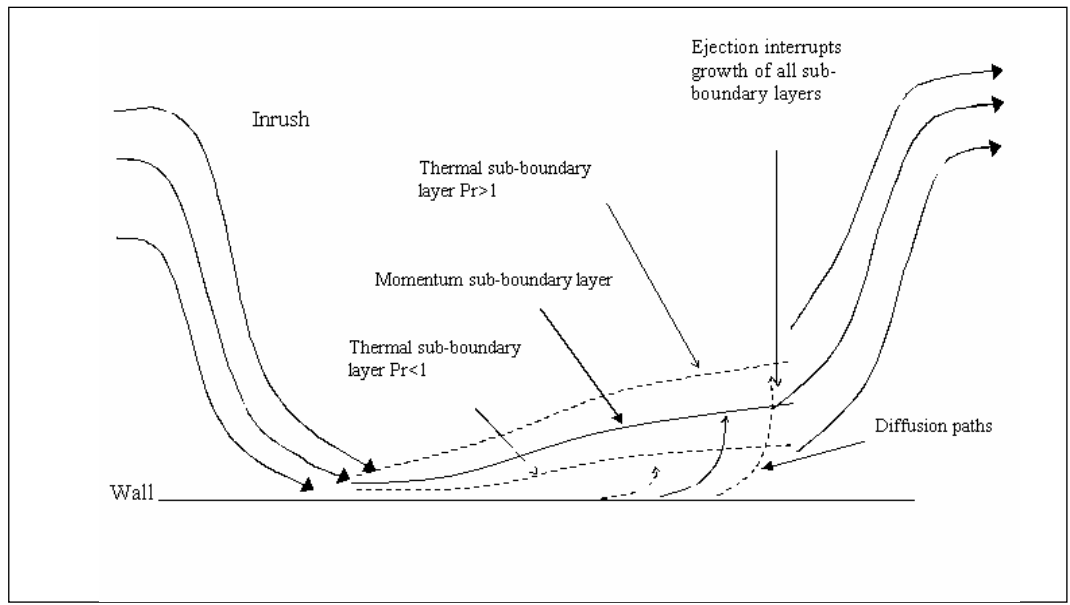

Figure 73  Contours of velocity and temperature in the wall layer.

In my view there is only one defining time scale for the wall layer $t_v$ because it defines the moment when the ejection occurs and therefore sets the life time of the low speed streak. Both the momentum and thermal content of the wall fluid are regenerated because of the subsequent inrush from the main stream. The surface renewal theory would require two different agitation events: the ejections and other separate rushes of fluid into the wall during the sweep phase. I have never seen any publication giving evidence of eddies or streams penetrating through the hair pin vortices into the low-speed streaks. To understand the role of the time scale $t_h$ we consider two cases. When Pr>1, then $t_h < t_v$ and the thermal boundary layer will have reached its maximum thickness before the low speed streak is ejected from the wall. Then for high Pr the Nusselt number scales with $Pr^{1/3}$ as shown in section 9.2. But when $t_h > t_v$ the pocket of fluid at the wall is ejected before the diffusion of heat from the wall has reached its full potential and the heat transfer process is still unsteady state. Therefore the Nusselt number scales with $Pr^{1/2}$.

### 13.3.3 The paradox of eddy diffusivities and viscosities

We have already shown that Reynolds' analogy does not necessarily imply that $Pr_t = 1$ as often stated (e.g. Bird et al, 2002, p.410). There is however much more to explore. The eddy viscosity and diffusivity are abstractions devised for convenience and it is much more useful to examine the problem by expressing the temperature in a turbulent flow field in terms of two components following the method of Reynolds

$$\theta = \Theta + \Theta' \tag{286}$$

Then time averaging equation (283) gives

$$U \frac{\partial \rho C_p \Theta}{\partial x_i} = k \frac{\partial \Theta}{\partial x_i^2} + \frac{\partial \rho C_p \overline{U_i' \Theta'}}{\partial x_i} \tag{287}$$

where the terms $\overline{U_i'\Theta'}$ are called the turbulent heat fluxes in similarity to the Reynolds stresses.

  $\Theta$ long term averaged temperature
  $\Theta'$ difference between instantaneous and long term average temperatures

We may again decompose the temperature further as in section 2.1 by writing

$$\theta = \Theta + \widetilde{\Theta} + \theta' = \widetilde{\theta} + \theta' \tag{288}$$

Where

  $\widetilde{\theta}$ is the smoothed phase temperature
  $\theta'$ the fast temperature fluctuations, the difference between the instantaneous and smoothed phase temperature, which are treated as periodic
  $\widetilde{\Theta}$ the difference between the smoothed phase temperature and the long term average

Substituting equation (288) into (283) and integrating over the period $t_v$ brings out new terms not evident in equation (287) but what do they represent physically?

$$\frac{\partial}{\partial t}\rho C_p \tilde{\theta} = k\frac{\partial^2 \tilde{\theta}}{\partial x_i^2} - \frac{\partial \rho C_p \tilde{u}_i \tilde{\theta}}{\partial x_i} - \frac{\partial \rho C_p \overline{u_i' \theta'}}{\partial x_i} \qquad (289)$$

Can the terms $\overline{u_i' \theta'}$ lead to an effect similar to the streaming function described in sections 3 and 4? To answer this question let us consider a hypothetical experiment where an unsteady laminar boundary layer is generated without small periodic velocity disturbances. This can be done for example by towing a plate steadily through a stagnant liquid for a period t. At the same time, let us apply a fluctuating heat load onto the liquid. Let us further repeat this exercise many times, each time changing the velocity and the duration of the towing so that the period varies between 0 and $t_v$. Then the local instantaneous velocity is $u_i = \tilde{u}_i + u_i'$ as in equation (8) where $\tilde{u}_i$ is the average of all the runs and $u_i'$ a fluctuating component. In this experiment the fluctuating component $u_i'$ is however not periodic. The temperature is also given by equation (288) but the term $\theta'$ is periodic. In this situation, the terms $u_i' \theta'$ are periodic and disappear upon averaging and only the terms $\tilde{u}_i \tilde{\theta}_i$ remain. Thus fast fluctuations in temperature alone cannot create a streaming effect and cannot destabilize a laminar flow field. This phenomenon is created by periodic fluctuations of velocity. There is thus a fundamental difference in the mathematical behaviour of the Reynolds stresses and the turbulent heat fluxes. There is a limit to drawing analogies between heat, mass and momentum transfers in turbulence.

It must be recalled that the eddy viscosity and diffusivity are just mathematical abstractions to help us correlate turbulent transport; they are not real physical quantities. This was clearly demonstrated in sections 7 and 8 when we showed that the solution of order $\varepsilon^0$, the erf, correlated very well the velocity profile in the wall layer and gave good predictions of most of the statistics traditionally attributed to turbulence. This solution is based solely on viscous diffusion and does not need to call on the eddy viscosity concept. In fact the traditional Reynolds stresses in the wall layer can be obtained simply by time averaging the differences of the instantaneous smoothed phase velocity and its long time average. A similar proof can be made for the eddy diffusivity in the wall layer. Thus it does not appear convincing to argue for particular models of the Reynolds stresses in closure methods for CFDs by invoking physical arguments of structures.

Kays (1994) also independently concluded that the turbulent Prandtl number has no physical significance. I will go even further and propose that the turbulent Prandtl is well predicted by equation (279) and that more attention must be given to accurate estimates of the distributions of momentum and heat flux than they have in modelling of turbulent transport.

**13.3.4 Correlations for heat and mass transfer**

Following the method in section 9.1, we accept a general form of the log-law as a starting point. Since heat, mass and momentum are convected by the same and only ejection, we accept the Karman constant already used in section 9.1 as the slope of the log-law. Then an analysis similar to section 9.1 gives

$$\Theta^+ = \int_0^{y^+} \frac{(q/q_w)dy^+}{1+0.4y^+(q/q_w)\sqrt{\frac{E_h}{1+E_h}\left(1-e^{-11.2\left[\frac{y^+(1-y^+/R^+)}{\delta_h^+}\right]^2}\right)}} \quad (290)$$

and the solution can be closed theoretically by feeding back into equation (290) the solution from equation (129).

It is useful to consider the solution from an alternative angle to discuss important problems related to the development of analogies both in the boundary layer or penetration approaches. All modern analogies accept that there are two distinct regions in a turbulent thermal/concentration/momentum field. A wall layer and a region outside it, which obeys a log law outside the wall layer and some other law like Cole's law of the wake in the far field region e.g. (Antonia & Kim, 1991a, 1991b) for the moment, following Prandtl (op.cit.) and Karman (op.cit.), we assume that the effect of the far field profile on the predictions of heat, mass and momentum transfer from the wall can be neglected and that the log-law can be assumed to extend to the limit of the boundary layer. This simplification is quite acceptable for pipe flow. The difference between the traditional boundary layer and the penetration approaches is that the former treats the wall layer in terms of time-averaged steady-state parameters that must be modelled empirically whereas the latter views it in terms of an unsteady process and models its time scales (particularly the distribution of renewal scales) also empirically.

In terms of flow dynamics there is only one wall layer that is dominated by the solution of order $\varepsilon^0$ identified in sections 3 and 4 and detailed in sections 7.5 and 7.6. In my view, we cannot argue for separate wall layers for heat, mass and momentum in a hydrodynamic sense; we can only argue (section 13.3.2) that the two time scales in equations (237) and (238) indicate different depths of penetration of heat, mass and viscous momentum into this unsteady laminar sub-boundary layer as shown in the analysis of Trinh and Keey (1992), section 9.2. We can again show that many statistical parameters that are traditionally interpreted as illustrative of turbulent mass and heat transfer can be derived from averages of the solution of transfer in an unsteady state laminar sub-boundary layer. I have not spent time finalizing these graphs, to free the precious little time I can devote to this pursuit of personal interest to discuss more important issues.

The analysis of Trinh and Keey (1992) shows that the boundary layer and penetration theories are inter convertible. The major difference between my views and those I find in the literature, particularly as far as penetration theories are concerned, is that I do not believe in small eddies coming in from the bulk flow to renew the low speed streaks while they are attached to the wall: I find no single published experimental evidence to support that postulate. In my view, the entire wall layer is renewed by the inrush that sets up new low-speed streaks. When we substitute the time scale defined by equation (135) into the solution for the thermal sub-boundary layer (equation (132), we obtain

$$\sigma = \frac{\delta_h}{\delta_v} = \frac{2\sqrt{\alpha f(u) x / U_v}}{\delta_v} = \Pr^b \qquad (291)$$

where the index b is given by equation (141). The reader will see immediately that the effect of the Prandtl number on the parameter $\sigma$ is the same for the wall layer as for traditional laminar boundary layer flow (Schlichting 1960, p), an observation I made intuitively in my early work (Trinh, 1969). Equation (280) may be rearranged as

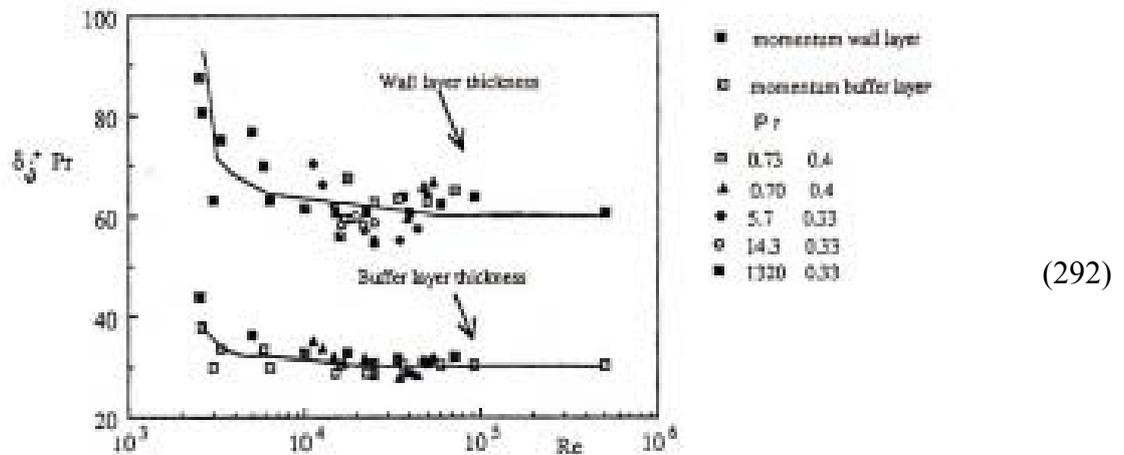

(292)

$$St = \frac{h}{\rho C_p U_v} = \frac{\tau_w}{\rho U_v^2} \Pr^{1-b} = \frac{f}{2} \Pr^{1-b}$$

and

$$\Theta_h^+ = U_v^+ \Pr^{1-b} \qquad (293)$$

where

$$\Theta^+ = \frac{(\Theta - \Theta_w)\rho C_p u_*}{q_w} \qquad (294)$$

Since

$$h = \frac{q_w}{\Theta - \Theta_w} \qquad (295)220)$$

Combining equations (64), (291) and (294) gives

$$\Theta_h^+ = \frac{\delta_h^+ \Pr}{4.16} \qquad (296)$$

We may now plot equation (296) against measurements of temperature/concentration profiles to give an estimate of the sub-boundary layer thickness for heat and mass diffusion. The results, shown in Figure 74 and should be compared with Figure 38 to appreciate the level of match between theory and experiment.

Figure 74 Normalised thickness of the thermal wall layer

The parameters in the log-law can now be estimated by forcing it though the point $\Theta_h^+, \delta_h^+$ given by equation (296) to give the full temperature, concentration profile (Figure 75)

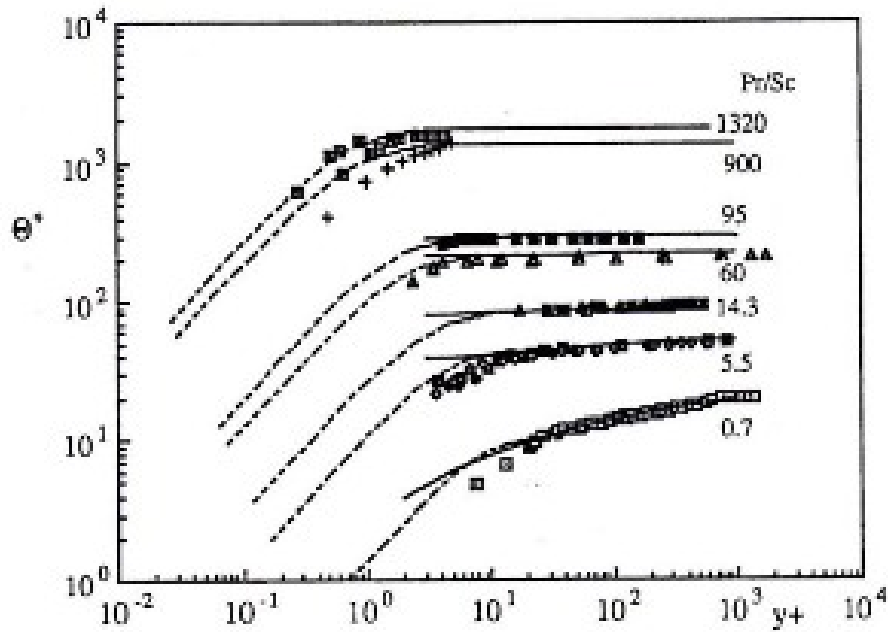

Figure 75. Temperature and concentration profiles predicted from Trinh 1992. Data for Pr=0.7 (Johnk & Hanratty, 1962), 5.5, 14.7(Gowen & Smith, 1967), 60(Neumann, 1968), 95(Janberg, 1970), Sc=700 (Lin, Moulton, & Putnam, 1953), 1320 (Flender & Hiby, 1981) also reported in (Kader, 1981)

At the pipe axis, $y^+ = R^+$, $\Theta^+ = \Theta_\infty^+$ and $R^+ = \dfrac{Re}{2}\sqrt{\dfrac{f}{2}}$ and the Nusselt number can be derived from the temperature profile by standard techniques (e.g. Schlichting, 1960). We may derive a multitude of simpler correlations by making a series of simplifying assumptions (Trinh, 1969). The simplest is to divide the temperature profile into a log law and an artificial thermal conductive 'sub-layer' where

$$\Theta^+ = y^+ \Pr \tag{297}$$

The intersection between these two lines occurs at the distance

$$y_{k,h}^+ = 11.8 \Pr^b \tag{298}$$

The parameter $y_{k,h}$ is the heat transfer equivalent of the Kolmogorov scale and further buttresses my argument that the Kolmogorov scale should not be interpreted as the size of the energy dissipating *eddies*. Are we going to argue then that there separate size eddies for the dissipation of heat, mass and momentum, a highly improbable physical concept, and how will that fit with the Reynolds analogy? The Kolmogorov scale in mass transfer is of the order of $y^+_{k,m} \approx 1$. The exponent b is obtained directly from the solutions of order $\varepsilon^0$ for heat, mass and momentum transfers and therefore are the same for laminar and turbulent flows. There is no need to determine it experimentally e.g. (Metzner & Friend, 1958).

And it can easily be shown (Trinh 1969) that this solution leads to

$$St = \frac{f/2}{1+11.8\sqrt{f/2}\left(Pr^{1-b}-1\right)} \quad (299)$$

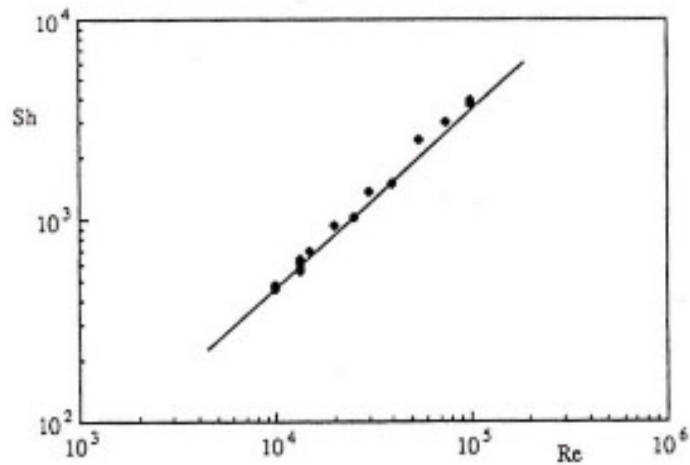

Figure 76  Sh vs Re Data Harriott and Hamilton (1965)

which becomes the Metzner-Friend analogy (1958) for high Prandtl numbers when $b \approx 1/3$. A more complex derivation that takes account of the buffer layer (Trinh, 1983, unpublished, section 13.3.8) similar to the Karman analogy (1939) leads to

$$St = \frac{\sqrt{f/2}}{2.5 l n \mathrm{Re}\sqrt{f} + 13.5 \mathrm{Pr}^{1-b} - 11.1 - 2.5 Ln \mathrm{Pr}^b - D(\mathrm{Pr})} \quad (300)$$

where

$$D(\mathrm{Pr}) = \Theta_\infty^+ - \Theta_b^+ \quad (301)$$

and $\Theta_b^+$ is the mixing cup temperature given by

$$\Theta_b^+ = \frac{1}{\pi R^{+2}} \int \Theta^+ U^+ 2\pi (R^+ - y^+) dy^+ \quad (302)$$

Equation (302) is shown against some literature data in Figure 77 for one Reynolds number.

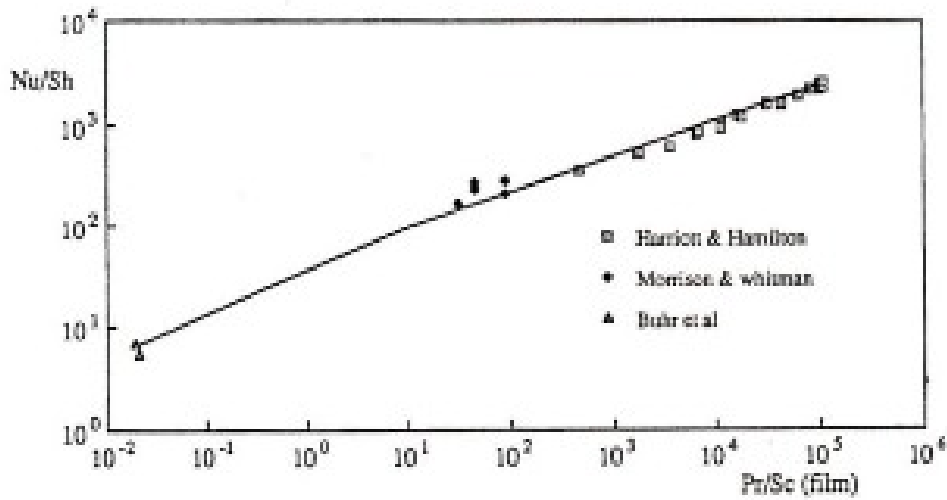

Figure 77 Predicted Nu/Sh vs Pr/Sc. Data of (Buhr, Carr, & Balzhiser, 1968; Harriott & Hamilton, 1965; Morrison and Whitman, 1928).

We note that the use of the solution of order $\varepsilon^0$ implies that the physical properties used for the Prandtl and Schmidt numbers must be evaluated at the temperatures and concentrations prevalent in the wall layer whereas the Reynolds number is usually based on the viscosity at the temperatures and concentrations of the bulk flow. When both the Prandtl and Reynolds number are evaluated at the bulk conditions, a correction must be made. In heat transfer, Sieder and Tate (1936) proposed a ratio of viscosities evaluated at the wall and bulk temperatures. Deissler (1955) took another approach and calculated directly the physical properties used in the Prandtl number at the temperatures of the wall layer. Harriott and Hamilton (1965) noted that the plot of Sherwood number against Reynolds number for mass transfer is usually 10% lower than the equivalent plot of Nusselt number for heat. I noted (Trinh 1969) that while the viscosity varied substantially with temperature, it is less dependent on the concentration of the diffusing species, particularly for the sparingly soluble solids normally used in mass transfer experiments. However the diffusivity changes rapidly with normal distance from the wall since the liquid is saturated at the wall but shows a very low concentration in the bulk flow. Thus when the Schmidt number was calculated from the value of diffusivity in the wall layer, the plots of Sherwood and Nusselt numbers coincided. The correction factor for mass transfer equivalent to Sieder and Tate's for heat transfer must be based on a ratio of diffusivities.

### 13.3.5 The Colburn analogy

We are not required to use the log-law to apply Reynolds' analogy, we can use the alternate power law representation of the velocity and temperature profile.

$$\frac{\Theta - \Theta_w}{\Theta_\infty - \Theta_w} = \left(\frac{y}{\delta_{h,t}}\right)^{p'} \qquad \frac{\Theta^+}{\Theta_\infty^+} = \left(\frac{y^+}{\delta_{h,t}^+}\right)^{p'} \qquad (303)$$

$$\frac{U}{U_\infty} = \left(\frac{y}{\delta_{m,t}}\right)^{p} \qquad \frac{U^+}{U_\infty^+} = \left(\frac{y^+}{\delta_{h,t}^+}\right)^{p} \qquad (304)$$

In general, I prefer the log-law because it is based on the similarity arguments of Millikan and as such can be viewed as more theoretical than the power law, which is an entirely empirical observation. Nonetheless, since there is still a debate as to which law should be used (Afzal, 2001, 2005; Zagarola & Smits, 1998) it is interesting to derive the analogy that Colburn (1933) simply stated empirically. I did this exercise to pass the time in the dark, one night when the power was cut-off (Trinh 1984)[12].

---

[12] In Saigon, the power was not available during the day for ordinary households throughout most of the 1980's and part of the 1970's after the communist victory but was turned on at 6 p.m. for 3 nights a week and after 10 for other days.

Power law profiles have a major weakness: they result in a zero velocity and temperature gradient at the wall and cannot be used to predict the rate of momentum and heat transport at the wall. We side step that problem by assuming that equations (303) and (304) do not apply straight to the wall and that a thin layer exists that is dominated by a diffusion process. Thus we forcing equation (304) through the Kolmogoroff point (11.8, 11.8) and obtain

$$\delta_{m,t}^+ = 11.8 \left( \frac{U_\infty^+}{11.8} \right)^{1/p} \tag{305}$$

Similarly we force equation (303) through the point $(11.8\,\mathrm{Pr}^{-b}, 11.8\,\mathrm{Pr}^{1-b})$ as we did for the power law in section 10.2

$$\delta_{h,t}^+ = 11.8\,\mathrm{Pr}^{-b} \left( \frac{\Theta_\infty^+}{11.8\,\mathrm{Pr}^{1-b}} \right)^{1/p'} \tag{306}$$

Then the ratio between the turbulent thermal and momentum boundary layers $\delta_{h,t}^+, \delta_{m,t}^+$ becomes

$$\sigma_t = \frac{\delta_{h,t}^+}{\delta_{m,t}^+} = \frac{\Theta_\infty^{+\,1/p'}}{U_\infty^{+\,1/p}} \mathrm{Pr}^{-b-\frac{1-b}{p'}} 11.8^{\frac{1}{p}-\frac{1}{p'}} \tag{307}$$

We now apply Reynolds' analogy by stating

$$p = p' \tag{308}$$

then
$$\sigma_t = \left( \frac{\Theta_\infty^+}{U_\infty^+} \right)^{1/p} \mathrm{Pr}^{\frac{b-bp-1}{p}} \tag{309}$$

But

$$\frac{\Theta_\infty^+}{U_\infty^+} = \frac{f/2}{St} \tag{310}$$

$$\sigma_t = \left(\frac{f/2}{St}\right)^p \Pr^{\frac{b-bp-1}{p}} \tag{311}$$

The Stanton number is obtained from the integral energy equation (Knudsen & Katz, 1958 p.420)

$$St = \frac{\partial}{\partial x} \int_0^{\delta_{h,t}} \frac{U}{U_\infty} \left[1 - \left(\frac{\Theta - \Theta_w}{\Theta_\infty - \Theta_w}\right)\right] dy \tag{312}$$

Substituting for (303) and (304)

$$St = \frac{\partial}{\partial x} \int_0^{\delta_{h,t}} \left(\frac{y}{\delta_{m,t}}\right)^p \left[1 - \left(\frac{y}{\delta_{h,t}}\right)^{p'}\right] dy \tag{313}$$

$$St = \frac{\partial}{\partial x}\left[\frac{p\delta_{m,t}}{(1+p)(1+p+p')} \sigma_t^{p+1}\right] = \left[\frac{p\sigma_t^{p+1}}{(2p+1)(p+1)}\right]\frac{\partial \delta_{m,t}}{\partial x} \tag{314}$$

Most experimental data show that the ratio $(f/2)/St$ is independent of $x$ (Reynolds claimed that it is unity, 1883) and we can take $\sigma_t$ to be independent of $x$. Then

$$St = \left(\frac{\delta_{m,t}}{x}\right)\left[\frac{p\sigma_t^{p+1}}{(2p+1)(p+1)}\right] \tag{315}$$

The boundary layer $\delta_{m,t}$ can be estimated from the integral momentum equation by standard techniques. The friction factor can be expressed as

$$f = \frac{\alpha}{\operatorname{Re}_g^\beta} \tag{316}$$

giving (Skelland, 1967)

$$p = \frac{\beta}{2 - \beta} \tag{317}$$

$$\frac{\delta_{m,t}}{x} = \left[\frac{\alpha(\beta+1)(\beta+2)}{2-\beta}\right]^{\frac{1}{\beta+1}} \left[\frac{\nu}{U_\infty x}\right]^{\frac{\beta}{\beta+1}} \tag{318}$$

$$\frac{\delta_{m,t}}{x} = \left[\frac{\alpha(1+3p)(1+2p)}{p}\right]^{\frac{1+p}{1+3p}} \left[\frac{\nu}{U_\infty x}\right]^{\frac{2p}{1+3p}} \tag{319}$$

Combining (311), (315) and (319), and rearranging gives

$$St = \frac{f}{2} \Pr^{(b-bp-1)\frac{p+1}{p+2}} \tag{320}$$

Putting $b = 1/3$ (high Schmidt numbers) and $p = 1/7$ (Blasius power law) gives

$$St = \frac{f}{2} \Pr^{-.63} \tag{321}$$

This derivation shows that the Colburn analogy applies best to external boundary layers and less to pipe flow where the radius R forces the thermal and momentum layers to the same thickness

### 13.3.6 The Calderbank and Moo-Young correlation

Many authors have attempted to use Kolmogorov's ideas of local isotropy to heat and mass transfer. Perhaps the most interesting work is due to Calderbank and Moo-Young (1962). They defined a "local isotropic Reynolds number as

$$\mathrm{Re}_k = \frac{d\sqrt{\overline{u'^2}}\rho}{\mu} \tag{322}$$

using Batchelor's relation for the fluctuating velocity over a distance $\lambda \leq d \leq L$

$$\sqrt{\overline{u'^2}} \approx \zeta^{2/3} \left(\frac{d}{\rho}\right)^{2/3} \tag{323}$$

$$\mathrm{Re}_k = \frac{d^{2/2} \zeta^{1/6} \rho^{1/3}}{\mu^{1/2}} \tag{324}$$

Using a standard formulation for the Sherwood number

$$Sh = A\operatorname{Re}_*^a Sc^b \tag{325}$$

And setting $b = 1/3$ Calderbank and Moo Young obtained

$$kSc^{2/3} = A\left(\frac{\zeta\mu}{\rho^2}\right)^{1/4} \tag{326}$$

The advantage of this correlation is that it is based on the power input per unit volume instead of the Reynolds number which is quite sensitive to the characteristic length of the geometry e.g. pipe diameter or impeller diameter in agitated vessels. Calderbank and MooYoung found that the value $A = .13$ correlated data over a wide range of flow geometries (Figure 78).

The Calderbank and Moo-Young relationship does not need to be empirical and can be derived from basic principles. Dimensional considerations suggest the following relation between the turbulent stress $u'^2$ the scale over which it is assessed $\lambda$ and energy per unit volume $\zeta$ as (Hinze, op.cit.)

$$\frac{\tau_t}{\rho} \approx u'^2 \approx \left(\frac{\zeta\lambda}{\rho}\right)^{2/3} \tag{327}$$

The turbulent stress cannot be readily predicted anywhere except at the Kolmogorov point $\lambda = y_K$ where it is equal to the viscous stress and therefore half of the total stress as discussed in section 7.6. Substituting equation (82) into ((327) gives

$$\frac{\tau_t}{\rho} \approx \left(\frac{\zeta\lambda_k}{\rho}\right)^{2/3} \approx \left(\frac{\zeta\mu}{\rho}\right)^{1/2} \tag{328}$$

$$\frac{\tau}{\rho} = 2\left(\frac{\zeta\mu}{\rho}\right)^{1/2} \tag{329}$$

We set approximately $\tau_w \approx \tau$ since $\lambda_k \ll \delta$ and substitute equation (329) into the mass transfer equivalent of equation (299)

$$St = \frac{f/2}{1+11.8\sqrt{f/2}(Sc^{1-b}-1)} \approx \frac{\sqrt{f/2}}{11.8(Sc^{1-b}-1)} \approx \frac{\sqrt{f/2}}{11.8Sc^{1-b}} \quad \text{for } Sc \gg 1 \tag{330}$$

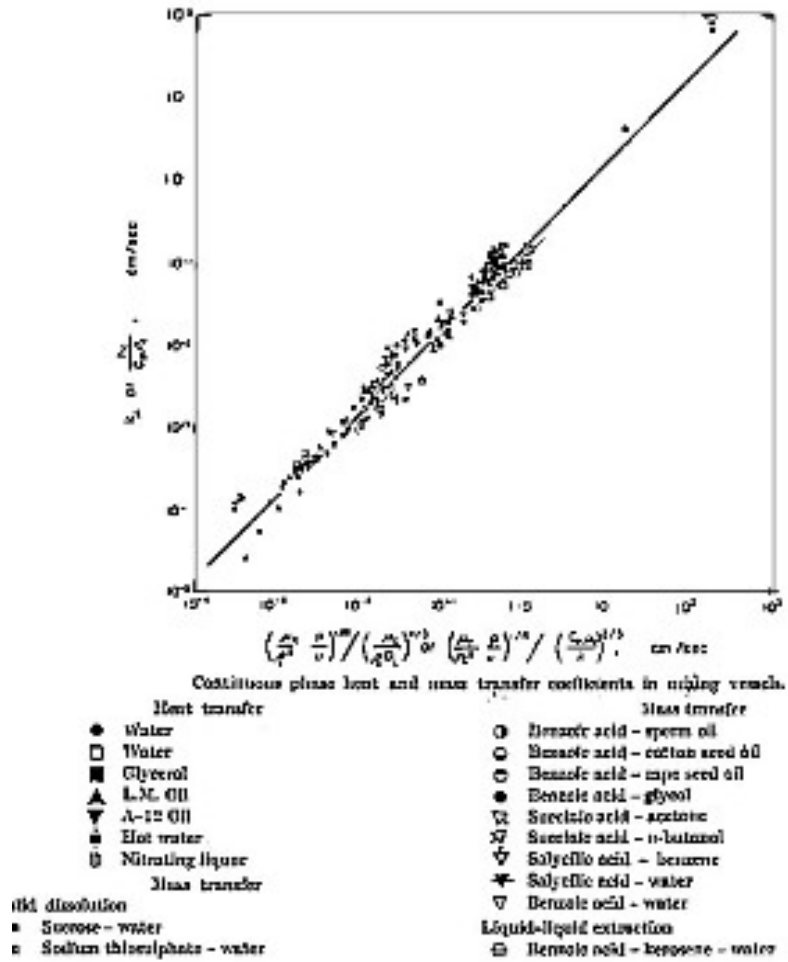

Figure 78 Calderbank and Moo-Young correlation for heat and mass transfer based on Power input per unit volume. The symbols and references to data source are given in the original Calderbank and Moo-Young paper.

Substituting for the mass transfer Stanton number St and setting $b = 1/3$ gives

$$St = \frac{k}{U} = \frac{\sqrt{\tau_w/\rho U^2}}{11.8 Sc^{2/3}} \tag{331}$$

$$kSc^{2/3} = 0.13 \left(\frac{\zeta \mu}{\rho}\right)^{1/4}$$

which is exactly the result of Calderbank and Moo-Young.

**13.3.7 Definition of viscosity in Non Newtonian flows, heat and mass transfer**

The solutions for turbulent heat and mass transfer in non-Newtonian fluids are easily obtained by changing the values of $U_v$ and $\delta_v$ to take account for the integration process. Alternatively, we can shift the Kolmogorov point according to equation (203). For non elastic fluids this gives rise to (Trinh 1969)

$$St = \frac{f/2}{1 + 11.8 \left(\frac{3n'+1}{4n'}\right)\sqrt{f/2}\left(\Pr^{1-b} - 1\right)} \tag{332}$$

which fits the experimental data of Metzner and Friend well. In viscoelastic flows, the wall layer edge and the Kolmogorov point are further shifted because the elastic forces dampen the fast fluctuations and delay the onset ejection but the zonal similarity profile is not affected (Figure 36). Thus one would now expand further the Kolmogorov scale using equation (226). Meyer (1966) could predict heat transfer in viscoelastic pipe flow once the velocity profile was measured, even though he mistakenly took the buffer layer as a new log-law mechanism as discussed in section 10.3.2.1 but such correlations are specific to the conditions of the velocity measurements. Attempts to generalize these correlations to other processing situations can lead to substantial discrepancies (e.g. Koskinen et al. op.cit.)

We conclude this discussion with a topic that has not been covered in detail: an examination of different definitions of viscosity which can add to our insight on turbulence. From early times, efforts to develop a correlation for turbulent pipe flows of non-Newtonian fluids centered around the definition of an effective viscosity that would collapse all non-Newtonian data onto the Newtonian curves obtained by Nikuradse e.g. (Alves, Boucher, & Pigford, 1952). The most widely accepted definition was proposed by Metzner and Reed (1955) who noted that the friction factor plots for non-Newtonian fluids was a family of lines with parameter $n'$ when the Reynolds number

$$\text{Re} = \frac{DV\rho}{\mu} \tag{57}$$

is expressed in terms of the apparent viscosity

$$\mu_a = \frac{\tau_w}{\dot{\gamma}_w} = K'\left(\frac{8V}{D}\right)^{n'-1}\left(\frac{3n'+1}{4n'}\right)^{n'-1} \tag{333}$$

But collapsed into a single line with the use of the "effective" viscosity term

$$\mu_e = K'\left(\frac{8V}{D}\right)^{n'-1}\left(\frac{3n'+1}{4n'}\right)^{n'} \tag{334}$$

Where

$$\mu_e = \mu_a\left(\frac{3n'+1}{4n'}\right) \tag{335}$$

However the use of the generalized Metzner-reed Reynolds number only collapses the friction factor curves in the laminar regime, not in the turbulent regime (Figure 55). This is because the different zones in a turbulent flow field are governed by different subsets of the NS equations and truly universal plots of pipe friction factors are better obtained by expressing the viscosity in terms of the instantaneous wall shear stress at the point of ejection (Figure 44) or by taking account of the apparent thickening of the wall layer arising from the integration process (Figure 68). The anomalies arising from these different definitions of the non-Newtonian viscosity term were never resolved by Metzner and his co-workers. For example, Bogue (op.cit.) found it more convenient to plot the velocity profiles in term of the apparent viscosity. Metzner and Friend retained the use of the Metzner-Reed Reynolds number in view of the widespread acceptance of the Dodge-Metzner correlation for pipe friction factor but could only correlate their heat transfer data by using the apparent viscosity for the Prandtl number.

$$\Pr = \frac{C_p \mu}{k} = \frac{\nu}{\alpha} \tag{336}$$

The effective viscosity also has to be redefined for each geometry e.g. flow past cylinders, sphere, between parallel plates.

In the present visualization the issue is most readily understood by looking at the Lagrangian equations (59) and (126). They indicate that the Prandtl number is introduced into solutions of heat transfer as a ratio of resistances to momentum and thermal diffusion and that the viscosity term of relevance must be defined in terms of the instantaneous shear stress and temperature. If we use the time-averaged wall shear stress, which is the easiest parameter to obtain experimentally, an integration coefficient is introduced as shown in section 10.1 which can be obtained from the apparent thickening of the wall layer if the definition in (333) is used or from a shift in the critical Reynolds number at the beginning of transition from laminar to turbulent flow if equation (334) is used (section 12). Thus the apparent viscosity is a better measure to use for the Prandtl and Schmidt numbers.

**13.3.8 Issues in the matching of Reynolds' analogy**

Figure 73 shows that the thicknesses for the sub-boundary layers for heat, mass and momentum are different and depend on the value of the Prandtl/Schmidt numbers. Because of that difference, an immediate question arises in matching the Reynolds analogy for the core region with a model for the diffusion dominated wall layer: should choose the matching point at the edge of the momentum sub-layer or at the edge of the thermal/diffusion sub-boundary layer?

The thickness of the diffusion sub-boundary layer at Sc=1320 is approximately $\delta_h^+ \approx 6$ whereas the thickness of the momentum sub-boundary layer is $\delta_\nu^+ = 64.7$. Thus there is a substantial region $6 < y^+ < 64.7$ where the profiles of velocity and concentration cannot possibly be similar and Reynolds' analogy cannot apply. Hanratty and his colleagues (op.cit.) have dedicated considerable effort to investigate the transfer of mass to turbulent flow at high Schmidt numbers and similarly argued that the assumption $E_h = E_\nu$ cannot hold near the wall. The error is particularly bad with analogies such as equation (275) because it assumes in this case that the diffusional sub-boundary layer is much thicker than it really is.

When the Prandtl number is smaller than unity, as in heat transfer to liquid metals, the diffusion of heat penetrates well beyond the boundary of the wall layer (Figure 73). The problem has always been treated separately from other transport analogies, which do not seem to apply to this particular case e.g. Seban and Shimazaki (1951), Lyon (1951). For $Pr = 0.01$, the thermal sub-boundary layer thickness is $\delta_h^+ = 647$ and there is a substantial region $64.7 < y^+ < 647$ where turbulent vortices can disturb the thermal diffusion process. Near the end of the sweep phase, the streaming flows represent real patches of fluid that penetrates in to the region $64.7 < y^+ < 647$ which is still dominated by thermal diffusion from the wall. This effect would be very similar to that of natural convection. The writer believes that this is the reason why many researchers in turbulent heat transfer to metals report a distortion of the temperature profile by natural convection e.g. (Buhr, et al., 1968).

We have seen in section 10.3.2.1 that a significant portion of the time-average error function near the edge of the wall layer can be approximated by a straight line in semi log-normal plots but with a different slope than the $1/\kappa$. The reader may recognise that line as Karman's description of a buffer layer but mistakenly interpreted, in my view, for a new Prandtl-Millikan logarithmic law of the wall for viscoelastic fluids. The Karman "buffer layer" concept can be useful but I want to stress that in using it I am still not convinced of the physical accuracy of Karman's interpretation of a transition between a purely (steady-state) laminar sub-layer proposed by Prandtl (op.cit.) and a turbulent core described by the log-law. I argue instead that the wall layer is well described by a process of unsteady viscous momentum into the flow along the wall but that part of its velocity profile can be approximated by a log-normal relationship.

Next to the wall, the time-averaged error function can be approximated by a linear relationship as postulated by Prandtl but in my estimation to a distance $y^+ = 4.5$ rather than 5. Again I am not using here Prandtl's visualisation of a steady state laminar sub-layer. Even in the sixties, Popovich and Hummel (1967) had observed a

197/218

linear relationship between velocity and distance next to the wall but the gradient was time dependent indicating that the even close to the wall the process was time-dependent.

We make use of these observations to investigate the case for $\Pr > 1$ by dividing the temperature/concentration field in three regions: a laminar sub-layer where diffusion predominates overwhelmingly, a log-law region and a buffer region where the relationship between $U^+$ or $\Theta^+$ and $y^+$ is no longer linear but where the Prandtl-Millikan log-law does not yet apply.

In the conductive/diffusive sub-layer

$$\Theta^+ = y^+ \Pr \tag{337}$$

$$U^+ = y^+ \tag{90)}$$

with thickness

$$\delta_{h,v}^+ = 4.5\,\Pr^{-a} \tag{338}$$

In the buffer layer $4.5\,\Pr^{-b} < y^+ < \delta_v^+$

$$\Theta^+ = A \ln y^+ + B \tag{339}$$

$$\frac{d\Theta^+}{dy^+} = \frac{A}{y^+} \tag{340}$$

Equation (273) can be rearranged as

$$E_h = \frac{dy^+}{d\Theta^+} - \frac{1}{\Pr} \tag{341}$$

At the edge of the conductive layer $y^+ = 4.5\Pr^{-b}$, $\Theta^+ = 4.5\Pr^{1-b}$ and $E_h = 0$ by definition. Then

$$A = 4.5\Pr^{1-b} \tag{342}$$

$$B = 4.5 \Pr{}^{1-b}\left[1 - \ln\left(4.5 \Pr{}^{-b}\right)\right] \tag{343}$$

$$\Theta^+ = 4.5 \Pr{}^{1-b}\left[1 + \ln\left(y^+ \Pr{}^{b}/4.5\right)\right] \tag{344}$$

It now remains to match equation (344) with the Prandtl-Millikan log-law. There are several matching points to choose from

1. the edge of the momentum wall layer

$$\left(y^+ = \delta_v^+ = 64.7, U_\infty^+ = 16, \Theta_v^+ = 4.5 \Pr{}^{1-b}\left[3.67 + b \ln \Pr\right]\right) \tag{345}$$

2. the edge of the thermal sub-boundary layer

$$\left(y^+ = \delta_h^+ = 64.7 \Pr{}^{-b}, U_h^+ = 16 - 2.5 b \ln \Pr, \Theta_h^+ = 4.5 \Pr{}^{1-b}\left[3.67 + \ln \Pr\right]\right) \tag{346}$$

3. the edge of the momentum (Karman) buffer layer

$$\left(y^+ = \delta_b^+ = 30, U_b^+ = 14, \Theta^+ = 4.5 \Pr{}^{1-b}\left[2.9 + b \ln \Pr\right]\right) \tag{347}$$

4. the edge of the thermal buffer layer

$$\left(y^+ = \delta_{h,b}^+ = 30 \Pr{}^{-b}, U_{h,b}^+ = 14 - 4.5 b \ln \Pr, \Theta_{h,b}^+ = 14 \Pr{}^{1-b}\right) \tag{348}$$

The reader may recall that these points are not blindly empirical; these points have clear physical connotations: $\delta_v^+$ is the maximum penetration of wall retardation by viscous momentum into the streamwise motion, $\delta_h^+$ is the maximum penetration of heat, $\delta_b^+$ is the result of time-averaging $\delta_v^+$, which is equivalent to statistical averaging with the assumption that low-speed streaks of all ages have and equal probability of passing the probe and $\delta_{h,b}^+$ is the time averaged value of $\delta_h^+$. The reader may note that the ratio of corresponding pairs of the momentum and thermal layers show the same exponent of the Prandtl number. This is because the averaging process does not involve the Prandtl number (see Figure 38 and Figure 63). Whichever matching point we choose the problem remains: there is no perfect match for the log-laws for velocity and temperature: we can only reduce the discrepancy. We present here the result of matching point 4. We apply Reynolds analogy by stating

$$\Theta_\infty^+ - \Theta_{h,b}^+ = U_\infty^+ - U_{h,b}^+ \tag{349}$$

Substituting for equation (348) gives equation (300).

## 14  Recapitulation

It is proposed that two types of Reynolds stresses must be distinguished respectively linked with the slow and fast variations of the instantaneous velocity. The fast velocity fluctuations give rise to "fast" Reynolds stresses through secondary fluid motions similar to the streaming process in laminar oscillating flow theory. This process illustrates how wave energy contained in the fast fluctuations can be transformed into kinetic energy for the secondary motion.

This secondary motion may be compared to intermittent jets in crossflow with respect to the mainstream as suggested also by Townsend (1970) and Grass (1976). The writer goes further than these previous authors by emphasising that the jets interact with the main stream to create a wake. This temporary breakdown of the mainstream is postulated as the source of the small scale turbulence. The log-law and the law of the wake regions are seen as the result of this disturbance of the original potential flow outside the laminar boundary layer and thus represent new layers added on top of the laminar boundary layer. The latter is identified with the wall layer in turbulent flow.

While the magnitude of the instantaneous Reynolds stresses are not large, except in the ejection phase of the bursting cycle, they play a major role in turbulence production through their disturbance to the mainstream.

## 15  Conclusion

Like the roads to Rome, there are many paths that can lead to an understanding of turbulence. The picture at the moment is still very diffuse and I claim, like Herbert (1988) a right to describe my own journey towards this goal. The reader will by now be aware that I see many of the traditional statistically deduced parameters used to describe turbulence as quite misleading in the sense that they all can be obtained by time averaging the contributions of a recurring transient laminar boundary layer. The central theme that I have presented is that the study of turbulence needs to begin with a more complex breakdown of the instantaneous velocity into many components, not just two as suggested by Reynolds. Substitution of this new expression for the instantaneous velocity into the NS equations highlighted the fact that there is more than one type of Reynolds stresses. I have only highlighted two, which I labelled slow and fast Reynolds stresses. The slow Reynolds stresses, which are captured in traditional measurements of turbulence parameters, are actually induced by the smoothed velocity of laminar circulation flow within passing coherent structures and only the fast Reynolds stresses can explain the mechanism of turbulence production. The reader by now will understand my reservations about past practice of making deductions from overall averaging and statistical analyses and may even forgive that silly joke about the statistics of China and India. The point needed to be made.

In the process, I have questioned and revised some iconic concepts such as the use of the Reynolds number as the sole criterion for transition between laminar and turbulent flows, the averaging of the NS equations into the Reynolds equations RANS as a starting point for turbulence studies, the physical interpretation of the parameters in the log law and of the Kolmogorov scale. These are concepts that I was raised on and there is no doubting the great contributions of Reynolds, Prandtl, Taylor and Kolmogorov to name only a few, to our present understanding of turbulence. I wish to state clearly my admiration for their genius. How else can we describe, for example, Reynolds' visual observations of eddies in 1883 by introducing a fluorescent dye and observing with an intermittent arc welder? I took another 80 years for Kline et al. to make visual observations of the wall layer process and that paper ushered another fifty years focus on coherent structures in turbulent flows! However, as heirs to the genial legacy of these elders we have a duty to continually modify and improve on the ideas that they develop within a context of very poor instrumentation and lack of detailed information.

It is a bold and foolhardy man that would attempt to revise such iconic concepts but then I was heir to a long tradition (4 centuries) of fighters and revolutionaries as my parents taught me and I wish to record here this debt. As the evidence mounted in this long, sometimes arduous and lonely, journey towards an understanding of turbulence, which remains a matter of personal interest rather than a career move or part of a job description, I felt the need to propose revisions to these concepts.

Will this new analysis lead to a formal general solution of the NS equations? I leave that to better mathematicians but I will highly recommend that the analysis be based on a multi-component expression for the instantaneous velocity rather than a lumped variable. I have chosen instead to concentrate on obtaining a picture by analysing linearised subsets on the NS equations that are more amenable to mathematical analysis and then connecting them through methods for matching asymptotic solutions with the understanding that these subsets pertain to domains of application, not flow regimes. In the process I have tried to illustrate the usefulness of theoretical approaches to the closure problems, of a more logical approach to non-Newtonian turbulent flows and of a new Lagrangian derivative along the path of scalar diffusing entities that greatly simplify the mathematical analysis. Hopefully, these tools will prove useful to colleagues interested in practical applications involving turbulence.

## 16    Literature

Adamorola, M. S., Sumner, D., & Bergstrom, D. J. (2007). Turbulent wake and vortex shedding for a stack partially immersed in a turbulent boundary layer. *Journal of Fluid Structures, 23*, 1189-1206.

Adrian, R. J., Meinhart, C. D., & Tomkins, C. D. (2000). Vortex organization in the outer region of the turbulent boundary layer. *Journal of Fluid Mechanics, 422*, 1-54.

Afzal, N. (2001). Power law and log law velocity profiles in fully developed turbulent pipe flow: equivalent relations at large Reynolds numbers. *Acta Mechanica, 151*(3-4), 171-183.

Afzal, N. (2005). Analysis of power law and log law velocity profiles in the overlap region of a turbulent wall jet. *Proceedings of the Royal Society A-Mathematical Physical and Engineering Sciences, 461*(2058), 1889-1910.

Akhavan, R., Kamm, R. D., & Saphiro, A. H. (1991). An investigation of transition to turbulence in bounded oscillatory Stokes flows. Part 2. Numerical simulations. *Journal of Fluid Mechanics, 225*, 423.

Alves, G. E., Boucher, D. F., & Pigford, R. R. (1952). *Chem. Eng. Prog, 48*, 385-393.

Andrade, E. N. (1931). On the circulation caused by the vibration of air in a tube. *Proc. Roy. Soc. London, A134*, 447.

Andrade, L. C. F., Petronílio, J. A., Edilsonde A. Maneschy, C., & Onofre de A. Cruz, D. (2007). The Carreau-Yasuda Fluids: a Skin Friction Equation for Turbulent Flow in Pipes and Kolmogorov Dissipative Scales. *J. of the Braz. Soc. of Mech. Sci. & Eng., XXIX*(2), 163-167.

Andreopoulos, J. (1989). Wind tunnel experiments on cooling tower plumes: part1 - in uniform crossflow. *Journal of Heat Transfer, 111*, 941948.

Antonia, R. A. (1980). *The organised motion in a turbulent boundary layer.* Paper presented at the 7th Australasian Hydraulics and Fluid Mechanics Conference, Brisbane.

Antonia, R. A., Bisset, D. K., & Browne, L. W. B. (1990). Effect of Reynolds number

on the topology of the organized motion in a turbulent boundary layer. *Journal of Fluid Mechanics, 213*, 267-286.

Antonia, R. A., Browne, L. W. B., & Bisset, D. K. (1989, 1989). *Effect of Reynolds number on the organised motion in a turbulent boundary layer" in "Near Wall Turbulence"*

Antonia, R. A., & Kim, J. (1991a). Reynolds stress and heat flux calculations in fully developed turbulent duct flow. *International Journal of Heat and Mass Transfer, 34*(8), 2013-2018.

Antonia, R. A., & Kim, J. (1991b). Turbulent Prandtl number in the near-wall region of a turbulent channel flow. *International Journal of Heat and Mass Transfer, 34*(7), 1905-1908.

Arcalar, M. S., & Smith, C. R. (1987a). A study of hairpin vortices in a laminar boundary layer. Part 2: hairpin vortices generated by fluid injection. *Journal of Fluid Mechanics, 175*, 43-83.

Arcalar, M. S., & Smith, C. R. (1987b). A study of hairpin vortices in a laminar boundary layer. Part I: hairpin vortices generated by fluid injection. *Journal of Fluid Mechanics, 175*.

Atkhen, K., Fontaine, J., & Wesfreid, J. E. (2000). Highly turbulent Couette-Taylor bubbly flow patterns. *Journal of Fluid Mechanics, 422*, 55-68.

Bakewell, H. P. J., & Lumley, J. L. (1967). Viscous sublayer and adjacent region in turbulent pipe flow. *Physics of Fluids, 10*, 1880-1889.

Baldwin, B. S., & Lomax, H. (1978). Thin layer approximation and algebraic model for separated turbulent flows. *AIAA paper 78-0257*.

Balmer, R. T., & Fiorina, M. A. (1980). Unsteady flow of an inelastic power law fluid in a circular tube. *Journal of Non-Newtonian Fluid Mechanics, 7*, 189.

Bandyopadhyay, P. R. (1986). REVIEW-Mean flow in turbulent boundary layers disturbed to alter skin friction. *Journal of Fluids Engineering, 108*, 127-140.

Bandyopadhyay, P. R. (2005). Trends in Biorobotic Autonomous Undersea Vehicles. *IEEE Journal of Oceanic Engineering, 30*(1), 109-139.

Bandyopadhyay, P. R., Henoch, C., Hrubes, J. D., Semenov, B. N., Amirov, A. I., Kulik, V. M., et al. (2005). Experiments on the effects of aging on compliant coating drag reduction. *Physics of Fluids, 17*(8), 085104.

Batchelor, G. K. (1960). *The Theory of Homogeneous turbulence* (Students' edition ed.). London: Cambridge University Press.

Bayley, F. J., Owen, J. M., & Turner, A. B. (1972). *Heat transfer*. Great Britain: Nelson and Sons ltd.

Bayly, B. J., Orszag, A. A., & Herbert, T. (1988). Instability mechanisms in shear flow transition. *Annual Review of Fluid Mechanics, 20*, 359.

Bechert, D. W., & Hage, W. (2006). Drag reduction with riblets in nature and engineering In R. Liebe (Ed.), *Flow Phenomena in Nature Volume 2*: WIT Press.

Bennett, A. (2006). *Lagrangian Fluid Dynamics*. Cambridge: Cambridge University Press.

Biferale, L., Calzavarini, E., Lanotte, A. S., Toschi, F., & Tripiccione, R. (2004). Universality of anisotropic turbulence. *Physica a-Statistical Mechanics and Its Applications, 338*(1-2), 194-200.

Bird, R. B., Stewart, W. E., & Lightfoot, E. N. (1960). *Transport Phenomena*. New York: Wiley and Sons.

Bird, R. B., Stewart, W. E., & Lightfoot, E. N. (1979). *Transport phenomena, 2nd ed*. New York: Wiley and Sons.

Black, T. J. (1969). Viscous drag reduction examined in the light of a new model of wall turbulence. In W. C.S. (Ed.), *Viscous Drag Reduction*. New York: Plenum Press.

Blackwelder, R. F., & Kaplan, R. E. (1976). On the Wall Structure of the Turbulent Boundary Layer. *Journal of Fluid Mechanics, 94*, 577-594.

Blackwelder, R. F., & Kovasznay, L. S. (1972). Time scales and correlations in a turbulent boundary layer. *Physics of Fluids, 15*, 1545.

Blasius, H. (1908). Grenzschichten in Flüssigkeiten mit kleiner Reibung. *Z. Math. u. Phys., 56*, 1-37, also NACA TM 1256.

Blasius, H. (1913). *Mitt. Forschungsarb., 131*, 1.

Blom, J., & deVries, D. A. (1968). *On the values of the turbulent Prandtl number*. Paper presented at the Third All-Union Heat and Mass Transfer Conference.

Bogard, G., & Tiederman, W. G. (1986). Burst detection with single point velocity measurements. *Journal of Fluid Mechanics, 162*, 389.

Bogue, D. C. (1961). *Velocity profiles in turbulent non-Newtonian pipe flow, Ph.D. Thesis*. University of Delaware.

Bogue, D. C., & Metzner, A. B. (1963). Velocity Profiles in Turbulent Pipe Flow - Newtonian and Non-Newtonian Fluids. *Industrial & Engineering Chemistry Fundamentals, 2*(2), 143.

Boussinesq, J. (1877). Théorie de l'écoulement tourbillant. *Mémoires Présentés a l'Académie des Sciences, 23*, 46.

Bowen, R. L. (1961). Scale-up for non-Newtonian fluid flow: Part 4, Designing turbulent-flow systems. *Chem. Eng. Prog, 68*(15), 143-150.

Bradshaw, P. (1971). *An introduction to Turbulence and Its Measurements*. Oxford: Pergammon Press.

Bradshaw, P., Launder, B. E., & Lumley, J. L. (1991). Collaborative testing of turbulence models: progress report. *Journal of Fluids Engineering, 113*, 3.

Brandstater, A., & Swinney, H. L. (1987). Strange attractors in weakly turbulent Couette-Taylor flow. *Phys. Rev. A, 35*, 2207-2220.

Broadwell, J. E., & Mungal, M. G. (1991). Large scale structures and molecular mixing. *Physics of Fluids, A3 (5)*, 1193.

Brown, D. (2003). *The Da Vinci code*: Doubleday.

Brown, G. L., & Thomas, A. S. W. (1977). Large structure in a turbulent boundary layer. *Phys. Fluid Suppl., 10*, S243.

Buhr, H. O., Carr, A. D., & Balzhiser, R. E. (1968). Temperature profiles in liquid metals and the effect of superimposed free convection in turbulent flow. *Int. J. Heat and Mass Transfer, 11*, 641.

Burger, E. D., Chorn, L. G., & Perkins, T. K. (1980). Studies of drag reduction conducted over a broad range of pipeline conditions when flowing Prudhoe Bay crude oil. *Journal of Rheology, 24*, 603-626.

Bushnell, D. M. (2003). Aircraft drag reduction - a review. *Journal of aerospace engineering, Vol. 217 Part G*, 1.

CAI, S. P., JIN, G. Y., & Yang, L. (2008). Drag reduction effect of coupling flexible tubes with turbulent flow. *Journal of Hydrodynamics, 20*(1), 96-100.

Calderbank, P. H. (1967). Mass transfer. In V. W. Uhl & J. B. Gray (Eds.), *Mixing. Theory and Practice*. (Vol II, pp. 1-114). New York: Academic Press.

Calderbank, P. H., & MooYoung, M. B. (1962). The continuous phase heat and mass transfer properties of dispersions. *Chem. Eng. Sci., 16*, 39.

Campbell, J. A., & Hanratty, T. J. (1983). Turbulent velocity fluctuations that control mass transfer at a solid boundary layer. *AIChE J., 29*, 215.

Camussi, R. (2002). Coherent structure identification from wavelet analysis of particle image velocimetry data. *Experiments In Fluids, 32*(1), 76-86.

Camussi, R., Guj, G., & Stella, A. (2002). Experimental study of a jet in a crossflow at very low Reynolds number. *Journal of Fluid Mechanics 454*, 113-144.

Cantwell, B. J. (1981). Organised Motion in Turbulent Flow. *Annual Review of Fluid Mechanics, 13*, 457.

Carlier, J., & Stanislas, M. (2005). Experimental study of eddy structures in a turbulent boundary layer using particle image velocimetry. *Journal of Fluid Mechanics, 535*, 143-188.

Carriere, Z. (1929). *J. Phys. Radium, 10*, 198.

Caton, F., Janiaud, B., & Hopfinger, E. J. (1999). Primary and secondary Hopf bifurcation in stratified Taylor-Couette flow. *Phys. Review letters, 82*(23), 4647-4650.

Caton, F., Janiaud, B., & Hopfinger, E. J. (2000). Stability and bifurcations in stratified Taylor-Couette flow. *Journal of Fluid Mechanics 419*, 93-124.

Cecebi, T., & Smith, A. M. O. (1974). *Analysis of turbulent boundary layers*.

Chan, D. T. L., Lin, A. M., & Kennedy, J. F. (1976). Entrainment and Drag Forces of Deflected Jets. *J. Hydraulics Div.*, 615.

Chassaing, P., George, J., Claria, A., & Sananes, F. (1974). Physical characteristics of subsonic jets in a cross-stream. *Journal of Fluid Mechanics, 62*, 41-64.

Chen, C. H. P., & Blackwelder, R. F. (1978). Large scale motion in a turbulent boundary layer: a study using temperature contamination. *Journal of Fluid Mechanics, 89*, 1.

Choi, K. S., Yang, X., Clayton, B. R., Glover, E., J., Atlar, M., Semenov, B. N., et al. (1997). Turbulent Drag Reduction Using Compliant Surfaces. *Proceedings: Mathematical, Physical and Engineering Sciences, 453*(1965), 2229-2240.

Chu, C. C., & Falco, R. E. (1988). Vortex Rings/Viscous Wall Layer Interaction Model of the Turbulence Production Process near Walls. *Experiments in Fluids, 6*, 305-315.

Churchill, S. W. (1996). A Critique of Predictive and Correlative Models for Turbulent Flow and Convection. *Ind. Eng. Chem. Res., 35*, 3122-3140.

Churchill, S. W. (1997). New Simplified Models and Formulations for Turbulent Flow and Convection. *AIChE Journal, 43*(5), 1125-1140.

Churchill, S. W. (2002). A Reinterpretation of the Turbulent Prandtl Number. *Ind. Eng. Chem. Res., 41*, 6393-6401.

Cigada, A., Malavasi, S., & Vanali, M. (2006). Effects of an asymetrical confined flow on a rectangular cylinder. *J.Fluids Structs., 22*, 213-227.

Clapp, R. M. (1961). Turbulent heat transfer in pseudoplastic nonNewtonian fluids. *Int. Developments in Heat Transfer, ASME, Part III, Sec. A*, 652.

Clauser, F. H. (1954). Turbulent Boundary layers in Adverse Pressure Gradients. *J.Aeoraunt. Sci., 21*, 91-108.

Colburn, A. P. (1933). A method of correlating forced convection heat transfer data and a comparison with fluid friction. *Trans. AIChE, 29*, 174.

Coles, D. E. (1956). The law of the wake in the turbulent boundary layer. *Journal of Fluid Mechanics, 1*, 191.

Coles, D. E. (1965). Transition in circular Couette flow. *Journal of Fluid Mechanics, 21*, 385-425.

Comte-Bellot, G., & Corrsin, S. (1966). The use of a contraction to improve the isotropy of a grid generated turbulence. *Journal of Fluid Mechanics, 139*, 67-95.

Corino, E. R., & Brodkey, R. S. (1969). A Visual investigation of the wall region in turbulent flow. *Journal of Fluid Mechanics, 37*, 1.

Cortelezzi, L., & Karagozian, A. R. (2001). On the formation of the counter-rotating vortex pair in transverse jets. *Journal of Fluid Mechanics 446*, 347-373.

Cruz, D. O. A., Batista, F. N., & Bortolus, M. (2000). A Law of the Wall Formulation for Recirculating Flows. *J. Braz. Soc. Mech. Sci. , 22*(1).

Danckwerts, P. V. (1951). Significance of liquid film coefficients in gas absorption. *Ind. Eng. Chem., 43*, 1460.

Darby, R., & Chang, H. D. (1984). Generalized correlation for friction loss in drag reducing polymer solutions. *AIChE Journal, 30*(2), 274-280.

Davies, C., Carpenter, P. W., Ali, R., & Lockerby, D. A. (2006). *Disturbance development in boundary layers over compliant surfaces.* Paper presented at the IUTAM Symposium on Laminar-Turbulent Transition.

Deissler, R. G. (1955). *Analysis of turbulent heat transfer, mass transfer, and friction in smooth tubes at high Prandtl and Schmidt numbers*.

Devenport, W. J., & Sutton, E. P. (1991). Near-wall behaviour of separated and reattaching flows. *AIAA JOURNAL, 29*, 25-31.

Dodge, D. W. (1959). *Turbulent flow of non-Newtonian fluids in smooth round tubes, Ph.D. Thesis*. US: University of Delaware.

Dodge, D. W., & Metzner, A. B. (1959). Turbulent Flow of Non-Newtonian Systems. *AICHE Journal, 5*(2), 189-204.

Dorfman, L. A. (1963). *Hydrodynamic resistance and the heat loss of rotating solids* Edinburgh: Oliver and Boyd.

Doshi, M. R., & Gill, W. N. (1970). A note on the mixing length theory of turbulent flow. *AIChE J., 16*, 885.

Driest, v. E. R. (1956). On Turbulent Flow Near a Wall. *J. Aero. Sci, 23*, 1007.

Dryden, H. L. (1934). *Boundary layer flow near flat plates.* Paper presented at the Fourth Intern. Congress for Appl. Mech., Cambridege, England.

Dryden, H. L. (1936). *Airflow in the boundary layer near flat plates*: NACA, Rep. 562.

Du, Y., & Karniadakis, G. E. (2000). Suppressing Wall Turbulence by Means of a Transverse Travelling Wave *Science 288*(5469), 1230 - 1234.

Dubief, Y., Terrapon, V. E., Shaqfeh, E. S. G., Moin, P., & Lele, S. K. (2004). On the coherent drag-reducing turbulence-enhancing behaviour of polymers in wall flows. *Journal of Fluid Mechanics, 514*, 271-280.

Dvorak, V. (1874). *Ann. Phys. Lpz., 151*, 634.

Eckelmann, H. (1974). The structure of the viscous sublayer and the adjacent wall region in a turbulent channel flow. *Journal of Fluid Mechanics, 65*, 439.

Einstein, H. A., & Li, H. (1956). The viscous sublayer along a smooth boundary. *J. Eng. Mech. Div. ASCE 82(EM2) Pap. No 945*.

Ekman, V. W. (1910). On the change from steady to turbulent motion of liquids. *Ark. f. Mat. Astron. och Fys., 6*(12).

Elmore, W. C., & Heald, M. A. (1969). *The physics of waves.* New York: McGraw-Hill.

Falco, R. E. (1977). Coherent motions in the outer region of turbulent boundary layers. *Physics of Fluids Suppl., 20*, S124.

Falco, R. E. (1991). A coherent structure model of the turbulent boundary layer and its ability to predict Reynolds number dependence. *phil. Trans. R. Soc. Lond.A, 336*, 103-129.

Faraday, M. (1831). On a Peculiar Class of Acoustical Figures; and on Certain Forms

Assumed by Groups of Particles upon Vibrating Elastic Surfaces *Phil. Trans. R. Soc. Lond. , 121*, 299-340.

Fiedler, H. E. (1988). Coherent structures in turbulent flows. *Prog. Aero. Sci, 25*, 231.

Fife, P., Wei, T., Klewicki, J., & McMurtry, P. (2005). Stress gradient balance layers and scale hierarchies in wall-bounded turbulent flows. *Journal of Fluid Mechanics, 532*, 165-189.

Flender, J. F., & Hiby, J. W. (1981). Investigation of solid/liquid mass transfer by a photometric method. *Ger. Chem. Eng, 4*, 370-379.

Flory, P. J. (1971). *Principles of polymer chemistry*. Ithaca, New York: Cornell University Press.

Fortuin, J. M. H., Musschenga, E. E., & Hamersma, P. J. (1992). Transfer processes in turbulent pipe flow described by the ERSR model. *AIChE J., 38*, 343.

Frater, K. R. (1967). Accoustic streaming in an elastico-viscous fluid. *Journal of Fluid Mechanics, 30*(4), 689-697.

Fric, T. F., & Roshko, A. (1994). Vortical structure in the wake of a transverse jet. *Journal of Fluid Mechanics, 279*, 1-47.

Frisch, U. (1995). *Turbulence, the legacy of A.N. Kolmogorov*: Cambridge University Press.

Frisch, U., Sulem, P., & Nelkin, M. (1978). A simple dynamical model of intermittent fully developed turbulence. *Journal of Fluid Mechanics, 87*, 719-736.

Fukagata, K., Kern, S., Chatelain, P., Koumoutsakos, P., & Kasagi, N. (2008). Evolutionary optimization of an anisotropic compliant surface for turbulent friction drag reduction. *J. Turbulence 9*(35), 1-17.

Fukuda, K., Tokunaga, J., Takashi, N., Nakatani, T., Iwasaki, T., & Kunitake, Y. (2000). Frictional drag reduction with air lubricant over a super-water-repellent surface. *J Mar Sci Technol 5*, 123-130.

Gad-el-Hak, M., Davis, S. H., McMurray, J. T., & Orszag, S. A. (1983). On the stability of the decelerating laminar boundary layer. *Journal of Fluid Mechanics*, 297-323.

Gad-el-Hak, M., & Hussain, A. K. M. (1986). Coherent structures in a turbulent boundary layer. Part 1:generation of artificial bursts. *Physics of Fluids, 29*, 2124.

Garner, F. H., & Keey, R. B. (1958). Mass-transfer from single solid spheres—I : Transfer at low Reynolds numbers *Chem. Eng. Sci, 9*(2-3).

Gasljevic, K., Aguilar, G., & Matthys, E. F. (2001). On two distinct types of drag-reducing fluids, diameter scaling, and turbulent profiles. *Journal of Non-Newtonian Fluid Mechanics 96* 405-425.

Gatski, T. B., & Rumsey, C. L. (2002). Linear and nonlinear eddy viscosity models. In B. E. Launder & N. D. Sandham (Eds.), *Closure strategies for turbulent and transitional flows*: Cambridge Academic press.

Gavrilakis, S. (1992). Numerical Simulation of Low-Reynolds Number Flow Through a Straight Square Duct. *Journal of Fluid Mechanics, 244*, 101-129.

Gorla, R. S. P., & Madden, P. E. (1984). A variational approach to nonsteady nonNewtonian flow in a circular pipe. *Journal of Non-Newtonian Fluid Mechanics, 16*, 251.

Gorman, M., & Swinney, H. L. (1979). Visual observation of the second characteristic mode in a quasiperiodic flow. *Physical Review Letters, 43*, 1871–1875.

Gowen, R. A., & Smith, J. W. (1967). The effect of the Prandtl number on temperature profiles for heat transfer in turbulent pipe flow. *Chem. Eng. Sci.,*

*22*, 1701-1711.

Grass, A. J. (1971). Structural features of turbulent flow over rough and smooth boundaries. *Journal of Fluid Mechanics, 50*, 223.

Gyr, A., & Bewersdorff, H. W. (1995). *Drag reduction of turbulent flows by additives*. Dorrecht, Netherlands: Kluwer Academy.

Haidari, A. H., & Smith, C. R. (1994). The generation and regeneration of single hairpin vortices. *Journal of Fluid Mechanics, 139*.

Hanjalic, K., & Jakirlic, S. (2002). Second-moment turbulence closure modelling. In B. E. Launder & N. D. Sandham (Eds.), *Closure strategies for turbulent and transitional flows*: Cambridge University Press.

Hanjalić, K., & Jakirlić, S. (2002). Second-moment turbulence closure. In B. E. Launder & N. D. Sandham (Eds.), *Closure strategies for turbulent and transitional flows*: Cambridge University Press.

Hanks, R. W., & Dadia, B. H. (1971). Theoretical Analysis of Turbulent Flow of Non-Newtonian Slurries in Pipes. *AICHE Journal, 17*(3), 554-&.

Hanks, R. W., & Ricks, B. L. (1975). Transitional and turbulent pipe flow of pseudoplastic fluids. *Journal of Hydronautics, 9*, 39-44.

Hanratty, T. J. (1956). Turbulent exchange of mass and momentum with a boundary. *AIChE J., 2*, 359.

Hanratty, T. J. (1989). A conceptual model of the viscous wall layer. In S. J. Kline & N. H. Afgan (Eds.), *Near wall turbulence* (pp. 81-103). New York: Hemisphere.

Harriott, P., & Hamilton, R. M. (1965). Solid-liquid mass transfer in turbulent pipe flow. *Chem. Eng. Sci., 20*, 1073.

Head, M. R., & Bandhyopadhyay, P. (1981). New aspects of turbulent boundary layer structure. *Journal of Fluid Mechanics, 107*, 297.

Hedley, T. B., & Keffer, J. F. (1974). Some turbulent/non-turbulent properties of the outer intermittent region of a boundary layer. *Journal of Fluid Mechanics, 64*, 645.

Herbert, T. (1988). Secondary instability of boundary layers. *Annual Review of Fluid Mechanics, 20*, 487.

Herrero, J., Grau, F. X., Grifoll, J., & Giralt, F. (1991). A near wall k-ε formulation for high Prandtl number heat transfer. *International journal of Heat and Mass Transfer, 34*(3), 711-721.

Higbie, R. (1935). Rate of absorption of a gas into a still liquid during short periods of exposure. *Transactions AIChE, 31*, 365.

Hino, M., Kawashiwayanagi, M., Nakayama, A., & Hara, T. (1983). Experiments on the turbulence statistics and the structure of a reciprocating oscillatory flow. *Journal of Fluid Mechanics, 131*, 363.

Hinze, J. O. (1959). *Turbulence* (2nd ed.). New York: McGraw-Hill

Holmes, P., Lumley, J. L., & Berkooz, G. (1998). *Turbulence, Coherent Structures, Dynamical Systems and Symmetry*: Cambridge University Press.

Hoyt, J. W. (1972). Effect of additives on fluid friction. *Transactions of ASME. Journal of Basic Engineering, 94*, 258-285.

Incropera, F. P., Dewitt, D. P., Bergman, T. L., & Lavine, A. S. (2007). *Fundamentals of Heat and Mass Transfer* (6th ed.). New York: John Wiley and Sons.

Janberg, K. (1970). Etude éxperimentale de la distribution des températures dans les films visqueux, aux grands nombres de Prandtl. *Int. J. Heat Mass transfer, 13*, 1234.

Jeong, J., & Hussain, F. (1995). On The Identification of a Vortex. *Journal of Fluid*

*Mechanics, 285*, 69-94.

Jeong, J., Hussain, F., Schoppa, W., & Kim, J. (1997). Coherent structures near the wall in a turbulent channel flow. *Journal of Fluid Mechanics, 332*, 185-214.

Jimenez, X., & Pinelli, A. (1999). The autonomous cycle of near-wall turbulence. *Journal of Fluid Mechanics, 389*, 335-359.

Johansson, A. V., Alfredson, P. H., & Eckelmann, H. (1987). On the evolution of shear layer structures in nearwall turbulence. In G. ComteBellot & J. Mathieu (Eds.), *Advances in Turbulence* (pp. 383). Berlin, Heidelberg.: Springer

Johansson, A. V., Alfresson, P. H., & Kim, J. (1991). Evolution and dynamics of shear-layer structures in near-wall turbulence. *Journal of Fluid Mechanics, 224*, 579-599.

Johnk, R. E., & Hanratty, T. J. (1962). Temperature profiles for turbulent flow of air in a pipe. *Chemical Engineering Science, 17*, 867.

Kader, B. A. (1981). Temperature and concentration profiles in fully turbulent boundary layers. *Int. J. Heat Mass transfer, 24*, 1541.

Karman, v. T. (1934). Turbulence and skin friction. *J. Aeronaut. Sci., 1*, 1-20.

Karman, v. T. (1939). The analogy between fluid friction and heat transfer. *Trans. ASME, 61*, 705.

Kawaguchi, Y., Li, F. C., Yu, B., & Wei, J. J. (2007). Turbulent drag reduction with surfactant additives- Basic research and application to an air conditioning system *NEW TRENDS IN FLUID MECHANICS RESEARCH. Proceedings of the Fifth International Conference on Fluid Mechanics, Aug.15-19, 2007, Shanghai, China* (pp. 29): Tsinghua University Press & Springer.

Kays, W. M. (1994). Turbulent Prandtl number. Where are we? *Journal of Heat Transfer, Transactions ASME 116*(2), 284-295.

Keffer, J. F., & Baines, W. D. (1963). The round jet in a crosswind. *Journal of Fluid Mechanics, 15*, 31.

Kelso, R. M., & Smits, A. J. (1995). Horseshoe vortex system resulting from the interaction between a laminar layer and a transverse jet. *Physics of Fluids A, 7*, 153-158.

Khishinevskii, M. K., & Kornienko, T. S. (1967). Heat and mass exchange between a smooth tube and turbulent liquid flow. *Theoretical Foundations of Chemical Engineering, 1*, 255.

Kim, H. T., Kline, S. J., & Reynolds, W. C. (1971). The Production of the Wall Region in Turbulent Flow. *Journal of Fluid Mechanics, 50*, 133.

Kim, J., Moin, P., & Moser, R. (1987). Turbulent Statistics in Fully Developed Channel Flow at Low Reynolds Number. *Journal of Fluid Mechanics, 177*.

Klebanoff, P. S. (1954). *Characteristics of turbulence in a boundary layer with zero pressure gradient*: NACA TN 3178.

Kline, S. J., Reynolds, W. C., Schraub, F. A., & Runstadler, P. W. (1967). The structure of turbulent boundary layers. *Journal of Fluid Mechanics, 30*, 741.

Knudsen, J. G., & Katz, D. L. (1958). *Fluid Dynamics and Heat Transfer*. New York McGraw-Hill.

Kolmogorov, A. N. (1941a). Dissipation of energy in the locally isotropic turbulence. *Dokl.Akad.Nauk SSSR, 32*, 1. Also Proc.Roy.Soc. London A (1991) 1**434**, 1915-1917.

Kolmogorov, A. N. (1941b). The local structure of turbulence in incompressible flow for very large Reynolds number. *C. R. Acad. Sci. U.S.S.R., 30*, 4.also Proc.Roy.Soc.London A (1991) 1**434**, 1999-1913.

Koskinen, J., Manninen, M., Pättikangas, T., Leinonen, T., Denifl, P., Pöhler, H., et al. (2005). *Controlling turbulence with drag reducing agents in stirred tank reactors* Paper presented at the 7th World Congress of Chemical Engineering

Kreplin, H. P., & Eckelmann, H. (1979). Propagation of perturbations in the viscous sublayer and adjacent wall region. *Journal of Fluid Mechanics, 95*, 305.

Krothapalli, A., Lourenco, L., & Buchlin, J. M. (1990). Separated flow upstream of a jet in a cross flow. *AIAA JOURNAL, 28*, 414-420.

Lamont, J. C., & Scott, D. S. (1970). An eddy cell model of mass transfer into the surface of a turbulent fluid. *AIChE Journal, 16*(513-519).

Landahl, M. T. (1973). *Drag reduction by polymer addition*. Paper presented at the Proceeding 13th International Congress Theoretical Applied Mechanics. Moscow.Springer.

Laufer, J. (1954). *The structure of turbulence in a fully developed pipe*: NACA TN No 2945.

Launder, B. E., & Sandham, N. D. (Eds.). (2002). *Closure strategies for transitional and turbulent flows*: Cambridge University Press.

Launder, B. E., & Shima, N. (1989). Second-moment closure for the nearwall sublayer:development and application. *AIAA JOURNAL, 27*, 1319.

Launder, B. E., & Spalding, D. B. (1974). The numerical computation of turbulent flows. *Computer Methods in Applied Mechanics and Engineering, 3*, 2182.

Laurien, E., & Kleiser, L. (1989). Numerical Simulation of Boundary-layer Transition and Transition Control. *Journal of Fluid Mechanics, 199*, 403-440.

Lawn, C. J. (1971). The Determination of the Rate of Dissipation in Turbulent Pipe Flow. *Journal of Fluid Mechanics, 48*, 477-505.

Lee, T., Fisher, M., & Schwarz, W. H. (1995). Investigation of the effects of a compliant surface on boundary-layer stability. *Journal of Fluid Mechanics Digital Archive, 288*(-1), 37-58.

Lesieur, M. (2008). *Turbulence in Fluids* (4th revised and enlarged edition ed.). Dordrecht: Springer.

Letelier, M. F., & Leutheusser, H. J. (1967). Skin friction in unsteady laminar pipe-flow. *J. Basic Eng. TASME, 89*, 847.

Levich, V. G. (1962). *Physico-Chemical Hydrodynamics*. New York: Prentice-Hall.

Liaw, G. C., Zakin, J. L., & Patterson, G. K. (1971). Effect of molecular characteristics of polymer on drag reduction. *AIChE Journal, 17*, 391-397.

Lin, C. S., Moulton, R. W., & Putnam, G. L. (1953). Mass transfer between solid wall and fluid streams: Mechanism and eddy distribution relationships in turbulent flow. *Ind. Eng. Chem., 45*, 636.

Liu, N. S., Shamroth, S. J., & McDonald, H. (1991). Reciprocal Interactions of Hairpin-shaped Vortices and a Boundary Layer. *AIAA JOURNAL, 29*, 720.

Loughlin, K. F., Abul-Hamayel, M. A., & Thomas, L. C. (1985). The surface rejuvenation theory of wall turbulence for momentum, heat and mass transfer: application to moderate and high Schmidt (Prandtl) numbers. *AIChE J., 31*, 1614-1620.

Luchak, T. S., & Tiederman, W. G. (1987). Timescale and structure of ejections and bursts in turbulent channel flows. *Journal of Fluid Mechanics, 174*, 529-552.

Lumley, J. (1969). Drag reduction by additives. *Annual Review of Fluid Mechanics, 1*, 367-384.

Lumley, J. L. (1973). Drag reduction in turbulent flow by polymer additives. *Journal of Polymer Science and Macromolecules Review 7, ed. Peterlin A., Interscience*, 263.

Lyon, R. N. (1951). Heat transfer at high fluxes in confined spaces. *Chem. Eng.*

*Prog., 47*, 75.

Malhotra, A., & Kang, S. S. (1984). Turbulent Prandtl number in circular pipes. *Int. J. Heat and Mass Transfer, 27*(11), 2158-2161.

Mankbadi, R. R. (1992). Dynamics and control of coherent structure in turbulent jets. *Appl Mech Rev, 45*(6), 219-247.

Martinelli, R. C. (1947). Heat transfer to molten metals. *Trans. ASME, 69*, 947.

Mathpati, C. S., & Joshi, J. B. (2007). Insight into Theories of Heat and Mass Transfer at the Solid-Fluid Interface Using Direct Numerical Simulation and Large Eddy Simulation. *Ind. Eng. Chem. Res., 46*, 8525-8557.

McComb, W. D. (1991). *The Physics of Turbulence*. Oxford: Clarendon Press.

McComb, W. D., & Rabie, L. H. (1979). Development of local turbulent drag reduction due to non uniform polymer concentration. *Physics of Fluids, 22*, 183-185.

McCormick, M. E., & Bhattacharyya, R. (1973). Drag reduction of a submersible hull by electrolysis

*Naval Engineers Journal, 85*(2), 11-16.

McEligot, D. M., Pickett, P. E., & Taylor, M. F. (1976). Measurement of wall region turbulent Prandtl numbers in small tubes. *Int. J. Heat Mass Transfer, 19*, 799.803.

McEligot, D. M., & Taylor, M. F. (1996). The turbulent Prandtl number in the near-wall region for Low-Prandtl-number gas mixtures. *Int. J. Heat Mass Transfer., 39*(6), 1287--1295.

McLeod, N., & Ponton, J. W. (1977). A model for turbulent transfer processes at a solid-fluid boundary. *Chem. Eng. Sci., 32*(483).

McMahon, H. M., Hester, D. D., & Palfery, J. G. (1971). Vortex shedding from a turbulent jet in a cross-wind. *Journal of Fluid Mechanics, 48*, 73-80.

McNaugton, K. (2008). The search for big eddy. from http://www.geos.ed.ac.uk/abs/research/micromet/Current/teal/

Meek, R. L., & Baer, A. D. (1970). The Periodic viscous sublayer in turbulent flow. *AIChE J., 16*, 841.

Meek, R. L., & Baer, A. D. (1973). Turbulent heat transfer and the periodic viscous sublayer in turbulent flow. *AIChE J., 16*, 1385.

Meinhart, C. D., & Adrian, R. J. (1995). On the existence of uniform momentum zones in a turbulent boundary layer. *Phys. Fluid 7*, 694-696.

Merkle, C., & Deutsch, S. (1989). Microbubble drag reduction. In M. Gad-el-Hak (Ed.), *Frontiers in Experimental Fluid Mechanics. Lecture Notes in Engineering* (Vol. 46, pp. 291): Springer.

Metzner, A. B., & Friend, W. L. (1958). Theoretical analogies between heat, mass and momentum transfer and modifications for fluids of high Prandtl/Schmidt numbers. *Can. J. Chem. Eng., 36*, 235.

Metzner, A. B., & Park, M. G. (1964). Turbulent flow characteristics of viscoelastic fluids. *Journal Fluid Mechanics, 20*, 291.

Metzner, A. B., & Reed, J. C. (1955). Flow of Non-Newtonian Fluids-Correlation of the Laminar, Transition and Turbulent Flow Regimes. *AIChE J., 1*, 434.

Meyer, W. A. (1966). A correlation of the friction characteristics for turbulent flow of dilute viscoelastic non-Newtonian fluids in pipes. *AIChE J., 12*, 522.

Millikan, C. B. (1939). A critical discussion of turbulent flows in channels and circular tubes. *Appl. Mech. Proc. Int. Congr. 5th*, 386.

Mohanarangam, K., Cheung, S. C. P., Tu, J. Y., & Chen, L. (2009). Numerical simulation of micro-bubble drag reduction using population balance model

*Ocean Engineering, 36*, 863-872.

Morrison, F. H., & Whitman, W. G. (1928). Heat transfer for oils and water in pipes. *Industrial and Engineering Chemistry, 20*(3), 234-240.

Moussa, Z. M., Trischka, J. W., & Eskinazi, S. (1977). The near field in the mixing of a round jet with a cross-stream. *Journal of Fluid Mechanics, 80*, 49-80.

Myska, J., & Zakin, J. L. (1997). Differences in the Flow Behaviours of Polymeric and Cationic Surfactant Drag-Reducing Additives. *Ind. Eng. Chem. Res. , 36*, 5483-5487.

Na, Y., & Hanratty, T. J. (2000). Limiting behaviour of turbulent scalar transport close to a wall. *International Journal of Heat and Mass Transfer 43*, 1749-1758.

National Research Council. (1997). Submarine platform technology. In Technology for the United States Navy and Marine Corps, 2000-2035: Becoming a 21st century force (Vol. 6, pp. 85-114). Washington, DC: National Academic Press.

Neumann, I. C. (1968). Transfert de chaleur en regime turbulent pour les grands nombres de Prandtl. *Inform. Aéraul. et Therm., 5*, 4-20.

Nieuwstadt, F. T. M., & Den Toonder, J. (2001). Drag reduction by additives: a review. In A. Soldati & R. Monti (Eds.), *Turbulence structure and motion* (pp. 269-316). New York: Springer Verlag.

Nikuradse, J. (1932). Gesetzmäßigkeit der turbulenten Strömung in glatten Rohren. *Forsch. Arb. Ing.-Wes. N0. 356*.

Nikuradse, J. (1933). Stromungsgesetz in rauhren Rohren. *VDIForschungshefte*, 361.

Nychas, S. G., Hershey, H. C., & Brodkey, R. S. (1973). A visual study of turbulent shear flow. *Journal of Fluid Mechanics, 61*, 513.

Obot, N. T. (1993). The frictional law of corresponding states: its origin and applications. *Trans IChemE, 71 Part A*, 3-10.

Offen, G. R., & Kline, S. J. (1974). Combined dye streak and hydrogen bubble visual observations of a turbulent boundary layer. *Journal of Fluid Mechanics, 62*, 223.

Oishi, Y., Murai, Y., Tasaka, Y., & Yasushi, T. (2009). *Frictional Drag reduction by wavy advection of deformable bubbles* Paper presented at the The 6th International Symposium on Measurement Techniques for Multiphase Flows.Journal of Physics: Conference Series volume 147, 012020.

Orlando news. (2004). New Speedo 'Shark' Swimsuit Is World's Fastest. from *www.clickorlando.com/news/2912946/detail.html*

Orr, W. M. F. (1907). The stability or instability of the steady motions of a perfect liquid and a viscous fluid. *Proc. Roy. Ir. Acad., A27*, 689.

Otis, D. R. (1985). Laminar startup flow in a pipe. *ASME J. Appl. Mech., 52*, 706.

Patankar, S. V. (1980). *Numerical Heat Transfer and Fluid Flow*. New York: McGraw-Hill.

Patel, V. C., Rodi, W., & Scheuerer, G. (1985). Turbulence models for nearwall and low Reynolds number flows: a review. *AIAA JOURNAL, 23*, 1308.

Patience, G. S., & Methrotra, A. N. (1989). Laminar Start Up Flow in Short Pipe Lengths. *Can. J. Chem. Eng., 67*, 883.

Peridier, V. J., Smith, F. T., & Walker, J. D. A. (1991). Vortex-induced Boundary-layer Separation. Part 1. The Unsteady Limit Problem Re. *Journal of Fluid Mechanics, 232*, 99-131.

Pinho, F. T., & Whitelaw, J. H. (1990). Flow Of Non-Newtonian Fluids In A Pipe. *Journal of Non-Newtonian Fluid Mechanics, 34*(2), 129-144.

Pletcher, R. H. (1988). Progress in Turbulent Forced Convection. *J. Heat Transfer, 110*, 1129.

Polhausen, K. (1921). Zur näherungsweisen Integration der Differentialgleichung der laminaren Reibungschicht. *ZAMM, 1*, 252-268.

Popovich, A. T., & Hummel, R. L. (1967). Experimental study of the viscous sublayer in turbulent pipe flow. *AICHE Journal, 13*(5), 854-860.

Popovich, A. T., & Hummel, R. L. (1967). Experimental study of viscous sublayer in turbulent pipe flow. *AIChE Journal, 13*(5), 854.

Prandtl, L. (1904). Uber Flussigkeitsbewegung bei sehr kleiner Reibung. *Proc. 3rd. Int. Math. Kongr. Heidelberg*.

Prandtl, L. (1910). Eine Beziehung zwischen Wärmeaustausch und Strömungswiderstand in Flüssigkeiten. *Phys. Z., 11*, 1072.

Prandtl, L. (1925). Uber die ausggebildete Turbulenz. *ZAMM, 5*, 136.

Prandtl, L. (1935). The Mechanics of Viscous Fluids. In D. W.F (Ed.), *Aerodynamic Theory III* (pp. 142). Berlin: Springer.

Pratte, B. D., & Baines, W. D. (1967). Profiles of the turbulent jet in a cross flow. *J.Hydronaut.Div.ASCE, 92*, 53-64.

Price, S. J., Sumner, D., Smith, J. G., Leong, K., & Paidoussis, M. P. (2002). Flow visualisation around a circular cylinder near to a plane wall. *J.Fluids Structs., 16*(2), 175-191.

Rao, K. N., Narasimha, R., & Narayanan, M. A. B. (1971). Bursting in a turbulent boundary layer. *Journal of Fluid Mechanics, 48*, 339.

Rashidi M, Hetsroni G, & S., B. (1991). Mechanisms of heat and mass-transport at gas-liquid interfaces. *International Journal of Heat and Mass Transfer, 7*, 1799–1810.

Rayleigh, L. (1880). On the stability of certain fluid motions. *Proc. London Math. Soc., 11*, 57.

Rayleigh, L. (1884). On the circulation of air observed in Kundt's tubes and on some allied accoustical problems. *Phil.Trans.Roy.Soc.Lond, 175*, 1-21.

Reichardt, H. (1943). *NACA TM 1047*.

Reichardt, H. (1961). In J. P. Hartnett (Ed.), *Recent Advances in Heat and Mass Transfer* (pp. 223): McGraw=Hill.

Reichardt, H. (1971). *MPI für Strömungsforschung*: Göttingen, Rep. No 6a/1971.

Reichart, H. (1963). In L. A. Dorfman (Ed.), *Hydrodynamic resistance and the heat loss of rotating bodies* (pp. 178). Edinburgh: Oliver and Boyd.

Reneaux, J. (2004). *Overview on drag reduction technologies for civil transport aircraft*. Paper presented at the European Congress on Computational Methods in Applied Sciences and Engineering-ECCOMAS 2004.

Reynolds, O. (1874). On the extent and action of the heating surface for steam boilers. *Proc. Manchester Lit. Phil. Soc., 14*, 7-12.

Reynolds, O. (1883). An experimental investigation of the circumstances which determine whether the motion of water shall be direct or sinuous, and of the law of resistances in parallel channels. *Phil Trans Roy Soc London, 174*, 935-982.

Reynolds, O. (1895). On the dynamical theory of incompressible viscous fluids and the determination of the criterion. *Philosophical Transactions of the Royal Society of London A., 186*, 123-164.

Richardson, L. F. (1922). *Weather prediction by numerical process*: Cambridge University Press.

Riley, N. (1967). *J. Inst. Math. Appl., 3*, 419.

Robinson, S. K. (1991). Coherent Motions in the Turbulent Boundary Layer. *Annual Review of Fluid Mechanics, 23*, 601.
Rodi, W. (1980). *Turbulence Models and Their Application in Hydraulics*. Delft, the Netherlands: International Association of Hydraulic Research Section, University of Karlsruhe.
Ruckenstein, E. (1968). A generalised penetration theory for unsteady convective mass transfer *Chem. Eng. Sci., 23*, 363.
Ruckenstein, E. (1987). Analysis of transport phenomena. *Advances in Chem. Eng., 13*, 1.
Saffman, P. G. (1981). Vortex interactions and coherent structures in turbulence. In R. Meyer (Ed.), *Transition and turbulence*: Academic press.
Sanders, W. C., Winkel, E. S., Dowling, D. R., Perlin, M., & Ceccio, S. (2006). Bubble friction drag reduction in a high-Reynolds-number flat-plate turbulent boundary layer. *Journal of Fluid Mechanics, 552*, 353-380.
Savory, E., Toy, N., McGuirk, J. J., & Sakellariou, N. (1990,). *An experimental and numerical study of the velocity field associated with a jet in a crossflow.* Paper presented at the Engineering Turbulence Modelling and Experiments, Proc. Int. Symp. Eng. Turbulence Modelling and Experiments.
Schiller, L. (1922). Untersuchungen uber laminare und turbulente Stromung. *Orsc. Ing. Wes. Heft*, 428.
Schlichting, H. (1932). Berechnung ebener periodischer Grenzschichtstromungen. *Phys.Z., 33*, 327-335.
Schlichting, H. (1933). Zur Enstehung der Turbulenz bei der Plattenströmung. *ZAMM, 13*, 171-174.
Schlichting, H. (1935). Amplitudenverleitung und Energiebilanz der kleinen Störungen bei der Plattenströmung. *Nachr. Ges. Wiss. Göttingen, Math. Phys. Klasse, Fachgruppe I, 1*, 47.
Schlichting, H. (1960). *Boundary Layer Theory* (4th Ed. ed.). New York. McGraw-Hill.
Schlichting, H. (1979). *Boundary layer theory. Seventh edition.* (Seventh edition ed.): McGraw-Hill.
Schlinder, W. G., & Sage, B. H. (1953). Velocity distribution between parallel plates. *IEC, 45*(12), 2636.
Schneck, D. J., & Walburn, F. J. (1976). Pulsatile blood flow in a channel of small exponential divergence Part II: Steady streaming due to the interaction of viscous effects with convected inertia. *Journal of Fluids Engineering*, 707.
Schoppa, W., & Hussain, F. (2000). Coherent structure dynamics in near-wall turbulence. *Fluid Dynamics Research, 26*(2), 119-139.
Schoppa, W., & Hussain, F. (2002). Coherent structure generation in near-wall turbulence. *Journal of Fluid Mechanics, 453*, 57-108.
Schubauer, G. B. (1954). Turbulent processes as observed in boundary layers and pipes. *J. Appl. Phys., 25*, 188-196.
Schubauer, G. B., & Skramstad, H. K. (1943). *Laminar boundary layer oscillations and transition on a flat plate*: NACA, Rep. No 909
Schultz-Grunow, F. (1941). NACA TM 986

Schwab, J. R. (1998). Turbulence modeling. In V. K. Garg (Ed.), *Applied computational fluid dynamics* (pp. 75-115): Dekker.
Seban, R. A., & Shimazaki, T. T. (1951). Heat transfer to a fluid in a smooth pipe with walls at a constant temperature. *Trans. ASME, 73*, 803.

Senecal, V. E., & Rothfus, R. R. (1953). Transition flow of fluids in smooth tubes. *Chem. Eng. Prog., 49*, 533.

Seyer, F. A., & Metzner, A. B. (1967). Turbulent flow of viscoelastic fluids. *The Canadian Journal of Chemical Engineering, 45*, 121.

Shaw, D. A., & Hanratty, T. J. (1964). Fluctuations in the Local Rate of Turbulent Mass Transfer to a Pipe Wall. *AIChE J., 10*(4), 475-482.

Shaw, D. A., & Hanratty, T. J. (1977). Influence of Schmidt number on the fluctuations of turbulent mass transfer to a wall. *AIChE J., 23*, 160.

Sibilla, S., & Beretta, C. P. (2005). Near-wall coherent structures in the turbulent channel flow of a dilute polymer solution. *Fluid Dynamics Research, 3*, 183.

Sideman, S., & Pinczewski, W. (1975). Turbulent heat and mass transfer at interfaces. In C. Gutfinger (Ed.), *Topics in Transport Phenomena*: Hemisphere Publishing House, Wiley

Sieder, E. N., & Tate, G. E. (1936). Heat Transfer and. Pressure Drop of Liquids in Tubes. *Industrial. Engineering Chemistry, 28*(28), 1429.

Skelland, A. H. P. (1967). *Non-Newtonian Flow and Heat transfer*. New York: John Wiley and Sons.

Smith, C. R., & Walker, J. D. A. (1995). Turbulent wall-layer vortices. *Fluid Mech. Applic., 30*.

Smith, C. R., Walker, J. D. A., Haidari, A. H., & Sobrun, U. (1991). On the Dynamics of near-wall Turbulence. *Phil. Trans. R. Soc. Lond., A336*, 131-175.

Sommerfield, A. (1908). *Ein Beitrag zur hydrodynamischen Erklarung der turbulenten Flussigkeisbewegung*. Paper presented at the Atti Congr. Int. Math. 4th Rome.

Son, S. M., & Hanratty, T. J. (1967). Limiting relation for the eddy diffusivity close to a wall. *AIChE J., 13*, 689-696.

Spalart, P. R. (1988). Direct Simulation of a Turbulent Boundary Layer up to $Re_{\grave{e}}$=1410. *Journal of Fluid Mechanics, 187*.

Spalart, P. R., & Allmaras, S. R. (1994). A one-equation turbulence model for aerodynamic flows. *La Recherche Aérospatiale 1*, 5-21.

Spalding, D. B. (1961). A single formula for the law of the wall. *ASME J. Appl. Math., 28*(3), 455-458.

Spina, E. F., & Smits, A. J. (1987). Organised structures in a compressible, turbulent boundary layer. *Journal of Fluid Mechanics, 182*, 85.

Squire, C. F. (1971). *Waves in physical systems*. Englewood Cliffs, New Jersey: Prentice-Hall.

Squire, H. B. (1942). *Heat transfer calculations for aeorofoils. ARC RM 1986 also in Schlichting "Boundary Layer Theory (1979) p.305*.

Stokes, G. G. (1851). On the effect of the internal friction of fluids on the motion of pendulums. *Camb. Phil. Trans., IX*, 8.

Strouhal, V. (1878). Über eine besondere Art der Tonerregung. *Ann. Phys. und Chemie, New series, 5*, 216-251.

Stuart, J. T. (1966). Double boundary layers in oscillatory viscous flow. *Journal of Fluid Mechanics, 24*, 673.

Suponitsky, V., Cohen, J., & Bar-Yoseph, P. Z. (2005). The generation of streaks and hairpin vortices from a localized vortex disturbance embedded in unbounded uniform shear flow. *Journal of Fluid Mechanics, 535*, 65-100.

Swearingen, J. D., & Blackwelder, R. F. (1987). The Growth and Breakdown of Streamwise Vortices in the presence of a Wall. *Journal of Fluid Mechanics, 182*, 255-290.

Szymanski, P. (1932). Quelques solutions exactes des équations de l'hydrodynamique de fluide visqueux dans le cas d'un tube cylindrique. *J. des Mathematiques Pures et Appliquées, 11 series 9*, 67.

Takeda, Y. (1999). Quasi-periodic state and transition to turbulence in a rotating Couette system. *Journal of Fluid Mechanics, 389*, 81-99.

Tanaka, H., & Yabuki, H. (1986). Laminarisation and reversion to turbulence of low Reynolds number flow through a converging to constant area duct. *Journal of Fluids Engineering, 108*, 325-330.

Tardu, S. F. (1995). Coherent structures and riblets. *Appl. Sci. Res., 54*, 349-385.

Taylor, G. I. (1921). *Proc.London Math Soc., 20*, 126.

Taylor, G. I. (1923). Stability of a viscous liquid contained between two rotating cylinders. *Phil. Trans. Roy. Soc. London. Series A, 223*, 289-343.

Taylor, G. I. (1935). The statistical theory of turbulence, Parts II-V. *Proc. Roy. Soc. London, Ser. A, 135*, 421.

Tennekes. (1966). Wall region in turbulent shear flow of non-Newtonian fluids. *Physics of Fluids, 9*, 872.

Tetlionis, D. M. (1981). Unsteady Viscous Flow. *SpringerVerlag, New York*.

Theodore, L. (1971). *Transport phenomena for Engineers*. Scranton: International Texbook Co.

Thomas, A. D. (1960). Heat and momentum transport characteristics of non-Newtonian aqueous Thorium oxide suspensions. *AIChEJ, 8*.

Thomas, L. C., & Al-Sharif, M. M. (1981). An integral analysis for heat transfer in turbulent incompressible boundary layer flow. *Journal of Heat Transfer-Transactions of the ASME, 103*, 772-777.

Thomas, L. C., & Fan, L. T. (1971). A model for turbulent transfer in the wall region. *J. Eng. Mech. Div., 97*, 169.

Tollmien, W. (1929). Über die Entstehung der Turbulenz. l. Mitteilung. *Nachr. Ges. Wiss. Göttingen, Mth. Phys. Klasse*, 21.

Toms, B. A. (1949). *Some observations on the flow of linear polymer solutions through straight tubes at large Reynolds numbers*. Paper presented at the Proc. Int. Cong. on Rheology, Vol II, p.135.North Holland Publishing Co.

Townsend, A. A. (1970). Entrainment and the structure of turbulent flow. *Journal of Fluid Mechanics, 41*(part 1), 13-46.

Townsend, A. A. (1979). Flow patterns of large eddies in a wake and in a boundary-layer *Journal of Fluid Mechanics, 95*(DEC), 515-537.

Toyoda, K., & Hiramoto, M. (2006). Effect of streamwise vortices on characteristics of jets. *JSME International Journal Series B, 49*(4), 884-889.

Trefethen, L. N., Trefethen, A. E., Reddy, S. C., & Driscoll, T. A. (1993). Hydrodynamic stability without eigenvalues. *Science, 261*.

Trinh, K. T. (1969). *A boundary layer theory for turbulent transport phenomena, M.E. Thesis,* New Zealand: University of Canterbury.

Trinh, K. T. (1992). *Turbulent transport near the wall in Newtonian and non-Newtonian pipe flow, Ph.D. Thesis*. New Zealand: University of Canterbury.

Trinh, K. T. (1993). *Power law relationships for viscous non-Newtonian fluids*. Paper presented at the Proceedings of the 6th APCChE Congress and 21st Chemeca Conference.

Trinh, K. T. (1994). *Similarity between the Wall Layers of Newtonian and Non-Newtonian Fluids in Turbulent Flow*. Paper presented at the IPENZ Conference.

Trinh, K. T. (1996). *Structural implications of Karman's universal constant in*

*turbulent shear flows*. Paper presented at the IPENZ Annual conference.

Trinh, K. T. (1999). *Pipe Flow of a Power Law Fluid: A Case Study for a Theory of Non-Newtonian Turbulence* Paper presented at the Chemeca, Newcastle.

Trinh, K. T. (2002). *A New Partial Derivative In Transport Dynamics, paper 881*. Paper presented at the 9th APCChE Congress and Chemeca 2002.

Trinh, K. T. (2005). Milk Powder Technology Course Notes. Massey University.

Trinh, K. T. (2005a). *A zonal similarity analysis of wall-bounded turbulent shear flows*. Paper presented at the 7th World Congress of Chemical Engineering: Proceedings.

Trinh, K. T. ( 2005b). *Structural significance of Karman's universal constant*. Paper presented at the 7th World Congress of Chemical Engineering: Proceedings.

Trinh, K. T., & Keey, R. B. (1992a). A Modified Penetration Theory and its Relation to Boundary Layer Transport. *Trans. IChemE, ser. A, 70*, 596-603.

Trinh, K. T., & Keey, R. B. (1992b). A Time-Space Transformation for Non-Newtonian Laminar Boundary Layers. *Trans. IChemE, ser. A, 70*, 604-609.

Tsinober, A. (2001). *An Informal Introduction to Turbulence*. London: Kluwer Academic Publishers.

Turney, D. E., & Banerjee, S. (2008). Transport phenomena at interfaces between turbulent fluids. *AIChE Journal, 54*(2), 344-349.

Virk, P. S. (1975). Drag reduction fundamentals. *AICHE Journal, 21*, 625-656.

Virk, P. S., Mickley, H. S., & Smith, K. A. (1970). The ultimate asymptote and mean flow structure in Toms'phenomena. *Journal of Applied Mechanics-Transactions of the ASME, 37*, 488.

Walker, J. D. A. (1978). The Boundary Layer due to a Rectilinear Vortex. *Proc. R. Soc. Lond., A359*, 167-188.

Walker, J. D. A., Abbott, D. E., Scharnhorst, R. K., & Weigand, G. G. (1989). Wall-layer Model for the Velocity profile in Turbulent Flows. *AIAA JOURNAL, 27*, 140.

Wang, L., Olsen, M. G., & Vigil, R. D. (2005). Reappearance of azimuthal waves in turbulent Taylor-Couette flow at large aspect ratio. *Chem. Eng. Sci., 60*, 555-5568.

Wark, C. E., & Nagib, H. M. (1991). Experimental investigation of coherent structures in boundary layers. *Journal of Fluid Mechanics, 230*, 183-208.

Wei, T., & Willmarth, W. W. (1989). Reynolds-number Effects on the Structure of a Turbulent Channel Flow. *Journal of Fluid Mechanics, 204*, 57-95.

Wereley, S. T., & Lueptow, R. M. (1998). Spatio-temporal character of non-wavy and wavy Taylor-Couette flow. *Journal of Fluid Mechanics, 364*, 59-80.

White, C. M., & Mungal, M. G. (2008). Mechanics and prediction of turbulent drag reduction with polymer additives. *Annual Review of Fluid Mechanics, 40*, 235-256.

White, C. M., Somandepalli, V. S. R., & Mungal, M. G. (2004). The turbulence structure of drag reduced boundary layer flow. *Experiments in Fluids, 36*, 62-69.

White, F. M., Christoph, G. H., & Lessmann, R. C. (1973). Calculation of turbulent heat transfer and skin friction. *AIAA Journal, 11*, 1046-1052.

Willmarth, W. W., & Lu, S. S. (1972). Structure of the Reynolds stress near the wall. *Journal of Fluid Mechanics, 55*, 65.

Wilson, K. C., & Thomas, A. D. (1985). A New Analysis of the Turbulent-Flow of Non-Newtonian Fluids. *Canadian Journal of Chemical Engineering, 63*(4), 539-546.

Yoo, S. S. (1974). *Heat transfer and friction factors for non-Newtonian fluids in turbulent flow, PhD thesis*. US: University of Illinois at Chicago Circle.

Zagarola, M. V., Perry, A. E., & Smits, A. J. (1997). Log laws or power laws: The scaling in the overlap region. *Physics of Fluids, 9*(7), 2094-2100.

Zagarola, M. V., & Smits, A. J. (1998). Mean-flow scaling of turbulent pipe flow. *Journal of Fluid Mechanics, 373*, 33-79.

Zanoun, E.-S., & Durst, F. (2003). Evaluating the law of the wall in two-dimensional fully developed turbulent channel flows. *Physics of Fluids, 15*(10), 3079-3089.

Zhang, L. H., & Swinney, H. L. (1985). Nonpropagating oscillatory modes in Couette-Taylor flow. *Phys. Rev. A, 31*, 1006-1009.